\documentclass[graybox,envcountchap]{svmono}

\usepackage{mathptmx}
\usepackage{helvet}
\usepackage{courier}
\usepackage{type1cm}         

\usepackage{makeidx}         
\usepackage{graphicx}        
\usepackage{multicol}        
\usepackage[bottom]{footmisc}

\usepackage{amssymb}  
\usepackage{amsmath, amsfonts}
\usepackage{textcomp}

\usepackage[normalem]{ulem}  

\newcommand{\non}{\nonumber\\}

\newcommand{\be}{\begin{equation}}
\newcommand{\ee}{\end{equation}}
\newcommand{\bea}{\begin{eqnarray}}
\newcommand{\eea}{\end{eqnarray}}
\newcommand{\ba}[1]{\begin{array}{#1}}
\newcommand{\ea}{\end{array}}

\newcommand{\bm}[1]{\mbox{\boldmath${#1}$}}
\newcommand{\uq}{\hat{\mathbf{q}}} 
\newcommand{\uk}{\hat{\mathbf{k}}}
\newcommand{\up}{\hat{\mathbf{p}}}
\newcommand{\hk}{\hat{k}}

\newcommand{\hp}{\hat{p}}
\newcommand{\vg}{\bm{\gamma}}

\newcommand{\Tr}{{\rm Tr}}

\makeindex  

\begin{document}

\author{Andreas Schmitt}
\title{Introduction to superfluidity}
\subtitle{Field-theoretical approach and applications}
\date{July 31, 2014}
\maketitle

\frontmatter

\preface
This course is about the theory of low-energy and high-energy, non-relativistic and relativistic, bosonic and fermionic superfluidity and 
superconductivity. 
Does that sound too much? Well, one important point of the course will be to show that these things are not as diverse as they might seem: the mechanism behind 
and the basic phenomenological properties of superfluidity are the same whether applied to ``ordinary'' low-energy superfluids or to more ``exotic'' superfluids in 
high-energy physics; non-relativistic and relativistic treatments may look quite different at first sight, but of course the former is only a limit case of the latter;
bosonic and fermionic superfluids can be continuously connected in some sense; and, once you understand what a superfluid is, it is very easy to understand 
what a superconductor is and vice versa.  

The motivation for this course arose from my own research in high-energy physics where certain kinds of superfluids and superconductors
are predicted in ultra-dense nuclear and quark matter. These are ``stellar superfluids'', since they are likely to occur in the interior of compact stars. 
Working on stellar superfluids, it was natural to learn about more down-to-earth superfluids which are firmly established experimentally.
Therefore, this course is interesting for researchers who are in a similar situation like myself, who have some background in 
high-energy physics and want to learn about superfluidity, explained in a field-theoretical language they are used to. 
I believe that the course is also insightful for researchers with a background in condensed matter physics who are interested in high-energy applications of 
their field and a relativistic field-theoretical formalism they usually do not employ. And, most importantly, this course is intended for 
advanced undergraduate students, graduate students, and researchers who simply want to understand what superfluidity is and what its applications  in modern physics are. 

Readers unfamiliar with quantum field theory might find some of the chapters challenging, even though I have tried to present most of the calculations in a 
self-contained way. When this was not possible, I have mentioned suitable references where the necessary elements of field theory are explained. 
However, not all of the chapters rely on field-theoretical methods. For instance, the course starts with an introduction to superfluid helium that 
can easily be understood with basic knowledge of statistical physics and thermodynamics. Most of the chapters that do
employ quantum field theory aim at a microscopic description of superfluids, i.e., the degrees of freedom of the theory are the bosons that condense or the 
fermions that form Cooper pairs. In this sense, the course is in large parts about the fundamental mechanisms behind superfluidity. But, I will 
emphasize the connection to phenomenology throughout the course and do not want the reader to get lost in technical details. For instance, I will show 
in a simple setting how a microscopic quantum field theory can be connected to the phenomenological two-fluid model of a superfluid. 

Despite the pompous announcement in the first sentence, this is a course that can be taught in about one semester. Therefore, it can only deal with 
a few selected aspects of superfluidity. This selection has been based on the aim to convey the underlying microscopic physics of superfluidity, on pedagogical
considerations, and of course is also, to some extent, a matter of taste. As a result of this subjective selection, there are many important aspects 
that I will not, or only marginally, discuss, such as vortices in a rotating superfluid, dissipative effects, or observable signatures of 
stellar superfluids. Literature that can be consulted for such topics and for further reading in general is given at the end of the introduction and throughout the text.

These lecture notes are based on a course that I taught at the Vienna University of Technology in the winter semester 2011/2012 and in the summer semester 2013.
I would like to thank all participants for numerous questions and many lively discussions that have improved my understanding of superfluidity. 
I am grateful to Mark Alford, Karl Landsteiner, S.\ Kumar Mallavarapu, David M\"{u}ller, Denis Parganlija, Florian Preis, Anton Rebhan, and Stephan Stetina  
for many helpful comments and discussions. This work has been supported by the Austrian science 
foundation FWF under project no.~P23536-N16 and by the NewCompStar network, COST Action MP1304.

\vspace{\baselineskip}
\begin{flushright}\noindent
Vienna, April 2014 \hfill {\it Andreas Schmitt}\\
\end{flushright}

\mainmatter

\tableofcontents

\chapter{Introduction}
\label{sec:intro}

\section{Setting the stage: what is a superfluid?}

Superfluidity was first observed in liquid helium. The key experiment was the study of flow through a thin capillary, and the key 
observation was that the fluid flows without friction. Hence the name superfluid. What is behind this phenomenon? Does it only occur in liquid helium? 
If not, where else? To generalize the 
specific observation of frictionless flow, we notice that in order to observe a flow, something is transported through the capillary. In liquid helium, 
we can say that mass is transported. We may also say that helium atoms are transported. This does not make a difference, neither 
the total mass of the liquid nor the total number of helium atoms is changed during the experiment. Both are conserved quantities. In relativistic systems,
mass is not a conserved quantity in general. So, if we call the mass, or better, the number of helium atoms, a 
``charge'', we can say that superfluidity is frictionless transport of a conserved charge. Formulated in this way, we can ask whether there are other systems 
where some other conserved charges show a dissipationless flow. 

Before we do so, let us stay with superfluid helium for a moment. It turns out that the frictionless flow is not its only spectacular property. For instance, if we try to 
rotate it, it will develop vortices, quasi-one-dimensional strings whose number is proportional to the externally imposed angular momentum. 
The existence of vortices is, besides the frictionless flow, another clear signature of superfluidity. Furthermore, one finds that the specific heat shows a 
peculiar behavior at a certain temperature. This is the temperature below which helium becomes superfluid and above which it behaves like a normal fluid. 
Therefore, a superfluid is a phase of a given system below a certain critical temperature at which a phase transition happens. What is the nature of this
phase transition and how can we describe it theoretically? For the case of liquid helium, more precisely for liquid $^4$He, 
the answer is Bose-Einstein condensation, where 
the helium atoms occupy a single quantum state, forming a ``condensate''. This phase transition can be characterized 
in terms of symmetries of the system, and we can make the connection to the conserved charge introduced above: in the superfluid phase of a system, a symmetry
of the system that is associated with a conserved charge is spontaneously broken. This does not mean that the total charge is no longer conserved. Roughly speaking, 
it means that charge can be deposited into or extracted from the condensate. 

Historically, superfluidity was discovered some time after superconductivity.
The similarity of the two words is no coincidence. A superconductor is also a phase of a system in which a charge is transported without dissipation. 
In a metal or alloy, this charge is electric charge, and it is electric resistance that becomes unmeasurably small. Also for the vortices, there is an analogue 
in a superconductor: flux tubes, quasi-one-dimensional objects in which a magnetic field can penetrate the superconductor. And, as for the 
superfluid, there is a critical temperature above which superconductivity is lost. So what is the actual difference between a superfluid and a superconductor
from the theoretical point of view? Electric charge is associated to a gauge symmetry. This is a local symmetry, i.e., a symmetry that allows for different
transformations at different points of space-time. A superconductor can be said to break a local symmetry spontaneously, while a superfluid breaks a 
global symmetry spontaneously. This statement shows the theoretical similarity of the two phenomena, but also emphasizes their only fundamental difference. 

Of course, a superconducting metal is different from superfluid helium in many aspects. For instance, electrons are fermions, while helium atoms are bosons
(for now, I am talking about $^4$He). Therefore, it cannot be Bose-Einstein condensation that leads to the phase transition in an electronic
superconductor, at least not in a direct way. The key mechanism is Cooper pairing at the Fermi surface due to an effectively attractive interaction between the 
electrons. We shall discuss both Bose-Einstein condensation and Cooper pairing in a field-theoretical framework and will also show that they can be continuously connected.
It is important to remember that the fundamental difference between a superfluid and a superconductor is given by the above distinction of symmetries, and {\it not} by 
the bosonic vs.\ fermionic nature of the underlying microscopic physics. Even though the best known superfluid is bosonic and the best known superconductor
is fermionic, there are also fermionic superfluids and bosonic superconductors.

Now that we have a rough theoretical concept for superfluidity and pointed out its similarity to 
superconductivity, we may ask whether there are systems with different characteristic energy (temperature) scales that show superfluidity and/or 
superconductivity. The exciting answer is that there is superfluidity all over the energy scale: while the critical temperature of superfluid $^4$He is about 2.2 K,
there are experiments with ultra-cold atomic gases that become superfluid at temperatures of the order of $10^{-7}\, {\rm K}$. On the other end of the scale, 
there are superfluids in 
dense nuclear and quark matter. Astrophysical observations indicate that the critical temperature for neutron matter is of the order of $10^8\,{\rm K}$, 
while theoretical estimates predict quark matter to be superfluid for temperatures up to about $10^{11}\,{\rm K}$. Thus, superfluidity can occur for systems whose typical 
energies are separated by about 18 orders of magnitude! Superfluidity and superconductivity of nuclear or quark matter has never been observed in the laboratory because 
there are currently no experiments that are able to create the necessary conditions. Therefore, our only current ``laboratories'' are compact stars, and it is 
an exciting topic of current research to combine theoretical predictions with signatures in astrophysical data to support or rule out the existence of high-energy 
superfluids. One of these high-energy superfluids is very illustrative regarding the above notion of superfluidity and superconductivity in terms of broken symmetries: 
quarks carry electric and color charges (local symmetries) {\it and} 
baryon number charge (global symmetry). Therefore, a given phase of quark matter where Cooper pairs form is a color superconductor 
and/or an electric superconductor and/or a baryonic superfluid, depending on the pairing pattern in which the quarks pair. For example, 
the theoretically best established phase, the so-called color-flavor locked phase, is a color superconductor and a superfluid, but not an electric superconductor. 

Finally, let me comment on the use of a relativistic treatment in large parts of this course. Advantages of this approach are its generality -- the non-relativistic
case can always be obtained as a limit -- and its formal rigor, but where are relativistic effects in superfluids 
of phenomenological relevance? Clearly, in the low-energy systems such as liquid helium and ultra-cold atomic gases,
a non-relativistic framework is appropriate, and we shall work in this framework when we discuss these systems explicitly. Relativistic effects in 
high-energy superfluids become important when the mass of the constituents is small compared to their kinetic energy. 
A neutron superfluid is sometimes treated non-relativistically because of the relatively large neutron mass; however, at high densities in the core of the star, 
the Fermi momentum of the neutrons becomes comparable with the neutron mass and relativistic effects have to be taken into account. 
Quark superfluidity in compact stars clearly has to be treated relativistically because of the light masses of the quarks 
(only three-flavor quark matter is of phenomenological interest, the three heavy flavors are not relevant at densities present in the interior of a compact star). 

Since this course combines non-relativistic with relativistic chapters, some slight inconsistencies in the notation are unavoidable. For instance, the same symbol
for the chemical potential can denote slightly different quantities because the rest mass is usually absorbed in the chemical potential in non-relativistic 
treatments. I have tried to make such inconsistencies clear at the point where they occur to avoid confusion. For consistency,  I have decided to work in natural 
units of particle physics throughout the course, even in the nonrelativistic chapters, i.e., Planck's constant divided by $2\pi$, the speed of light, 
and the Boltzmann constant are set to one, $\hbar=c=k_B=1$. In the relativistic calculations, I denote four-vectors by capital letters, for instance the 
four-momentum $K=(k_0,{\bf k})$, and the modulus of the three-vector by $k =  |{\bf k}|$. At nonzero temperature, $k_0=-i\omega_n$, where $\omega_n$ are the 
Matsubara frequencies, $\omega_n=2n\pi T$ for bosons and $\omega_n=(2n+1)\pi T$ 
for fermions, with $n\in {\mathbb Z}$ and the temperature $T$. (As an exception, I use small letters for four-vectors in Sec.\ \ref{sec:derivation} for the sake of 
compactness.) The convention for the Minkowski metric is $g^{\mu\nu}=(1,-1,-1,-1)$.

\section{Plan of the course and further reading}

The course starts with an introduction to the physics of superfluid helium, chapter \ref{sec:he4}. This is done on a macroscopic level, i.e., we do not 
discuss a microscopic theory on the level of the helium atoms. The goal of this chapter is to become familiar with the phenomenology 
of a superfluid, and to introduce the basic concepts of superfluid hydrodynamics, in particular the two-fluid model. Most parts of this chapter and further 
details can be found in the textbooks \cite{khal,pines,landau1,annett}. 

In chapter \ref{sec:phi4} we discuss superfluidity in a microscopic framework, 
a bosonic field theory. We introduce important theoretical concepts such as spontaneous symmetry breaking and the Goldstone theorem, which are discussed in generality
in standard field theory textbooks such as Refs.\ \cite{srednicki,pokorsky}, and discuss the field-theoretical version of concepts introduced in chapter \ref{sec:he4},
for instance the superfluid velocity and the dispersion of the Goldstone mode. In some parts of this chapter and several other instances in this course 
I will make use of elements of thermal quantum field theory which are explained in the textbooks \cite{kapusta,lebellac} and in my own lecture notes \cite{thermal}.

Chapter \ref{sec:reltwo} connects the previous two chapters by discussing the relativistic generalization of the 
two-fluid model and by showing how the field-theoretical results of chapter \ref{sec:phi4} give rise to that model. 
More details about the covariant two-fluid formalism can be found in the research papers \cite{1982PhLA...91...70K,1982ZhETF..83.1601L,carter89,1992PhRvD..45.4536C};
for an extended version of the field-theoretical derivation of the two-fluid model presented here, see Ref.\ \cite{Alford:2012vn}. 

In chapter \ref{sec:cooper} we switch to 
fermionic systems and discuss the field-theoretical derivation of the mean-field gap equation. Parts of this chapter are based on Ref.\ \cite{Pisarski:1999av}, and 
there is plenty of literature about the analogous derivation in the non-relativistic context, see for instance the textbooks \cite{fetter,tinkham}. At the end
of the chapter, we use the general result to discuss some examples of fermionic superfluids and superconductors. This discussion is more or less restricted 
to solving the gap equation, for more extensive discussions I refer the reader to the specific literature such as Refs.\ \cite{vollhardt} (superfluid $^3$He), 
\cite{Alford:2007xm} (color-superconducting quark matter), or \cite{Page:2006ud,arXiv:1001.3294,2013arXiv1302.6626P} (astrophysical aspects of 
superfluids/superconductors in nuclear and quark matter).

The goal of chapter \ref{sec:meissner1} is to point out that a gauge boson in a system with spontaneously broken gauge symmetry acquires a mass.
In our context, this is the Meissner mass, which is responsible for the Meissner effect in a superconductor. The calculation of the Meissner mass for a 
fermionic superconductor is worked out in detail, for related research papers in the context of quark matter see for instance 
Refs.\ \cite{Rischke:2000qz,Rischke:2000ra,Schmitt:2003aa,Alford:2005qw}. Discussions of the Meissner effect in an ordinary superconductor 
can be found in many textbooks such as Ref.\ \cite{tinkham}. 

In chapter \ref{sec:BCSBEC} we discuss the BCS-BEC crossover, a crossover from a weakly coupled fermionic
system that forms Cooper pairs according to the Bardeen-Cooper-Schrieffer (BCS) theory \cite{bcs} to a Bose-Einstein condensation (BEC) of di-fermions. Since this 
crossover has been observed in the laboratory with ultra-cold fermionic gases, our theoretical discussion will be embedded into this context and we will work in 
a non-relativistic framework. I will only touch the basic points of this crossover, much more can be found in reviews such 
as \cite{giorgini,ketterle,2010AnPhy.325..233L,2012LNP...836.....Z}, for relativistic discussions see for instance 
Refs.\ \cite{Nishida:2005ds,Abuki:2006dv,Deng:2006ed,2009NuPhA.823...83G,He:2013gga}. 

In chapter \ref{sec:lowenergy} we come back to 
a relativistic fermionic superfluid and discuss the Goldstone mode by introducing fluctuations on top of the mean-field approximation of 
chapter \ref{sec:cooper}. In a way, this is the superfluid counterpart to chapter \ref{sec:meissner1} where a superconductor, exhibiting a massive
gauge boson instead of a Goldstone mode, is discussed. Related works in recent research are Refs.\ \cite{2008PhRvA..77b3626D,Gubankova:2008ya,2011AnPhy.326..193S} 
in the context of a Fermi gas and Ref.\ \cite{Fukushima:2005gt} in the context of color-flavor locked quark matter. 

In the final chapter \ref{sec:mismatch} we discuss the field-theoretical calculation of the free energy of a Cooper-paired system with mismatched Fermi surfaces. 
This situation, originally discussed in the context of an electronic superconductor \cite{Chandrasekhar:1962,Clogston:1962}, has gained a lot of interest in recent 
research and has applications in the fields of ultra-cold atoms \cite{2006Sci...311..492Z,2006Sci...311..503P,sheehy}, dense quark matter 
\cite{Alford:2007xm,Alford:2002kj,Rajagopal:2005dg}, and even in the context of chiral symmetry breaking in a strong magnetic field at nonzero baryon chemical potential 
\cite{Preis:2012fh}.

\chapter{Superfluid helium}
\label{sec:he4}

Helium was first liquefied in 1908 by H.\ Kamerlingh Onnes who cooled it below the liquid/gas transition temperature of 4.2 K\footnote{In this chapter, 
helium is always synonymous to $^4$He, which is bosonic. I will write $^4$He only when I want to emphasize the bosonic nature. The fermionic 
counterpart $^3$He can also become superfluid, see Sec.\ \ref{sec:examples}.}. 
Later, in 1927, M.\ Wolfke and W.H.\ Keesom realized that there is another phase transition at lower
temperatures, around 2.17 K. This phase transition had manifested itself in a discontinuity of the specific heat, whose curve as a function of temperature resembles 
the Greek letter $\lambda$, and thus the transition point was called $\lambda$ point. The two phases of liquid helium were termed ``helium I'' and ``helium II''.
The remarkable superfluid properties of liquid helium below the 
$\lambda$ point, helium II, were experimentally established by P.\ Kapitza in 1938 \cite{1938Natur.141...74K}, 
and independently by J.F.\ Allen and A.D.\ Misener in the same year \cite{1938Natur.141...75A}. 
Kapitza had set up an experiment with two cylinders that were connected by a thin tube with a thickness of $0.5\,\mu{\rm m}$. Only below the $\lambda$ point, helium was 
flowing easily through the tube, suggesting a strikingly low viscosity.  
Kapitza received the Nobel prize for this discovery in 1978 (interestingly, together with
Penzias and Wilson who received it for the completely unrelated discovery of the cosmic microwave background radiation). Kapitza coined the term 
``superfluidity'' in his paper of 1938, having some intuition about a deep connection to superconductivity. This is remarkable because, 
although superconductivity had been observed much earlier in 1911, a microscopic understanding was only achieved much later in 1957.
Only then, with the help of the microscopic theory of Bardeen, Cooper, and Schrieffer it was possible to appreciate the deep connection between electronic 
superconductivity and superfluidity in $^4$He. For the case of a bosonic superfluid such as $^4$He, the theoretical background of Bose-Einstein 
condensation was already known since 1924 \cite{1924ZPhy...26..178B,einstein}, 
and F.\ London proposed shortly after the discovery that helium undergoes a Bose-Einstein condensation 
\cite{1938Natur.141..643L}. Other early theoretical developments, such as the phenomenological two-fluid model, were put forward 
by L.\ Landau \cite{1941PhRv...60..356L} and L.\ Tisza \cite{1938Natur.141..913T}. More details about the interesting history of the discovery of superfluidity 
can be found in  Refs.\ \cite{balibar,griffin}.

\begin{figure} [t]
\begin{center}
\includegraphics[width=0.75\textwidth]{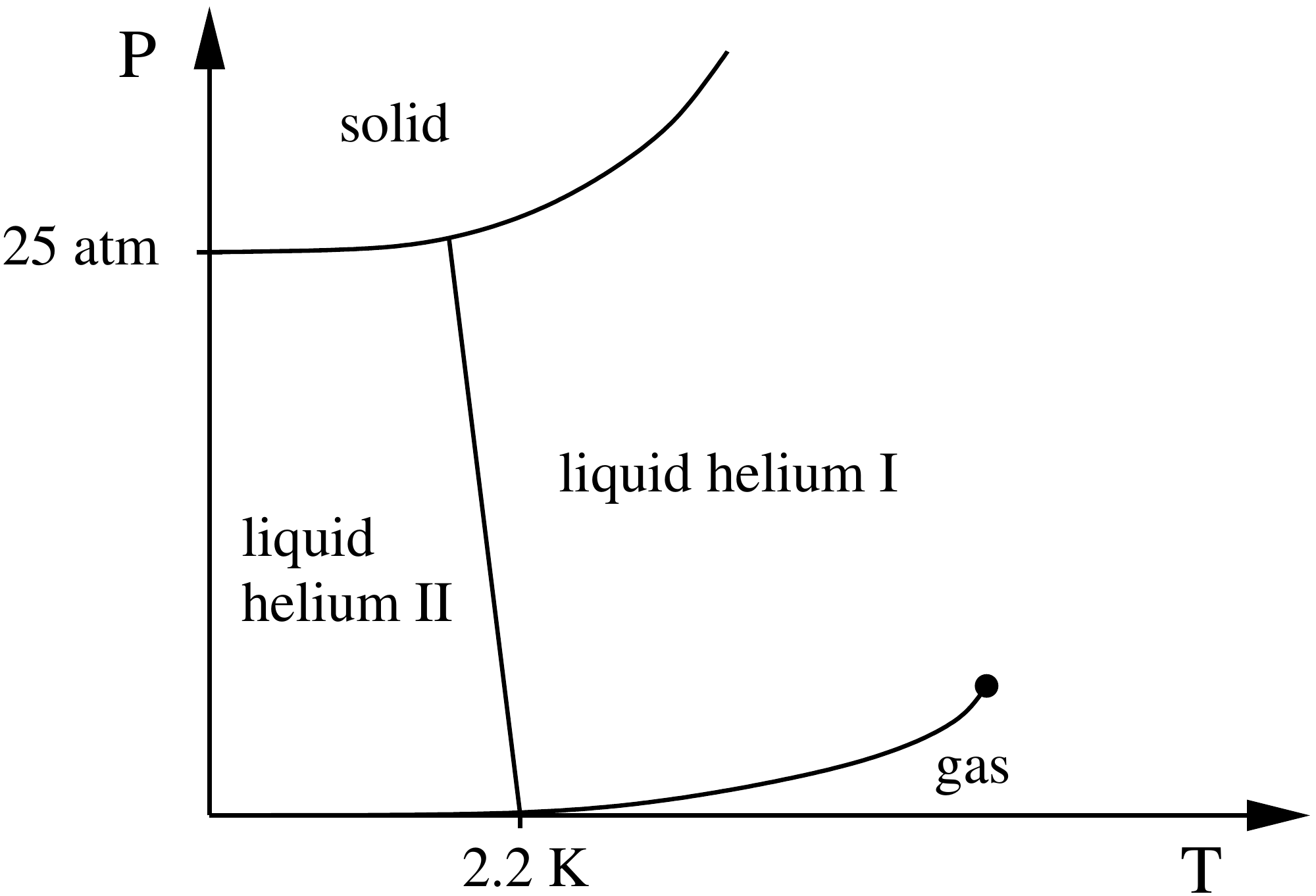}
\hspace{1cm}
\caption{Schematic phase diagram of $^4$He in the plane of pressure $P$ and temperature $T$. Below a certain pressure, helium remains liquid for arbitrarily
small temperatures, allowing for a superfluid phase below a critical temperature, sometimes called $\lambda$-temperature. 
Superfluid and normal fluid phases are denoted by helium II and helium I, respectively. This terminology has historical origin and was given to the 
two phases after the discovery of the phase transition, but before the discovery of superfluidity of helium II.
}
\label{fighe4}
\end{center}
\end{figure}

What is special about helium, i.e., why can it become superfluid at low temperatures? Superfluidity is a quantum effect, so if we were to invent 
a liquid that becomes superfluid, one thing we would have to make sure is that it remains a liquid for very small temperatures, because only there
quantum effects become important. Helium is special in this sense. All other elements solidify at some point when they are cooled down. The reason is that 
the kinetic energy of their atoms becomes sufficiently small at small temperatures to confine the atoms within their lattice sites. For very small
temperatures, the kinetic energy is solely given by the zero-point motion. It turns out that the zero-point motion for helium atoms is sufficiently large  
 to prevent them from forming a solid. Only under strong pressure does helium solidify. Although hydrogen atoms are lighter, their inter-atomic attractive 
interactions are much stronger, so hydrogen does solidify. The phase diagram of helium is shown in Fig.\ \ref{fighe4}.

\section{Landau's critical velocity}
\label{sec:vc}

To explain why helium can be superfluid, we need to explain why it transports charge (here: mass, or helium atoms) without friction. 
The most important ingredient is the Bose-Einstein condensate. It carries charge and can flow without losing energy. The excitations on top of the 
condensate potentially lead to dissipation. Landau came up with a very general argument that 
results in a condition for these excitations in order to allow for superfluidity: let us consider a superfluid moving through a capillary with velocity 
${\bf v}_s$. In the rest frame of the fluid (where the capillary moves with velocity $-{\bf v}_s$), let the energy of such an excitation and its corresponding 
momentum be $\epsilon_p>0$ and ${\bf p}$. Now, in the rest frame of the capillary, the energy of the fluid is given 
by the kinetic energy $E_{\rm kin}$ plus the energy of the elementary excitations, transformed into the new frame\footnote{Here, in the context 
of superfluid helium, we change frames by a Galilei transformation, and do not use the more general Lorentz transformation. Later we shall discuss relativistic 
excitations whose transformation reduces in the low-velocity limit to Eq.\ (\ref{epsgalilei}), see Eq.\ (\ref{epsflow}) and discussion below that equation.},
\be \label{epsgalilei}
E = E_{\rm kin} + \epsilon_p + {\bf p}\cdot{\bf v}_s \, .
\ee
The fluid loses energy through dissipation if  
\be
\epsilon_p + {\bf p}\cdot{\bf v}_s < 0 \, . 
\ee
The left-hand side can only be negative if its minimum is negative, \mbox{$\epsilon_p - p v_s < 0$}. Consequently, the system transports charge without dissipation
for velocities smaller than the critical velocity
\be \label{vc}
v_c = \min_p\,\frac{\epsilon_p}{p} \, .
\ee
This simple argument by Landau is of fundamental importance for the understanding of a superfluid. A direct consequence is that systems where 
$\min_p\,\frac{\epsilon_p}{p} = 0$ cannot be superfluid since then $v_c=0$ and an arbitrarily small velocity would result in dissipation. 
We can write the minimum of $\epsilon_p/p$ as the solution of
\be
0 = \frac{\partial}{\partial p}\frac{\epsilon_p}{p} \;\; \Rightarrow \;\; \frac{\partial \epsilon_p}{\partial p} = \frac{\epsilon_p}{p} \, .
\ee
For a given point on the curve $\epsilon_p$ we are thus asking whether the slope of the curve is identical to the slope of a straight line
from the origin through the given point. Or, in other words, to check the superfluidity of a system, take 
a horizontal line through the origin in the $\epsilon_p$-$p$ plane and rotate it upwards. If you can do so by a finite amount before touching 
the dispersion curve, the system supports superfluidity. 
The slope of the line at the touching point is the critical velocity according to Landau above which 
superfluidity is destroyed. In particular, any gapless dispersion with slope zero in the origin must lead to dissipation for any 
nonzero velocity. It is important to remember that the criterion for superfluidity is not only a requirement for the excitations of the system. 
Otherwise one might incorrectly conclude that a free gas of relativistic particles with dispersion $\epsilon_p = \sqrt{p^2+m^2}$ is a superfluid. The criterion 
rather requires a nonzero critical velocity $v_c$ {\it and} the existence of a condensate. Without a condensate, there is nothing to transport the charge without 
friction.

We shall see later that Bose-Einstein condensation is always accompanied by a gapless mode $\epsilon_{p=0}=0$ due to the Goldstone theorem, and this 
gapless mode is called Goldstone mode. One might think that the Goldstone mode can very easily be excited. And this is true in some sense. For instance, due to the 
gaplessness, such a mode becomes populated for arbitrarily small temperatures. Landau's argument, however, shows that even a gapless mode is sufficiently 
difficult to excite by forcing the fluid to move through a capillary: if for instance the dispersion of the Goldstone mode is linear, $\epsilon_p \propto p$, 
the mode is gapless but Landau's critical velocity is nonzero, and in fact identical to the slope of the Goldstone mode. Typically, the slope of a 
Goldstone mode is indeed linear for small momenta. This is true for instance in superfluid helium. On the other hand, if we had $\epsilon_p \propto p^2$
for small momenta, the slope of the dispersion at the origin would be zero and as a consequence $v_c=0$. 


\section{Thermodynamics of superfluid helium}
\label{sec:thermohelium}

\begin{figure} [t]
\begin{center}
\includegraphics[width=0.75\textwidth]{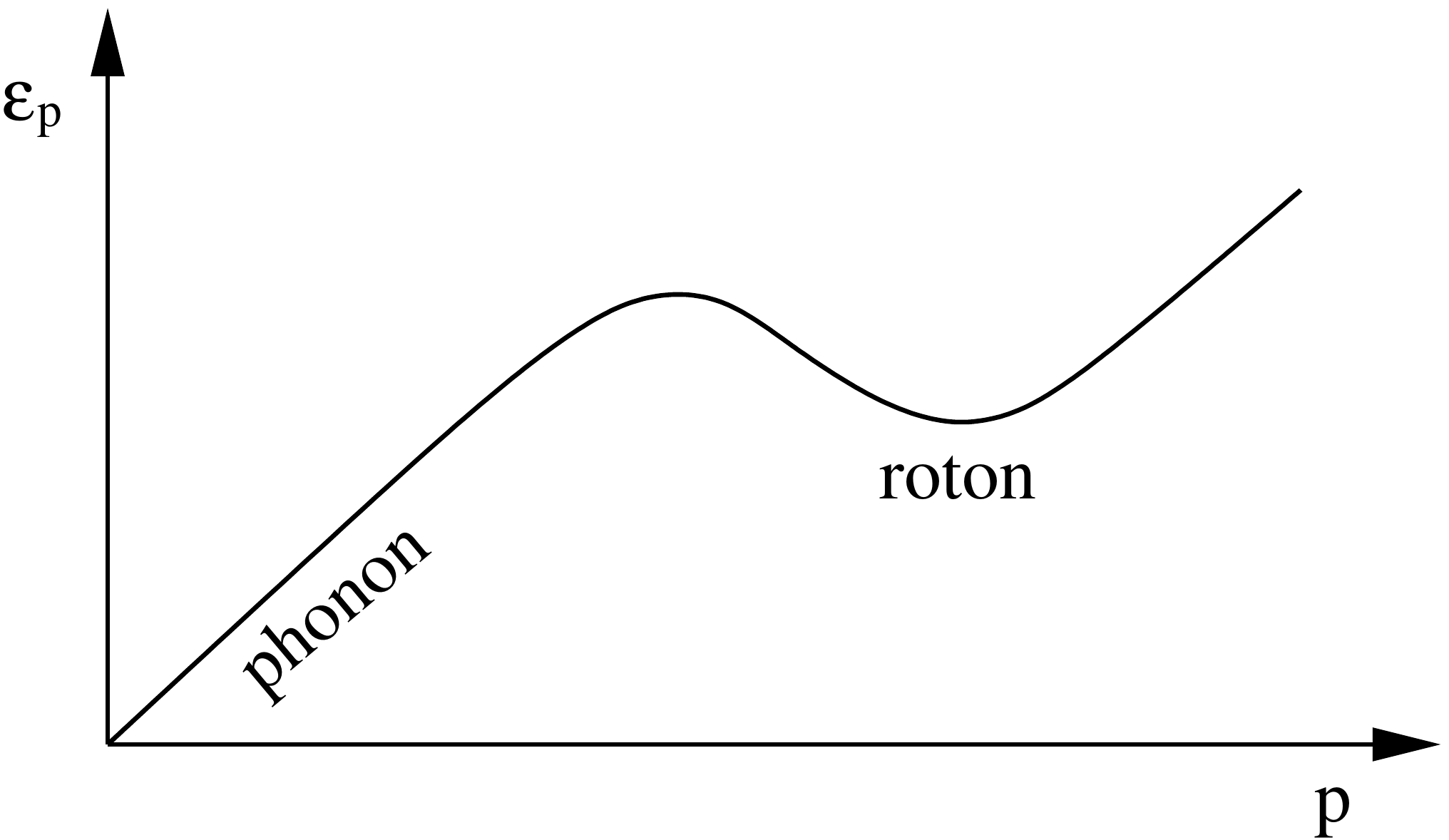}
\caption{Schematic plot of the Goldstone dispersion for superfluid helium. This mode is often modelled in terms of two different modes, the phonon and the roton, 
see Fig.\ \ref{figphrot1}}
\label{figphonon}
\end{center}
\end{figure}

\begin{figure} [t]
\begin{center}
\includegraphics[width=0.75\textwidth]{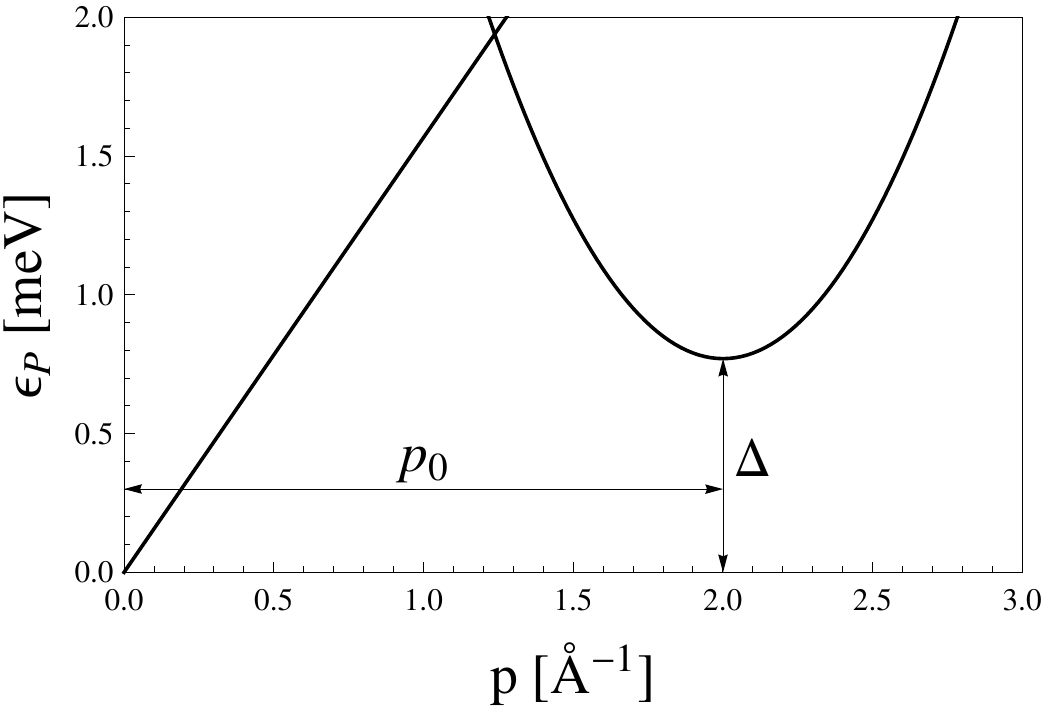}
\caption{Dispersions for phonons and rotons from Eqs.\ (\ref{phononrotondisp}) 
with parameters $m=1.72\times 10^{-24}\,{\rm g}$, $p_0=2.1\times 10^{-19}\,{\rm g}\,{\rm cm}\,{\rm s}^{-1}$, 
$\Delta = 8.9\,{\rm K}$, $c= 238\,{\rm m}\,{\rm s}^{-1}$. 
}
\label{figphrot1}
\end{center}
\end{figure}

While the existence of a Goldstone mode and the linearity at small $p$ are very general
features, the details of the complete dispersion of this mode depend on the details of the interactions in a given system. In superfluid helium, it turns
out that the mode has a dispersion of the form shown in Fig.\ \ref{figphonon}. For low energies, it can effectively be described by two 
different excitations, one accounting for the linear low-momentum part -- this is called the ``phonon'' -- and one accounting for the 
vicinity of the local minimum at a finite value of $p$ -- this is called the ``roton''. We can write these two dispersions as
\begin{subequations} \label{phononrotondisp}
\bea
\epsilon_p &=& c p \qquad\qquad\qquad\quad  \mbox{(``phonon'')} \, , \\
\epsilon_p &=& \Delta + \frac{(p-p_0)^2}{2m} \qquad \mbox{(``roton'')} \, . 
\eea
\end{subequations}
with parameters $c$, $\Delta$, $p_0$, $m$, whose values are specified in Fig.\ \ref{figphrot1}.

Let us first compute some of the thermodynamic properties given by the Goldstone mode. We start from the general expression for the pressure, 
\be \label{PTint}
P = -T\int\frac{d^3{\bf p}}{(2\pi)^3}\ln\left(1-e^{-\epsilon_p/T}\right) = \frac{1}{3}\int\frac{d^3{\bf p}}{(2\pi)^3}\, p 
\frac{\partial \epsilon_p}{\partial p} f(\epsilon_p) \, ,
\ee
where, in the second step, we have used partial integration, where $T$ is the temperature, and where 
\be
f(\epsilon_p)=\frac{1}{e^{\epsilon_p/T}-1} 
\ee
is the Bose distribution function. 

Consequently, the phonon contribution to the pressure is
\be \label{Pph}
P_{\rm ph} = \frac{c}{6\pi^2}\int_0^\infty dp\,\frac{p^3}{e^{cp/T}-1} = \frac{T^4}{6\pi^2 c^3}
\underbrace{\int_0^\infty dy\frac{y^3}{e^y-1}}_{\displaystyle{\pi^4/15}} = 
\frac{\pi^2T^4}{90c^3} \, .
\ee
If the dispersion were linear for all $p$, this result would be valid for any $T$. However, the dispersion is linear only for 
small $p$. Since the corrections to the linear behavior become important at larger temperatures, this result cannot be trusted for all $T$.
(Obviously, the critical temperature for superfluidity is another, absolute, limit above which this result is inapplicable). 
The result for the pressure is similar to the Stefan-Boltzmann pressure of blackbody radiation. The reason is that a photon has the same linear dispersion as the 
superfluid phonon, but moves with the speed of light. Thus, we recover the Stefan-Boltzmann pressure if we set $c$ equal 
to the speed of light and multiply the result by 2 because a photon has two degrees of freedom.

We can now compute the entropy and the specific heat per unit volume from the usual thermodynamic definitions,
\be \label{sph}
s_{\rm ph} = \frac{\partial P_{\rm ph}}{\partial T} = \frac{2\pi^2T^3}{45 c^3} \, ,
\ee
and 
\be \label{cVph}
c_{V,{\rm ph}} = T\frac{\partial s_{\rm ph}}{\partial T} = \frac{2\pi^2T^3}{15 c^3} = 3s_{\rm ph}\, , 
\ee
where the subscript $V$ indicates that the specific heat is computed at fixed volume (as opposed to fixed pressure).

The calculation of the roton contribution is a bit more complicated,
\be
P_{\rm rot} = \frac{1}{6\pi^2m}\int_0^\infty dp\, p^3\frac{p-p_0}{e^{\epsilon_p/T}-1} \, .
\ee
In general, this integral has to be solved numerically. Here we proceed by making the assumption $T\ll \Delta$, such that we can approximate 
\be
\frac{1}{e^{\frac{\Delta}{T}+\frac{(p-p_0)^2}{2mT}}-1} \simeq e^{-\Delta/T}e^{-\frac{(p-p_0)^2}{2mT}} \, , 
\ee
and thus 
\be
P_{\rm rot} \simeq \frac{e^{-\Delta/T}}{6\pi^2m} \int_0^\infty dp\,p^3(p-p_0)e^{-\frac{(p-p_0)^2}{2mT}} \, .
\ee
This expression shows that the contribution of the rotons is exponentially suppressed for temperatures much smaller than $\Delta$. 
To obtain the subleading temperature dependence, we introduce the new integration variable $y=(p-p_0)/\sqrt{2mT}$,
\bea \label{Prot}
P_{\rm rot} &=& \frac{e^{-\Delta/T}T(2mT)^{3/2}}{3\pi^2} \int_{-\frac{p_0}{\sqrt{2mT}}}^\infty dy\,y \left(y+\frac{p_0}{\sqrt{2mT}}\right)^3 e^{-y^2}\non[2ex]
&\simeq&
 \frac{e^{-\Delta/T}Tp_0^2(2mT)^{1/2}}{\pi^2} \underbrace{\int_{-\infty}^\infty dy\,y^2 e^{-y^2}}_{\displaystyle{\sqrt{\pi}/2}}\non[2ex]
&=&\sqrt{\frac{m}{2\pi^3}}\,p_0^2 \,T^{3/2} e^{-\Delta/T}
 \, ,
\eea
where we have assumed $T\ll p_0^2/(2m)$. With the parameters given in Fig.\ \ref{figphrot1} we have $p_0^2/(2m)\simeq 93\,{\rm K}$, 
i.e., since we already have
assumed that $T$ is much smaller than $\Delta \simeq 8.9 \,{\rm K}$, $T$ is also much smaller than $p_0^2/(2m)$. Again we may compute entropy and specific heat,
\be
s_{\rm rot} \simeq \sqrt{\frac{m}{2\pi^3}}\, \frac{p_0^2\Delta}{T^{1/2}} e^{-\Delta/T} \, ,
\ee
and
\be
c_{V,{\rm rot}} 
\simeq \sqrt{\frac{m}{2\pi^3}}\,\frac{p_0^2\Delta^2}{T^{3/2}} e^{-\Delta/T} \, ,
\ee
where we have neglected terms of higher order in $T/\Delta$.

\section{Two-fluid model}
\label{sec:2fluid}

The hydrodynamics of a superfluid is often described within a so-called two-fluid model, suggested by Tisza \cite{1938Natur.141..913T}
and Landau \cite{1941PhRv...60..356L} shortly after the discovery of superfluidity. 
A priori, this was a purely phenomenological description. We shall discuss later how it emerges as a kind of effective theory from a microscopic description. 
In the two-fluid picture, the system is formally divided into two fluids, 
the {\it superfluid} and {\it normal fluid}, which interpenetrate each other. 
The superfluid component consists of the condensate, 
while the 
normal component contains the elementary excitations, i.e., the phonon and roton excitations in the case of superfluid helium. 
This picture suggests that at zero temperature there is only a superfluid. Then, upon heating up the system, the normal fluid will start to appear
and become more and more dominant until the superfluid completely vanishes at and above the critical temperature. Originally, the model served to explain the ``viscosity
paradox'' which had appeared from two apparently contradicting behaviors of superfluid helium: damping times of the oscillations of a torsion pendulum in liquid 
helium suggested a viscosity \cite{1935RSPSA.151..342W}, in apparent contrast to the dissipationless flow through a thin capillary \cite{1938Natur.141...74K}. 
In the two-fluid picture
it is only the superfluid component that can flow through the thin tube while the pendulum sees both fluid components, i.e., the excitations of the normal 
fluid were responsible for the damping of the pendulum.  The model predicts the existence of a second sound mode, see Sec.\ \ref{sec:sounds}, which 
was indeed observed after the two-fluid picture was suggested.

The flow of the system is described by two fluids with independent velocity fields.  
The momentum density ${\bf g}$ receives contribution from both fluids,
\be \label{gboth}
{\bf g} = \rho_s {\bf v}_s +\rho_n {\bf v}_n \, ,
\ee
where ${\bf v}_s$ and ${\bf v}_n$ are the velocities of the superfluid and the normal fluid, respectively, and $\rho_s$ and $\rho_n$
are the superfluid and normal-fluid mass densities, such that the total mass density is 
\be
\rho=\rho_n+\rho_s \, .
\ee 
To compute the normal-fluid density, we consider the rest frame of the superfluid, in which the normal fluid is moving with 
velocity ${\bf w} \equiv  {\bf v}_n-{\bf v}_s$. In this frame, the momentum density only receives a contribution from the normal fluid and 
is given by $\rho_n{\bf w}$. We can also express the momentum density of the normal fluid as 
\be \label{g02}
\rho_n{\bf w} = \int\frac{d^3{\bf p}}{(2\pi)^3}\,{\bf p}\,f(\epsilon_p-{\bf p}\cdot{\bf w}) \, , 
\ee
where we have taken into account that the distribution function of the elementary excitations depends on the relative velocity between the two 
fluids.  As in the previous subsections, $\epsilon_p$ is the dispersion of the elementary excitations measured in the superfluid rest frame. In particular,
we recover the Galilei transformed excitations from Eq.\ (\ref{epsgalilei}) for ${\bf v}_n=0$. 
Multiplying both sides of Eq.\ (\ref{g02}) with ${\bf w}$, we obtain an expression for the normal-fluid density,
\bea \label{rhon2}
\rho_n &=& \frac{1}{w}\int\frac{d^3{\bf p}}{(2\pi)^3}\,\hat{\bf w}\cdot{\bf p}\,f(\epsilon_p-{\bf p}\cdot{\bf w}) \, ,
\eea
where $\hat{\bf w}\equiv {\bf w}/w$. It is important to realize that the concept of normal-fluid and superfluid densities only makes sense in the presence of a 
(at least infinitesimal) relative velocity. In general, $\rho_s$ and $\rho_n$ are functions of this relative velocity. For many applications one is 
interested in the limit of small relative velocities. To compute $\rho_n$ in the limit $w\to 0$, we insert the Taylor expansion
\be
f(\epsilon_p-{\bf p}\cdot{\bf w}) = f(\epsilon_p) -{\bf p}\cdot{\bf w}\left.\frac{\partial f}{\partial\epsilon_p}\right|_{{\bf w}=0} + {\cal O}(w^2) 
\ee
into Eq.\ (\ref{rhon2}). The integral over the first term of this expansion vanishes, and we obtain 
\bea \label{rhonw0}
\rho_n({\bf w}\to0) &=& -\int\frac{d^3{\bf p}}{(2\pi)^3} ({\bf p}\cdot\hat{\bf w})^2 \frac{\partial f}{\partial\epsilon_p}\non[2ex]
&=&-\frac{1}{3}\int\frac{d^3{\bf p}}{(2\pi)^3} p^2 \frac{\partial f}{\partial\epsilon_p} \non[2ex]
&=& \frac{1}{3T}\int\frac{d^3{\bf p}}{(2\pi)^3} \frac{p^2 e^{\epsilon_p/T}}{(e^{\epsilon_p/T}-1)^2} \, .
\eea
Since we have modelled the  dispersion of the Goldstone mode by two separate excitations, we can compute
their contribution to the normal-fluid density separately. Let us start with the phonon contribution.

With the $z$-axis of our coordinate system pointing in the direction of ${\bf w}$ and $x=\cos\theta$ with $\theta$ being the 
angle between ${\bf w}$ and the momentum ${\bf p}$, the phonon contribution from Eq.\ (\ref{rhon2}) is 
\bea
\rho_{n,{\rm ph}} &=& \frac{1}{4\pi^2 w} \int_{-1}^1 dx\,x \int_0^\infty dp\,\frac{p^3}{e^{p(c-wx)/T}-1}  \allowdisplaybreaks \non[2ex]
&=& 
\frac{T^4}{4w\pi^2}\underbrace{\int_{-1}^1 dx \frac{x}{(c-wx)^4}}_{\displaystyle{\frac{8cw}{3(c^2-w^2)^3}} \; \mbox{for $w<c$}}
\int_0^\infty dy\,\frac{y^3}{e^y-1} \allowdisplaybreaks \non[2ex]
&=& \frac{2\pi^2T^4}{45c^5}\left(1-\frac{w^2}{c^2}\right)^{-3} \, .
\eea
The condition $w<c$ is necessary to ensure superfluidity: we can repeat the argument for Landau's critical velocity from Sec.\ \ref{sec:vc}, replacing the rest frame 
of the capillary with the rest frame of the normal fluid. This shows that the relative velocity $w$ has an upper limit given by Eq.\ (\ref{vc}) above which 
dissipation sets in. In the absence of rotons, this limit would be given by $c$. As Fig.\ \ref{figphrot1} shows, the presence of the rotons only decreases 
the limit. 

For small relative velocities $w$ we find
\be \label{rhophs}
\rho_{n,{\rm ph}} = \frac{2\pi^2T^4}{45c^5} \left[1+{\cal O}\left(\frac{w^2}{c^2}\right)\right] 
\simeq \frac{s_{\rm ph}T}{c^2} \, ,
\ee
where the result for the entropy density (\ref{sph}) has been used. One can check that the same $w\to 0$ result is obtained by directly using Eq.\ (\ref{rhonw0}).

For the roton contribution we find for small temperatures
\bea
\rho_{n,{\rm rot}} &\simeq& \frac{e^{-\Delta/T}}{4\pi^2w}\int_0^\infty dp\,p^3e^{-\frac{(p-p_0)^2}{2mT}}\int_{-1}^1 dx \,xe^{\frac{pwx}{T}}\non[2ex]
&=& \frac{Te^{-\Delta/T}}{2\pi^2w^2}\int_0^\infty dp\,p^2e^{-\frac{(p-p_0)^2}{2mT}}\left(\cosh\frac{pw}{T}-\frac{T}{wp}\sinh\frac{pw}{T}\right)
\non[2ex]
&\simeq&\frac{Te^{-\Delta/T}p_0^2}{2\pi^2w^2}\left(\cosh\frac{p_0w}{T}-\frac{T}{wp_0}\sinh\frac{p_0w}{T}\right)
\int_{-p_0}^\infty dq\,e^{-\frac{q^2}{2mT}} \non[2ex]
&\simeq& \sqrt{\frac{m}{2\pi^3}}\frac{T^{3/2}e^{-\Delta/T}p_0^2}{w^2}
\left(\cosh\frac{p_0w}{T}-\frac{T}{wp_0}\sinh\frac{p_0w}{T}\right) \, .
\eea
In the limit $w\to 0$ this becomes 
\be \label{rhorotappr}
\rho_{n,{\rm rot}} = \sqrt{\frac{m}{2\pi^3}}\frac{e^{-\Delta/T}p_0^4}{3T^{1/2}}\left[1+{\cal O}\left(\frac{p_0^2w^2}{T^2}\right)\right] 
\simeq \frac{p_0^2}{3T^2}P_{\rm rot} \, .
\ee
Again, we can check that this result is obtained from the general expression (\ref{rhonw0}). This is left as a small exercise to the reader. 

\begin{figure} [t]
\begin{center}
\includegraphics[width=0.75\textwidth]{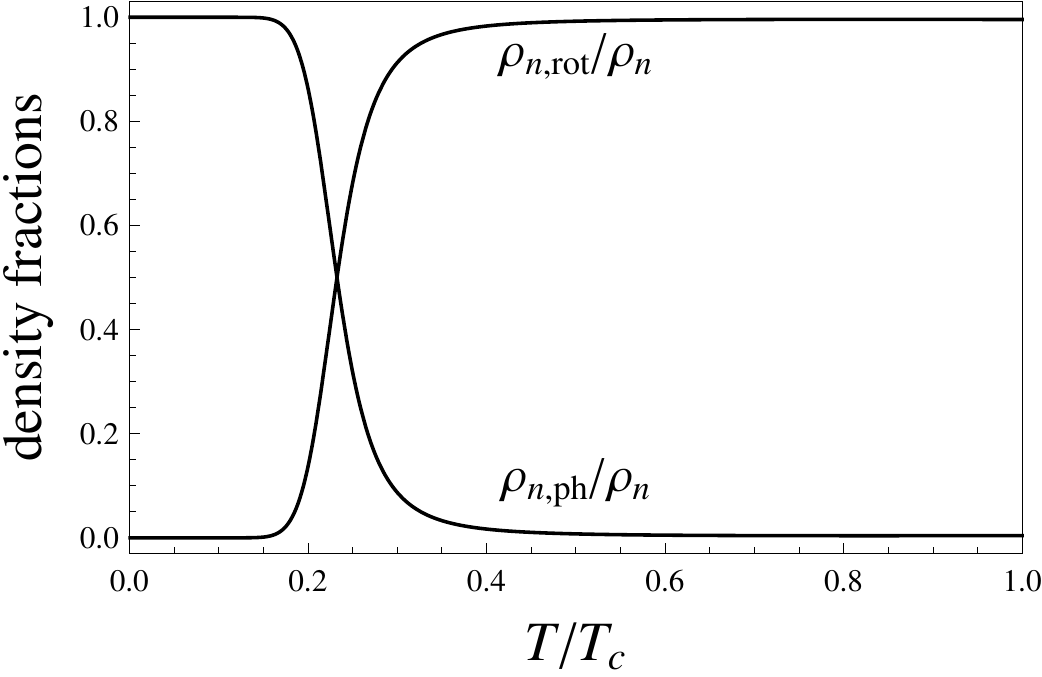}
\caption{Contributions of phonons and rotons to the normal-fluid density for all temperatures up to the critical temperature in the absence of a relative flow,  
${\bf w}=0$, with the parameters of Fig.\ \ref{figphrot1}.  
}
\label{figrhophrhorot}
\end{center}
\end{figure}
\begin{figure} [t]
\begin{center}
\includegraphics[width=0.75\textwidth]{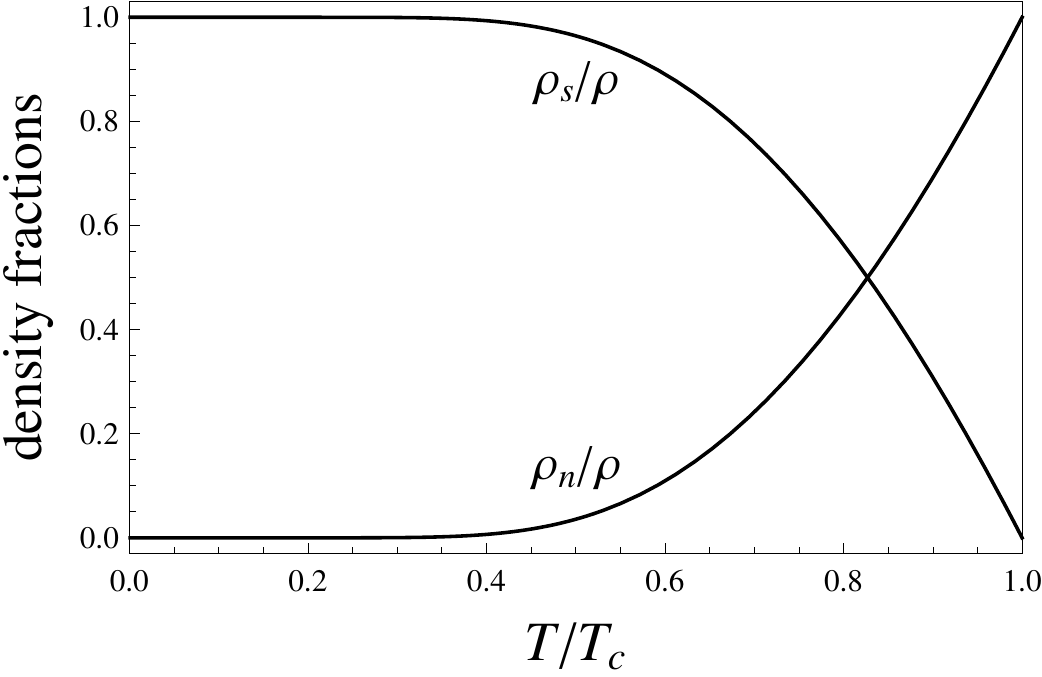}
\caption{Normal-fluid and superfluid density fractions in the absence of a relative flow,  ${\bf w}=0$,
with the parameters of Fig.\ \ref{figphrot1} and the total mass density $\rho = 0.147\,{\rm g}\,{\rm cm}^{-3}$. 
For all temperatures above $T_c$, we have $\rho_n=\rho$, $\rho_s=0$. 
}
\label{figphrot2}
\end{center}
\end{figure}

We can now compute the total normal-fluid density from the two separate contributions. 
Together with a given total density $\rho$, this allows us to compute the density fractions for superfluid and normal-fluid components for all temperatures
up to the critical temperature. Since at the critical temperature $T_c$ all mass sits in the normal fluid, we can compute $T_c$ by solving
$\rho = \rho_{n,{\rm ph}}(T_c) + \rho_{n,{\rm rot}}(T_c)$ numerically for $T_c$. For the limit $w\to 0$ and with $\rho = 0.147\,{\rm g}\,{\rm cm}^{-3}$ one obtains 
$T_c\simeq 2.47\,{\rm K}$. This number is obtained by using the full temperature dependence of $\rho_n$: the phonon contribution (\ref{rhophs}) is exact
for all temperatures (under the assumption that the phonon dispersion continues linearly for all momenta), while for the roton contribution 
the full expression (\ref{rhonw0}) has been used, including a numerical momentum integral. (Using the low-temperature approximation (\ref{rhorotappr}) gives a slightly 
larger critical temperature.) The discrepancy to the actual value of the critical temperature of $T_c\simeq 2.17\,{\rm K}$ is due to the model assumption of 
separate phonon and roton excitations, which differs from the correct quasiparticle spectrum, see Fig.\ \ref{figphrot1}. This difference is important for large 
temperatures. Within the given model, we show the phonon and roton contributions to the normal-fluid density in Fig.\ \ref{figrhophrhorot}, 
and the superfluid and normal-fluid density fractions in Fig.\ \ref{figphrot2}.

\section{First and second sound}
\label{sec:sounds}

One interesting consequence of the two-fluid model is the existence of two sound modes. 
The second sound mode was first observed in superfluid helium \cite{peshkov1946determination,1947PhRv...71..600L}, for a nice popular article about second sound 
and its significance for developments in the theory of superfluid helium see Ref.\ \cite{donnelly}. Much more recently, 
second sound was also measured in an ultra-cold Fermi gas \cite{2013arXiv1302.2871S}.  
Before we can discuss first and second sound, we need to discuss some hydrodynamics. We shall give a very brief introduction to 
single-fluid hydrodynamics before we add a second fluid in order to describe the superfluid. Here we shall only discuss ideal, i.e., dissipationless, hydrodynamics. 
This is sufficient for the discussion of first and second sound, which can propagate non-dissipatively. If you are interested in a much more detailed account of 
hydrodynamics, see for instance Ref.\ \cite{landauhydro}.

\subsection{Single-fluid hydrodynamics}
\label{sec:single}

We shall start from the relativistic form of hydrodynamics and then take the non-relativistic limit. 
The conservation equations for charge and (four-)momentum are 
\be \label{dmuTmunu}
\partial_\mu T^{\mu\nu} = 0 \,, \qquad \partial_\mu j^\mu  =  0 \, ,
\ee
where  
\be \label{jnv}
j^\mu = nv^\mu
\ee
is the current associated with the conserved charge, and 
\be \label{Tmunu}
T^{\mu\nu} = (\epsilon + P) v^\mu v^\nu  - g^{\mu\nu} P 
\ee
is the stress-energy tensor for an ideal fluid. Here, $n$, $\epsilon$, and $P$ are number density, energy density, and pressure, measured in the rest frame of the fluid, 
and $g^{\mu\nu}={\rm diag}(1,-1,-1,-1)$ is the metric tensor. Moreover,   
\be
v^\mu = \gamma(1,{\bf v})  
\ee
is the four-velocity of the fluid, expressed in terms of the three-velocity ${\bf v}$ and the Lorentz factor $\gamma = (1-v^2)^{-1/2}$. Here, 
$v^2$ denotes the square of the modulus of the three velocity. This form of the four-velocity ensures
\be
v_\mu v^\mu = 1.
\ee
Even though we have omitted the arguments, in general all quantities of course depend on space-time, i.e., ${\bf v}= {\bf v}({\bf x},t)$ etc. 

The various components of the stress-energy tensor are 
\begin{subequations}
\bea
T^{00} &=& \frac{\epsilon+Pv^2}{1-v^2} \, , \\[2ex]
T^{0i} &=& T^{i0}=\frac{\epsilon+P}{1-v^2}\,v_i \, , \\[2ex]
T^{ij} &=& \frac{\epsilon+P}{1-v^2}v_iv_j + \delta_{ij}P \, ,
\eea
\end{subequations}
where $v_i$ are the components of the three-velocity ${\bf v}$, not of the four-velocity. 
In particular, the stress-energy tensor is symmetric. We define the rest frame of the fluid by ${\bf v}=0$, i.e., $v^\mu=(1,0,0,0)$. In this particular frame,
the stress-energy tensor assumes the simple form 
\be \label{Tmunu0}
T^{\mu\nu} = \left(\begin{array}{cccc} \epsilon &0&0&0\\0&P&0&0\\0&0&P&0\\0&0&0&P\end{array}\right) \, .
\ee
As a simple exercise, one can check that the general stress-energy tensor (\ref{Tmunu}) can be obtained via a Lorentz transformation
from the stress-energy tensor (\ref{Tmunu0}). In a single-fluid system with uniform fluid velocity, it is obviously always possible to choose a global frame in which 
the three-velocity vanishes. This is the rest frame of the fluid. 
This is not possible in a two-fluid system, even if the two velocities of the two fluids are uniform. In that case, 
one may choose to work in the rest frame of one of the fluids, and the direction of the velocity of the other fluid will necessarily break rotational invariance.

In order to take the non-relativistic limit, we introduce the rest mass density $\rho$ via the temporal component of the four-current,
\be
\rho= mj^0  \, .
\ee
With Eq.\ (\ref{jnv}), this means that $mn = \rho\sqrt{1-v^2}$, i.e., $\rho\sqrt{1-v^2}$ is the rest mass density in the fluid rest frame, while $\rho$
is the rest mass density in the frame where the fluid moves with velocity ${\bf v}$. Eventually, after having derived the non-relativistic limit, the mass density 
$\rho$ will be assumed to be frame independent. 

The spatial components of Eq.\ (\ref{jnv}) now give
\be
m{\bf j} = \rho {\bf v} \, .
\ee
Next, we need the non-relativistic version of the stress-energy tensor. To this end, we introduce the non-relativistic energy density $\epsilon_0$
in the fluid rest frame by separating the rest energy,
\be
\epsilon = \rho\sqrt{1-v^2}+\epsilon_0 \, .
\ee
Neglecting terms of order $v^4$, we can write
\bea
T^{00} &\simeq& \epsilon+(\epsilon+P)v^2 \non[2ex]
&\simeq& \rho+\epsilon_0 +\left(\frac{\rho}{2}+\epsilon_0+P\right)v^2  \, . 
\eea
We now remove the rest mass density $\rho$ and assume that $\epsilon_0 + P \ll \rho$ in the kinetic term to 
obtain the non-relativistic version
\be
T^{00}_{\rm non-rel.} = \epsilon_0+\frac{\rho v^2}{2} \, ,
\ee
which contains the energy density in the fluid rest frame plus a kinetic term which has the usual non-relativistic form.  We proceed analogously for the
other components. First, we write 
\bea
T^{0i} &\simeq& (\epsilon + P)(1+v^2)v_i \non[2ex]
&\simeq& \rho v_i + \left(\epsilon_0 + \frac{\rho v^2}{2}+P\right)v_i \, , 
\eea
where, in the ${\cal O}(v^3)$ terms, we have again neglected $\epsilon_0+P$ compared to $\rho$.
Then, we define the {\it momentum density} by 
\be
g_i\equiv T^{0i}_{\rm non-rel.} = \rho v_i \, , 
\ee
and the {\it energy flux} by
\be
q_i\equiv T^{i0}_{\rm non-rel.} = \left(\epsilon_0 + \frac{\rho v^2}{2}+P\right)v_i \, .
\ee
As a consequence, in the non-relativistic version the stress-energy tensor is not symmetric, $T^{0i}_{\rm non-rel.}\neq T^{i0}_{\rm non-rel.}$. Finally, 
\bea
T^{ij} &\simeq& (\epsilon +P)v_iv_j+\delta_{ij}P\non[2ex]
&\simeq& (\rho+\epsilon_0 +P)v_iv_j+\delta_{ij}P \, , 
\eea
and, again using $\epsilon_0+P\ll\rho$, we define the non-relativistic stress tensor  
\be
\Pi_{ij}\equiv T^{ij}_{\rm non-rel.} = \rho v_iv_j+\delta_{ij}P \, .
\ee
We are now prepared to formulate the conservation equations (\ref{dmuTmunu}) in the non-relativistic limit, 
\begin{subequations} \label{hydro1}
\bea
\frac{\partial \rho}{\partial t} + \nabla\cdot {\bf g} &=& 0 \, , \label{drho}\\[2ex]
\frac{\partial\epsilon}{\partial t} + \nabla\cdot {\bf q} &=& 0 \, , \label{de}\\[2ex]
\frac{\partial g_i}{\partial t} + \partial_j\Pi_{ji} &=& 0 \, , \label{dg}
\eea
\end{subequations}
where we have defined the energy density\footnote{The $\epsilon$ defined here is {\it not} the relativistic
$\epsilon$ used above. But since for the rest of the chapter we shall work in the non-relativistic framework, this slight abuse of notation should not cause any 
confusion.}
\be
\epsilon \equiv  \epsilon_0+\frac{\rho v^2}{2} 
\ee
(remember that $\epsilon_0$ is the energy density in the rest frame of the fluid).
The first equation is the current conservation $\partial_\mu j^\mu=0$ multiplied by $m$, the second equation is the $\nu=0$ component of the four-momentum 
conservation $\partial_\mu T^{\mu\nu}=0$,  and the third equation is the $\nu=i$ component of the four-momentum conservation. 
In summary, we repeat the definitions of the non-relativistic quantities that appear in these equations,
\be \label{gqPi}
{\bf g} = \rho{\bf v} \, , \qquad {\bf q} = \left(\epsilon + P\right){\bf v} \,, \qquad \Pi_{ij} = \rho v_iv_j + \delta_{ij}P \, . 
\ee
Using these definitions, we can also bring the hydrodynamic equations in the following form\footnote{Dissipative effects are included by 
adding the following terms to the energy flux and the stress tensor (the momentum density remains unchanged),
\be \label{gqPidiss}
q_i = (\epsilon + P)v_i +v_j\delta\Pi_{ij}+Q_i  \,, \qquad \Pi_{ij} = \rho v_iv_j + \delta_{ij}P +\delta\Pi_{ij}\, , \nonumber
\ee
where
\be
{\bf Q}\equiv -\kappa\nabla T \, , \qquad \delta\Pi_{ij} \equiv 
 -\eta\left(\partial_iv_j-\partial_jv_i-\frac{2}{3}\delta_{ij}\nabla\cdot {\bf v}\right)-\zeta\delta_{ij}\nabla
\cdot{\bf v} \, , \nonumber
\ee
with the thermal conductivity $\kappa$, the shear viscosity $\eta$, and the bulk viscosity $\zeta$. 
In the presence of dissipation, the entropy current 
is no longer conserved, i.e., the right-hand side of Eq.\ (\ref{entropycons}) is not zero. With dissipative terms, the Euler equation (\ref{euler})
is known as the Navier-Stokes equation. Existence and smoothness of general solutions to the Navier-Stokes equation (and also to the Euler equation) 
are an unsolved problem in mathematical physics and its solution is worth a million dollars, 
see {\it http://www.claymath.org/millenium-problems/navier-stokes-equation}.} 
\begin{subequations} \label{hydroeqs2}
\bea
\frac{\partial \rho}{\partial t} +\nabla\cdot(\rho{\bf v}) &=& 0 \, , \label{drho2} \allowdisplaybreaks \\[2ex]
\frac{\partial s}{\partial t} +\nabla\cdot(s{\bf v}) &=& 0 \, , \label{entropycons}\allowdisplaybreaks\\[2ex]
\frac{\partial {\bf v}}{\partial t} + ({\bf v}\cdot\nabla){\bf v} &=& -\frac{\nabla P}{\rho} \qquad 
\mbox{(Euler equation)} \, . \allowdisplaybreaks \label{euler}
\eea
\end{subequations}
The continuity equation for the mass current (\ref{drho2}) is simply copied from above; Eq.\ (\ref{euler}) is straightforwardly obtained by inserting 
Eqs.\ (\ref{gqPi}) into Eq.\ (\ref{dg}) and using the continuity equation (\ref{drho2}). To derive Eq.\ (\ref{entropycons}) -- which is a continuity equation for the 
entropy current -- start from Eq.\ (\ref{de}) and write 
\bea \label{dedt}
0&=& \frac{\partial}{\partial t}\left(\epsilon_0+\frac{\rho v^2}{2}\right) + \nabla\cdot \left(\epsilon_0 +\frac{\rho v^2}{2} + P\right){\bf v} \non[2ex]
&=& \frac{\partial\epsilon_0}{\partial t} + \frac{v^2}{2}\frac{\partial\rho}{\partial t} + \rho {\bf v}\cdot \frac{\partial {\bf v}}{\partial t} \non[2ex]
&& +(\mu\rho+Ts)\nabla\cdot {\bf v} +{\bf v}\cdot\nabla \epsilon_0 + \nabla\cdot \left(\frac{\rho v^2}{2}{\bf v}\right) + {\bf v}\cdot\nabla P \, ,
\eea
where we have used the relation 
\be
\epsilon_0+P=\mu\rho+Ts \, , 
\ee
where $s$ is the entropy density and $\mu$ the chemical potential\footnote{In this non-relativistic context, we work with the 
chemical potential per unit mass $\mu$, which has the same units as a velocity squared (i.e., it is dimensionless if the speed of light is set to one). 
In the relativistic treatment, starting in chapter \ref{sec:phi4}, $\mu$ will denote the chemical potential per unit charge, which has the same units as energy.}. 
Now we remember the thermodynamic relations
\be \label{thermo1}
d\epsilon_0 = \mu \,d\rho+Tds \, ,
\ee
and 
\be \label{dP}
dP = \rho \,d\mu + s\,dT \, . 
\ee
These two thermodynamic relations reflect the fact that $\epsilon_0$ and $P$ are related via two Legendre transforms with respect to the pairs 
$(T,s)$ and $(\mu,\rho)$. We shall need Eq.\ (\ref{dP}) later. Here we make use of Eq.\ (\ref{thermo1}) which we insert into Eq.\ (\ref{dedt}) to obtain
\bea
0 &=& \mu\frac{\partial\rho}{\partial t} + T\frac{\partial s}{\partial t} + (\mu\rho+Ts)\nabla\cdot {\bf v} + \mu{\bf v}\cdot\nabla\rho + 
T{\bf v}\cdot\nabla s \non[2ex]
&& +\underbrace{\frac{v^2}{2}\frac{\partial\rho}{\partial t} + \rho {\bf v}\cdot \frac{\partial {\bf v}}{\partial t}
+ \nabla\cdot \left(\frac{\rho v^2}{2}{\bf v}\right)+{\bf v}\cdot\nabla P}_{=0} \allowdisplaybreaks\non[2ex] 
&=& T\left[\frac{\partial s}{\partial t}+\nabla\cdot(s{\bf v})\right] \, , 
\eea
where we have used the continuity equation (\ref{drho2}) twice and the Euler equation (\ref{euler}). The result is the entropy 
conservation (\ref{entropycons}). The entropy current is only conserved in the absence of dissipation.

\subsection{Two-fluid hydrodynamics}
\label{sec:subsub2fluids}

In view of the two-fluid model discussed in Sec.\ \ref{sec:2fluid}, we have to modify the single-fluid hydrodynamics because each of the 
fluid components of the superfluid acquires its own, independent velocity field. Let us distinguish two (local) reference frames in the following way. 
Imagine a superfluid flowing through a tube. Then, our first reference frame is the frame where the tube is at rest and where superfluid 
and normal fluid have velocities ${\bf v}_s$ and ${\bf v}_n$, respectively. The second reference frame is the one where the superfluid
is at rest, i.e., the tube moves with velocity $-{\bf v}_s$ and the superfluid and normal fluid move with velocities 
zero and ${\bf v}_n-{\bf v}_s$, respectively. We denote quantities in the superfluid rest frame with a subscript 0 and quantities in the rest frame of the tube
without additional subscript.

In the rest frame of the tube, the momentum density is given by the sum of both fluids, as already stated in Eq.\ (\ref{gboth}). In the superfluid
rest frame, the momentum density is only given by the normal fluid which has mass density $\rho_n$ and which moves with velocity  ${\bf v}_n-{\bf v}_s$.
Consequently, 
\begin{subequations}
\bea
{\bf g} &=& \rho_n{\bf v}_n + \rho_s{\bf v}_s \, , \qquad {\bf g}_0 = \rho_n({\bf v}_n-{\bf v}_s)  \\[2ex]
\Rightarrow \qquad {\bf g} &=& {\bf g}_0 + \rho{\bf v}_s   \, , \label{gal1}
\eea 
\end{subequations}
where $\rho=\rho_n + \rho_s$ is the total mass density, as above.
The stress tensor in the two frames reads 
\begin{subequations}
\bea
\Pi_{ij} &=& \rho_n v_{ni}v_{nj} + \rho_s v_{si} v_{sj} + \delta_{ij} P \, , \non[2ex] 
\Pi_{0ij}&=& \rho_n (v_{ni}-v_{si})(v_{nj}-v_{sj})  + \delta_{ij} P  \label{stress1}\\[2ex]
\Rightarrow \qquad \Pi_{ij} &=& \Pi_{0ij} + \rho v_{si}v_{sj} + v_{si}g_{0j}+v_{sj}g_{0i}  \, .\label{gal2}
\eea
\end{subequations}
For completeness, although we shall not need this in the following, let us also write down the energy density and the energy density current 
in the two frames. We have 
\begin{subequations}
\bea
\epsilon &=&\epsilon_n + \epsilon_s + \frac{\rho_nv_n^2}{2} +\frac{\rho_sv_s^2}{2} \, , \qquad \epsilon_0 = 
\epsilon_n + \epsilon_s + \frac{\rho_n({\bf v}_n-{\bf v}_s)^2}{2} \hspace{0.5cm} \\[2ex]
\Rightarrow \qquad \epsilon &=& \epsilon_0 + {\bf v}_s\cdot{\bf g}_0  + \frac{\rho v_s^2}{2}  \, ,\label{gal3}
\eea
\end{subequations}
where 
\be \label{epsnepss}
\epsilon_n = -P_n + \mu \rho_n + Ts \, , \qquad \epsilon_s = -P_s + \mu \rho_s  
\ee 
are the energy densities of normal fluid and superfluid, measured in their respective rest frames. Analogously, $P_n$ and $P_s$ are the pressures of the
normal fluid and superfluid, and with $P=P_n+P_s$ the relations (\ref{epsnepss}) imply $\epsilon_n+\epsilon_s = P+\mu\rho +Ts$. 
In the absence of a normal fluid, we have $\epsilon_s=\epsilon_0$, which makes the connection to the notation of the previous subsection. 
In Eqs.\ (\ref{epsnepss}) we have used that only the normal fluid carries entropy. 

Finally, for the energy flux we find
\begin{subequations}
\bea
{\bf q} &=& \left(\epsilon_n + P_n + \frac{\rho_nv_n^2}{2}\right){\bf v}_n + \left(\epsilon_s + P_s + \frac{\rho_sv_s^2}{2}\right){\bf v}_s
\, , \non[2ex]
\quad {\bf q}_0 &=& \left[\epsilon_n + P_n + \frac{\rho_n({\bf v}_n-{\bf v}_s)^2}{2}\right]({\bf v}_n-{\bf v}_s) \\[2ex]
\Rightarrow \qquad q_i &=& q_{0i}+ \left(\epsilon_0 + {\bf v}_s\cdot{\bf g}_0  + \frac{\rho v_s^2}{2}\right) v_{si} +\frac{v_s^2}{2}g_{0i}+v_{sj}\Pi_{0ij} 
\, .\label{gal4}
\eea
\end{subequations}
Eqs.\ (\ref{gal1}), (\ref{gal2}), (\ref{gal3}), and (\ref{gal4}) are the Galilei transforms of momentum density, stress tensor, energy density, and energy flux 
from the superfluid rest frame into the corresponding quantities in the rest frame of the tube. Notice that they are expressed solely in terms 
of quantities measured in the superfluid rest frame and the superfluid velocity. Since the relative velocity between the two reference frames is given by ${\bf v}_s$, 
the normal-fluid velocity ${\bf v}_n$ does not appear in the Galilei transform, as it should be.

We write the hydrodynamic equations in the rest frame of the tube as 
\begin{subequations} \label{hydro2}
\bea
&&\frac{\partial \rho}{\partial t} + \nabla\cdot {\bf g} = 0 \, , \\[2ex]
&&\frac{\partial s}{\partial t} +\nabla\cdot(s{\bf v}_n) = 0 \, , \\[2ex]
&&\frac{\partial {\bf g}}{\partial t} + {\bf v}_s(\nabla\cdot{\bf g}) + ({\bf g}\cdot\nabla){\bf v}_s + 
{\bf g}_0(\nabla\cdot{\bf v}_n) + ({\bf v}_n\cdot\nabla){\bf g}_0 +\nabla P = 0 \, . \label{dg2}
\eea
\end{subequations}
The first two equations have the same form as for the single fluid case, see Eqs.\ (\ref{hydroeqs2}), with $\rho$ and ${\bf g}$ now being the total mass and 
momentum densities, receiving contributions from both fluids, and the entropy density $s{\bf v}_n$ solely coming from the normal fluid. 
To derive Eq.\ (\ref{dg2}) from (\ref{dg}) one first easily checks that the stress tensor from Eq.\ (\ref{stress1}) can be written as 
\be \label{piij}
\Pi_{ij} = v_{sj}g_i + v_{ni}g_{0j}+\delta_{ij}P \, .
\ee
(Although not manifest in this form, the stress tensor is of course still symmetric.) Inserting Eq.\ (\ref{piij}) 
into Eq.\ (\ref{dg}) immediately yields Eq.\ (\ref{dg2}).

Before we turn to the sound modes we derive one more useful relation. Using the hydrodynamic equations, the thermodynamic
relations (\ref{thermo1}) and (\ref{dP}), and the explicit two-fluid form of $\epsilon$ and ${\bf q}$, a rather tedious calculation yields 
\bea \label{vs1}
\frac{\partial\epsilon}{\partial t}+\nabla\cdot{\bf q} &=& -\rho_s({\bf v}_n-{\bf v}_s)\left[\nabla\mu+({\bf v}_s\cdot\nabla){\bf v}_s
+\frac{\partial{\bf v}_s}{\partial t}\right] \non[2ex]
&& +\frac{({\bf v}_n-{\bf v}_s)^2}{2}\left[\frac{\partial\rho_s}{\partial t}+\nabla\cdot(\rho_s{\bf v}_s)\right] \, .
\eea
Now we use that the left-hand side of this equation is zero due to (the two-fluid version of) Eq.\ (\ref{de}) and neglect the term quadratic in the relative 
velocity ${\bf v}_n-{\bf v}_s$ on the right-hand side to obtain the following relation for the superfluid velocity,
\be\label{vs2}
({\bf v}_s\cdot\nabla){\bf v}_s+\frac{\partial{\bf v}_s}{\partial t}=-\nabla\mu \, .
\ee

\subsection{Sound modes}
\label{sec:sound}

Imagine both fluid components to be at rest and the system to be in thermodynamic equilibrium. A sound wave is a (small) oscillation
in the thermodynamic quantities like entropy, pressure etc and in the velocities of the two fluids. We thus imagine adding small deviations from equilibrium to the 
thermodynamic quantities like $s({\bf x},t) = s_0+\delta s({\bf x},t)$, $P({\bf x},t) = P_0+\delta P({\bf x},t)$ etc and small deviations (from zero) to the velocities, 
${\bf v}_n({\bf x},t) = \delta{\bf v}_n({\bf x},t)$ and ${\bf v}_s({\bf x},t) = \delta{\bf v}_s({\bf x},t)$. Here, the 
subscript 0 denotes thermodynamic equilibrium. In general, one might also compute the sound modes in the presence of a relative velocity of the two fluids, i.e., 
one may choose nonzero values of ${\bf v}_{n,0}$ and ${\bf v}_{s,0}$. Here we restrict ourselves to the isotropic situation ${\bf v}_{n,0}={\bf v}_{s,0}=0$.

Since we are interested in small deviations from equilibrium, we neglect terms quadratic in the deviations, for instance 
\bea
\nabla\cdot {\bf g} &=& (\rho_{n,0}+\delta\rho_n)\nabla\cdot\delta{\bf v}_n + (\rho_{s,0}+\delta\rho_s)\nabla\cdot\delta{\bf v}_s 
+\delta{\bf v}_n\cdot\nabla\delta\rho_n +\delta{\bf v}_s\cdot\nabla\delta\rho_s \non[2ex]
&\simeq&\rho_{n,0} \nabla\cdot \delta{\bf v}_n +\rho_{s,0}\nabla\cdot \delta{\bf v}_s \, .
\eea
The linearized hydrodynamic equations (\ref{hydro2}) thus become 
\begin{subequations} \label{hydroapprox}
\bea
\frac{\partial \rho}{\partial t} + \rho_n\nabla\cdot {\bf v}_n +\rho_s\nabla\cdot {\bf v}_s  &\simeq& 0 \, , \label{rhoapprox}\\[2ex]
\frac{\partial s}{\partial t} +s \nabla\cdot {\bf v}_n &\simeq& 0 \, , \label{sapprox}\\[2ex]
\rho_n\frac{\partial {\bf v}_n}{\partial t}+\rho_s\frac{\partial {\bf v}_s}{\partial t} +\nabla P &\simeq& 0 \, , \label{gapprox} 
\eea
\end{subequations}
and Eq.\ (\ref{vs2}) simplifies to
\be
\frac{\partial {\bf v}_s}{\partial t}+ \nabla\mu \simeq 0 \, . \label{vsapprox}
\ee
A usual sound wave is a density oscillation and is described by a wave equation that relates a spatial second derivative to a temporal second derivative. 
Due to the presence of two fluids, we will now obtain a second wave equation for the entropy. The two wave equations are 
\begin{subequations}
\bea
\frac{\partial^2\rho}{\partial t^2} &=& \Delta P \, , \label{wave1}\\[2ex]
\frac{\partial^2 S}{\partial t^2} &=& \frac{S^2\rho_s}{\rho_n}\Delta T \, , \label{wave2}
\eea
\end{subequations} 
where $S$ is the entropy per unit mass, such that $s=\rho S$. The equations are derived as follows. 
Eq.\ (\ref{wave1}) is immediately obtained by taking the time derivative of Eq.\ (\ref{rhoapprox}) and the divergence of Eq.\ (\ref{gapprox}). 
Eq.\ (\ref{wave2}) requires some more work. From the thermodynamic relation (\ref{dP})
we obtain $\nabla P = \rho\nabla\mu+s\nabla T$. Inserting $\nabla P$ from Eq.\ (\ref{gapprox}) and 
$\nabla\mu$ from Eq.\ (\ref{vsapprox}) into this relation, taking the divergence on both sides, and keeping only terms linear in the deviations
from equilibrium yields
\be \label{div}
\rho_n\frac{\partial}{\partial t} \nabla\cdot ({\bf v}_n-{\bf v}_s) \simeq -s\Delta T \, .
\ee
In order to replace the divergence on the right-hand side of this equation we observe
\bea
\frac{\partial S}{\partial t} &=& \frac{1}{\rho}\frac{\partial s}{\partial t} - \frac{S}{\rho}\frac{\partial \rho}{\partial t} \non[2ex]
&=&-S\nabla\cdot {\bf v}_n+\frac{S}{\rho}(\rho_n\nabla\cdot{\bf v}_n+\rho_s\nabla\cdot {\bf v}_s) \non[2ex]
&=& -\frac{S\rho_s}{\rho}\nabla\cdot({\bf v}_n-{\bf v}_s) \, , 
\eea
where, in the second step, we have used Eqs.\ (\ref{rhoapprox}) and (\ref{sapprox}). Inserting this result into Eq.\ (\ref{div})
and again using the linear approximation yields the second wave equation (\ref{wave2}). 

Next, we solve the wave equations. We take $T$ and $P$ as independent variables, such that $S$ and $\rho$ are functions of $T$ and $P$,
\be
\delta S = \frac{\partial S}{\partial T}\delta T + \frac{\partial S}{\partial P}\delta P  \, , \qquad  \
\delta \rho = \frac{\partial\rho}{\partial T}\delta T  + \frac{\partial\rho}{\partial P}\delta P  
 \, , 
\ee
where all derivatives are evaluated in equilibrium. Inserting this into Eqs.\ (\ref{wave1}) and (\ref{wave2}) yields to linear order in $\delta T$, $\delta P$,
\begin{subequations}
\bea
\frac{\partial\rho}{\partial P}\frac{\partial^2\delta P}{\partial t^2} + \frac{\partial\rho}{\partial T}\frac{\partial^2\delta T}{\partial t^2}
&=& \Delta (\delta P) \, , \\[2ex]
\frac{\partial S}{\partial P}\frac{\partial^2\delta P}{\partial t^2} + \frac{\partial S}{\partial T}\frac{\partial^2\delta T}{\partial t^2}
&=& \frac{S^2\rho_s}{\rho_n} \Delta (\delta T) \, .
\eea
\end{subequations}
The deviations from equilibrium are assumed to be harmonic oscillations, $\delta P({\bf x},t)=\delta P_0e^{-i(\omega t-kx)}$, 
$\delta T({\bf x},t)=\delta T_0e^{-i(\omega t-kx)}$, where the amplitudes $\delta P_0$, $\delta T_0$ are constant in time and space, and where $\omega$ and $k$
are frequency and wave number of the oscillation. Without loss of generality, we have chosen the sound waves to propagate in the $x$-direction.
We define the sound velocity
\be
u = \frac{\omega}{k} \, , 
\ee
such that the wave equations become
\begin{subequations}
\bea
\left(u^2\frac{\partial\rho}{\partial P}-1\right)\delta P_0 + u^2 \frac{\partial\rho}{\partial T}\delta T_0 &=& 0 \,  , \\[2ex]
u^2\frac{\partial S}{\partial P}\delta P_0 + \left(u^2\frac{\partial S}{\partial T}-\frac{S^2\rho_s}{\rho_n}\right)\delta T_0 &=& 0 \, .
\eea
\end{subequations}
For this system of equations to have nontrivial solutions, we must require the determinant to vanish,
\be \label{polyu41}
u^4 |J_f(T,P)|  
- u^2 \left(\frac{\partial \rho}{\partial P}\frac{S^2\rho_s}{\rho_n}+\frac{\partial S}{\partial T}\right)
+\frac{S^2\rho_s}{\rho_n} = 0 \, ,
\ee
where $|J_f(T,P)|$ is the determinant of the Jacobian matrix of the function $f(T,P)\equiv (S(T,P),\rho(T,P))$. 
The Jacobian matrix of $f$ is 
\be
J_f(T,P) \equiv \frac{\partial(S,\rho)}{\partial (T,P)} = \left(\begin{array}{cc} \displaystyle{\frac{\partial S}{\partial T}} &
\displaystyle{\frac{\partial S}{\partial P}} \\[2ex] \displaystyle{\frac{\partial \rho}{\partial T}} &
\displaystyle{\frac{\partial \rho}{\partial P}}\end{array}\right) \, .
\ee
Now remember that the derivative of the inverse function $f^{-1}(S,\rho)=(T(S,\rho),P(S,\rho))$ is given by 
the inverse of the Jacobian matrix of $f$, $J_{f^{-1}}[f(T,P)] = [J_f(T,P)]^{-1}$. Therefore,
\be
J_{f^{-1}}(S,\rho) \equiv \frac{\partial(T,P)}{\partial (S,\rho)} = \left(\begin{array}{cc} \displaystyle{\frac{\partial T}{\partial S}} &
\displaystyle{\frac{\partial T}{\partial \rho}} \\[2ex] \displaystyle{\frac{\partial P}{\partial S}} &
\displaystyle{\frac{\partial P}{\partial \rho}}\end{array}\right) = \frac{1}{|J_f(T,P)|} 
\left(\begin{array}{cc} \displaystyle{\frac{\partial \rho}{\partial P}} &
-\displaystyle{\frac{\partial S}{\partial P}} \\[2ex] -\displaystyle{\frac{\partial \rho}{\partial T}} &
\displaystyle{\frac{\partial S}{\partial T}}\end{array}\right) \, .\nonumber
\ee
Consequently, from the diagonal elements of this matrix equation we read off
\be \label{diagel}
\frac{\partial T}{\partial S} = \frac{1}{|J_f(T,P)|}\frac{\partial \rho}{\partial P} \, , \qquad \frac{\partial P}{\partial \rho} = \frac{1}{|J_f(T,P)|}
\frac{\partial S}{\partial T} \, .
\ee
After dividing Eq.\ (\ref{polyu41}) by $|J_f(T,P)|$ we can use these relations to write 
\be \label{polyu42}
u^4 - u^2 \left(\frac{\partial T}{\partial S} \frac{S^2\rho_s}{\rho_n} + \frac{\partial P}{\partial \rho}\right) +
\frac{1}{|J_f(T,P)|} \frac{S^2\rho_s}{\rho_n}=0 \, . 
\ee
Up to now we have worked with the function $f(T,P)=[S(T,P),\rho(T,P)]$ and its inverse $f^{-1}(S,\rho)=[T(S,\rho),P(S,\rho)]$. In Eq.\ (\ref{polyu42}), 
derivatives of $f^{-1}$ appear, and thus the derivatives of $T$ with respect to $S$ and of $P$ with respect to $\rho$ are obviously taken at fixed $\rho$ and $S$, 
respectively. We further simplify the polynomial for $u$ as follows. Take the first component of $f^{-1}$, $T(S,\rho)$, and read it as a function of $S$ only, with
a fixed $\rho$. Inversion of this function then yields a function $S(T,\rho)$. Now do the same with the second component of $f$, $\rho(P,T)$, i.e., invert
this function at a fixed $T$. This yields
\begin{subequations} \label{TSrhoP}
\bea
\frac{\partial T}{\partial S} &=& \left(\frac{\partial S}{\partial T}\right)^{-1}= \frac{T\rho}{c_V} \;\;\;\; \mbox{at fixed}\;\rho \, , \label{TSrhoP1}\\[2ex]
\left(\frac{\partial \rho}{\partial P}\right)^{-1} &=& \frac{\partial P}{\partial \rho} \;\;\;\; \mbox{at fixed}\;T \, , 
\eea
\end{subequations}
with the definition for the specific heat per unit mass  
\be
\frac{c_V}{\rho} = T\frac{\partial S}{\partial T}  \, ,  
\ee
where $c_V$ is the specific heat per unit volume, and the derivative is taken at fixed $\rho$. With the help of Eqs.\ (\ref{TSrhoP}) we write the first relation 
of Eq.\ (\ref{diagel}) as
\be \label{Jfinv}
\frac{1}{|J_f(T,P)|} = \frac{T\rho}{c_V}\frac{\partial P}{\partial \rho} \, , 
\ee
with the derivative taken at fixed $T$. Now we insert Eqs.\ (\ref{TSrhoP1}) and (\ref{Jfinv}) into the polynomial (\ref{polyu42}) to obtain
\be
u^4 -u^2 \left(\frac{\rho S^2T\rho_s}{c_V\rho_n} + \frac{\partial P}{\partial \rho} \right) 
+  \frac{\rho S^2T\rho_s}{c_V\rho_n}\frac{\partial P}{\partial \rho}=0 \, . 
\ee
The two derivatives of $P$ with respect to $\rho$ appearing here were not identical originally because the derivatives are taken at fixed $S$ and at fixed $T$. 
However, we approximate these derivatives to be equal, which is equivalent to approximating the specific heat at constant pressure $c_P$ by the specific heat at 
constant volume $c_V$. This approximation turns out to be a good approximation for superfluid helium.

The resulting equation has the simple structure $u^4 - u^2(a+b)+ab=0$ with solutions $u^2=a,b$. Consequently, the two positive solutions for $u$ are  
\begin{subequations} \label{u1u2he}
\bea
u_1&=& \sqrt{\frac{\partial P}{\partial \rho}} \, , \\[2ex]
u_2 &=& \sqrt{\frac{\rho S^2T\rho_s}{c_V\rho_n}} = 
\sqrt{\frac{s^2T\rho_s}{\rho c_V\rho_n}}\, .
\eea
\end{subequations}
These are the velocities of first and second sound. Since we have not worked with a relative velocity ${\bf v}_n-{\bf v}_s$ between normal fluid and 
superfluid (except for the small oscillations that constitute the sound waves), the sound velocities are pure numbers, i.e., they do not depend on the direction
of propagation. 

\begin{figure} [t]
\begin{center}
\includegraphics[width=0.85\textwidth]{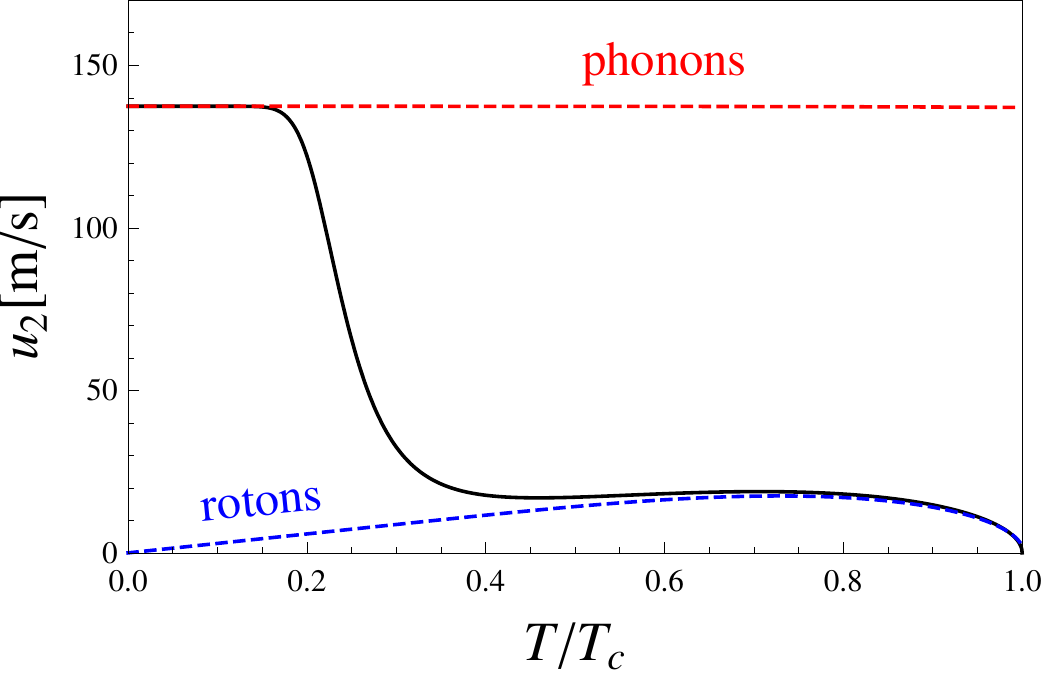}
\caption{(Color online) Speed of second sound $u_2$ as a function of temperature 
in superfluid helium from Eq.\ (\ref{u1u2he}) with the parameters given in Fig.\ \ref{figphrot1}. The dashed lines are obtained by only 
taking into account phonon and roton contributions. 
}
\label{figsounds}
\end{center}
\end{figure}

At low temperatures, as we shall see in the next section in a microscopic model, 
\be \label{Prhoc}
\frac{\partial P}{\partial \rho} \simeq  c^2 \, , 
\ee
i.e., the speed of first sound is given by the slope of the Goldstone dispersion, $u_1(T\to 0)=c$. For the speed of second sound, we may use 
the results from our thermodynamic calculations in Sec.\ \ref{sec:thermohelium}. At low temperatures, the roton contribution is irrelevant, and we use 
$c_{V,{\rm ph}}=3s_{\rm ph}$ and $\rho_{n,{\rm ph}} = s_{\rm ph}T/c^2$, see Eqs.\ (\ref{cVph}) and (\ref{rhophs}), respectively.
We may also approximate $\rho_s\simeq \rho$. Inserting all this into our expression for $u_2$ we find
\be \label{u2T0}
u_2(T\to 0) = \frac{c}{\sqrt{3}} = \frac{u_1(T\to 0)}{\sqrt{3}}\, .
\ee
The full temperature dependence of $u_2$, within the present phonon/roton model, is shown in Fig.\ \ref{figsounds}. Remember that second sound is only possible 
due to the presence of the second fluid. Therefore, it is easy to understand that $u_2$ goes to zero at the critical temperature, because at that point $\rho_s\to 0$
and the system becomes a single-fluid system. Interestingly, the behavior at small temperatures is different. Had we set $T=0$ exactly, there would have been no
normal fluid and thus no second sound. However, starting with two fluids and then taking the limit $T\to 0$ leads to a nonzero speed of second sound. 
The figure also shows the characteristic behavior of $u_2$ due to the presence of the phonons and rotons. As expected from the discussion above, see 
in particular Fig.\ \ref{figrhophrhorot}, the phonons dominate at low temperatures, $T\lesssim 0.25\,T_c$, while the rotons dominate for all larger temperatures 
below $T_c$. 
This characteristic behavior is special for helium, and superfluids that have no rotons show a different behavior. In contrast, the ratio of first
and second sound at low temperatures given in Eq.\ (\ref{u2T0}) is more universal because it only depends on the linear behavior of the Goldstone mode at 
small momenta. If you are interested in recent theoretical studies about sound waves in superfluids, for instance in the context of superfluid atomic gases or 
relativistic superfluids, see Refs.\ \cite{Alford:2012vn,2009PhRvA..80e3601T,2010NJPh...12d3040H,Alford:2013koa,Schmitt:2013nva}.

\chapter{Superfluidity in quantum field theory}
\label{sec:phi4}

The theoretical treatment of superfluid helium used in the previous section was phenomenological in the sense that the microscopic degrees of freedom, the helium
atoms, never appeared in our description. We took it as given that there is a gapless excitation, and we modelled its form in terms of phonons and rotons,
if you wish because experiments tell us so. In particular, the Bose condensate only appeared in a very indirect way: by fixing the total density and having a 
model for the normal fluid, we could compute the superfluid density. We shall now take a more microscopic approach. We shall start from a theory for the 
degrees of freedom that form a Bose condensate and attempt to gain a more fundamental understanding of the characteristic properties of a superfluid. 
The following general concepts, which shall be made more precise in this chapter, will play a central role:
\begin{itemize}

\item {\it $U(1)$ symmetry}: this is the simplest continuous symmetry, given by one real parameter; the Lagrangian of our model will be invariant under this symmetry, 
and its so-called spontaneous breaking is a necessary condition for superfluidity.

\item {\it conserved charge}: a conserved charge is a consequence of the $U(1)$ symmetry via Noether's theorem, and it is essential for all superfluids
because this charge is transported by a superflow.

\item {\it Bose-Einstein condensation}: in the present context of a bosonic superfluid, Bose-Einstein condensation
is just another way of saying that the $U(1)$ symmetry is spontaneously broken.

\item {\it spontaneous symmetry breaking}: the ground state, i.e., the Bose-Einstein condensate, in a superfluid is not invariant with respect to transformations 
of the original symmetry of the Lagrangian of the system; this is called spontaneous symmetry breaking.

\item {\it Goldstone mode}: if the spontaneously broken symmetry is global, as it is the case in this chapter, 
a massless mode arises for all temperatures below the critical temperature; we shall compute the dispersion of this Goldstone mode explicitly. 

\item {\it symmetry restoration \& critical temperature}: all superfluids/superconductors we discuss exist at sufficiently small temperatures, and there 
is a certain critical temperature where the condensate has melted or, in other words, the ground state has become symmetric under the full symmetry of the Lagrangian.

\end{itemize}

\section{Lagrangian and conserved charge}

We start from the following Lagrangian 
${\cal L}$ for a complex scalar field $\varphi(X)$ depending on space-time, $X\equiv (x_0,{\bf x})$\footnote{As mentioned in the introduction, 
I denote space-time and momentum four-vectors 
by capital letters $X$, $Y$, $\ldots$ and $K$, $Q$, $\ldots$. This leaves the small letters $x=|{\bf x}|$ and $k=|{\bf k}|$ for the moduli of the three-vectors.},
\be \label{lagr}
{\cal L} = \partial_\mu\varphi^*\partial^\mu\varphi -m^2|\varphi|^2 -\lambda |\varphi|^4 \, .
\ee
The Lagrangian describes spin-0 bosons with mass $m$ which interact repulsively with each other with a coupling constant $\lambda>0$. 

We first observe that ${\cal L}$ is invariant under $U(1)$ rotations of the field,
\be
\varphi \to e^{-i\alpha} \varphi \, ,
\ee
with a constant $\alpha \in \mathbb{R}$. Since $\alpha$ is constant one talks about a {\it global} transformation or a {\it global} symmetry, as opposed to a {\it local}
symmetry where $\alpha$ would be allowed to depend on space-time.

In order to account for Bose-Einstein condensation, we need to separate the condensate from the fluctuations. This is done by writing 
\be \label{shift}
\varphi(X) \to \phi(X) + \varphi(X) \, , 
\ee
where $\phi(X)$ is the condensate and $\varphi(X)$ are the fluctuations. The point of this decomposition is that the fluctuations are a 
dynamical field, i.e., we perform a functional integration over $\varphi$ and $\varphi^*$ when we compute the partition function, 
while the condensate is a classical field, which here we shall 
determine from the Euler-Lagrange equations. A priori we do not know whether Bose-Einstein condensation will occur, so in a sense Eq.\ (\ref{shift}) is an ansatz,
and later it will turn out whether $\phi(X)$ is nonzero and which value it assumes. 

Remember from the textbook treatment of (non-relativistic) Bose-Einstein condensation that the condensate describes a macroscopic occupation of the bosons in the ground 
state of the system, usually the zero-momentum state. Analogously, in field theory, we may Fourier decompose the field and separate the state with zero four-momentum 
$K=(k_0,{\bf k})$. More generally, if we want to allow for a nonzero superfluid velocity ${\bf v}_s$, condensation takes place in a state with nonzero four-momentum, 
say $P$. Therefore, a uniform condensate that moves with a constant velocity determined by $P$ can be written as $\phi(X)=\varphi(P)e^{-i P\cdot X}$, while the 
fluctuations then are $\varphi(X)=\sum_{K\neq P}e^{-iK\cdot X}\varphi(K)$. At this point, however, we do not need to make any assumptions for $\phi(X)$, 
although later we shall mostly talk about a uniform condensate, or, even simpler, about a condensate at rest, $P=0$. 

We write the complex condensate in terms of its modulus $\rho$ and its phase $\psi$, 
\be
\phi(X) = \frac{\rho(X)}{\sqrt{2}} e^{i\psi(X)} \, .
\ee
The fluctuations $\varphi(X)$ will later be needed to compute the dispersion of the Goldstone mode. As a first
step, we neglect them. In this case, the Lagrangian only depends on the classical field, 
\be
{\cal L} = {\cal L}^{(0)} + \mbox{fluctuations} \, , 
\ee
where
\be \label{lagr0}
{\cal L}^{(0)} = \frac{1}{2}\partial_\mu\rho\partial^\mu\rho +\frac{\rho^2}{2}(\partial_\mu\psi\partial^\mu\psi -m^2) - \frac{\lambda}{4}\rho^4 \, .
\ee
Next we write down the equations of motion for $\rho$ and $\psi$. Notice that the phase $\psi$ only appears through its space-time derivative, and thus
the Euler-Lagrange equations are
\begin{subequations} \label{eullag}
\bea
0&=& \frac{\partial{\cal L}}{\partial\rho} - \partial_\mu\frac{\partial{\cal L}}{\partial(\partial_\mu\rho)} \\[2ex]
0&=&\partial_\mu\frac{\partial{\cal L}}{\partial(\partial_\mu\psi)} \, .
\eea
\end{subequations}
From this general form we compute 
\begin{subequations} \label{eoms}
\bea
\Box\rho &=& \rho(\sigma^2-m^2-\lambda\rho^2) \, , \label{eom1} \\[2ex]
\partial_\mu(\rho^2\partial^\mu\psi) &=& 0  \, , \label{eom2} 
\eea
\end{subequations}
where we have abbreviated
\be
\sigma\equiv \sqrt{\partial_\mu\psi\partial^\mu\psi} \, .
\ee
The second equation of motion (\ref{eom2}) is nothing but the continuity equation for the conserved current.
We know from Noether's theorem that a system with a continuous symmetry has a conserved current\footnote{It is therefore important that we 
consider a complex field; the same Lagrangian for a real scalar field has only a discrete $\mathbb{Z}_2$ symmetry $\varphi\to-\varphi$ and 
thus no conserved current.}: 

\bigskip
{\bf Noether's Theorem:} If the Lagrangian is invariant under transformations of a continuous global symmetry group there exists a four-current $j^\mu$
that obeys a continuity equation and a corresponding conserved charge $Q=\int d^3{\bf x}\,j^0$.

\bigskip
 
In our case, the current assumes the 
form\footnote{In the basis of $\varphi$ and $\varphi^*$, we have $j^\mu = i(\varphi\partial^\mu\varphi^*-\varphi^*\partial^\mu\varphi)$.} 
\be \label{noetherj}
j^\mu = \frac{\partial{\cal L}}{\partial(\partial_\mu\psi)} = \rho^2\partial^\mu\psi \, , 
\ee
such that Eq.\ (\ref{eom2}) is 
\be
\partial_\mu j^\mu = 0 \, .
\ee
This is one of the hydrodynamic equations introduced in Eqs.\ (\ref{dmuTmunu}). The second one, namely the conservation of the stress-energy tensor $T^{\mu\nu}$
can be derived as follows. We use the definition
\be \label{Tmunugrav}
T^{\mu\nu} = \frac{2}{\sqrt{-g}}\frac{\delta(\sqrt{-g}{\cal L})}{\delta g_{\mu\nu}} = 2\frac{\delta{\cal L}}{\delta g_{\mu\nu}} - g^{\mu\nu}{\cal L} \, ,
\ee
where $\sqrt{-g}\equiv \sqrt{-{\rm det}\,g^{\mu\nu}}$. It is applicable to general metric tensors $g^{\mu\nu}$, in particular to curved space-times, and is thus 
sometimes called the {\it gravitational} definition. Of course, here we are only interested in flat space-time, i.e., after taking the derivatives 
in Eq.\ (\ref{Tmunugrav}) we set $g^{\mu\nu}=(1,-1,-1,-1)$. The advantage of this definition is that the stress-energy tensor is manifestly 
symmetric, as opposed to the so-called {\it canonical} stress-energy tensor, which is conserved too, but in general not symmetric.

With our classical Lagrangian ${\cal L}^{(0)}$ we obtain 
\be \label{Tmunumicro}
T^{\mu\nu} = \partial^\mu\rho\partial^\nu\rho+\rho^2\partial^\mu\psi\partial^\nu\psi-g^{\mu\nu}{\cal L}^{(0)} \, .
\ee
Consequently, 
\be
\partial_\mu T^{\mu\nu} = \partial_\mu(\partial^\mu\rho\partial^\nu\rho+\rho^2\partial^\mu\psi\partial^\nu\psi)-\partial^\nu{\cal L}^{(0)} \, .
\ee
We compute
\bea
\partial^\nu {\cal L}^{(0)} &=& \frac{\partial{\cal L}^{(0)}}{\partial\rho}\partial^\nu \rho + \frac{\partial{\cal L}^{(0)}}{\partial(\partial_\mu\rho)}
\partial^\nu\partial_\mu \rho +\frac{\partial{\cal L}^{(0)}}{\partial(\partial_\mu\psi)}
\partial^\nu\partial_\mu \psi  \non[2ex]
&=& \partial^\nu \rho \left[\frac{\partial{\cal L}^{(0)}}{\partial\rho}-\partial_\mu\frac{\partial{\cal L}^{(0)}}{\partial(\partial_\mu\rho)}\right]
-\partial^\nu\psi\partial_\mu\frac{\partial{\cal L}^{(0)}}{\partial(\partial_\mu\psi)} \non[2ex]
&&+\partial_\mu\left[\frac{\partial{\cal L}^{(0)}}{\partial(\partial_\mu\rho)}\partial^\nu \rho
+\frac{\partial{\cal L}^{(0)}}{\partial(\partial_\mu\psi)}\partial^\nu\psi\right] \non[2ex]
&=& \partial_\mu(\partial^\mu\rho\partial^\nu\rho+\rho^2\partial^\mu\psi\partial^\nu\psi) \, ,
\eea
where we have used the equations of motion (\ref{eullag}). We thus have 
\be
\partial_\mu T^{\mu\nu} = 0 \, ,
\ee
as expected. Just as for the current, this conservation is also a consequence of Noether's theorem. In this case, the associated symmetry is translational invariance. 

\section{Spontaneous symmetry breaking}

Let us now for simplicity assume that $\rho$ and $\partial^\mu\psi$ are constant in space and time. As a consequence, the current and the stress-energy
tensor also become constant, and the equations $\partial_\mu j^\mu = \partial_\mu T^{\mu\nu}=0$ are trivially fulfilled. Therefore, 
with this assumption we will not be able to discuss complicated hydrodynamics, but we shall be able to discuss the basic concepts of spontaneous symmetry breaking
and the physics of a uniform superflow.
 
In this case, the Lagrangian becomes
\be \label{treelevelU}
{\cal L}^{(0)} = -U \, , \qquad U = -\frac{\rho^2}{2}(\sigma^2-m^2)+\frac{\lambda}{4}\rho^4 \, ,
\ee
where $U$ is called tree-level potential. Since the second equation of motion (\ref{eom2}) is now trivially fulfilled we are left with the first one
(\ref{eom1}) to determine the condensate. This equation is equivalent to finding extremal points of $U$ with respect to $\rho$,
\be
0 = \frac{\partial U}{\partial \rho} = \rho(\sigma^2-m^2-\lambda\rho^2) \, ,
\ee
which has the solutions 
\be \label{sols}
\rho= 0 \, , \qquad \rho^2 = \frac{\sigma^2-m^2}{\lambda} \, .
\ee
Since $\rho\in \mathbb{R}$ and thus $\rho^2>0$, we see that we need $\sigma^2>m^2$ for the nontrivial solution to exist. 
In order to understand this condition, remember that relativistic Bose-Einstein condensation occurs when the 
chemical potential is larger than the mass\footnote{In the textbook example of a non-relativistic, noninteracting Bose gas, 
the chemical potential is always non-positive, and Bose-Einstein condensation occurs when the chemical potential is equal to zero.
This might be a bit confusing because the definition of the chemical potential in a non-relativistic treatment differs from the one 
in a relativistic treatment by the rest mass, $\mu_{\rm non-rel.}=\mu_{\rm rel.}-m$. In other words, in a non-relativistic description, you get your massive 
particles ``for free'', while in general you need to invest the energy $m$, which can be provided by the chemical potential.
With this difference in mind we understand that Bose-Einstein condensation at $\mu_{\rm non-rel.}=0$ corresponds to $\mu_{\rm rel.}=m$. More generally speaking, 
$\mu_{\rm rel.}=m$ induces a condensate with positive charge, and $\mu_{\rm rel.}=-m$ a condensate with negative charge. In the presence of interactions, here due 
to the $\varphi^4$ term, $|\mu_{\rm rel.}|$ is allowed to become larger than $m$. In this chapter, $\mu = \mu_{\rm rel.}$.}. 
This suggests that $\sigma$ plays the role of a chemical potential. Let us try to understand how this comes about. Usually, 
a chemical potential $\mu$ is introduced via ${\cal H}-\mu{\cal N}$ with the Hamiltonian ${\cal H}$ and the conserved charge density ${\cal N}=j^0$.
This is equivalent to adding the chemical potential to the Lagrangian as if it were the temporal component of a gauge field (see for instance Ref.\ \cite{thermal}), i.e., 
\be \label{Lmu}
{\cal L} = |(\partial_0-i\mu)\varphi|^2-|\nabla\varphi|^2 -m^2|\varphi|^2 -\lambda |\varphi|^4 \, .
\ee
One can show that this modified Lagrangian with kinetic term $|(\partial_0-i\mu)\varphi|^2-|\nabla\varphi|^2$ and a condensate with a trivial phase, 
$\psi=0$, is identical to the original Lagrangian with kinetic term $\partial_\mu\varphi\partial^\mu\varphi^*$ and a condensate 
with a time-dependent phase $\psi= \mu t$. The explicit check of this statement is left as an exercise, but it is not difficult to understand because it is 
closely related to a gauge transformation: in a $U(1)$ gauge theory, the transformation that gives a phase to the field is exactly compensated by the covariant 
derivative. We conclude that it does not matter whether we put the chemical potential directly into the Lagrangian or whether we introduce it through the phase of the 
condensate. Consequently, $\mu=\partial_0\psi$ is the chemical potential associated with the conserved charge. However, this is not exactly what we were expecting. 
We had conjectured that  $\sigma=\sqrt{\partial_\mu\psi\partial^\mu\psi}$ plays the role of a chemical potential. This is only identical to $\partial_0\psi$ if we 
set $\nabla\psi=0$. We shall discuss the case of a nonzero $\nabla\psi$ below and find that $\sigma$ is the chemical potential in the rest frame of the superfluid, 
while $\partial_0\psi$ is the chemical potential in the frame where the superfluid moves with a velocity determined by $\nabla\psi$. Here, we first discuss the 
simpler case $\nabla\psi=0$ where indeed $\mu=\sigma=\partial_0\psi$.  
\begin{figure} [t]
\begin{center}
\includegraphics[width=\textwidth]{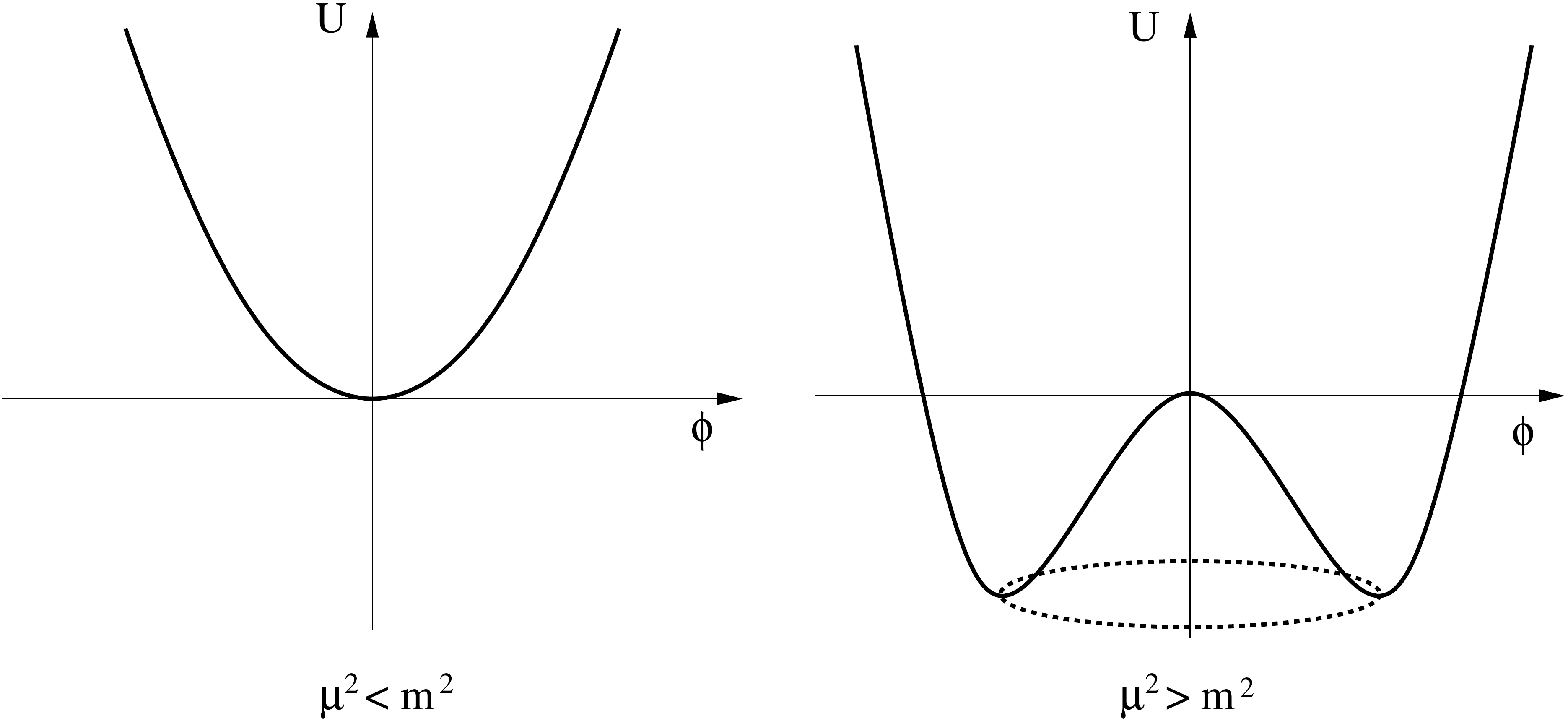}
\caption{
Illustration of the zero-temperature tree-level potential $U(\phi)$ for $\mu^2<m^2$ (left) and $\mu^2>m^2$ (right). In the latter case, 
the order parameter acquires a nonzero value at a fixed, but arbitrary value on the bottom circle of the potential, thus breaking the 
$U(1)$ symmetry ``spontaneously''. 
}
\label{figmexican}
\end{center}
\end{figure}

We illustrate the potential $U$ in Fig.\ \ref{figmexican}. 
For chemical potentials $|\mu|<m$, the minimum is at $\phi=0$, i.e., there is no condensation. For condensation, one needs a negative coefficient 
in front of the $\phi^2$ term, i.e., the modulus of the chemical potential must be larger than the 
mass, $|\mu|>m$. 
In this case, the potential has a ``Mexican hat'' or ``bottom of a wine bottle'' shape. (Since we consider a repulsive interaction for which 
$\lambda>0$ the potential is bounded from below, otherwise the system would be unstable.) 
The $U(1)$ symmetry of the Lagrangian is reflected in the rotationally symmetric wine bottle potential. The 
nontrivial minimum, at a given angle $\psi$ of the condensate, is not invariant under $U(1)$ because a $U(1)$ transformation rotates the condensate along 
the bottom of the wine bottle. This mechanism, where the Lagrangian has a symmetry which is not respected by the ground state, 
is called {\it spontaneous symmetry breaking}. The object that breaks the symmetry and which is zero in the symmetric phase, here the condensate $\phi$, 
is called the {\it order parameter}.  

Before we introduce a superflow into this picture, let us, as an aside, discuss another solution to the equations of motion. 
We consider a static situation with cylindrical symmetry, such that $\rho(X) = \rho({\bf x})$, $\psi(X)=\psi({\bf x})$, and 
${\bf x} = (r,\theta,z)$ in cylindrical coordinates. Then, with the above identification $\mu=\partial_0\psi$, the equations of motion (\ref{eoms}) read
\begin{subequations}
\bea
-\Delta\rho &=& \rho[\mu^2-(\nabla\psi)^2-m^2-\lambda\rho^2]\, ,  \label{vortex1}\\[2ex]
\nabla\cdot(\rho^2\nabla\psi) &=& 0 \, .\label{vortex2}
\eea
\end{subequations}
We now assume that the profile of the condensate does not depend on $z$, such that the problem becomes two-dimensional. If we move around the $r=0$ line
along a circle, the condensate $\phi$ must return to its original value since it must not be multi-valued. However, this does not require the 
phase to return to its original value, we are rather allowed to return to an integer multiple of $2\pi$ times the original value. This situation is borne out
in the ansatz $\psi = n\theta$ with $n\in {\mathbb N}$, and $\rho({\bf x}) = \rho(r)$. 
Here, $n$ is called the winding number since it indicates how many times the phase has wound around
the $U(1)$ circle in the internal space while going once around the $r=0$ line in position space. With this ansatz and the differential operators in cylindrical 
coordinates\footnote{We need
\bea
\nabla\psi &=& \frac{\partial\psi}{\partial r}\,{\bf e}_r + \frac{1}{r}\frac{\partial\psi}{\partial \theta}\,{\bf e}_\theta 
+\frac{\partial\psi}{\partial z}\,{\bf e}_z \, , \non[2ex]
\Delta\rho &=& \frac{1}{r}\frac{\partial}{\partial r}\left(r\frac{\partial\rho}{\partial r}\right) +\frac{1}{r^2}\frac{\partial^2\rho}{\partial\theta^2}
+\frac{\partial^2 \rho}{\partial z^2} \, . \nonumber
\eea},
Eq.\ (\ref{vortex2}) is automatically fulfilled. We can write the remaining equation (\ref{vortex1}) as
\be \label{vortexdiff}
\frac{1}{\eta}\frac{\partial}{\partial\eta}\left(\eta\frac{\partial R}{\partial \eta}\right)+\left(1-\frac{n^2}{\eta^2}\right)R-R^3 = 0 \, , 
\ee
where we have introduced the dimensionless function $R\equiv \rho/\rho_0$ and the dimensionless radial variable $\eta\equiv \sqrt{\lambda}\rho_0r$, where 
$\rho_0^2 = (\mu^2-m^2)/\lambda$ is the condensate squared in the homogeneous case, see Eq.\ (\ref{sols}). With the boundary conditions $R(\eta=0)=0$ and 
$R(\eta=\infty)=1$, Eq.\ (\ref{vortexdiff}) is an equation for the profile of a single straight-line vortex. In this form, it is identical to the equation obtained in the
non-relativistic context from the so-called Gross-Pitaevskii equation \cite{gross1961structure,pitaevskii1961vortex}. Vortices occur in rotating superfluids and 
are characterized 
by a vanishing condensate in their center. Much more could be said about them, but I refer the reader to the standard literature, see for instance
Refs.\ \cite{pines,annett,donnellybook}. 
As an exercise, you may solve the nonlinear differential equation (\ref{vortexdiff}) numerically and/or find analytical solutions close to the center of the 
vortex or far away from it.

\section{Superfluid velocity}

The solution of the equations of motion (\ref{sols}) does not fix the value of $\sigma$, i.e., we are free to choose the four-gradient
of the phase of the condensate. We have already argued that it is related to the chemical potential, at least in the case $\nabla\psi=0$. 
We shall now allow for a nonzero value of $\nabla\psi$ and discuss the meaning of the phase of the condensate with the help of the hydrodynamic
form of current and stress-energy tensor,
\begin{subequations}
\bea
j^\mu &=& n v^\mu \, , \label{jmu}\\[2ex]
T^{\mu\nu} &=& (\epsilon + P) v^\mu v^\nu -g^{\mu\nu} P \, . \label{Tmunu7}
\eea
\end{subequations}
We have already seen these expressions in Sec.\ \ref{sec:single}. To repeat, $n$, $\epsilon$, and $P$ are charge density, energy density, and pressure
in the rest frame of the superfluid. For the superfluid four-velocity $v^\mu$ we have by definition $v^\mu v_\mu=1$, and as for any four-velocity we can thus write
\be \label{vmuvs}
v^\mu = \gamma(1,{\bf v}_s) \, , \qquad \gamma = \frac{1}{\sqrt{1-v_s^2}} \, ,
\ee
where ${\bf v}_s$ is the superfluid three-velocity.
Contracting Eq.\ (\ref{jmu}) with $v_\mu$ and $j_\mu$, the two resulting equations yield
\be \label{nrhosigma}
n = \sqrt{j_\mu j^\mu} = \rho^2\sigma \, ,
\ee
where we have used $j^\mu = \rho^2\partial^\mu\psi$ from Eq.\ (\ref{noetherj}), and thus
\be \label{vmugradient}
v^\mu = \frac{j^\mu}{n} = \frac{\partial^\mu\psi}{\sigma} \, .
\ee
This is an important result that connects the macroscopic four-velocity to the microscopic phase of the condensate. 
With the help of Eq.\ (\ref{vmuvs}) we obtain the superfluid three-velocity\footnote{In the non-relativistic context, the superfluid velocity is related
to the phase of the condensate by the analogous relation
\be
{\bf v}_s = -\frac{\nabla\psi}{m} \, ,  \nonumber
\ee
with $m$ being the mass of the constituent particles of the condensate, for instance the mass of the helium atoms in superfluid helium. 
This relation shows that the superfluid flow is curl-free,
\be
\nabla\times {\bf v}_s = 0 \, .\nonumber
\ee
This property is responsible for the formation of vortices in a rotating superfluid since only the vortices, in whose center the condensate is zero, are
able to carry an externally imposed angular momentum.} 
\be \label{vs}
{\bf v}_s = -\frac{\nabla\psi}{\partial_0\psi} \, .
\ee
The minus sign appears because the 3-velocity ${\bf v}_s$ corresponds to the spatial components of the 
{\it contravariant} 4-vector $v^\mu$, while the operator $\nabla$ corresponds to the spatial components of the {\it covariant} 4-vector $\partial_\mu$, i.e., 
$\partial^\mu = (\partial_t,-\nabla)$. 

We now know how $n$ and $v^\mu$ can be expressed in terms of field-theoretical quantities. We may also determine $\epsilon$ and $P$. 
To this end, we first notice that they can be obtained by the following contractions from the stress-energy tensor,
\be
\epsilon = v_\mu v_\nu T^{\mu\nu} \, , \qquad P = -\frac{1}{3}(g_{\mu\nu}-v_\mu v_\nu)T^{\mu\nu} \, ,
\ee
where we have used $g_{\mu\nu}g^{\mu\nu} = g^{\mu}_{\;\;\;\mu} = 4$. 
Using the microscopic result for $T^{\mu\nu}$ from Eq.\ (\ref{Tmunumicro}) (with a constant $\rho$), this yields
\be \label{PL0}
P = {\cal L}^{(0)} \, , \qquad \epsilon = \rho^2\sigma^2 - P = \sigma n - P \, .
\ee
The latter relation, $\epsilon+P=\sigma n $, is a thermodynamic relation at $T=0$ from which we confirm that $\sigma$ plays the role of the chemical potential.
Since this relation holds in the rest frame of the superfluid ($\epsilon$, $n$, and $P$ are all measured in this frame), $\sigma$ is the chemical potential
measured in the superfluid rest frame. With the definition of the three-velocity (\ref{vs}) we can write $\sigma = \partial_0\psi\sqrt{1-{\bf v}_s^2}$, 
which shows that $\partial_0\psi$ is the chemical potential in the frame in which the superfluid has velocity ${\bf v}_s$. 
 
As a result of these ``translations'' of the microscopic, field-theoretical objects into macroscopic quantities we learn that the phase of the condensate plays a very 
interesting role; it determines the chemical potential as well as the superflow. These values are not fixed by the equations of motion, they rather are 
boundary conditions. Take for instance $\partial_0\psi$: we are free to choose how many times per unit time the condensate rotates around the $U(1)$ 
circle (= around the bottom of the Mexican hat). By this choice we fix the chemical potential. (Multiple) rotations around the $U(1)$ circle are called topological modes 
because closed paths 
on the circle can be topologically classified by their winding number; a closed path with winding number $n$ cannot be continuously deformed into 
a closed path with winding number $m\neq n$. These modes are different from small (harmonic) oscillations of the phase and the modulus of the condensate, 
which exist on top of the topological modes and  which determine the excitations of the system such as the Goldstone mode, see next subsection. 
Analogously to the rotations per unit time, we are also free to choose how many times per unit length
the phase winds around. This determines the superflow or, more precisely, the superfluid three-velocity ${\bf v}_s$. Consequently, one can picture a 
superflow microscopically
as a spiral whose axis points in the direction of the superflow and whose windings are in the internal $U(1)$ space. In this picture, denser windings correspond
to larger flow velocities. 

As we have discussed above, the time derivative of the 
phase corresponds to the temporal component of a gauge field, and we could have started with a Lagrangian containing this field with the same effect. The same holds
for the spatial components of course. It is left as an exercise to start from a Lagrangian with a covariant derivative $D^\mu = \partial^\mu - iA^\mu$ and show that 
this Lagrangian, together with a trivial phase of the condensate, leads to the same results,  with $\partial^\mu\psi$ replaced by $A^\mu$. 

Finally, we may compute charge density, energy density and pressure explicitly. With Eqs.\ (\ref{treelevelU}), (\ref{nrhosigma}), (\ref{PL0}), and the condensate 
$\rho$ from Eq.\ (\ref{sols}), we find
\begin{subequations}
\bea
n&=& \sigma\frac{\sigma^2-m^2}{\lambda} \, , \label{nPepsilon1}\\[2ex]
P &=& \frac{(\sigma^2-m^2)^2}{4\lambda} \, ,\\[2ex]
\epsilon &=& \frac{(3\sigma^2+m^2)(\sigma^2-m^2)}{4\lambda} \, .
\eea
\end{subequations}
With the help of these results we can write the energy density as a function of the pressure, 
\be \label{eps3P}
\epsilon = 3P+\frac{2m^2\sqrt{P}}{\sqrt{\lambda}} \, . 
\ee
This relation shows that the trace of the stress energy tensor, $T^\mu_{\;\;\;\mu}=\epsilon-3P$ is nonzero only in the presence of a mass parameter. 
If $m=0$, there is no energy scale in our Lagrangian, and the trace of the stress-energy tensor vanishes. We may use the relation (\ref{eps3P}) to compute
\be \label{epsP}
\frac{\partial\epsilon}{\partial P} = \frac{3\sigma^2-m^2}{\sigma^2-m^2} \, .
\ee
Remember that we have claimed in the previous chapter that $(\partial P/\partial \rho)^{1/2}$ is the slope of the Goldstone dispersion, see Eq.\ (\ref{Prhoc}).
As a consequence, this slope was identical to the speed of first sound at small temperatures. Here, the role of the mass density $\rho$ is played 
by the energy density $\epsilon$, and we will show in the next section that (\ref{epsP}) is indeed the inverse of the slope of the Goldstone dispersion (squared).

\section{Goldstone mode}

The excitation energies of the Bose-condensed system are computed by taking into account the fluctuations $\varphi(X)$ that we have introduced and then 
immediately dropped at the beginning of the chapter. You may think of these excitations in the following way. If the scalar field of our theory were non-interacting, 
the single-particle excitations would be given by 
$\epsilon_k^\pm = \sqrt{k^2+m^2}\mp\mu$, were the two different signs correspond to particles and anti-particles, which are distinguished according to their
$U(1)$ charge, say a particle carries charge +1, and an anti-particle carries charge $-1$. A nonzero chemical potential introduces an asymmetry
between particles and antiparticles. If for instance $\mu>0$ it takes more energy to excite anti-particles than particles. For any given momentum 
this energy difference is $2\mu$. Interactions usually change these excitation energies. If we switch on a small interaction, the dispersion relations usually do 
not change much, and often one can absorb the effect of the interaction into a new, effective mass of the excitations. These new single-particle excitations are then 
sometimes called quasiparticles, suggesting that the system with interactions again looks approximately like a system of non-interacting particles, but with 
modified dispersion relations. In the case of Bose-Einstein condensation, something more substantial happens to the system. Not only do the particles interact with 
each other, but there is a rearrangement of the ground state of the system. Nevertheless, we can still compute quasiparticle excitations.
The modification of the excitation energies goes beyond a small correction to the 
mass. There will rather be a qualitative difference compared to the uncondensed case. The most interesting effect will be the existence of a gapless mode, the Goldstone
mode. But also the nature of the excitations will change in the sense that the two excitation branches no longer carry charges +1 and $-1$. This is a consequence
of the spontaneous breaking of the symmetry. 

It is easy to see why in the Bose-condensed phase such a qualitative difference must occur. Let us assume that $\mu>0$ and start from the situation where 
$\mu<m$, i.e., there is no Bose condensate. Here, $m$ can be the mass of non-interacting bosons or an effective mass in which the effect of a weak interaction
is absorbed. We are in the vacuum with zero occupation of the spectrum, or at some given temperature with 
some thermal occupation of the spectrum. For $\mu>0$, the lowest excitation branch is $\epsilon_k^+ = \sqrt{k^2+m^2}-\mu$.  Now increase $\mu$. Obviously, when we 
approach $\mu=m$, the excitation energy is about to turn negative for small momenta. This indicates an instability of the system, which is resolved by the 
formation of a Bose-Einstein condensate. In this new ground state, instead of the excitation that is about to turn negative, there will be a new mode that remains 
gapless even if we keep increasing $\mu$. We now confirm this picture by doing the actual calculation.

A very direct way of computing the excitation energies is to start from the equations of motion (\ref{eoms}) and insert the following ansatz for modulus and phase of the 
condensate, $\rho\to \rho + \delta \rho\,e^{i(\omega t - {\bf k}\cdot{\bf x})}$, $\psi\to \psi + \delta \psi\,e^{i(\omega t - {\bf k}\cdot{\bf x})}$.
Linearizing the equations in the fluctuations $\delta \rho$ and $\delta \psi$ yields a condition for $\omega$ from which the excitation energies are computed.
I leave this calculation as an exercise and will present a different calculation that gives a more complete field-theoretical description of the system.

It is convenient to introduce the transformed fluctuation field $\varphi'(X)$ via
\be \label{vpprime}
\varphi'(X) = \varphi(X)e^{-i\psi(X)} \, ,
\ee
where $\varphi(X)$ is the fluctuation field from Eq.\ (\ref{shift}). The reason for this transformation is that in the new basis the tree-level propagator
will be diagonal in momentum space. We also introduce real and imaginary part of the transformed fluctuations,
\be \label{phi1phi2}
\varphi'(X) = \frac{1}{\sqrt{2}}[\varphi_1'(X) + i\varphi_2'(X)] \, . 
\ee
With these notations, the Lagrangian (\ref{lagr}) becomes 
\be
{\cal L} = {\cal L}^{(0)} + {\cal L}^{(1)}+ {\cal L}^{(2)}+ {\cal L}^{(3)}+ {\cal L}^{(4)} \, , 
\ee
with ${\cal L}^{(0)}$ from Eq.\ (\ref{lagr0}) and the fluctuation terms, listed by their order in the fluctuation from linear to quartic,
\begin{subequations} \label{L1L2L3L4}
\bea
{\cal L}^{(1)}&=& \varphi_1'\Big[\rho(\sigma^2 - m^2 -\lambda\rho^2)-\Box\rho\Big] \non[2ex]
&&-\frac{\varphi_2'}{\rho}\partial_\mu(\rho^2\partial^\mu
\psi) +\partial_\mu(\varphi_1'\partial^\mu\rho+\varphi_2'\rho\partial^\mu\psi) \, , \allowdisplaybreaks\\[2ex]
{\cal L}^{(2)}&=&\frac{1}{2}\Big[\partial_\mu\varphi_1'\partial^\mu\varphi_1'+\partial_\mu\varphi_2'\partial^\mu\varphi_2'
+(\varphi_1'^2+\varphi_2'^2)(\sigma^2-m^2) \non [2ex]
&& +\,2\partial_\mu\psi(\varphi_1'\partial^\mu\varphi_2'
-\varphi_2'\partial^\mu\varphi_1')-\lambda\rho^2(3\varphi_1'^2+\varphi_2'^2)\Big] \, , \label{L2}\allowdisplaybreaks\\[2ex]
{\cal L}^{(3)}&=& -\lambda\rho\varphi_1'(\varphi_1'^2+\varphi_2'^2) \, , \allowdisplaybreaks\\[2ex]
{\cal L}^{(4)}&=& -\frac{\lambda}{4}(\varphi_1'^2+\varphi_2'^2)^2 \, .
\eea
\end{subequations}
The terms linear in the fluctuations reduce to a total derivative term when we use the equations of motion (\ref{eoms}). Thus, they yield no contribution
to the action, assuming that the fields vanish at infinity. This is of course no surprise since computing the terms linear in the fluctuations is basically
a re-derivation of the Euler-Lagrange equations. Since we are only interested in the basic properties of the system, we shall only be interested 
in the terms quadratic in the fluctuations. We point out, however, that the condensate induces cubic interactions, even though we have started from 
only quartic interactions. 

From the quadratic terms ${\cal L}^{(2)}$ we determine the tree-level propagator whose poles give the excitation energies we are interested in.
To this end, we introduce the Fourier transformed field as 
\be \label{fourier}
\varphi_i'(X) = \frac{1}{\sqrt{TV}}\sum_{K} e^{-iK\cdot X} \varphi'_i(K) \, ,
\ee
where $T$ is the temperature, $V$ the volume, and $\varphi_i'(K)$ the (dimensionless) Fourier transform of $\varphi_i'(X)$. 
The scalar product in the exponential is formally taken with the Minkowski metric, $K\cdot X = k_0x_0-{\bf k}\cdot{\bf x}$. 
However, in the imaginary time formalism of thermal field theory, we have $x_0=-i\tau$ with  
$\tau\in[0,\beta]$, where $\beta=1/T$, and $k_0 = -i\omega_n$ with the bosonic Matsubara frequencies $\omega_n = 2\pi nT$, $n\in \mathbb{Z}$. 
Hence, \mbox{$K\cdot X =-(\omega_n\tau +{\bf k}\cdot {\bf x})$} is essentially a Euclidean scalar product. Imaginary time arises in thermal field theory due to the 
formal equivalence between the statistical partition function $Z=\Tr\, e^{-\beta\hat{H}}$ and a sum over transition amplitudes with identical initial and final states, 
$\langle\varphi|e^{-i\hat{H}t}|\varphi\rangle$. 
   
With the Fourier transform (\ref{fourier}) we compute the contribution to the action $S$ from the quadratic terms,
\bea \label{act2}
S^{(2)} &=&\int_0^{1/T}d\tau \int d^3{\bf x} \; {\cal L}^{(2)} \non[2ex]
&=& -\frac{1}{2}\sum_K[\varphi_1'(-K),\varphi_2'(-K)]
\frac{D^{-1}(K)}{T^2} \left(\begin{array}{c} \varphi_1'(K) \\ \varphi_2'(K) \end{array}\right)\, ,
\eea
where 
\be \label{nondiag}
D^{-1}(K) = \left(\begin{array}{cc}-K^2+m^2+3\lambda\rho^2-\sigma^2 & 
-2iK_\mu\partial^\mu\psi \\[1ex] 2iK_\mu\partial^\mu\psi & 
-K^2+m^2+\lambda\rho^2-\sigma^2 \end{array}\right) 
\ee
is the inverse tree-level propagator in momentum space. 

The thermodynamic potential density is defined as
\be \label{OmegaTZ}
\Omega = -\frac{T}{V}\ln Z \, , 
\ee
with the partition function
\be \label{Z}
Z = \int {\cal D}\varphi_1'{\cal D}\varphi_2' \, e^S \, . 
\ee
In our approximation, the action $S$ contains the contributions from ${\cal L}^{(0)}$ and ${\cal L}^{(2)}$. As in the previous section, we assume that
$\rho$ and $\partial_\mu\psi$ are constant. Therefore, the space-time integration over ${\cal L}^{(0)}$ is trivial and simply yields a factor $\frac{V}{T}$. Since
the resulting contribution does not depend on the dynamical fields $\varphi_1'$, $\varphi_2'$, also the functional integration is trivial. As a result, we
obtain a contribution $-\frac{V}{T} U$ to $\ln Z$, with the tree-level potential $U$ from Eq.\ (\ref{treelevelU}). The functional integration over 
$S^{(2)}$ can be done exactly because the fields $\varphi_1'$, $\varphi_2'$ only appear quadratically. 
As a result, we obtain the thermodynamic potential density at tree level, 
\bea \label{Omegatree}
\Omega &=& U + \frac{1}{2}\frac{T}{V}{\rm Tr}\ln \frac{D^{-1}(K)}{T^2} \, ,
\eea
where the trace is taken over the internal $2\times 2$ space and over momentum space. For the explicit evaluation of this expression we need  
to compute the zeros of the determinant of the inverse propagator. This determinant is a quartic polynomial in $k_0$
whose solutions in the presence of a superflow $\nabla\psi$ are very complicated. We shall come back to the general solutions in Sec.\ \ref{sec:connect}, 
where we discuss the connection of the field-theoretical results with the two-fluid formalism.
Here we proceed with the much simpler case $\nabla\psi=0$, where the dispersion relations are isotropic. 
In this case, the determinant has two zeros that we denote by $\epsilon_k^\pm$, plus their negatives. 
Consequently, we can write ${\rm det}\,D^{-1} = [k_0^2-(\epsilon_k^+)^2][k_0^2-(\epsilon_k^-)^2]$ and compute
\bea \label{lndet}
\frac{1}{2}\frac{T}{V}{\rm Tr}\ln \frac{D^{-1}(K)}{T^2} &=& \frac{1}{2}\frac{T}{V}\ln{\rm det} \frac{D^{-1}(K)}{T^2} \allowdisplaybreaks \non[2ex]
&=& 
\frac{1}{2}\frac{T}{V}\ln\prod_K\frac{[(\epsilon_k^+)^2-k_0^2][(\epsilon_k^-)^2-k_0^2]}{T^4} \allowdisplaybreaks\non[2ex]
&=&  \frac{1}{2}\frac{T}{V}\sum_K \left[\ln\frac{(\epsilon_k^+)^2-k_0^2}{T^2}+\ln\frac{(\epsilon_k^-)^2-k_0^2}{T^2}\right] \allowdisplaybreaks\non[2ex]
&=& \sum_{e=\pm}\int\frac{d^3{\bf k}}{(2\pi)^3}\left[\frac{\epsilon_k^e}{2}+T\ln\left(1-e^{-\epsilon_k^e/T}\right)\right]\, , 
\eea
where, in the last step, we have performed the summation over Matsubara frequencies (dropping a constant, infinite contribution)
and have taken the thermodynamic limit, $\frac{1}{V}\sum_{\bf k}
\to \int\frac{d^3{\bf k}}{(2\pi)^3}$. Comparing the second term of this expression with the familiar result from statistical mechanics shows that 
the poles of the propagator indeed are the excitation energies of the system. In chapter \ref{sec:he4}, we have worked with the pressure $P=-\Omega$ (\ref{PTint})
that has exactly the form of the second term in the last line of Eq.\ (\ref{lndet}).
In the present field-theoretical treatment there is an additional contribution which is divergent and has to be renormalized. 
It vanishes if we subtract the thermodynamic potential at $T=\mu=0$ to obtain the renormalized thermodynamic potential. 

First we are interested in the excitation energies themselves. Solving ${\rm det}\,D^{-1}=0$ with $\nabla\psi=0$ for $k_0$ yields 
\be \label{epspm}
\epsilon_k^\pm = \sqrt{k^2+m^2+2\lambda\rho^2+\mu^2\mp \sqrt{4\mu^2(k^2+m^2+2\lambda\rho^2)+\lambda^2\rho^4}} \, , 
\ee
where we have identified $\partial_0\psi = \mu$.
If we set the condensate to zero, $\rho=0$, we recover the dispersion relations of free bosons,
\be
\rho=0:\qquad \epsilon_k^\pm = \sqrt{k^2+m^2}\mp \mu \, , 
\ee
where the upper sign corresponds to particles, which carry $U(1)$ charge $+1$, and the lower sign to antiparticles, which carry $U(1)$ charge $-1$.
With the $T=0$ result for the condensate (\ref{sols}) the dispersions become
\be \label{GM1}
\epsilon_k^\pm=\sqrt{k^2+(3\mu^2-m^2)\mp\sqrt{4\mu^2k^2+(3\mu^2-m^2)^2}} \, .
\ee
We see that $\epsilon_k^+$ becomes gapless, i.e., $\epsilon_{k=0}^+=0$. This mode is called Goldstone mode and behaves linearly for small momenta, 
\be \label{GM2}
\epsilon_k^+ =  \sqrt{\frac{\mu^2-m^2}{3\mu^2-m^2}}\, k  + {\cal O}(k^3) \, .
\ee
Since this result holds only in the condensed phase, where $\mu^2>m^2$, the argument of the square root is positive. Comparing this result 
with Eq.\ (\ref{epsP}) shows that the slope of the low-momentum, zero-temperature dispersion of the Goldstone mode is identical to $(\partial P/\partial\epsilon)^{1/2}$
(remember that for $\nabla\psi=0$ we have $\sigma=\mu$). This, in turn, is the speed of first sound, as we know from Sec.\ \ref{sec:sound}. Consequently, 
in the low-temperature limit, the speed of first sound is identical to the slope of the Goldstone dispersion. 
 
\begin{figure} [t]
\begin{center}
\includegraphics[width=0.85\textwidth]{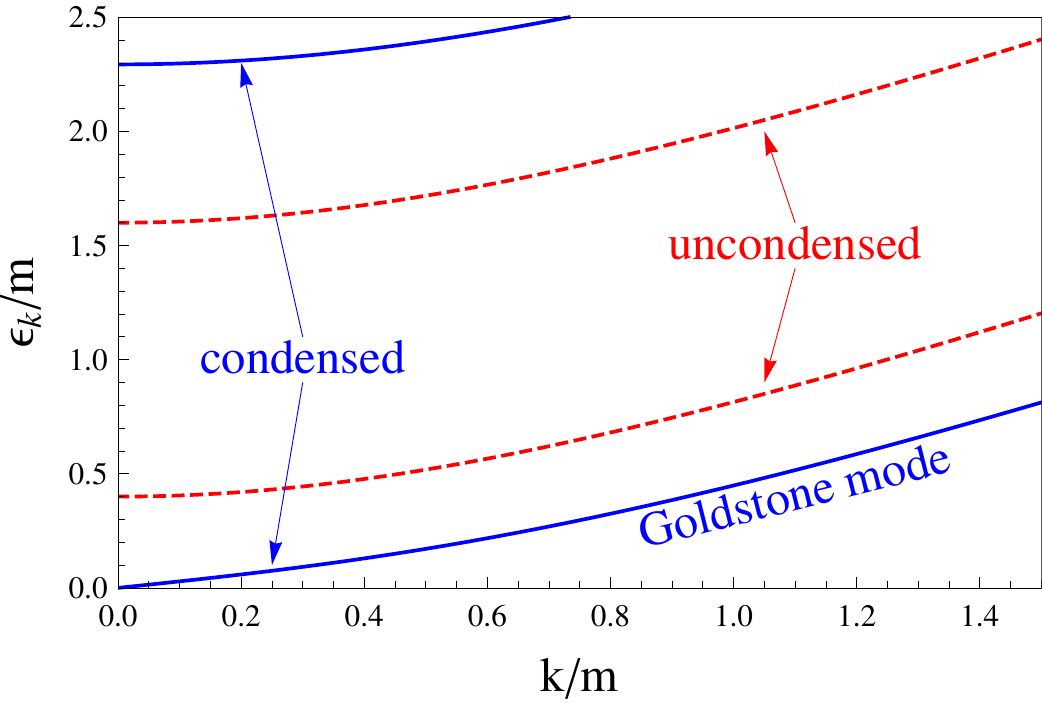}
\caption{(Color online) Single-particle excitation energies $\epsilon_k^\pm$ for the case without condensation (dashed curves, $|\mu|<m$, here $\mu=0.6 m$), 
and with condensation (solid curves, $|\mu|>m$, here $\mu=1.1 m$. In the condensed case, there is a gapless mode with linear behavior for small momenta,
the Goldstone mode.}
\label{figGoldstone}
\end{center}
\end{figure}

The second mode does have an energy gap and behaves quadratically for small momenta,  
\be
\epsilon_k^- = \sqrt{2}\sqrt{3\mu^2-m^2}+\frac{1}{2\sqrt{2}}\frac{5\mu^2-m^2}{(3\mu^2-m^2)^{3/2}}\,k^2 + {\cal O}(k^4)  \, .
\ee
This massive mode is sometimes called Higgs mode: 
in the Higgs mechanism of the standard model of particle physics, there is a condensate of the Higgs field, giving rise to a Goldstone mode (which, since the
spontaneously broken symmetry is a gauge symmetry, is not a physical degree of freedom) and a massive mode. The latter is nothing but the Higgs boson which has
recently been found at the Large Hadron Collider (LHC) \cite{2012PhLB..716....1A,2012PhLB..716...30C}. 
We plot the dispersions of both modes and compare them to the uncondensed case in Fig.\ \ref{figGoldstone}.

The gaplessness of one of the modes is a very general phenomenon for spontaneously broken global symmetries. 
Its existence is predicted by the 

\bigskip
{\bf Goldstone Theorem:} If a continuous global symmetry of the Lagrangian is spontaneously broken there exists 
a gapless mode. This mode is called Goldstone mode.

\bigskip
The proof of this theorem can be found in most textbooks about quantum field theory, see for instance
Ref.\ \cite{pokorsky}. It should be emphasized that for the existence of the gapless mode it is crucial that the broken symmetry is global. 
The case of a spontaneously broken local symmetry will be discussed in chapter \ref{sec:meissner1}.

In Lorentz invariant systems, one can also make a precise statement about the number of Goldstone modes: if the global symmetry group $G$ of the 
Lagrangian is spontaneously broken to a subgroup $H\subset G$, there exist ${\rm dim}\, G/H$ many gapless modes, i.e., there are as many Goldstone modes as broken 
generators. In our case, $G=U(1)$, $H={\bf 1}$, such that there is ${\rm dim}\, G/H=1$ Goldstone mode. Under some circumstances, this counting rule may be violated and 
the number of broken generators merely presents an upper limit for the number of Goldstone modes. The violation of the counting rule is closely related to the 
appearance of Goldstone modes with a quadratic, instead of a linear, low-energy dispersion relation, see 
Refs.\ \cite{1976NuPhB.105..445N,Miransky:2001tw,Brauner:2010wm,Watanabe:2011ec,2012PhRvL.108y1602W,2014arXiv1402.7066W} for more details.

There are many examples for systems with Goldstone modes, for instance

\begin{itemize}

\item {\it superfluid $^4$He:} in terms of symmetries, this case is identical to the present field-theoretical $\varphi^4$ model. The Goldstone mode can 
be modelled in terms of phonon and roton excitations, as discussed in detail in chapter \ref{sec:he4}. This example shows that symmetries and 
their spontaneous breaking predict the masslessness and the long-wavelength behavior (the phonon), but the complete dispersion of the Goldstone mode depends 
strongly on the details of the system (there is no roton-like behavior in our $\varphi^4$ model).

\item {\it (anti-)ferromagnetism:} in this case, rotational symmetry is broken, $SO(3) \to U(1)$. In a ferromagnet, there is one 
Goldstone mode with quadratic dispersion, called {\it magnon} (spin wave). This is an example where the number of Goldstone modes
is less than the number of broken generators. In an antiferromagnet, in contrast, there are two magnon degrees of freedom, both with linear 
dispersion. 

\item {\it chiral symmetry breaking in QCD:} for three quark flavors, the symmetry breaking pattern is $SU(3)_L\times SU(3)_R \to SU(3)_{R+L}$, where 
$SU(3)_L$ and $SU(3)_R$ are the groups of transformations in flavor space of fermions with left-handed and right-handed chirality, and $SU(3)_{R+L}$ transformations are 
joint rotations of left-handed and right-handed fermions. There are eight Goldstone modes, the meson octet $\pi^0$, $\pi^\pm$, $K^0$, $\bar{K}^0$, $K^\pm$, $\eta$; 
since the chiral symmetry is not an exact symmetry to begin with, these Goldstone modes have small masses and are therefore called pseudo-Goldstone 
modes.
 
\end{itemize}

It is instructive to write down the expression for the charge density of the condensed system. With $P=-\Omega$, Eq.\ (\ref{lndet}), and the usual 
thermodynamic definition for the charge density we compute
\bea
n &=& \frac{\partial P}{\partial \mu} =  -\sum_{e=\pm}\int\frac{d^3{\bf k}}{(2\pi)^3}\left[\frac{1}{2}+f(\epsilon_k^e)\right]
\frac{\partial \epsilon_k^e}{\partial \mu}
\, .
\eea
Let us ignore the temperature-independent (divergent) first term. For the uncondensed case, we have $\partial \epsilon_k^e/\partial \mu=-e$, and the charge density has 
the expected form: one particle and one antiparticle contribute one unit of charge each, with opposite signs. This is different in the condensed case, where particles
and antiparticles are replaced by the Goldstone and the massive mode. These quasiparticles do not have a well-defined charge (they are no eigenstates of the 
charge operator). This is reflected in the momentum-dependent factor $\partial \epsilon_k^e/\partial \mu$ which becomes complicated in the presence of a condensate. 
One can say that the quasiparticles are momentum-dependent combinations of the original modes. We will encounter an analogous situation in a fermionic system 
with Cooper pairing, see chapter \ref{sec:cooper}. In that case, the corresponding coefficients are called Bogoliubov coefficients and determine the 
mixing between fermions and fermion holes, see Eq.\ (\ref{nbog}).

\section{Symmetry restoration at the critical temperature}

Eqs.\ (\ref{Omegatree}) and (\ref{lndet}) give the thermodynamic potential for nonzero temperatures,
\be \label{omegavut}
\Omega = U + T\sum_{e=\pm}\int\frac{d^3{\bf k}}{(2\pi)^3} \ln\left(1-e^{-\epsilon_k^e/T}\right) \, , 
\ee
where we have subtracted the divergent vacuum  part. For a simple estimate of the temperature-dependent term we make use of the 
high-temperature expansion of the corresponding term of a non-interacting, uncondensed Bose gas \cite{kapusta,Haber:1981fg},
\bea \label{expand}
&&T\sum_{e=\pm}\int\frac{d^3{\bf k}}{(2\pi)^3} \ln\left[1-e^{-(\sqrt{k^2+M^2}-e\mu)/T}\right] \non[2ex]
&&= -\frac{\pi^2 T^4}{45} 
+\frac{(M^2-2\mu^2)T^2}{12} - \frac{(M^2-\mu^2)^{3/2}T}{6\pi} + \ldots 
\eea
Now, unfortunately, the dispersion relations of our condensed phase are more complicated. However, we may bring them into an appropriate form by 
applying the approximation
\be
\epsilon_k^\pm \simeq \sqrt{k^2 + m^2 + 2\lambda \rho^2} \mp\mu  \, , 
\ee 
which is obtained from neglecting the term $\lambda^2\rho^4$ in Eq.\ (\ref{epspm}), i.e., we assume the condensate to be small. 
In this approximation we lose the gaplessness of our Goldstone mode. But, since we are interested in large temperatures, where a small mass of the Goldstone mode
is negligible compared to the temperature, this is unproblematic. 

With this approximation and the help of Eq.\ (\ref{expand}) we can compute the potential,
\be \label{OmT}
\Omega\simeq \left(-\frac{\mu^2-m^2}{2}+\frac{\lambda T^2}{6}\right)\rho^2 +\frac{\lambda}{4}\rho^4 + {\rm const} \, , 
\ee
where ``const'' denotes terms that do not depend on $\rho$. We see that the effect of the temperature is
to make the quadratic term in $\rho$ less negative and eventually positive. Then, there is no nontrivial minimum for $\rho$ and the 
$U(1)$ symmetry of the ground state is restored. This is the phase transition to the non-superfluid phase were there is no condensate. The critical temperature
is the temperature where the prefactor of the quadratic term is zero,
\be
T_c^2 \simeq \frac{3(\mu^2-m^2)}{\lambda} \, .
\ee
We may use Eq.\ (\ref{OmT}) to compute the temperature-dependent condensate, 
\be \label{phiT}
\rho^2(T) \simeq \frac{\mu^2-m^2}{\lambda}-\frac{T^2}{3} = \rho^2_0\left(1-\frac{T^2}{T_c^2}\right) \, ,
\ee
with $\rho_0=\rho(T=0)$. We see that the condensate melts when $T$ approaches $T_c$. At $T=T_c$, the condensate vanishes and remains zero for 
$T>T_c$ (Eq.\ (\ref{phiT}) is only valid for $T<T_c$). Although there is no obvious problem with the temperature dependence of the condensate,
we need to keep in mind that we have employed a very crude high-temperature approximation. The crudeness becomes obvious 
when we consider the dispersion relation of the Goldstone mode. Inserting Eq.\ (\ref{phiT}) into Eq.\ (\ref{epspm}) and expanding for small $T$
shows that $\epsilon_{k=0}^+$ becomes imaginary,
\be
(\epsilon_{k=0}^+)^2 = -\frac{\lambda T^2}{3}\frac{\mu^2-m^2}{3\mu^2-m^2} \, .
\ee
This is an unphysical result because the Goldstone theorem holds for all temperatures and tells us that $\epsilon_{k=0}^+=0$. 
The reason for this problem is that we have ignored loop corrections which yield further $\lambda T^2$ terms. Here we were only interested in showing the 
melting of the condensate and restoration of the symmetry at $T_c$. We shall therefore not go into further details; see for instance Ref.\ \cite{thermal}
for an explanation how the one-loop self-energy corrects the unphysical excitation energies.

\chapter{Relativistic two-fluid formalism}
\label{sec:reltwo}

What is the relation between the relativistic field-theoretical approach of the previous chapter and the two-fluid formalism explained in 
chapter \ref{sec:he4}? The answer to this question is not obvious because the two-fluid formalism developed for superfluid helium is manifestly non-relativistic, as one
can see for example from the use of mass densities $\rho_s$, $\rho_n$. Since mass is not a conserved quantity, these densities have to be generalized in a 
relativistic
framework. Another way of saying this is that we would like to have a two-fluid formalism where we work with two four-currents (instead of the two
three-currents ${\bf j}_s$ and ${\bf j}_n$), which allow for a covariant formulation. This relativistic generalization of the two-fluid formalism 
was developed by I.M.\ Khalatnikov and V.V.\ Lebedev in 1982 \cite{1982PhLA...91...70K,1982ZhETF..83.1601L} and later, in a different formulation, 
by B.\ Carter \cite{carter89}. Both formulations are equivalent \cite{1992PhRvD..45.4536C,Andersson:2006nr}.

\section{Covariant formulation}

As a starting point, we recall the field-theoretical 
expression of the Noether current,
\be \label{noether2}
j^\mu = \frac{\partial {\cal L}}{\partial (\partial_\mu\psi)} \, ,
\ee
where $\psi$ is the phase of the condensate.
This definition relates the four-current $j^\mu$ with its conjugate momentum $\partial^\mu\psi$. Notice that $j^\mu$ has mass dimensions 3 while 
$\partial^\mu\psi$ has mass dimensions 1. We now generalize this concept to two currents.  
In the absence of dissipation there are two conserved currents, the charge current $j^\mu$ and the entropy current $s^\mu$. 
Therefore, the second current we use for our two-fluid formalism
is the entropy current. This decomposition into two currents is 
different from introducing superfluid and normal-fluid components of the charge current. We shall discuss later how this picture can be recovered from the 
covariant formalism. To construct the analogue of Eq.\ (\ref{noether2}) for the entropy current, we need to introduce another conjugate momentum which we call
$\Theta^\mu$. We shall see that this momentum is related to temperature: while, as we have seen, the temporal component of $\partial^\mu\psi$ is the chemical potential, 
the temporal component of $\Theta^\mu$ corresponds to temperature. A straightforward analogue of Eq.\ (\ref{noether2}) does not exist because 
a Lagrangian of a quantum field theory is not a priori equipped with such a four-vector $\Theta^\mu$. Let us therefore introduce a function $\Psi$,
which in some sense must correspond to the Lagrangian; its exact relation to field theory will be discussed later. The function $\Psi$ is called 
{\it generalized pressure} for reasons that will become clear below. We thus postulate
\be \label{dPsid}
j^\mu = \frac{\partial \Psi}{\partial (\partial_\mu\psi)} \, , \qquad s^\mu = \frac{\partial \Psi}{\partial \Theta_\mu}\, .
\ee
Since $\Psi$ is a Lorentz scalar, and the only four-vectors it depends on are $\partial^\mu\psi$ and $\Theta^\mu$, it must be a function of the Lorentz scalars
that can be built out of these four-vectors, namely $\sigma^2$, $\Theta\cdot\partial\psi$, $\Theta^2$ 
(here and in the following we use the notation \mbox{$\Theta\cdot\partial\psi = \Theta_\mu\partial^\mu\psi$}, $\Theta^2=\Theta_\mu\Theta^\mu$ 
for Minkowski scalar products, and 
$\sigma^2\equiv \partial_\mu\psi\partial^\mu\psi$, as defined in the previous chapter). Consequently, we can write 
$\Psi = \Psi(\sigma^2, \Theta\cdot\partial\psi, \Theta^2)$, and 
\be  \label{dPsi}
d\Psi = j_\mu d(\partial^\mu\psi) + s_\mu d\Theta^\mu \, .
\ee
We can now relate the two currents to the two conjugate momenta,
\begin{subequations}\label{js}
\bea
j^\mu &=&  \overline{B}\, \partial^\mu\psi + \overline{A}\, \Theta^\mu  \, ,
\label{js1} \\[2ex]
s^\mu &=&  \overline{A}\, \partial^\mu\psi + \overline{C}\, \Theta^\mu  \, ,
\label{js2}
\eea
\end{subequations}
where, using Eq.\ (\ref{dPsid}) and the chain rule, we have 
\be \label{entrainbar}
\overline{A}\equiv \frac{\partial \Psi}{\partial(\Theta\cdot\partial\psi)} \, , \qquad 
\overline{B}\equiv 2\frac{\partial \Psi}{\partial \sigma^2}
\,, \qquad \overline{C}\equiv 2\frac{\partial \Psi}{\partial\Theta^2} \, .
\ee
We see from Eqs.\ (\ref{js}) that the two currents are in general not four-parallel to their conjugate momenta, they rather may receive an admixture from the 
momenta that are associated with the other current. This effect of the interaction of the two currents is called entrainment, and the 
corresponding coefficient $\overline{A}$ is called {\it entrainment coefficient}.\footnote{The notation $\overline{A}$, $\overline{B}$, 
$\overline{C}$ is chosen because we write the inverse transformation, see Eqs.\ (\ref{psiTheta}), in terms of $A$, $B$, 
$C$, which is in accordance with Ref.~\cite{Carter:1995if} where $A$, $B$, $C$ were termed anomaly, bulk, and caloric
coefficients.}

With the help of the generalized pressure $\Psi$ we can formulate a generalized thermodynamic relation. Usually, energy density $\epsilon$ and pressure $P$ are related
via $\epsilon = - P + \mu n + T s$. We now introduce the {\it generalized energy density} $\Lambda$ via the generalized relation
\be \label{Lamgen}
\Lambda  =-\Psi + j\cdot\partial\psi + s\cdot\Theta \, . 
\ee
Analogously to usual thermodynamics, $\Psi$ and $\Lambda$ are related by a double Legendre
transform, i.e., while the pressure depends on the currents, the energy density depends on the conjugate momenta. Therefore, we have 
$\Lambda = \Lambda(s^2,j^2,s\cdot j)$ and 
\be \label{dLambda}
d\Lambda = \partial_\mu\psi\,dj^\mu + \Theta_\mu ds^\mu \, . 
\ee
The momenta can thus be written as a combination of the currents as follows,  
\begin{subequations}\label{psiTheta}
\bea
\partial^\mu\psi &=& \frac{\partial\Lambda}{\partial j_\mu} = B j^\mu + A s^\mu  \, ,\label{psiTheta1}\\[2ex]
\Theta^\mu &=& \frac{\partial \Lambda}{\partial s_\mu} = A j^\mu + C s^\mu  \, ,\label{Theta}
\eea
\end{subequations}
where
\be
A\equiv \frac{\partial \Lambda}{\partial (j\cdot s)} \, , \qquad B\equiv 2\frac{\partial \Lambda}{\partial j^2}
\,, \qquad C\equiv 2\frac{\partial \Lambda}{\partial s^2} \, .
\ee
The coefficients of the two transformations (\ref{js}) and (\ref{psiTheta}) are obviously related by a matrix inversion,
\be \label{AAbar}
\overline{C} = \frac{B}{BC-A^2} \, , \qquad \overline{B} = \frac{C}{BC-A^2} \, , 
\qquad \overline{A} = -\frac{A}{BC-A^2} \, .
\ee
The microscopic information of the system is contained in $\Psi$ and $\Lambda$, i.e., when we know one of these functions we can compute for instance the 
coefficients $A$, $B$, $C$, which in turn are needed to set up the (dissipationless) hydrodynamics of the system. In the literature,
$\Lambda$ is sometimes called master function, and the two approaches based either on the generalized energy density or the generalized pressure
are termed ``convective variational approach'' and "potential variational approach'', respectively \cite{1992PhRvD..45.4536C}.

For the hydrodynamic equations we need to know the stress-energy tensor. It can be written as
\be \label{tmunucomer}
T^{\mu\nu} = -g^{\mu\nu}\Psi +j^\mu\partial^\nu\psi + s^\mu\Theta^\nu \, . 
\ee
With this form and the relation (\ref{Lamgen}) we have $T^\mu_{\;\;\;\mu} = \Lambda -3\Psi$, which is a generalized way of saying that the trace of the 
stress-energy tensor
is given by $\epsilon-3P$. We can ``solve'' Eq.\ (\ref{tmunucomer}) for $\Psi$, i.e., write the generalized pressure in terms of various contractions of the 
stress-energy tensor. For instance, using contractions with $s^\mu$ and $\partial^\mu\psi$, we can write 
\be \label{Psicontract}
\Psi = \frac{1}{2}\left[\frac{s\cdot \partial\psi(s_\mu\partial_\nu\psi + s_\nu\partial_\mu\psi)-s^2\partial_\mu\psi\partial_\nu\psi
-\sigma^2s_\mu s_\nu}{(s\cdot \partial\psi)^2-s^2\sigma^2}-g_{\mu\nu}\right]T^{\mu\nu}   \, .
\ee
We shall use this form below to express $\Psi$ in terms of the pressures of the superfluid and normal-fluid components.

It is not obvious from the form (\ref{tmunucomer}) that the stress-energy tensor is symmetric. With the relations derived above, however, 
we can bring it into a manifestly symmetric form. We can for instance eliminate the two conjugate momenta, 
\be \label{Tsj}
T^{\mu\nu} = -g^{\mu\nu}\Psi +Bj^\mu j^\nu + Cs^\mu s^\nu +A(j^\mu s^\nu + s^\mu j^\nu) \, . 
\ee
Besides being manifestly symmetric, this form also confirms the interpretation of the entrainment coefficient: the term that couples the two currents is 
proportional to $A$. As an exercise, you can derive the following expressions for $A$, $B$, $C$ in terms of 
the stress-energy tensor, the generalized energy density, and the currents,
\begin{subequations}
\bea 
A &=&-\frac{j^\mu s^\nu T_{\mu\nu}-(j\cdot s)\Lambda}{(s\cdot j)^2-s^2j^2}  \, , \\[2ex] 
B &=&\frac{s^\mu s^\nu T_{\mu\nu}-s^2\Lambda}{(s\cdot j)^2-s^2j^2}  \, , \\[2ex] 
C &=&\frac{j^\mu j^\nu T_{\mu\nu}-j^2\Lambda}{(s\cdot j)^2-s^2j^2} \, .
\eea
\end{subequations}
Using the stress-energy tensor (\ref{tmunucomer}), we can also reformulate the hydrodynamic equations (\ref{dmuTmunu}). 
We compute 
\bea
0&=&\partial_\mu T^{\mu\nu} \non[2ex]
&=& \partial^\nu\psi\partial_\mu j^\mu+  j_\mu(\partial^\mu\partial^\nu\psi - \partial^\nu\partial^\mu\psi) \non[2ex]
&&+ \Theta^\nu \partial_\mu s^\mu 
+ s_\mu (\partial^\mu\Theta^\nu-\partial^\nu\Theta^\mu) \, ,
\label{dThydro}
\eea
where we have used $\partial^\nu \Psi = j_\mu\partial^\nu\partial^\mu\psi+s_\mu\partial^\nu\Theta^\mu$, which follows from Eq.~(\ref{dPsi}). 
Because of current conservation $\partial_\mu j^\mu=0$, the first term on the right-hand side vanishes. The second term vanishes 
obviously. Note that it vanishes due to the specific form of the momentum associated to the charge current. Had we started with a general 
two-fluid system (not necessarily a superfluid), this term would not be zero. As a consequence, we see that at zero temperature, where there is no entropy current, 
the energy-momentum conservation is automatically fulfilled. To derive the additional equations for nonzero temperature, we 
contract the two remaining terms with $s_\nu$. For $s\cdot \Theta\neq 0$ this yields the entropy conservation $\partial_\mu s^\mu = 0$. Consequently, 
we can write the conservation equations as 
\be
\partial_\mu j^\mu = 0 \, , \qquad \partial_\mu s^\mu = 0 \,, \qquad  s_\mu \omega^{\mu\nu}  = 0 \, ,
\ee
where we have introduced the {\it vorticity} 
\be
\omega^{\mu\nu} \equiv  \partial^\mu\Theta^\nu-\partial^\nu\Theta^\mu \, .
\ee

\section{Relation to the original two-fluid formalism}

The formalism that we have just explained looks different from the two-fluid formalism introduced in Sec.\ \ref{sec:2fluid}. In particular, it makes no reference
to superfluid and normal-fluid components. We now show that there is a translation from 
one formulation into the other, meaning that both descriptions are equivalent, see for instance Ref.\ \cite{Alford:2012vn} and appendix A of Ref.\ \cite{Herzog:2008he}. 
A decomposition of the conserved current into superfluid and normal-fluid 
contributions reads
\bea
j^\mu &=& n_nu^\mu + n_s v^\mu \non[2ex]
&=& \frac{n_n}{s} \, s^\mu + \frac{n_s}{\sigma}\, \partial^\mu\psi \, . \label{jmacroT} 
\eea
In the first step, we have written the current in terms of two velocities $u^\mu$ and $v^\mu$ and two charge densities $n_n$ and $n_s$, which correspond 
to superfluid and normal-fluid contributions. We have thus simply added a term of the same form to the current in a single-fluid system, see Eq.\ (\ref{jmu}). 
In the second step, we have written the superfluid four-velocity in terms of the gradient of the phase of the condensate, 
\be
v^\mu = \frac{\partial^\mu\psi}{\sigma} \, , 
\ee
as we know from our field-theoretical discussion, see Eq.\ (\ref{vmugradient}). Furthermore, we have defined the normal-fluid
velocity through the entropy current, 
\be
u^\mu = \frac{s^\mu}{s} \, , 
\ee
with $s\equiv (s^\mu s_\mu)^{1/2}$. We see that the decomposition (\ref{jmacroT}) is a ``mixed'' form
compared to the two descriptions discussed above: it neither uses the two currents nor the two momenta as its variables, but rather one current, namely $s^\mu$, 
and one momentum, namely $\partial^\mu\psi$. The somewhat more physical decomposition into superfluid and normal fluid goes along with a less natural 
decomposition in terms of currents and conjugate momenta. Note also that neither of the two currents $n_s v^\mu$ and $n_n u^\mu$ obeys a continuity equation.

We may decompose the stress-energy tensor in an analogous way,
\bea
T^{\mu\nu} &=& (\epsilon_n + P_n)u^\mu u^\nu -g^{\mu\nu} P_n  + (\epsilon_s+P_s) v^\mu v^\nu -g^{\mu\nu}P_s
 \, ,\label{tmunuson}
\eea
where, as in the non-relativistic approach in Sec.\ \ref{sec:subsub2fluids}, $\epsilon_s$ and $P_s$ are energy density and pressure of the superfluid,
measured in the superfluid rest frame, while $\epsilon_n$ and $P_n$ are energy density and pressure of the normal fluid, measured in the normal-fluid rest 
frame. First of all, it is instructive to insert this form of the stress-energy tensor into Eq.\ (\ref{Psicontract}), which yields
\bea 
\Psi &=& \frac{1}{2}\left[\frac{u\cdot v(u_\mu v_\nu + u_\nu v_\mu)-v_\mu v_\nu-u_\mu u_\nu}{(u\cdot v)^2-1}-g_{\mu\nu}\right]T^{\mu\nu} \non[2ex]   
&=& P_n+P_s \, ,
\eea
where $u_\mu u^\mu = v_\mu v^\mu =1$ has been used.
This result gives us an idea of the physical meaning of the generalized pressure: it is the sum of the two pressures of the two fluids, each measured in 
the rest frame of the respective fluid. Now that we have written the stress-energy tensor in terms of superfluid and normal-fluid velocities (\ref{tmunuson}),
we should ask how it is related to the stress-energy tensor (\ref{tmunucomer}) that is expressed in terms of conserved currents and their conjugate momenta. 
We had already derived a version that only depends on the two currents (\ref{Tsj}), now we need a form that depends on the current $s^\mu$ and the momentum 
$\partial^\mu\psi$. With the help of Eqs.\ (\ref{js}) and (\ref{psiTheta}) we find
\be \label{Tdpsis}
T^{\mu\nu} = -g^{\mu\nu}\Psi + \frac{1}B\partial^\mu\psi\partial^\nu\psi +\frac{1}{\overline{C}}s^\mu s^\nu \, .
\ee
We see that no mixed terms of the form  $s^\mu \partial^\nu\psi$ appear. This justifies in hindsight that we have not included terms of the form
$u^\mu v^\nu$ in Eq.\ (\ref{tmunuson}), i.e., we can indeed simply add the contributions of superfluid and normal-fluid in the stress-energy tensor.

Now we can translate the coefficients $A$, $B$, $C$ into more conventional thermodynamic quantities, including the superfluid and normal-fluid charge densities. 
First, with the help of
Eqs.\ (\ref{psiTheta}), we write the current $j^\mu$ in terms of $s^\mu$ and $\partial^\mu\psi$. Comparing the result with the current in the form 
(\ref{jmacroT}) yields
\be \label{AB1}
A = -\frac{\sigma n_n}{s n_s} \, , \qquad B = \frac{\sigma}{n_s} \, .
\ee
Next, we compare the two forms of the stress-energy tensor (\ref{tmunuson}) and (\ref{Tdpsis}). We have already identified $\Psi=P_n+P_s$. Then, the remaining 
superfluid contribution yields $1/B = (\epsilon_s+P_s)/\sigma^2$. With the thermodynamic relation in the superfluid rest frame $\epsilon_s+P_s = \sigma n_s$ 
(remember that $\sigma$ is the chemical potential measured in the superfluid rest frame and that there is no $Ts$ term in this relation because 
the superfluid does not carry entropy), one confirms the 
result for $B$ in Eq.\ (\ref{AB1}). The remaining normal-fluid contribution yields $1/\overline{C} = (\epsilon_n+P_n)/s^2$. This time, we employ
the thermodynamic relation in the normal-fluid rest frame $\epsilon_n+P_n = \mu n_n + T s$, use the expression for $\overline{C}$ from Eq.\ (\ref{AAbar}) and for 
$A$ and $B$ from Eq.\ (\ref{AB1}) to find
\be \label{C1}
C = \frac{\mu n_n + Ts}{s^2} + \frac{\sigma n_n^2}{s^2 n_s} \, .
\ee

\section{Connecting field theory with the two-fluid formalism}
\label{sec:connect}

The results of the previous two subsections provide the setup for the hydrodynamic description of a relativistic superfluid. However, we have not
yet seen how this setup can be derived from a microscopic theory or model. We shall make this connection in this section, using the low-temperature approximation
of the field theory discussed in chapter \ref{sec:phi4}. As above, we shall work in the dissipationless limit and only consider a uniform, i.e., stationary and 
spatially homogeneous, relative velocity between superfluid and normal fluid. For further literature connecting field theory with the hydrodynamics of a superfluid see
for instance Refs.\ \cite{Alford:2012vn,Carter:1995if,Son:2002zn,Comer:2002dm,Nicolis:2011cs}.

One of the fundamental quantities in the two-fluid formalism is the generalized pressure $\Psi$,
which usually is not part of a field-theoretical setup.  
In general, $\Psi$ should be identified with the effective action density of the microscopic theory \cite{Alford:2012vn,Alford:2013koa}, which, if evaluated
at the minimum, gives the thermodynamic pressure. More precisely, in single-fluid systems we can go to 
the rest frame of the fluid where $T_{ij} = {\rm diag}(P,P,P)$ and then have $\Psi=P$. In a two-fluid system, we cannot avoid an anisotropic stress-energy tensor.
When we go to the rest frame of one of the fluids we can write $T_{ij} = {\rm diag}(P_\perp,P_\perp,P_\parallel)$, and $\Psi$ is identical to the 
pressure perpendicular to the flow of the other fluid $\Psi=P_\perp$.
In addition to this identification, we also have to remember that $\Psi$ 
depends on Lorentz scalars, while the pressure in the field theory usually is written in terms of quantities that do change under Lorentz transformations, 
in our case the chemical potential $\mu$, the temperature $T$, and the superflow $\nabla\psi$. The thermal field theory calculation is usually performed
in the rest frame of the heat bath, which in our context is the rest frame of the normal fluid. Therefore, $\mu$, $T$ and $\nabla\psi$ are all measured in the 
normal-fluid rest frame. Note that $\nabla\psi$ will be treated as a thermodynamic variable, just like $\mu$ and $T$.

In the covariant two-fluid formalism, there are 2 independent four-vectors, and we have seen that they can either be chosen to be two currents, or two
momenta, or one current and one momentum. In any case, there are $2\times 4$ degrees of freedom. In the field-theoretical calculation, there are the 
thermodynamic variables $\mu$, $T$, and $\nabla\psi$, i.e., five degrees of freedom. The additional three parameters are hidden in the condition that in thermal 
field theory we are working in the normal-fluid rest frame, ${\bf s}=0$. How do we compute for instance the quantities $A$, $B$, $C$ from the 
microscopic physics? From Eqs.\ (\ref{AB1}) and (\ref{C1}) we see that they are given by ``usual'' thermodynamic quantities plus the superfluid density $n_s$.
(The normal-fluid density can be expressed in terms of the total charge density $n$ and the superfluid density, $n_n=n-\mu n_s/\sigma$.) 
To express $n_s$ in field-theoretical terms, we take the spatial components of Eq.\ (\ref{jmacroT}) in the normal-fluid rest frame, where ${\bf s}=0$, 
and contract both sides of the equation with $\nabla\psi$ to obtain
\be \label{nsdef}
n_s = -\sigma\frac{\nabla\psi\cdot{\bf j}}{(\nabla\psi)^2} \, .
\ee
The three-current ${\bf j}$ can be computed in field theory in a standard way since it is given by the derivative of the pressure with respect to $\nabla\psi$,
and we know how to compute the pressure in field theory. (Here, pressure refers to $P=-\Omega$, which is the transverse pressure $P_\perp$ in the stress-energy tensor.)
This expression for the superfluid density, together with Eqs.\ (\ref{AB1}) and (\ref{C1}), 
yields a recipe to compute $A$, $B$, $C$. If we are interested in small values of the superflow $\nabla\psi$ we can simplify the expression for $n_s$. 
To this end, we expand the pressure,
\be
P = P(\nabla\psi=0) + \frac{1}{2}\partial_i\psi\partial_j\psi\,\left[\frac{\partial^2 P}{\partial(\partial_i\psi)(\partial_j\psi)}\right]_{\nabla\psi=0} + \ldots\, .
\ee
Here we have dropped the linear term which is equivalent to assuming that the current vanishes for $\nabla\psi=0$. Then, after rewriting the derivatives
with respect to $\partial_i\psi$ in terms of derivatives with respect to $|\nabla\psi|$, we obtain 
\be \label{nszerosf}
\mbox{zero superflow:}\qquad n_s = -\mu \left(\frac{\partial^2 P}{\partial|\nabla\psi|^2}\right)_{\nabla\psi=0} \, .
\ee
To summarize, from the field-theoretical point of view, $\partial_0\psi=\mu$, $\partial_i\psi=-\mu v_{si}$, and $\Theta_0=T$ are thermodynamic variables in the 
grand canonical ensemble; $s_i=0$ is given due to the ``natural'' frame of the calculation, and $j_0=n$, $j_i$, and $s_0=s$ are thermodynamic equilibrium 
quantities that can be computed via usual thermodynamic definitions. Only the three-vector $\Theta_i$ is uncommon in usual thermodynamics. By writing down
the spatial components of Eqs.\ (\ref{psiTheta}) we find for this vector
\be \label{bftheta}
\Theta_i = -\frac{A}{B}\partial_i\psi = \frac{n_n}{s}\partial_i\psi \, .
\ee
This identification of the various four-vectors relies on our assumptions of homogeneity and vanishing dissipation. In general hydrodynamics,
the currents and conjugate momenta are of course not given by uniform equilibrium quantities. 

In the following, we compute some of these quantities explicitly to illustrate the translation between field theory and two-fluid hydrodynamics.

\subsection{Goldstone mode and superfluid density}

For a field-theoretical derivation of the two-fluid model it is crucial to include nonzero temperatures $T$ (otherwise there is only a single fluid) and 
a nonzero superflow $\nabla\psi$. The latter is necessary to compute the superfluid density which describes the response of the system to the superflow.
In other words, even if we are eventually interested in the case of vanishing superflow, we need to include an infinitesimal superflow to compute the superfluid
density, for instance via Eq.\ (\ref{nszerosf}). However, we shall be a bit more ambitious and keep a finite $\nabla\psi$ in our calculation. It turns out that within
the low-temperature approximation relatively simple results can be obtained even in the presence of a superflow.
 
We start from the pressure that we have derived in chapter \ref{sec:phi4}, see Eq.\ (\ref{omegavut}),
\be \label{Psuper}
P(\mu,T,\nabla\psi)  = - U - T\sum_{e=\pm}\int\frac{d^3{\bf k}}{(2\pi)^3} \ln\left(1-e^{-\epsilon_{\bf k}^e/T}\right) \, , 
\ee
where the $T=0$ contribution is the negative of the tree-level potential, evaluated at the minimum $\lambda \rho^2 = \sigma^2-m^2$, 
\be
-U = \frac{(\sigma^2-m^2)^2}{4\lambda} \, .
\ee 
The quasiparticle excitations $\epsilon_{\bf k}$ in the presence of a nonzero $\nabla\psi$ are complicated, as discussed in chapter \ref{sec:phi4}, and in general 
it is best to proceed numerically. However, if we restrict ourselves to low temperatures, we
get away with the following simplification. At sufficiently low $T$, the massive mode $\epsilon_{\bf k}^-$ becomes irrelevant, and for the Goldstone mode
$\epsilon_{\bf k}^+$ we only need to keep the low-energy dispersion, which is linear in the momentum $k$. 
The quasiparticle excitations are computed from the determinant of the inverse propagator (\ref{nondiag}),
\bea
0 &=& {\rm det}\,D^{-1} \non[2ex]
&=& k_0^4 - 2k_0^2[k^2+3\sigma^2-m^2+2(\nabla\psi)^2] -8k_0\mu\,{\bf k}\cdot\nabla\psi \non[2ex]
&& +k^2[k^2+2(\sigma^2-m^2)]-4({\bf k}\cdot\nabla\psi)^2 \, ,
\eea
where we have ordered the contributions according to the powers of $k_0$, and where we have inserted the $T=0$ solution for the condensate. 
Since we work at low temperatures we use the ansatz $k_0=c(\uk)k$ to determine the (now angular dependent) slope of the Goldstone dispersion $c(\uk)$. 
After inserting this ansatz into the determinant, the terms quadratic in $k$ yield a simple quadratic equation for $c(\uk)$ that has the following solution 
(plus one unphysical negative solution),
\be \label{epsflow}
\epsilon_{\bf k}^+ \simeq \frac{\sqrt{(\sigma^2-m^2)[3\sigma^2-m^2+2(\nabla\psi)^2\sin^2\theta]}-2\mu |\nabla\psi|\cos\theta}{3\sigma^2-m^2+2(\nabla\psi)^2} \, k \, , 
\ee
where $\theta$ is the angle between $\nabla\psi$ and the quasiparticle momentum ${\bf k}$. 
It is instructive to take the non-relativistic 
limit of this expression. In this case, the superfluid velocity ${\bf v}_s = -\nabla\psi/\mu$ is much smaller than the speed of light, $v_s\ll 1$. Moreover, the 
mass $m$ is large. Since we are in the condensed phase, $m$ must always be smaller than the chemical potential, and thus $1-m^2/\mu^2$ has to be small. 
Expanding in this smallness parameter as well as in the superfluid velocity, yields 
$\epsilon_{\bf k}^+ = \epsilon_k^+ + {\bf k}\cdot{\bf v}_s$, with $\epsilon_k^+$ being the (linear) Goldstone dispersion in the absence of a superflow (\ref{GM2}). 
As a limit, we have thus reproduced the Galilei transformed dispersion from our discussion of Landau's critical velocity, see Eq.\ (\ref{epsgalilei}) in 
chapter \ref{sec:he4}. The dispersion (\ref{epsflow}) is the generalization to a 
Lorentz transformed excitation energy. We can also see this by starting from the inverse propagator in the absence of a superflow,
\be \label{propnonflow}
D^{-1}(K) = \left(\begin{array}{cc}-K^2+2(\mu^2-m^2) & \;\;
-2ik_0\mu \\[1ex] 2ik_0 \mu & \;\;
-K^2 \end{array}\right) \, .
\ee
Here, $K^2$ and $m^2$ are Lorentz scalars, while a Lorentz transformation acts on $k_0$ and $\mu$,
\be
\mu \to \mu\sqrt{1-v_s^2} \, , \qquad k_0 \to \frac{k_0-{\bf k}\cdot{\bf v}_s}{\sqrt{1-v_s^2}} \, . 
\ee
Inserting this into the inverse propagator (\ref{propnonflow}), we obtain 
the inverse propagator in the presence of a superflow (\ref{nondiag}). Therefore, the dispersion (\ref{epsflow}) is the Lorentz transformed dispersion,
where the Lorentz transformation is performed from the superfluid rest frame into the normal-fluid rest frame. Remember that ${\bf v}_s$ is the superfluid 
velocity measured in the normal-fluid rest frame. So, (\ref{epsflow}) is the dispersion measured in the normal-fluid rest frame
in the presence of a moving superfluid, for which we measure the velocity ${\bf v}_s$. As discussed in the context of Landau's critical velocity, the negativity
of $\epsilon_{\bf k}^+$ indicates the breakdown of superfluidity through dissipative processes. It is left as an exercise to compute the critical velocity 
from Eq.\ (\ref{epsflow}). Remember that in chapter \ref{sec:he4} our argument made use of the presence of a capillary, which introduced a second frame besides the 
rest frame of the superfluid. Now, at nonzero temperatures, the normal fluid provides such a second frame. Therefore, dissipation would also occur in 
the case of an infinite system without any walls of a capillary.

Even though the superflow introduces a non-trivial angular integration, we can evaluate the pressure (\ref{Psuper}) at low temperatures in an analytical form. We write
the nonzero-temperature part as
\bea
&&T\int\frac{d^3{\bf k}}{(2\pi)^3} \ln\left[1-e^{-c(\uk)k/T}\right] = \frac{T^4}{2\pi^2}\int\frac{d\Omega}{4\pi}\frac{1}{c^3(\uk)}
\int_0^\infty dy\,y^2\ln(1-e^{-y}) \non[2ex]
&&= -\frac{\pi^2 T^4}{90}\frac{[3\sigma^2-m^2+2(\nabla\psi)^2]^3}{(\sigma^2-m^2)^{3/2}}\frac{1}{2}
\int_{-1}^1 dx\,\frac{1}{(\sqrt{\alpha-\beta x^2}-\gamma x)^3} \, ,
\eea
where we have used 
\be
\int_0^\infty dy\,y^2\ln(1-e^{-y}) = -\frac{\pi^4}{45} \, , 
\ee
and where we have abbreviated 
\be
\alpha\equiv 3\sigma^2-m^2+2(\nabla\psi)^2 \, , \qquad \beta\equiv 2(\nabla\psi)^2 \, , \qquad \gamma\equiv \frac{2\mu|\nabla\psi|}{\sqrt{\sigma^2-m^2}} \, .
\ee
Now, with 
\be
\frac{1}{2}\int_{-1}^1 dx\,\frac{1}{(\sqrt{\alpha-\beta x^2}-\gamma x)^3} = \frac{(\alpha-\beta)^{3/2}}{\alpha(\alpha-\beta-\gamma^2)^2} \, , 
\ee
and adding the $T=0$ contribution, we find
\be \label{Pmsigma}
P \simeq \frac{(\sigma^2-m^2)^2}{4\lambda} +\frac{\pi^2 T^4}{90}\frac{(3\sigma^2-m^2)^{3/2}(\sigma^2-m^2)^{1/2}}{[\sigma^2-m^2-2(\nabla\psi)^2]^2} \, .
\ee
From this expression for the pressure, we can derive all thermodynamic quantities up to the given order in temperature. Since we have kept the 
full dependence on the superflow, we should read the pressure as a function of the thermodynamic variables $T$, $\mu$, and $\nabla\psi$. In particular, 
we obtain the charge current by taking the derivative with respect to $\nabla\psi$ and from the result we can compute the superfluid density, see definition
(\ref{nsdef}). It is straightforward to compute the result for arbitrary values of the superflow. Here, for compactness, we give the result for vanishing superflow,
\be
\mbox{zero superflow:} \qquad n_s \simeq \frac{\mu(\mu^2-m^2)}{\lambda} - \frac{\pi^2 T^4}{45}\frac{\mu(3\mu^2-m^2)^{1/2}(12\mu^2-m^2)}{(\mu^2-m^2)^{5/2}} \, .
\ee  
As expected, the superfluid density decreases with increasing temperature. 
For \mbox{$T=0$}, the superfluid density should be identical to the total charge density $n=\frac{\partial P}{\partial\mu}$. 
To check this, remember that 
$n_s$ is the superfluid density measured in the superfluid rest frame, while $n$ is the total charge density measured in the normal-fluid rest frame. 
Therefore, we need to multiply 
$n_s$ by the Lorentz factor $\mu/\sigma=1/\sqrt{1-v_s^2}$ to obtain\footnote{In our approximation,  the superfluid density at $T=0$ is solely given by the condensate, 
$n_s = \sigma \rho^2$. Therefore, {\it superfluid density} and {\it condensate density} are identical. In general, in an interacting system, the 
condensate density is smaller than the superfluid density \cite{annett,2001cond.mat..1299S,oai:arXiv.org:cond-mat/0305138}, see also Ref.\ \cite{2011PPN....42..460Y}
for a discussion about the inequivalence of superfluid and condensate densities.}
\be
n(T=0) = \frac{\mu}{\sigma} n_s(T=0) \, ,
\ee
where $n(T=0)$ has already been computed above, see Eq.\ (\ref{nPepsilon1}).
The normal-fluid density (for all temperatures) is then given by $n_n=n-\mu/\sigma\,n_s$. 

Our approximation is only valid for small temperatures and can thus not be used to compute the superfluid and normal-fluid densities up to the critical temperature, 
as we have done for superfluid helium in 
chapter \ref{sec:he4}, see Fig.\ \ref{figphrot2}. It is instructive to point out the difference between the two calculations: our calculation for helium was
purely phenomenological, not based on a microscopic model. We simply made an assumption (motivated from experiment) for the excitation energies and were thus
able to compute the properties of the normal fluid. From these results, together with fixing the total mass density, we obtained the superfluid 
density ``for free'', without even having to talk about the condensate in a microscopic sense. In the present field-theoretical calculation, we do know the 
microscopic physics.  In our calculation within the 
grand canonical ensemble, we can in principle compute ``everything'' as a function of chemical potential, temperature, and superfluid velocity. Working at fixed chemical 
potential is more natural from the field-theoretical point of view, but of course we could always switch to the canonical ensemble since we know the relation 
between the chemical potential and the charge density. With a lot more effort, one can extend the present field-theoretical model to all temperatures
below the critical temperature \cite{Alford:2013koa}.

\subsection{Generalized pressure and sonic metric}

We have seen that the covariant two-fluid formalism is built upon the generalized pressure $\Psi$, which contains the microscopic information
of the system. We have already mentioned that this is not an object usually encountered in field theory. We may thus ask if we can, at least a posteriori,
construct the generalized pressure from the results we have now obtained. Since $\Psi$ depends on the Lorentz scalars $\sigma^2$, $\Theta^2$, and 
$\partial\psi\cdot\Theta$, we must express the frame-dependent quantities $\mu$, $T$, and $\nabla\psi$ in terms of these Lorentz scalars. To this end, we write

\begin{subequations} \label{sigthet}
\bea
\sigma^2 &=& \mu^2-(\nabla\psi)^2 \, , \\[2ex]
\Theta^2 &=& T^2 - \frac{A^2}{B^2}(\nabla\psi)^2 \, , \\[2ex]
\partial\psi\cdot\Theta &=& \mu T -\frac{A}{B}(\nabla\psi)^2 \, , 
\eea
\end{subequations}
where $T=\Theta^0$, $\mu=\partial^0\psi$, and Eq.\ (\ref{bftheta}) have been used. Next, we have to compute $A$ and $B$. As Eq.\ (\ref{AB1}) shows,
this requires to compute the superfluid and normal-fluid densities and the entropy. This can be done straightforwardly with the help of the 
pressure (\ref{Pmsigma}). We give the results for the ultra-relativistic case $m=0$ and leave the more general case with a nonzero $m$ as an exercise,
\begin{subequations}\label{nsnns}
\bea
n_s &=& \frac{\sigma^3}{\lambda} - \frac{4\pi^2T^4}{5\sqrt{3}\mu}\frac{(1-v_s^2)^{3/2}}{(1-3v_s^2)^3} \, ,\\[2ex] 
n_n &=& n-\frac{\mu}{\sigma}n_s = \frac{4\pi^2 T^4}{5\sqrt{3}\mu}\frac{(1-v_s^2)^2}{(1-3v_s^2)^3} \, , \\[2ex]
s &=& \frac{\partial P}{\partial T} = \frac{2\pi^2 T^3}{5\sqrt{3}}\frac{(1-v_s^2)^2}{(1-3v_s^2)^2} \, .
\eea
\end{subequations}
We see that the temperature dependent terms 
diverge for $v_s=\frac{1}{\sqrt{3}}$. This indicates that our temperature expansion breaks down for superfluid velocities close to this critical value,
and we can trust the results only for sufficiently small $v_s$. In fact, this critical value is nothing but Landau's critical velocity, as one can 
check with the help of the quasiparticle excitation (\ref{epsflow}).  
 
We now insert Eqs.\ (\ref{nsnns}) into Eqs.\ (\ref{AB1}) and the result into Eqs.\ (\ref{sigthet}). Then, we solve the resulting equations for $T$, $\mu$ and 
$\nabla\psi$, and insert the result back into the pressure (\ref{Pmsigma}). Since we have expanded the pressure up to order $T^4$ we have to discard all higher-order terms
that we have generated in this calculation. To this end, we take into account that $\Theta^2\propto T^2$ and $\partial\psi\cdot\Theta\propto T$, while $\sigma^2$ does
not depend on $T$. As a result, we can write the generalized pressure as 
\be \label{PsiT4}
\Psi[\sigma^2,\Theta^2,\partial\psi\cdot\Theta] \simeq \frac{\sigma^4}{4\lambda} + \frac{\pi^2}{90\sqrt{3}}\left[\Theta^2+2\frac{(\partial\psi\cdot\Theta)^2}{\sigma^2}
\right]^2 \, . 
\ee
Once we have the pressure in this form, we can compute for instance the coefficients $\overline{A}$, $\overline{B}$, $\overline{C}$ from their original 
definition (\ref{entrainbar}). It is left as an exercise to perform this calculation and to re-express the resulting expressions in terms of the 
field-theoretical variables $T$, $\mu$, and ${\bf v}_s$. 

The pressure (\ref{PsiT4}) can be rewritten in the following way,
\be 
\Psi[\sigma^2,\Theta^2,\partial\psi\cdot\Theta] \simeq \frac{\sigma^4}{4\lambda} 
+ \frac{\pi^2}{90\sqrt{3}}({\cal G}^{\mu\nu} \Theta_\mu \Theta_\nu)^2 \, ,
\ee
where we have introduced the so-called {\it sonic metric} (sometimes also called {\it acoustic metric}) \cite{Carter:1995if,unruh,2005LRR.....8...12B,Mannarelli:2008jq}
\be \label{sonic}
{\cal G}^{\mu\nu} = g^{\mu\nu} + \left(\frac{1}{c^2}-1\right)v^\mu v^\nu  \, , 
\ee
with $c=\frac{1}{\sqrt{3}}$ and the Minkowski metric $g^{\mu\nu}$. This result suggests that the superfluid velocity effectively introduces 
a curved space. Remember that photons in Minkowski space have light-like four-momenta, i.e., their four-momenta are null vectors with respect to the 
Minkowski metric. In analogy, phonons have four-momenta that are null vectors with respect to the sonic metric, which means
\be
{\cal G}^{\mu\nu}K_\mu K_\nu = 0 \, .
\ee
Using the explicit form of the sonic metric (\ref{sonic}) and solving this null condition for the energy $k_0$ gives 
\be \label{sonicdisp}
k_0 = \frac{\sqrt{(1-v_s^2)[3(1-v_s^2)+2v_s^2\sin^2\theta]}+2v_s\cos\theta}{3-v_s^2}\, k \, .
\ee
This is the low-energy dispersion relation for the Goldstone mode that we have derived above from the pole of the propagator: setting 
$m=0$ in Eq.\ (\ref{epsflow}) gives exactly the relation (\ref{sonicdisp}). (For the sake of a consistent notation, $\theta$ is still the angle between 
$\nabla\psi$ and ${\bf k}$, which means it is the angle between $-{\bf v}_s$ and ${\bf k}$ because of $\mu{\bf v}_s=-\nabla\psi$.)

\chapter{Fermionic superfluidity: Cooper pairing}
\label{sec:cooper}

We have seen in the previous chapters that a necessary condition for superfluidity is the formation of a Bose-Einstein condensate. It seems that this restricts
superfluidity to bosonic systems. However, also fermionic systems can become superfluid. In order for a fermionic system to develop a Bose-Einstein condensate, 
it has to form some kind of bosonic states. The mechanism that provides these states is Cooper pairing, which we shall discuss in this section. 
Cooper pairing is a very generic phenomenon because any Fermi surface is unstable if there is an arbitrarily small attractive interaction between the fermions.
This instability manifests itself in a new ground state, in which pairs of fermions are created at the Fermi surface. 
Simply speaking, fermionic systems undergo an intermediate step, Cooper pairing, before they are ``ready'' to condense and thus become a superfluid or a superconductor.
Before we discuss this mechanism in detail in a field-theoretical approach, I give a brief overview over the history of fermionic 
superfluids and superconductors. 

The most prominent system where Cooper pairing takes place is an electronic superconductor. Superconductivity was first 
observed in mercury by H.\ Kamerlingh Onnes in 1911 \cite{1911KNAB...14..113K}, see Ref.\ \cite{vandelft} for an interesting historical account. 
He observed a vanishing resistivity below a temperature of $T_c = 4.2\, {\rm K}$ 
(by the way, he cooled his system with liquid helium). In 1933, W.\ Meissner and R.\ Ochsenfeld
discovered that a superconductor expels an externally applied magnetic field \cite{1933NW.....21..787M}, now called the Meissner effect (or Meissner-Ochsenfeld effect). 
Although some properties of superconductors
could be described with phenomenological models, for instance the Ginzburg-Landau model in 1950, it took almost 50 years until the microscopic BCS
theory and thus Cooper pairing of electrons was formulated. The original paper appeared in 1957 \cite{bcs}, and J.\ Bardeen, L.\ Cooper, and J.\ Schrieffer were 
awarded the 
Nobel Prize in 1972. One obstacle in the understanding of electronic superconductors is the origin of the attractive interaction. 
This is provided by the lattice of ions, more precisely by its excitations. These excitations, called phonons, correspond to the Goldstone mode associated to the 
spontaneous breaking of translational invariance through the lattice. Although two electrons have the same charge and thus appear to repel 
each other, the exchange of phonons gives rise to a net attractive interaction. It is crucial that the repulsive force is screened in the 
crystal by the positively charged ions. Important progress had been made in the years before BCS regarding the quantitative understanding
of the electron-phonon interaction, in particular by H.\ Fr\"{o}hlich in 1950 \cite{1950PhRv...79..845F}. 
The next major discovery in the field of electronic superconductors was the discovery of 
high-temperature superconductors in 1986 by J.G.\ Bednorz and K.A.\ M\"{u}ller \cite{1986ZPhyB..64..189B} (Nobel Prize 1987). 
They found a material that shows superconducting properties for temperatures
below $T\simeq 35\, {\rm K}$. By now, superconductors with critical temperatures as large as $T_c\simeq 130\, {\rm K}$ have been observed. The mechanism 
behind high-$T_c$ superconductivity is still a matter of current research and cannot be understood within the weak-coupling methods discussed in this chapter. 

Once the mechanism of Cooper pairing had been understood, the question arose which other systems may show this effect. The next one that was established 
experimentally is superfluid $^3$He which, in contrast to $^4$He, is a fermionic system. In this case, there is an attractive interaction 
between the helium atoms (more precisely between quasi-fermions whose properties in the liquid are significantly different from helium atoms in vacuum)
in the spin-triplet, $p$-wave channel. As a consequence, a very rich structure of conceivable order parameters exists, allowing for several superfluid phases in 
$^3$He. The superfluid phase transition of $^3$He was first observed 
by D.\ Lee, P.\ Osheroff, and R.\ Richardson in 1972 \cite{1972PhRvL..28..885O}, who received the Nobel Prize in 1996 \cite{RevModPhys.69.645,RevModPhys.69.667}. 
In fact, two different transitions at $2.6\, {\rm mK}$ and $1.8\, {\rm mK}$ were observed,
owing to two different superfluid phases of $^3$He, called {\it A phase} and {\it B phase}.

Most recently, Cooper pairing has been observed in ultra-cold Fermi gases. Superfluidity was directly observed by vortex formation in a gas of $^6$Li atoms in 2005 
by W.\ Ketterle and his group \cite{2005Natur.435.1047Z}. In this case, the critical temperature is about \mbox{$200\,{\rm nK}$}. We will say much more about these atomic
systems in chapter \ref{sec:BCSBEC}.

There are also systems in high-energy physics where Cooper pairing is expected, but experimental evidence is very difficult to establish. For instance, it is assumed that 
neutrons are superfluid and protons superconducting in the interior of neutron stars, first suggested by N.N.\ Bogoliubov in 1958 \cite{bogol} and 
A.B.\ Migdal in 1959 \cite{1959NucPh..13..655M}. Indications that this is indeed the case come from various astrophysical 
observations; for instance from observed pulsar glitches, sudden jumps in the rotation frequency of the star that are attributed to a sudden ``un-pinning'' 
of superfluid vortices in the inner crust of the star; or from the cooling behavior of the star: recent measurements that show an unexpectedly rapid cooling over 
about 10 years are attributed to a superfluid phase transition, suggesting a critical temperature for neutron superfluidity of $T_c\simeq 5.5\cdot 10^8\,{\rm K}$
\cite{oai:arXiv.org:1011.6142,Shternin:2010qi}. 

Moreover, quark matter may form a {\it color superconductor} in the deconfined phase of Quantum Chromodynamics (QCD). 
If the density in the interior of compact stars is sufficiently large, color superconductivity may also be of astrophysical relevance. The attractive force between 
quarks in ultra-dense matter (as well as the one for nucleons in dense nuclear matter) is provided by the strong interaction, i.e., by QCD. In this sense, 
color superconductivity is a very fundamental form of Cooper pairing, the attractive interaction is directly provided by a fundamental interaction, and no lattice of
ions is needed like in an electronic superconductor. The possibility of Cooper pairing of quarks was already mentioned in 1969 \cite{Ivanenko:1969gs}, 
before the theory of QCD was even established. Pioneering work on color superconductivity was 
done in the late seventies \cite{Barrois:1977xd,Frautschi:1978rz,Barrois:1979pv,Bailin:1979nh}, and many possible phases of color superconductivity were discussed by 
D.\ Bailin and A.\ Love in 1984 \cite{Bailin:1979nh}. Only much later, in 1998, it was realized that the energy 
gap in quark matter may be large enough to be important for the phenomenology of compact stars \cite{Son:1998uk}. The critical temperature of
color superconductors depends strongly on the specific pairing pattern, and can be as large as about $10^{11}\, {\rm K}$ 
(since quarks have more quantum numbers than electrons, a multitude of phases with quark Cooper pairing is conceivable). For a review about color superconductivity,  
see Ref.\ \cite{Alford:2007xm}.

We shall come back to some of these examples of Cooper pairing in Sec.\ \ref{sec:examples}, after having derived the BCS gap equation in a relativistic  
field-theoretical calculation. For the main arguments that lead to Cooper pairing it does not
matter much whether one starts with relativistic fermions with dispersion $\epsilon_k=\sqrt{k^2+m^2}-\mu$ or from non-relativistic fermions with 
dispersion $\epsilon_k = \frac{k^2}{2m}-\mu$\footnote{Again, notice that the chemical potential for the non-relativistic dispersion, although written with the 
same symbol as the one for the relativistic case, includes the rest mass; for $m\gg k$ we have  
$\sqrt{k^2+m^2}-\mu_{\rm rel.} \simeq m + \frac{k^2}{2m} -\mu_{\rm rel.} = \frac{k^2}{2m} -\mu_{\rm non-rel.}$}. 
The reason is that (weak-coupling) Cooper pairing is a Fermi surface phenomenon and {\it at the 
Fermi surface} both dispersions are linear. We can expand the dispersion for momenta close to the Fermi momentum $k_F$
\be \label{fermivel}
\epsilon_k \simeq v_F(k-k_F) \, , 
\ee
with the {\it Fermi velocity}
\be \label{vF}
v_F \equiv \left. \frac{\partial \epsilon_k}{\partial k}\right|_{k=k_F} =  \left\{\begin{array}{ccc} \displaystyle{\frac{k_F}{m}} & \;\;\mbox{non-relativistic}\;\;, 
& k_F=\sqrt{2\mu m}
\\[2ex] \displaystyle{\frac{k_F}{\mu}}  & \mbox{relativistic}, & k_F=\sqrt{\mu^2-m^2} \\[2ex] 1 & \;\;\mbox{ultra-relativistic}\;\;, & k_F=\mu \end{array} \right. \, .
\ee
Consequently, all three cases show qualitatively the same behavior at the Fermi surface, only the slope of the linear dispersion is different.

\section{Derivation of the gap equation}
\label{sec:derivation}

In this section we shall derive the gap equation for the energy gap in the quasiparticle spectrum of a Cooper-paired system. The main result of the derivation 
on the following pages is Eq.\ (\ref{gapeq1}).

\subsection{Lagrangian}

We consider a theory that contains fermions which interact via boson exchange, think for instance of electrons that interact via phonon exchange, 
or quarks that interact via exchange of gluons. Our Lagrangian thus assumes the following form,
\be
{\cal L} = {\cal L}_{\rm fermions} + {\cal L}_{\rm bosons} + {\cal L}_{\rm interactions} \, .
\ee
Here,  
\be
{\cal L}_{\rm fermions} = \overline{\psi}(i\gamma^\mu\partial_\mu+\gamma^0\mu-m)\psi
\ee
is the free fermionic part with the four-spinor $\psi$, $\overline{\psi}=\psi^\dag\gamma^0$, the chemical potential $\mu$ and the mass $m$. This form of the 
fermionic Lagrangian holds for a single fermion species. Therefore, we cannot really apply the following to quark matter, where there are $N_f N_c$ many 
species, with $N_f=3$ and $N_c=3$ being the numbers of flavors and colors (for applications in compact stars, only up, down and strange quarks are relevant). 
Nevertheless, even when you are interested in Cooper pairing in quark matter, it is 
instructive to go through the single-flavor, single-color calculation before adding the complication of multiple fermion species. 

The bosonic Lagrangian for a real, scalar boson with mass $M$ is
\be
{\cal L}_{\rm bosons} = \frac{1}{2}\partial_\mu\varphi\partial^\mu\varphi - \frac{1}{2}M^2\varphi^2 \, ,
\ee
and for the interaction we write 
\be
{\cal L}_{\rm interactions} = -g\overline{\psi}\psi\varphi \, ,
\ee
where $g>0$ is the coupling constant.
We have kept the structure of this term as simple as possible, using a Yukawa-type interaction, but our main arguments will also hold for more complicated interactions.
For instance, the quark-gluon interaction in QCD has a much richer structure,
\be
-g\overline{\psi}\psi\varphi \to  -g \overline{\psi}_\alpha\gamma^\mu  T_a^{\alpha\beta} \psi_\beta A_\mu^a \, ,
\ee
with the gluon fields $A_\mu^a$, the Gell-Mann matrices $T_a$ ($a=1,\ldots 8$), and color indices $1\le \alpha,\beta\le 3$.

The partition function is 
\be
Z = \int{\cal D}\overline{\psi}{\cal D}\psi{\cal D}\varphi \, e^S \, , 
\ee
with the action\footnote{In order to avoid very space-consuming expressions, we denote the four-vectors in this section 
with small letters: $x, y, \ldots$ for space-time vectors, $k,q,\ldots$ for four-momenta. In Sec.\ \ref{sec:quasi} 
we go back to capital letters, as in all other chapters.}  
\bea \label{actionXY}
S &=& \int_x{\cal L} 
= \int_{x,y}\left[ \overline{\psi}(x) G_0^{-1}(x,y) \psi(y)  - \frac{1}{2} \varphi(x)D^{-1}(x,y)\varphi(y)\right] \non[2ex]
&& \hspace{1cm}- g  \int_x \overline{\psi}(x)\psi(x)\varphi(x) \, ,
\eea
where we have abbreviated the space-time integral
\be \label{intxdef}
\int_x \equiv \int_0^{1/T} d\tau \int d^3{\bf x} \, , 
\ee
and where  
\be \label{fermtree}
G_0^{-1}(x,y) = \delta(x-y)(i\gamma^\mu\partial_\mu + \gamma^0\mu - m)  
\ee
is the inverse fermionic tree-level propagator, and $D^{-1}(x,y)$ the inverse bosonic propagator, whose specific form is not relevant for now.

The first step is to integrate out the bosonic fields. To this end, we use  
\be \label{xiAB}
\frac{1}{2}\xi^T A^{-1}\xi + b^T\xi = -\frac{1}{2} b^T A b + \frac{1}{2}\xi'^T A^{-1}\xi' \, , 
\ee
where $\xi' = \xi+Ab$ with a symmetric matrix $A$ and vectors $\xi$ and $b$, such that all four matrix products in the relation result in scalars.
We apply this relation to the last two terms of Eq.\ (\ref{actionXY}), i.e., we identify $\xi\to \varphi(x)$, 
$A\to D(x,y)$, $b\to g\overline{\psi}(x)\psi(x)$, and the matrix products in Eq.\ (\ref{xiAB}) are products in position space. 
As a result, the terms linear and quadratic in the original bosonic fields can be rewritten as terms constant and 
quadratic in the new, shifted bosonic fields. We can thus easily integrate over the shifted fields to obtain 
\be
Z = Z_{\rm bosons} \int{\cal D}\overline{\psi}{\cal D}\psi \, e^{S'} \, , 
\ee
with a bosonic partition function $Z_{\rm bosons}$ that is irrelevant for our purpose, and the new fermionic action
\be
S' = \int_{x,y}\left[ \overline{\psi}(x) G_0^{-1}(x,y) \psi(y)  + \frac{g^2}{2} 
\overline{\psi}(x)\psi(x)D(x,y)\overline{\psi}(y)\psi(y) \right] \, .
\ee
The interaction term proportional to $g^2$ is composed of two elementary Yukawa interactions: it contains two incoming fermions, the propagator of the 
exchanged boson, and two outgoing fermions. At each elementary vertex, there is a Yukawa coupling $g$, hence there is a $g^2$ for the total process. 

\subsection{Mean-field approximation}

We shall now try to find an approximation for this interaction term. The goal will be to write the product of two fermion spinors as its expectation value plus 
fluctuations around this value. The expectation value of the two fermions will correspond to  a condensate of fermion 
pairs. In a way, we are looking for an analogue of our ansatz for the Bose-Einstein condensate in the bosonic field theory, see Eq.\ (\ref{shift}). 

For a di-fermionic condensate, there are two options. First, there might be a condensate of fermion-antifermion pairs. In that case, one may 
proceed rather straightforwardly since the scalar $\overline{\psi}\psi$ is the relevant object. The physics described by such a condensate is for instance
chiral symmetry breaking in QCD. However, here we are not interested in this condensate. Cooper pairs in a superfluid or a superconductor 
are fermion-fermion pairs, not fermion-antifermion pairs. In this case, the object that ``wants'' to obtain an expectation value cannot simply be written 
as $\psi\psi$: think of $\psi$ as a column vector and $\overline{\psi}$ as a 
row vector in Dirac space, then $\overline{\psi}\psi$ is a scalar, $\psi\overline{\psi}$ a $4\times 4$ matrix, but the products $\psi\psi$ and 
$\overline{\psi}\,\overline{\psi}$ are not defined. In other words, we would like to have a fermion which is described by a row vector. This is 
done by introducing the so-called charge-conjugate spinor $\psi_C$, such that a Cooper pair of fermions can be written as $\psi\overline{\psi}_C$
and a Cooper pair of anti-fermions as $\psi_C\overline{\psi}$. (Cooper pairing of anti-fermions will play no role in the physical systems we discuss but 
it is convenient to introduce it too.)
The details of this procedure are as follows. With the charge-conjugation matrix 
$C=i\gamma^2\gamma^0$ we define\footnote{Here and
in the following we need a few properties of the Dirac matrices, such as $(\gamma^0)^\dagger = \gamma^0$, $(\gamma^i)^\dagger = -\gamma^i$, and 
$\{\gamma^\mu,\gamma^\nu\}=2g^{\mu\nu}$ (consequently, $\gamma^0$ anti-commutes with $\gamma^i$). Moreover, $(\gamma^0)^2=1$, $(\gamma^i)^2=-1$. 
In the Dirac representation, we have
\be
\gamma^0 = \left(\begin{array}{cc} 1 & 0 \\ 0 & -1\end{array}\right) \, , \qquad \gamma^i = \left(\begin{array}{cc} 0 & \sigma_i \\ -\sigma_i& 0\end{array}\right)
\, , \qquad \gamma^5 \equiv i\gamma^0\gamma^1\gamma^2\gamma^3 = \left(\begin{array}{cc} 0 & 1 \\ 1 & 0\end{array}\right) \, , \nonumber
\ee
with the Pauli matrices $\sigma_i$, which are defined as
\be
\sigma_1 = \left(\begin{array}{cc} 0 & 1 \\ 1 & 0 \end{array}\right) \, , \qquad \sigma_2 = \left(\begin{array}{cc} 0 & -i \\ i & 0 \end{array}\right) \, , 
\qquad \sigma_3 = \left(\begin{array}{cc} 1 & 0 \\ 0 & -1 \end{array}\right) \, .\nonumber
\ee}
\be\label{psiC}
\psi_C \equiv  C\overline{\psi}^T   \, , 
\ee
which implies 
$\overline{\psi}_C = \psi^TC$, $\psi = C\overline{\psi}^T_C$, $\overline{\psi} = \psi_C^TC$. Here, $\overline{\psi}_C$ is understood as
first charge-conjugating, then taking the Hermitian conjugate and multiplying by $\gamma^0$. For instance, the first relation is obtained as 
$\overline{\psi}_C = (C\overline{\psi}^T)^\dag\gamma^0 = (C\gamma^0\psi^*)^\dag\gamma^0 = \psi^T \gamma^0C^\dag\gamma^0 = -\psi^T\gamma^0C\gamma^0 = \psi^TC$.
Since 
\be
\overline{\psi}_C\psi_C = \psi^TCC\overline{\psi}^T=-\psi^T\overline{\psi}^T = (\overline{\psi}\psi)^T = \overline{\psi}\psi \, , 
\ee
where $C=-C^{-1}$ and the Grassmann property of the fermion spinor have been used, we can write 
\bea
&&\overline{\psi}(x)\psi(x)\overline{\psi}(y)\psi(y) = 
\frac{1}{2}[\overline{\psi}_C(x)\psi_C(x)\overline{\psi}(y)\psi(y)+\overline{\psi}(x)\psi(x)\overline{\psi}_C(y)\psi_C(y)] \non[2ex]
&&=-\frac{1}{2}\Tr[\psi_C(x)\overline{\psi}(y)\psi(y)\overline{\psi}_C(x)+\psi(x)\overline{\psi}_C(y)\psi_C(y)\overline{\psi}(x)]
\, ,
\eea 
where the trace is taken over Dirac space. Again, the minus sign arises since the fermion field is a Grassmann variable.
Now we can separate the appropriate di-fermion expectation values,
\begin{subequations}
\bea
\psi_C(x)\overline{\psi}(y) &=& \langle\psi_C(x)\overline{\psi}(y)\rangle -[\langle\psi_C(x)\overline{\psi}(y)\rangle - \psi_C(x)\overline{\psi}(y)] \, , \\[2ex]
\psi(y)\overline{\psi}_C(x) &=& \langle\psi(y)\overline{\psi}_C(x)\rangle  -[\langle\psi(y)\overline{\psi}_C(x)\rangle - \psi(y)\overline{\psi}_C(x)] \, , 
\eea
\end{subequations}
and consider the square brackets as small fluctuations. Neglecting terms quadratic in these fluctuations, we
derive (the few lines of algebra for the derivation is left as an exercise)
\bea
&&\int_{x,y} D(x,y) \overline{\psi}(x)\psi(x)\overline{\psi}(y)\psi(y) = \int_{x,y} D(x,y) \Tr[ \langle\psi_C(x)\overline{\psi}(y)\rangle
\langle\psi(y)\overline{\psi}_C(x)\rangle] \non[2ex]
&& - \int_{x,y}D(x,y) \Tr[\langle\psi_C(x)\overline{\psi}(y)\rangle \psi(y)\overline{\psi}_C(x) + \langle\psi(x)\overline{\psi}_C(y)\rangle
\psi_C(y)\overline{\psi}(x)] \, ,
\eea
where we have assumed the boson propagator to be symmetric in position space, $D(x,y)=D(y,x)$. The first term does not depend on the fermion fields
and thus we can pull it out of the functional integral,  
\be \label{Z2}
Z = Z_{\rm bosons} Z_0 \int{\cal D}\overline{\psi}{\cal D}\psi \, e^{S''} \, ,
\ee
with 
\be \label{Z0}
Z_0 \equiv \exp\left\{\frac{g^2}{2}\int_{x,y}D(x,y)\Tr[ \langle\psi_C(x)\overline{\psi}(y)\rangle
\langle\psi(y)\overline{\psi}_C(x)\rangle]\right\} \, . 
\ee
In the following derivation of the gap equation, $Z_0$ will play no role. However, this contribution is important for the 
thermodynamic potential. One can also derive the gap equation by minimizing the thermodynamic potential with respect to the gap; in this case, $Z_0$ has to be kept. 
We shall come back to this term when we go beyond the mean-field approximation in chapter \ref{sec:lowenergy}, see Eq.\ (\ref{LMF}).

In Eq.\ (\ref{Z2}) we have abbreviated the new action 
\be \label{action2}
S'' = \int_{x,y}\left\{ \overline{\psi}(x) G_0^{-1}(x,y) \psi(y)  + \frac{1}{2}[\overline{\psi}_C(x)\Phi^+(x,y)\psi(y) + 
\overline{\psi}(x)\Phi^-(x,y)\psi_C(y)]\right\} \, , 
\ee
where we have defined
\begin{subequations} \label{Phipm}
\bea
\Phi^+(x,y) &\equiv & g^2 D(x,y)\langle\psi_C(x)\overline{\psi}(y)\rangle \, , \label{gapplus}\\[2ex]
\Phi^-(x,y) &\equiv & g^2 D(x,y)\langle\psi(x)\overline{\psi}_C(y)\rangle \, .
\eea
\end{subequations}
It is easy to check that $\Phi^+$ and $\Phi^-$ are related via
\be \label{DD}
\Phi^-(y,x) = \gamma^0[\Phi^+(x,y)]^\dag\gamma^0 \, .
\ee

\subsection{Nambu-Gorkov space}

All effects of the interaction are now absorbed into $\Phi^\pm$, and our new action $S''$ is quadratic in the fields, i.e, we can perform the 
functional integral. To this end, let us go to momentum space by introducing the Fourier transforms of the fields, 
\begin{subequations}
\bea
\psi(x) &=& \frac{1}{\sqrt{V}}\sum_k e^{-i k\cdot x} \psi(k) \, , \qquad \overline{\psi}(x) = \frac{1}{\sqrt{V}}\sum_k e^{ik\cdot x} \overline{\psi}(k) \, , 
\label{fourier1}\\[2ex]
\psi_C(x) &=& \frac{1}{\sqrt{V}}\sum_k e^{-i k\cdot x} \psi_C(k) \, , \qquad \overline{\psi}_C(x) = \frac{1}{\sqrt{V}}\sum_k e^{ik\cdot x} \overline{\psi}_C(k) \, . 
\label{fourier2}
\eea
\end{subequations}
The normalization factor containing the three-volume $V$ is chosen such that the Fourier-transformed fields are dimensionless ($\psi(x)$ has mass dimensions 3/2). 
The temporal component of the four-momentum $k=(k_0,{\bf k})$ is given by the fermionic Matsubara frequencies, $k_0=-i\omega_n$, with $\omega_n = (2n+1)\pi T$.
There is some freedom in the choice of the signs in the exponentials of the charge-conjugate fields. They are chosen deliberately to lead to an 
action diagonal
in momentum space, see Eq.\ (\ref{diagk}). By charge-conjugating both sides of the first relation in 
Eq.\ (\ref{fourier1}) and comparing the result with the first relation in Eq.\ (\ref{fourier2}) we see that the given choice implies 
$\psi_C(k)=C\overline{\psi}^T(-k)$ and, analogously, $\overline{\psi}_C(k)=\psi^T(-k)C$. Hence, in momentum space, charge conjugation includes a 
sign flip of the four-momentum, while in position space we have $\psi_C(x)=C\overline{\psi}^T(x)$.

For the Fourier transformation of $\Phi^\pm$ we assume translational invariance, $\Phi^\pm(x,y)= \Phi^\pm(x-y)$, to write 
\be \label{fourierD}
\Phi^\pm(x-y) = \frac{T}{V}\sum_k e^{-ik\cdot (x-y)}\Phi^\pm(k) \, . 
\ee
With Eq.\ (\ref{DD}), this implies $\Phi^-(k) = \gamma^0[\Phi^+(k)]^\dag\gamma^0$.
We can now insert the Fourier decompositions into the interaction terms of the action (\ref{action2}). We find 
\bea \label{diagk}
\int_{x,y}\overline{\psi}_C(x)\Phi^+(x-y)\psi(y)  
&=& \frac{1}{T}\sum_k  \overline{\psi}_C(k)\Phi^+(k)\psi(k) \, , 
\eea
where we have used
\be
\int_x e^{-ik\cdot x} = \frac{V}{T} \delta_{k,0} \, .
\ee
Eventually, we want to consider $\overline{\psi}_C(k)$, $\psi_C(k)$ as independent integration variables, in addition to the variables $\overline{\psi}(k)$, $\psi(k)$. 
We can rewrite the integration in terms of an integral over all four variables 
by restricting ourselves to four-momenta in one half of the full momentum space, 
\bea \label{measure}
{\cal D}\overline{\psi}{\cal D}\psi &=& \prod_kd\overline{\psi}(k)\,d\psi(k) \allowdisplaybreaks \non[2ex]
&=& \prod_{k>0}d\overline{\psi}(k)\,d\overline{\psi}(-k)d\psi(k)\,d\psi(-k) \allowdisplaybreaks \non[2ex]
&=&{\cal N}\prod_{k>0}d\overline{\psi}(k)\,d{\psi}_C(k)d\psi(k)\,d\overline{\psi}_C(k) \, , 
\eea
with an irrelevant constant ${\cal N}$ which arises from the change of integration variables. 
In the integrand, we divide the sum over $k$ in Eq.\ (\ref{diagk}) into a sum over $k>0$ and a sum over $k<0$ and show that both sums are the same, 
\bea \label{help1}
\sum_{k<0}  \overline{\psi}_C(k)\Phi^+(k)\psi(k) &=& \sum_{k>0}  \overline{\psi}_C(-k)\Phi^+(-k)\psi(-k) \non
&=&  \sum_{k>0}  \psi^T(k)C\Phi^+(-k)C\overline{\psi}_C^T(k) \non
&=&  \sum_{k>0}  \left[\psi^T(k)C\Phi^+(-k)C\overline{\psi}_C^T(k)\right]^T \non
&=&  - \sum_{k>0}  \overline{\psi}_C(k)C[\Phi^+(-k)]^TC\psi(k) \non
&=& \sum_{k>0}  \overline{\psi}_C(k)\Phi^+(k)\psi(k) \, , 
\eea
where, in the last step, we have used $C[\Phi^+(-k)]^TC=-\Phi^+(k)$, which can be seen as follows. In position space we have
\bea
\int_{x,y}\overline{\psi}_C(x)\Phi^+(x-y)\psi(y) &=& -\int_{x,y}\psi^T(y)[\Phi^+(x-y)]^T\overline{\psi}_C^T(x) \non[2ex]
&=& -\int_{x,y}\overline{\psi}_C(x)C[\Phi^+(y-x)]^TC\psi(y) \, , 
\eea
i.e., $\Phi^+(x-y)=-C[\Phi^+(y-x)]^TC$. With the Fourier transform (\ref{fourierD}) this yields 
$C[\Phi^+(-k)]^TC=-\Phi^+(k)$. 

Using Eq.\ (\ref{help1}) in Eq.\ (\ref{diagk}), and analogously for $\overline{\psi}(x)\Phi^-(x,y)\psi_C(y)$,  yields the interaction part of the action in momentum space 
\bea \label{interaction}
&&\frac{1}{2}\int_{x,y}[\overline{\psi}_C(x)\Phi^+(x,y)\psi(y) + \overline{\psi}(x)\Phi^-(x,y)\psi_C(y)]\non[2ex]
&&= \frac{1}{T}\sum_{k>0} [ \overline{\psi}_C(k)\Phi^+(k)\psi(k)+ \overline{\psi}(k)\Phi^-(k)\psi_C(k)] \, .
\eea
Finally, we need to write the tree-level contribution in terms of fermions and charge-conjugate fermions. With the definition of the tree-level propagator in 
position space (\ref{fermtree}) and the Fourier transformed fields from Eq.\ (\ref{fourier1}) we obtain  
\bea
\int_{x,y}\overline{\psi}(x) G_0^{-1}(x,y) \psi(y) &=& \frac{1}{T}\sum_k\overline{\psi}(k)(\gamma^\mu k_\mu+\mu\gamma^0-m)\psi(k) \, .
\eea
Again, we need to divide the sum over $k$ into two sums, one over $k>0$ and one over $k<0$ and rewrite the latter as
\bea
\sum_{k<0}\overline{\psi}(k)(\gamma^\mu k_\mu+\mu\gamma^0-m)\psi(k) &=& \sum_{k>0}\overline{\psi}(-k)(-\gamma^\mu k_\mu+\mu\gamma^0-m)\psi(-k) \non
&=& \sum_{k>0}\psi_C^T(k)C(-\gamma^\mu k_\mu+\mu\gamma^0-m)C\overline{\psi}_C^T(k) \non
&=& \sum_{k>0}\left[\psi_C^T(k)C(-\gamma^\mu k_\mu+\mu\gamma^0-m)C\overline{\psi}_C^T(k)\right]^T \non
&=& -\sum_{k>0}\overline{\psi}_C(k)C(-\gamma_\mu^T k^\mu+\mu\gamma^0-m)C\psi_C(k) \non
&=& \sum_{k>0}\overline{\psi}_C(k)(\gamma^\mu k_\mu-\mu\gamma^0-m)\psi_C(k) \, , 
\eea
where $C\gamma_\mu^TC = \gamma_\mu$ has been used. Consequently,
\bea \label{tree2}
&&\int_{x,y}\overline{\psi}(x) G_0^{-1}(x,y) \psi(y) \non[2ex] = 
&&\frac{1}{T}\sum_{k>0} \Big\{\overline{\psi}(k)[G_0^+(k)]^{-1}\psi(k)+\overline{\psi}_C(k)[G_0^-(k)]^{-1}\psi_C(k)\Big\} \, , \label{treeG}
\eea
with the propagators for fermions and charge-conjugate fermions in momentum space,
\be \label{G0rel}
[G_0^\pm(k)]^{-1} = \gamma^\mu k_\mu \pm \gamma_0\mu - m \, .
\ee
With the new integration measure (\ref{measure}), the interaction part (\ref{interaction}) and the tree-level part (\ref{tree2}),
the partition function (\ref{Z2}) can be written in the following compact way,
\be \label{ZNZZ}
Z = {\cal N}Z_{\rm bosons}Z_0\int{\cal D}\overline{\Psi}\,{\cal D}\Psi\,\exp\left[\sum_{k>0}\overline{\Psi}(k)\frac{{\cal S}^{-1}(k)}{T}
\Psi(k)\right] \, . 
\ee
Here we have abbreviated the integration measure,  
\be
{\cal D}\overline{\Psi}\,{\cal D}\Psi \equiv \prod_{k>0}d\overline{\psi}(k)\,d{\psi}_C(k)d\psi(k)\,d\overline{\psi}_C(k) \, , 
\ee
introduced the new spinors
\be
\Psi \equiv \left(\begin{array}{c} \psi \\ \psi_C \end{array}\right) \, , \qquad \overline{\Psi} \equiv (\overline{\psi}, \overline{\psi}_C) \, , 
\ee
and the inverse propagator 
\be \label{fullSinv}
{\cal S}^{-1}(k) = \left(\begin{array}{cc} [G_0^+(k)]^{-1} & \Phi^-(k) \\[2ex] \Phi^+(k) & [G_0^-(k)]^{-1} \end{array}\right) \, .
\ee
The two-dimensional space that has emerged from the introduction of charge-conjugate spinors is called {\it Nambu-Gorkov space}. 
Together with the $4\times 4$ structure of Dirac space, ${\cal S}^{-1}$ is an $8\times 8$ matrix.

\subsection{Gap equation}

We can write Eq.\ (\ref{fullSinv}) in the form of a Dyson-Schwinger equation, where the inverse propagator is decomposed into a non-interacting part ${\cal S}_0^{-1}$
and a self-energy $\Sigma$,
\be \label{DysonS}
{\cal S}^{-1} = {\cal S}_0^{-1} + \Sigma \, , 
\ee
with 
\be \label{inversetree}
{\cal S}_0^{-1} = \left(\begin{array}{cc} [G_0^+]^{-1} & 0  \\ 0 & [G_0^-]^{-1} \end{array}\right) \, , \qquad  
\Sigma = \left(\begin{array}{cc} 0  & \Phi^- \\ \Phi^+ & 0  \end{array}\right) \, .
\ee
The propagator ${\cal S}$ is computed by inverting the matrix (\ref{fullSinv}),
\be \label{SGF}
{\cal S} = \left(\begin{array}{cc} G^+ &  F^-  \\ F^+  & G^- \end{array}\right) \, , 
\ee
with 
\begin{subequations}
\bea
G^\pm&\equiv& \left([G_0^\pm]^{-1}-\Phi^\mp G_0^\mp \Phi^\pm\right)^{-1} \, , \label{Gpm}\\[2ex]
F^\pm &\equiv& -G_0^\mp\Phi^\pm G^\pm \, .
\eea
\end{subequations}
This form of the propagator is easily verified by computing ${\cal S}^{-1}{\cal S}=1$. 
The off-diagonal elements $F^\pm$ of the full propagator are called {\it anomalous propagators}.
They describe the propagation of a fermion that is converted into a charge-conjugate fermion or vice versa. This is possible through 
the Cooper pair condensate which can be thought of as a reservoir of fermions and fermion-holes. This is another way of saying that 
the symmetry associated to charge conservation is spontaneously broken. More formally speaking, $\Phi^\pm$ is not invariant 
under $U(1)$ rotations: with the definition (\ref{Phipm}) and the transformation of the fermion spinor $\psi \to e^{-i\alpha}\psi$ we have
\be \label{Phitransform}
\Phi^\pm \to e^{\pm 2i\alpha}\Phi^\pm \, .
\ee
Consequently, $\Phi^\pm$ transforms non-trivially under symmetry transformations of the Lagrangian. This is just like the Bose condensate in our discussion of bosonic 
superfluidity. In fact, here the order parameter $\Phi^\pm$ is invariant under multiplication of both fermion spinors with $-1$, $\alpha=\pi$ in Eq.\ (\ref{Phitransform}),
and thus there is a residual subgroup $\mathbb{Z}_2$. This is a difference to the bosonic case, where the residual group was the trivial group, which only contains 
the unit element.

\begin{figure} [t]
\begin{center}
\includegraphics[width=0.85\textwidth]{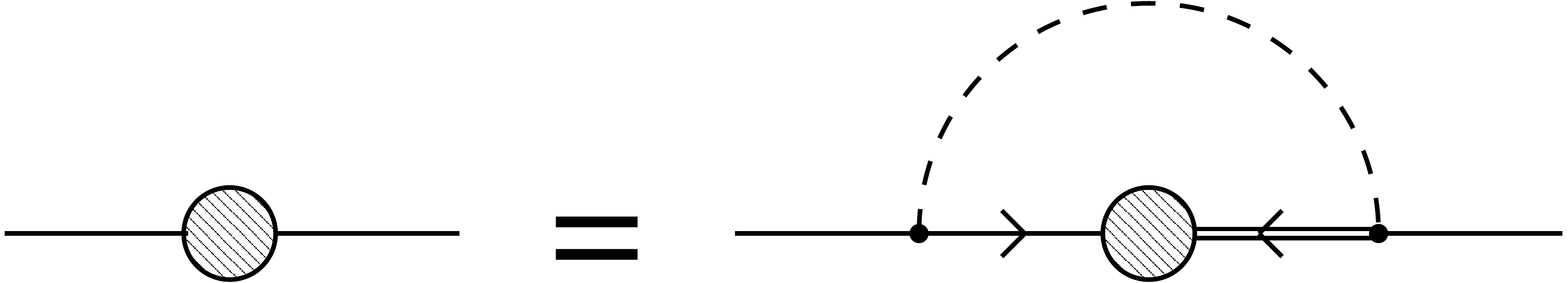}
\caption{Diagrammatic version of the gap equation (\ref{gapeq1}). The hatched circle is the gap function $\Phi^+$, the dashed line is the boson propagator $D$, while
the single and double lines represent the tree-level propagator $G_0^-$ and the full propagator $G^+$. The opposite charges in these two fermionic propagators
is indicated by the different directions of the arrows. The loop on the right-hand side contains the 
anomalous propagator $F^+ = -G_0^-\Phi^+ G^+$.   
}
\label{figgapeq}
\end{center}
\end{figure}

To derive the gap equation, we notice that the fermionic propagator ${\cal S}$ is, on the one hand, given by Eq.\ (\ref{SGF}). On the other hand, 
we require ${\cal S}$ to have the usual form of a propagator, extended to Nambu-Gorkov space, 
\be
{\cal S}(x,y) = -\langle\Psi(x)\overline{\Psi}(y)\rangle  = -\left(\begin{array}{cc} \langle\psi(x)\overline{\psi}(y)\rangle & \langle\psi(x)\overline{\psi}_C(y)\rangle \\[2ex]
 \langle\psi_C(x)\overline{\psi}(y)\rangle &  \langle\psi_C(x)\overline{\psi}_C (y)\rangle\end{array}\right) \, .
\ee
For instance, taking the lower left component, this implies $F^+(x,y) = -\langle\psi_C(x)\overline{\psi}(y)\rangle$ (we could also consider the upper
right component, the resulting gap equation would be equivalent).
Inserting this relation into Eq.\ (\ref{gapplus}) yields $\Phi^+(x,y)=-g^2D(x,y)F^+(x,y)$, which becomes in Fourier space 
\bea
\frac{T}{V}\sum_p e^{-ip\cdot(x-y)}\Phi^+(p) &=& -g^2\frac{T^2}{V^2}\sum_{q,k}e^{-i(q+k)\cdot(x-y)}D(q)F^+(k) \non[2ex]
&=& -g^2\frac{T^2}{V^2}\sum_{p,k}e^{-ip\cdot(x-y)}D(p-k)F^+(k) \, , 
\eea
where again we have assumed translational invariance, and where, in the second step, we have introduced the new summation variable $p=q+k$. 
We can now compare the coefficients of the Fourier series in $p$ to 
obtain
\be \label{gapeq1}
\Phi^+(p) = -g^2\frac{T}{V}\sum_kD(p-k)F^+(k) \, .
\ee
This is the gap equation, which is shown in diagrammatic form in Fig.\ \ref{figgapeq}. 
Since $F^+(k)$ contains the gap function $\Phi^+(k)$, the gap equation is an integral equation for the gap function.

\section{Quasiparticle excitations}
\label{sec:quasi}

Next we need to compute the various components of the propagator in Nambu-Gorkov space explicitly. This
is necessary for solving the gap equation, but even before doing so we will learn something about the structure of the 
fermionic quasiparticles. On general grounds we expect a Goldstone mode, i.e., a bosonic quasiparticle, due to the spontaneous breaking of the $U(1)$ symmetry. 
We shall discuss this mode in chapter \ref{sec:lowenergy}, and the absence of this mode in the case of a gauge symmetry 
in chapter \ref{sec:meissner1}. Here we focus on the fermionic excitations and the solution of the gap equation.

In the following, we restrict ourselves for simplicity to ultra-relativistic fermions, $m=0$. Including a mass renders the 
calculation more complicated, but the essential physics will be captured already in the massless case.  
It is convenient to express the inverse tree-level propagators for massless fermions in terms of energy projectors,
\bea \label{G0m1}
[G_0^\pm]^{-1} &=& \gamma^\mu K_\mu \pm \gamma^0\mu \non[2ex]
&=& \sum_{e=\pm}[k_0\pm(\mu-ek)]\gamma^0\Lambda_k^{\pm e} \, , 
\eea
with
\be
\Lambda_k^e \equiv \frac{1}{2}\left(1+e\gamma^0\vg\cdot\uk\right) \, .
\ee
It is easy to check that $\Lambda_k^+$ and $\Lambda_k^-$ form a complete set of orthogonal projectors,  
\be
\Lambda_k^++ \Lambda_k^- = 1\,, \qquad \Lambda_k^+ \Lambda_k^- = 0 \, , \qquad \Lambda_k^e \Lambda_k^e = \Lambda_k^e \, .
\ee
One benefit of this formulation is that inversion becomes very simple. A matrix of the form $A=\sum_i a_i{\cal P}_i$ with a 
complete set of orthogonal projectors ${\cal P}_i$ and scalars $a_i$, has obviously the inverse $A^{-1}=\sum_i a_i^{-1} {\cal P}_i$. The only small 
difference in our case is the additional matrix $\gamma^0$. But, because $\gamma^0$ and $\Lambda_k^e$ obey the simple commutation relation 
$\gamma^0\Lambda_k^e = \Lambda_k^{-e}\gamma^0$, we easily find 
\be \label{G0}
G_0^\pm = \sum_e\frac{\gamma^0\Lambda_k^{\mp e}}{k_0\pm(\mu-ek)} \, .
\ee
Next, we use the following ansatz for the gap matrix,
\be \label{Phiansatz}
\Phi^\pm(K) = \pm \Delta(K)\gamma^5 \, ,
\ee
with a gap function $\Delta(K)$ which is assumed to be real. Remember that $\Phi^+$ and $\Phi^-$ are related via $\Phi^-=\gamma^0(\Phi^+)^\dagger\gamma^0$, i.e., 
once we make the ansatz $\Phi^+=\Delta\gamma^5$, we obtain the expression for $\Phi^-$. There are various possible Dirac structures of the gap matrix.
The ansatz (\ref{Phiansatz}) respects the overall anti-symmetry of the Cooper pair with respect to exchange of the two fermions and 
corresponds to even-parity, spin-singlet pairing, where fermions of the same chirality form Cooper pairs, see Refs.\ \cite{Pisarski:1999av,Bailin:1983bm}
for a detailed discussion and a complete study of all possible Dirac structures.  
We shall compute the value of the gap $\Delta$ with the help of the gap equation in the next subsection for the 
case where $\Delta(K)$ is constant. First we discuss some properties of the superfluid system for which the actual value of $\Delta$ is not relevant. 
 
\begin{figure} [t]
\begin{center}
\includegraphics[width=0.8\textwidth]{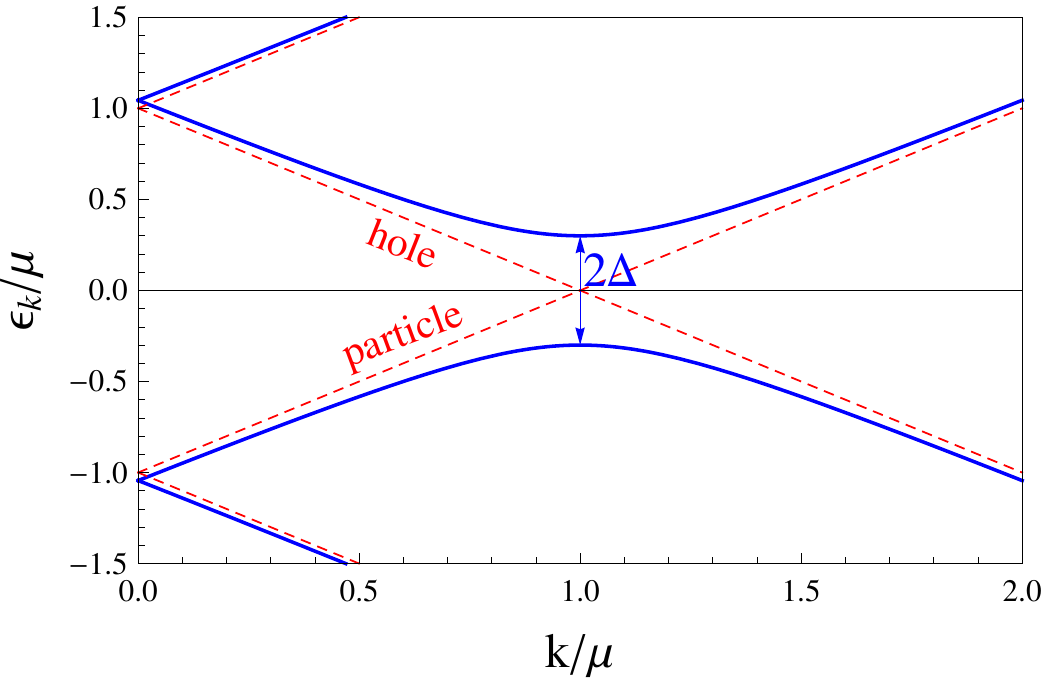}
\caption{(Color online) Fermionic quasiparticle dispersions (solid lines) in the presence of Cooper pairing, which introduces an energy gap $2\Delta$.
The thin dashed lines are the particle and hole dispersions for the unpaired case. Cooper pairing leads to a mixing of particle and hole states. 
}
\label{figgap}
\end{center}
\end{figure}

Inserting Eqs.\ (\ref{G0m1}), (\ref{G0}), (\ref{Phiansatz}) into the expressions for the propagator and charge-conjugate propagator (\ref{Gpm}) 
and using that $\gamma^5$ anti-commutes with the other Dirac matrices, $\{\gamma^5,\gamma^\mu\}=0$, as well as $(\gamma^5)^2=1$, yields
\bea \label{Gexp}
G^\pm(K) &=& \left\{ \sum_e\left[k_0\pm(\mu-ek)-\frac{\Delta^2}{k_0\mp(\mu-ek)}\right]\gamma^0\Lambda_k^{\pm e}\right\}^{-1} \non[2ex]
&=& \sum_e \frac{k_0\mp (\mu-ek)}{k_0^2-(\epsilon_k^e)^2}\,\gamma^0\Lambda_k^{\mp e} \, ,
\eea
with 
\be \label{epskeequiv}
\epsilon_k^e \equiv  \sqrt{(\mu-ek)^2+\Delta^2} \, .
\ee
The anomalous propagators become
\be \label{Fexp}
F^\pm(K) = \pm \sum_e\frac{\Delta(K)\gamma^5\Lambda_k^{\mp e}}{k_0^2-(\epsilon_k^e)^2} \, .
\ee
We see that all components of the Nambu-Gorkov propagator have the same poles, $k_0=\pm \epsilon_k^e$. These are the excitation energies for 
quasi-particles ($e=+$) and quasi-antiparticles ($e=-$) (both for the upper sign) and quasi-holes ($e=+$) and quasi-anti-holes ($e=-$)
(both for the lower sign). We now see that $\Delta$ is indeed an energy gap in the quasiparticle spectrum, see Fig.\ \ref{figgap}. This energy gap
is the reason that there is frictionless charge transport in a fermionic superfluid. 
 
We can gain some further insight into the nature of the Cooper-paired system by computing the 
charge density and the occupation numbers. To this end, we start from the pressure $P$, which is defined as 
\be
P=\frac{T}{V}\ln Z \, .
\ee
With the partition function $Z$ from Eq.\ (\ref{ZNZZ}) we have 
\be \label{Ptrln}
P = \frac{1}{2}\frac{T}{V}\sum_K\Tr\ln{\cal S}^{-1} + \frac{T}{V}\ln Z_0 + \frac{T}{V}\ln Z_{\rm bosons} \, , 
\ee
where the trace in the first term is taken over Nambu-Gorkov and Dirac space. We have evaluated the functional integral in the partition function, 
which is formally the same as for non-interacting fermions. Remember that the additional degree of freedom from the charge-conjugate fermions had resulted 
in restricting the momentum sum to one half space, $K>0$. In equation (\ref{Ptrln}) we sum over all $K$, but have taken care of this
overcounting by multiplying by $\frac{1}{2}$. We are interested in the charge density $n$, which is the derivative of $P$ with respect to the chemical potential $\mu$. 
Therefore, the contribution of $Z_0$, which does not depend explicitly on $\mu$, and the bosonic part are irrelevant for the following, and the charge density becomes 
\bea \label{nNG}
n &=& \frac{1}{2}\frac{T}{V}\sum_K\Tr\left[{\cal S}\frac{\partial {\cal S}^{-1}}{\partial \mu}\right] \non[2ex]
&=&\frac{1}{2}\frac{T}{V}\sum_K\Tr[\gamma^0(G^+-G^-)] \, ,
\eea
where, in the second step, the explicit form of ${\cal S}^{-1}$ (\ref{fullSinv}) has been used and the trace over Nambu-Gorkov space has been performed. 
Inserting the propagators from Eq.\ (\ref{Gexp}) and using
$\Tr[\Lambda_k^e]=2$ yields
\bea
n &=& -2\frac{T}{V}\sum_K\sum_e\frac{\mu-ek}{k_0^2-(\epsilon_k^e)^2} \non[2ex]
&=& 2\sum_e\int\frac{d^3{\bf k}}{(2\pi)^3}\frac{\mu-ek}{2\epsilon_k^e}
\tanh\frac{\epsilon_k^e}{2T} \, . 
\eea
Here we have performed the sum over fermionic Matsubara frequencies, $k_0=-i\omega_n$, with $\omega_n = (2n+1)\pi T$, 
\be \label{Matsusimple}
T\sum_{k_0}\frac{1}{k_0^2-a^2} = -\frac{1}{2a}\tanh\frac{a}{2T} \, ,
\ee
and taken the thermodynamic limit $\frac{1}{V}\sum_{\bf k}\to \int\frac{d^3{\bf k}}{(2\pi)^3}$. 
Now, with $\tanh\frac{x}{2T} = 1-2f(x)$, where 
\be
f(x) = \frac{1}{e^{x/T}+1}  
\ee
is the Fermi distribution 
function, we can write the result as 
\bea \label{nbog}
n &=& 2\sum_e e \int\frac{d^3{\bf k}}{(2\pi)^3}\left[\frac{1}{2}\left(1-\frac{k-e\mu}{\epsilon_k^e}\right)
+\frac{k-e\mu}{\epsilon_k^e}f(\epsilon_k^e)\right]\non[2ex]
&=&2\sum_e e \int\frac{d^3{\bf k}}{(2\pi)^3}\Big\{ |u_k^e|^2 f(\epsilon_k^e) + |v_k^e|^2 [1-f(\epsilon_k^e)]\Big\}
\, ,
\eea
with
\be
|u_k^e|^2 \equiv \frac{1}{2}\left(1+\frac{k-e\mu}{\epsilon_k^e}\right) \, , \qquad  
|v_k^e|^2 \equiv \frac{1}{2}\left(1-\frac{k-e\mu}{\epsilon_k^e}\right) \, .
\ee
The second line of Eq.\ (\ref{nbog}) shows that the quasiparticles are mixtures of fermions with occupation $f$ and and fermion-holes with 
occupation $1-f$, where the  
mixing coefficients are the so-called {\it Bogoliubov coefficients} $|u_k^e|^2$ and $|v_k^e|^2$ with $|u_k^e|^2+ |v_k^e|^2 =1$.

We may first check that this expression reduces to the usual charge density of free fermions if we set the gap to zero. In this case,
$\epsilon_k^e = |k-e\mu|$ and we find 
\bea
n_{\Delta=0} &=& 2\sum_e e \int\frac{d^3{\bf k}}{(2\pi)^3}\left[\Theta(e\mu-k)+{\rm sgn}\,(k-e\mu)\,f(|k-e\mu|)\right] \non[2ex]
&=&2\sum_e e \int\frac{d^3{\bf k}}{(2\pi)^3}f(k-e\mu)
\, ,
\eea
as expected. To see the second step, consider the momentum integral over the two intervals $[0,e\mu]$ and $[e\mu,\infty]$ separately and use $1-f(x)=f(-x)$.

\begin{figure} [t]
\begin{center}
\includegraphics[width=0.8\textwidth]{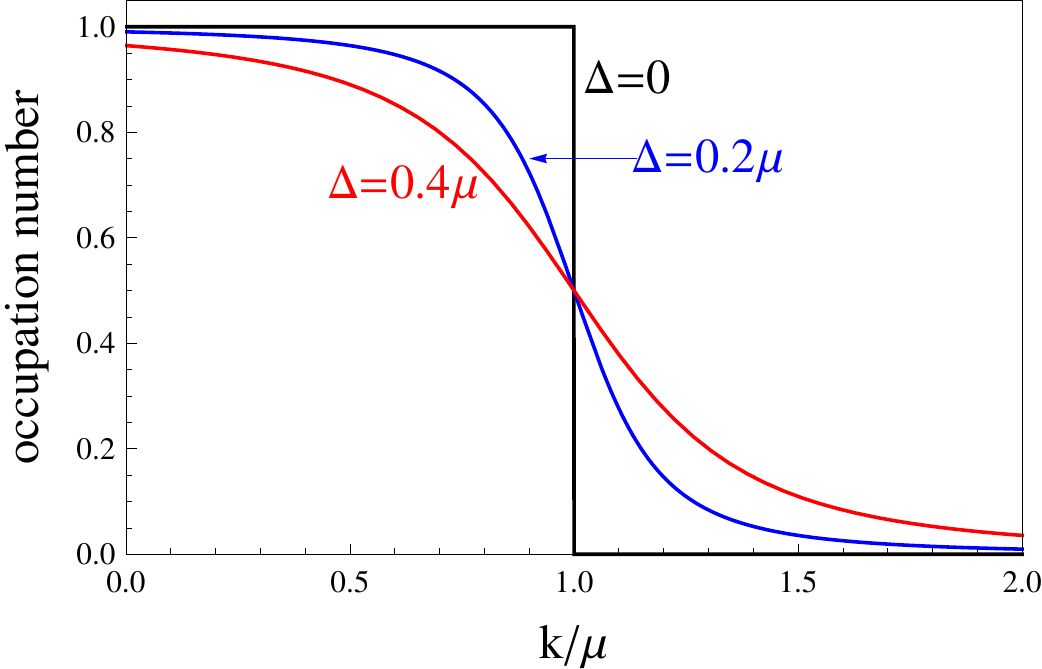}
\caption{(Color online) Zero-temperature fermion occupation number from Eq.\ (\ref{occupy}) for various values of the energy gap $\Delta$. 
}
\label{figocc}
\end{center}
\end{figure}

Finally, let us take the zero-temperature limit of Eq.\ (\ref{nbog}). Since $\epsilon_k^e>0$, we have 
$f(\epsilon_k^e)\to \Theta(-\epsilon_k^e)=0$ at zero temperature, and thus, neglecting the contribution
of the antiparticles,
\be \label{occupy}
n_{T=0}\simeq 2\int\frac{d^3{\bf k}}{(2\pi)^3} \frac{1}{2}\left[1-\frac{k-\mu}{\sqrt{(k-\mu)^2+\Delta^2}}\right] \, .
\ee
We plot the integrand, i.e., the occupation number in the presence of a gap, in Fig.\ \ref{figocc}. We see that the gap has a similar
effect as a nonzero temperature: the sharp Fermi surface of the non-interacting system becomes a smeared surface in the superfluid system.

\section{Solving the gap equation}
\label{sec:BCSgap}

\begin{figure} [t]
\begin{center}
\includegraphics[width=0.75\textwidth]{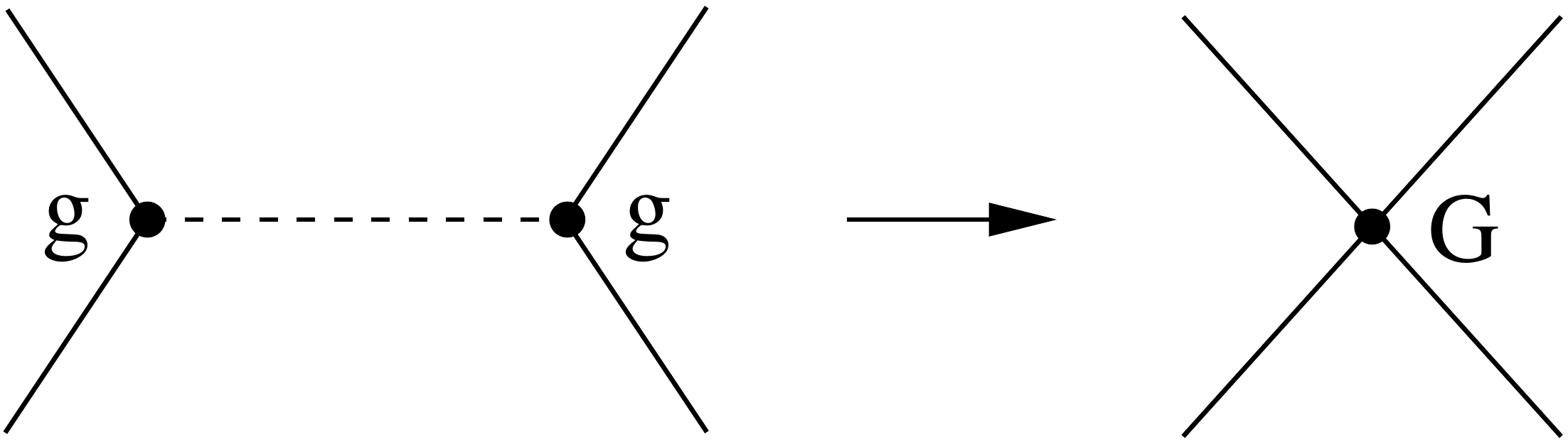}
\caption{General interaction via boson exchange and fundamental coupling $g$ (left) and point-like approximation with effective coupling $G$ (right). The
point-like four-fermion interaction is used in the solution of the gap equation in Sec.\ \ref{sec:BCSgap}. Long-range effects of the interaction
become important for instance in QCD, where the dashed line would correspond to a gluon propagator, see Sec.\ \ref{sec:quarkmatter}.
}
\label{figpointlike}
\end{center}
\end{figure}

Inserting the ansatz for the gap matrix (\ref{Phiansatz}) and the anomalous propagator (\ref{Fexp}) into the gap equation (\ref{gapeq1}),
we obtain
\be
\Delta(P)\gamma^5 = -g^2\frac{T}{V}\sum_K D(P-K) \frac{\Delta(K)\gamma^5\Lambda_k^{-}}{k_0^2-\epsilon_k^2} \, ,
\ee
where we have neglected the antiparticle contribution and abbreviated $\epsilon_k\equiv \epsilon_k^+$.
To get rid of the matrix structure, we multiply both sides of the equation with $\gamma^5$ and take the trace over Dirac 
space, 
\be
\Delta(P) = -\frac{g^2}{2}\frac{T}{V}\sum_KD(P-K) \frac{\Delta(K)}{k_0^2-\epsilon_k^2} \,,  
\ee
where we have used $\Tr[\Lambda_k^{e}]=2$.
Now let us assume that the interaction between the fermions is point-like, i.e., 
the inverse boson propagator can be approximated by the boson mass squared, $D^{-1}(Q) = -Q^2+M^2 \simeq M^2$.
In this case, $\Delta(P)$ becomes independent of $P$ and after performing the Matsubara sum we obtain
\be \label{gapeq2}
\Delta = G  \int\frac{d^3{\bf k}}{(2\pi)^3}\frac{\Delta}{2\epsilon_k}\tanh\frac{\epsilon_k}{2T} \, ,
\ee
with the effective coupling constant 
\be
G = \frac{g^2}{2 M^2} \, . 
\ee
Note that, while $g$ is dimensionless, $G$ has mass dimensions $-2$. The approximation of the interaction via exchange of a boson by 
a four-fermion interaction\footnote{Had we only been interested in the gap equation with four-fermion interaction, we could have put this simpler interaction 
term into our Lagrangian from the beginning. In fact, we shall do so when we discuss fluctuations around the BCS mean-field solution in chapter \ref{sec:lowenergy},
see Eq.\ (\ref{NJLLag}). The resulting model is called Nambu-Jona-Lasinio (NJL) model \cite{Nambu:1961tp,Nambu:1961fr}, and has been 
used for instance as a simplified description of Cooper pairing in quark matter \cite{Buballa:2003qv}.} is shown in Fig.\ \ref{figpointlike}. 

Let us first discuss the solution of Eq.\ (\ref{gapeq2}) for zero temperature, where $\tanh\frac{\epsilon_k}{2T}=1$. We also assume that 
the interaction is only nonzero for fermions in a small vicinity around the Fermi surface $[\mu-\delta,\mu+\delta]$ with 
\be\label{delta}
\Delta_0\ll\delta\ll\mu \, ,
\ee
where $\Delta_0\equiv \Delta(T=0)$. This assumption corresponds to the weak coupling limit because Pauli blocking does not allow for any scattering 
processes of fermions deep in the Fermi sea. The stronger the coupling, the more fermions in the Fermi sea become relevant.
Within this approximation, the gap equation becomes
\be \label{gapeqT0}
\Delta_0 \simeq \frac{\mu^2 G}{2\pi^2}\int_0^\delta d\xi\,\frac{\Delta_0}{\sqrt{\xi^2+\Delta_0^2}} \, , 
\ee
where we have approximated $dk\,k^2\simeq dk\,\mu^2$, introduced the new integration variable $\xi = k-\mu$, and then have used the symmetry of the 
integrand with respect to $\xi\to -\xi$, such that we can restrict ourselves to the interval $\xi\in [0,\delta]$ and multiply the result by 2. Obviously, $\Delta_0=0$
is one solution of the equation. To find the nontrivial solution, we divide both sides of the equation by $\Delta_0$. Then we see that there must be
a nonzero $\Delta_0$ for any coupling $G>0$, no matter how small: if $\Delta_0$ were zero, there would be a logarithmic divergence from the lower boundary,
which corresponds to the Fermi surface. This is the essence of the instability towards Cooper pairing. 
The reason for this infrared divergence is that the integral has essentially become one-dimensional due to the restriction
to momenta within a small vicinity around the Fermi surface. Therefore, a formal way of saying why the Fermi surface is unstable with respect to the formation 
of a Cooper pair condensate is that, at weak coupling, there is an effective dimensional reduction of the dynamics of the system from 3+1 to 1+1 dimensions. 

It is now easy to compute $\Delta_0$. With 
\be
\int \frac{d\xi}{\sqrt{\xi^2+\Delta_0^2}} = \ln\left[2\left(\xi+\sqrt{\xi^2+\Delta_0^2}\right)\right] \, , 
\ee
we find 
\be \label{gapT0}
\Delta_0 \simeq 2\delta \exp\left(-\frac{2\pi^2}{G\mu^2}\right) \, .
\ee
This is the famous result for the BCS gap. It shows the dependence of the energy gap on the coupling: at weak coupling (only in this regime is our mean-field
approximation valid, and only in this regime are we allowed to restrict ourselves to a small vicinity of the Fermi surface) the energy gap is exponentially 
suppressed, with the fermion-boson coupling $g$ appearing quadratically in the exponential, $G\propto g^2$.
This result is non-perturbative because there is no Taylor expansion around $G=0$. We can also see the non-perturbative nature from the structure of the 
gap equation itself, see for instance the diagrammatic form in Fig.\ \ref{figgapeq}: the gap appearing in the loop on the right-hand side of the equation 
is itself determined by a loop that contains the gap etc. Therefore, we effectively resum infinitely many diagrams rather than computing diagrams up to a 
fixed power in the coupling constant. 
The BCS gap equation is thus a nice example to illustrate that taking the weak coupling limit does not necessarily allow for a perturbative 
calculation.


We may also use the gap equation to compute the critical temperature $T_c$ for the superconducting phase transition. In BCS theory, this phase transition 
is of second order, i.e., the gap vanishes continuously at the critical point. 
Therefore, we may use the gap equation and imagine we are sitting at a point just below the critical temperature. At this point, the gap is still nonzero, and we may 
divide Eq.\ (\ref{gapeq2}) by $\Delta$. Then we can take the limit $\Delta\to 0$ in the resulting equation to obtain an equation for the critical temperature,
\be
1 \simeq \frac{G\mu^2}{2\pi^2} \int_0^\delta \frac{d\xi}{\xi}\tanh\frac{\xi}{2T_c} \, .
\ee
With the new integration variable $z=\xi/(2T_c)$ and after integration by parts we obtain
\bea
\frac{2\pi^2}{G\mu^2} &=& \ln z\, \tanh z\Big|_0^{\delta/(2T_c)} -\int_0^{\delta/(2T_c)} dz\, \frac{\ln z}{\cosh^2z} \non[2ex]
&\simeq & \ln\frac{\delta}{2T_c} - \underbrace{\int_0^\infty dz\,\frac{\ln z}{\cosh^2z}}_{\displaystyle{-\gamma+\ln\frac{\pi}{4}}}  \,  ,
\eea
where $\gamma\simeq 0.577$ is the Euler-Mascheroni constant, and, in the second step,
we have used that $\delta\gg T_c$. This assumption is justified because $T_c$ will turn out to be of the same order as $\Delta_0$, and we have already 
assumed that $\delta \gg \Delta_0$. 
Solving the resulting equation for $T_c$ and using Eq.\ (\ref{gapT0}) yields
\be \label{TcD0}
T_c = \frac{e^\gamma}{\pi} \Delta_0 \simeq 0.57\, \Delta_0 \, ,
\ee
i.e., the critical temperature of a BCS superfluid or superconductor is about half the zero-temperature gap. (Amusingly, $\gamma$ and $e^\gamma/\pi$ have almost the same 
numerical value.) As an exercise, you may solve the gap equation
numerically for all temperatures below $T_c$. 

\section{Examples}
\label{sec:examples}

In the beginning of the chapter we have mentioned several systems that exhibit Cooper pairing. Our derivation of the gap equation has been done in 
a more or less specific setting that cannot account for the details of all these systems. Nevertheless, our gap equation is sufficiently generic that 
we can now, with very simple modifications, discuss various physical systems separately.

\subsection{Electronic superconductor}

In a superconducting metal or alloy, the fermions that form Cooper pairs are electrons and their dispersion is non-relativistic, i.e., we have to 
replace our relativistic quasiparticle dispersion by
\be
\epsilon_k = \sqrt{\xi_k^2 +\Delta^2} \, , \qquad \xi_k \equiv \frac{k^2}{2m}-\mu \, .
\ee
The interaction is given by the exchange of phonons, i.e., in general one has to take into account the specific structure of the phonon propagator.
However, as an approximation, we can keep the structure of the gap equation (\ref{gapeq2}), where a point-like interaction is assumed. The cutoff $\delta$ that we have 
introduced above, is now given by the Debye frequency $\omega_D$, which is a natural cutoff frequency determined by the ion crystal. Consequently, 
at zero temperature we can write 
\bea
1 &=& G\int\frac{d^3{\bf k}}{(2\pi)^3}\frac{1}{2\sqrt{\xi_k^2+\Delta_0^2}} \non[2ex]
&=& G\int_{-\omega_D}^{\omega_D}d\xi\int\frac{d^3{\bf k}}{(2\pi)^3}\delta(\xi-\xi_k)\frac{1}{2\sqrt{\xi_k^2+\Delta_0^2}} \non[2ex]
&=& G\int_{-\omega_D}^{\omega_D} d\xi\,\frac{N(\xi)}{2\sqrt{\xi^2+\Delta_0^2}} \, , 
\eea
with the density of states 
\be
N(\xi) \equiv \left[\frac{k^2}{2\pi^2}\left(\frac{\partial \xi_k}{\partial k}\right)^{-1}\right]_{k=k(\xi)} \, , 
\ee
where $k(\xi)$ is the solution of $\xi=\xi_k$. Again, we use that the integral is dominated by a small vicinity around
the Fermi surface to write 
\be
1\simeq G\,N(0)\int_0^{\omega_D}\frac{d\xi}{\sqrt{\xi^2+\Delta_0^2}} \, , \qquad N(0) = \frac{k_Fm}{2\pi^2} \, ,
\ee
with the Fermi momentum $k_F = \sqrt{2m\mu}$. Thus, in complete analogy to above,
\be
\Delta_0 \simeq 2\omega_D\exp\left[-\frac{1}{G\,N(0)}\right] \, . 
\ee
This derivation shows that the factor $\mu^2/(2\pi^2)$, which appears in the exponential of the 
relativistic version, Eq.\ (\ref{gapT0}), is nothing but the density of states at the Fermi surface. In both non-relativistic
and ultra-relativistic cases we can express the density of states at the Fermi surface in the universal form $k_F^2/(2\pi^2 v_F)$, with the 
Fermi velocity $v_F$ introduced in Eq.\ (\ref{fermivel}). This means that the smaller the Fermi velocity the larger the density of states at the 
Fermi surface and thus the larger the energy gap $\Delta$.

\subsection{Anisotropic superfluid}

In the situation discussed so far, the order parameter for superfluidity breaks an internal $U(1)$ spontaneously, but not rotational invariance. 
There are systems, however, where rotational symmetry is spontaneously broken by a Cooper pair condensate. One example is 
superfluid $^3$He, where the order parameter is a $3\times 3$ matrix in the space of spin and angular momentum. In this case, various different 
phases are conceivable, characterized by different residual symmetry groups \cite{vollhardt}. One of these phases is the so-called {\it A phase}, where the energy 
gap turns out to 
be anisotropic in momentum space. Another example for anisotropic Cooper pairing is quark matter where quarks of the same flavor form Cooper pairs. 
In this case, Cooper pairs carry nonzero total angular momentum, and phases not unlike the ones in superfluid $^3$He have been 
predicted \cite{Schafer:2000tw,Alford:2002rz,Schmitt:2004et}. 

It is beyond the scope of this course to go into the details of these systems. We rather model a system with an anisotropic gap 
by choosing the following ansatz for the gap matrix that has a preferred direction, say the 3-direction,
\be \label{Phianiso}
\Phi^\pm = \Delta \hat{k}_3 \, . 
\ee
In our approach there is no reason why the ground state should be anisotropic, i.e., Eq.\ (\ref{Phianiso}) is a solution to the gap equation, 
but if we were to compute the free energy of the corresponding phase, we would find it to be larger than the one of the isotropic phase, i.e., the anisotropic
phase would be disfavored. Nevertheless, 
we shall compute the relation between the critical temperature and the zero-temperature gap and will see that it is modified compared to the standard BCS 
relation (\ref{TcD0}). This modification is applicable to the more complicated scenarios mentioned above where the anisotropic phase {\it is} favored.

With the ansatz (\ref{Phianiso}), we first determine the dispersion relation which, repeating the calculation that leads to Eq.\ (\ref{epskeequiv}),
turns out to be 
\be
\epsilon_k = \sqrt{(k-\mu)^2+\hat{k}_3^2\Delta^2} \, .
\ee
This dispersion shows that there are directions in momentum space where the quasifermions are ungapped. More precisely, the gap 
function $\hat{k}_3\Delta$ has a nodal line at the equator of the Fermi sphere. The anomalous propagators become
\be
F^\pm = -\sum_{e=\pm}\frac{\Delta \hat{k}_3\Lambda_k^{\mp e}}{k_0^2-(\epsilon_k^e)^2} \, , 
\ee
and thus the gap equation reads
\be
\Delta\hat{p}_3 = 2G\frac{T}{V}\sum_K\frac{\Delta \hat{k}_3\Lambda_k^-}{k_0^2-\epsilon_k^2} \, .
\ee
To deal with the angular dependence, we multiply both sides 
with $\hat{p}_3\Lambda_p^+$, take the trace over Dirac space and take the angular average with respect to the direction of the 
external vector ${\bf p}$,
\be
\langle \hat{p}_3^2\rangle_p = G\int\frac{d^3{\bf k}}{(2\pi)^3}\frac{\hat{k}_3\left\langle\hat{p}_3\uk\cdot\up\right\rangle_p}{2\epsilon_k} \tanh\frac{\epsilon_k}{2T}
\, , 
\ee
where $\langle  - \rangle_p \equiv \int\frac{d\Omega_p}{4\pi}$. We have performed the Matsubara sum, divided both sides of the equation by $\Delta$, and 
dropped the first term of the Dirac trace $\Tr[\Lambda_k^{-}\Lambda_p^+]= 1-\uk\cdot\up$ which vanishes upon angular integration. 

It is left as an exercise to work out the details of evaluating this gap equation at $T=0$ and at the critical point. As a result, one finds
a modified relation between $\Delta_0$ and $T_c$,
\bea
T_c &=& \frac{e^\gamma}{\pi}\Delta_0 \exp\left[\frac{1}{2}\frac{\langle \hat{k}_3^2 \ln \hat{k}_3^2\rangle}{\langle \hat{k}_3^2\rangle}\right] \non[2ex] 
&=& \frac{e^\gamma}{\pi}e^{-\frac{1}{3}}\Delta_0 \simeq 0.717 \frac{e^\gamma}{\pi}\Delta_0\, .
\eea
In this anisotropic case, the meaning of $\Delta_0$ is of course a bit different: the energy gap at zero temperature is $\hat{k}_3\Delta_0$, i.e., $\Delta_0$ 
is the maximal gap, and for all directions in momentum space except for the 3-direction the actual gap is smaller. 

\subsection{Color superconductor}
\label{sec:quarkmatter}

Cooper pairing in quark matter is called color superconductivity because a quark-quark Cooper pair carries color
charge and thus breaks the color gauge group spontaneously, in analogy to an electronic superconductor where the Cooper pairs carry electric charge. For instance,
a Cooper pair of a red and a blue quark carries color charge anti-green (because we know that a baryon composed of a red, blue, and green 
quark is color neutral). At sufficiently large densities, quarks are weakly interacting because of asymptotic freedom, which is a fundamental property 
of QCD \cite{Gross:1973id,Politzer:1973fx}.
In this case, the attractive interaction between quarks is provided by one-gluon exchange. Therefore, in generalization of Eq.\ (\ref{gapeq1}), 
the QCD gap equation can be written as \cite{Alford:2007xm}
\be
\Phi^+(P) = g^2\frac{T}{V}\sum_K\gamma^\mu T_a^T F^+(K)\gamma^\nu T_b D_{\mu\nu}^{ab}(P-K) \, , 
\ee
where $g$ is the QCD coupling constant, $T_a$ ($a=1,\ldots,8$) the Gell-Mann matrices, and $D_{\mu\nu}^{ab}$ the gluon propagator.
The main differences to the cases discussed so far are $(i)$ the larger number of fermionic degrees of freedom (color \& flavor) and $(ii)$ the specific form 
of the gluonic interaction.

Regarding point $(i)$, $\Phi^+$ is not only a matrix in Dirac space but also in color and flavor space, i.e., it is a $4N_cN_f\times 4N_cN_f$ matrix. 
Therefore, by choosing an ansatz for the gap matrix, one has to ``guess'' which quarks pair with which other quarks. It is beyond the scope of this course to 
discuss the various possible pairing patterns. We only mention the most symmetric pairing pattern in which {\it all} quarks are involved in pairing. The 
resulting phase is called color-flavor locked (CFL) phase \cite{Alford:1998mk} and is the ground state of three-flavor QCD at sufficiently large densities.
The reason for the name is that the CFL phase is invariant under simultaneous color and flavor transformations, i.e., color and flavor degrees of freedom 
become ``locked'' in a certain sense.

Ignoring all complications from the color-flavor structure, we are still left with point $(ii)$, the effect of the fundamental QCD interaction. 
We shall not go into the technical details of this point, see for instance Refs.\ \cite{Alford:2007xm,arXiv:1001.3294,Son:1998uk,Pisarski:1999tv} for 
a complete discussion, but we briefly point out the main effect of the interaction. 
Instead of Eq.\ (\ref{gapeqT0}) the zero-temperature QCD gap equation becomes
\be
\Delta_p \simeq \frac{g^2}{18\pi^2}\int_0^{\delta} d(k-\mu)\,\frac{\Delta_k}{\epsilon_k}\frac{1}{2}
\ln\frac{b^2\mu^2}{|\epsilon_k^2-\epsilon_p^2|} \, ,
\ee
where the gap depends on three-momentum, and where $b\equiv 256\pi^4[2/(N_fg^2)]^{5/2}$. The different structure arises from the specific
form of the gluon propagator; more precisely, from the long-range interaction mediated by Landau-damped magnetic gluons. 
It has a crucial consequence for the dependence of the gap on the coupling. One finds for the zero-temperature value of the 
weak-coupling gap at the Fermi surface $k=\mu$,
\be
\Delta_0 \simeq  2b\mu\exp\left(-\frac{3\pi^2}{\sqrt{2}g}\right) \, .
\ee
Consequently, the QCD gap is parametrically larger than the BCS gap because of the different power of the fermion-boson coupling constant, 
$e^{-{\rm const}/g}$ vs.\ $e^{-{\rm const}/g^2}$. 

One may also use the QCD gap equation to compute the critical temperature of color superconductivity \cite{Pisarski:1999tv,Brown:1999aq}. 
Even in the presence of long-range gluonic interactions, the BCS relation (\ref{TcD0}) between the critical temperature 
and the zero-temperature gap may still hold. Whether it actually holds, depends on the specific pairing pattern \cite{Schmitt:2002sc}. 
We have already seen that it 
can be violated in an anisotropic phase. It can also be violated if the quasiparticles have different energy gaps. This situation occurs in the CFL phase,
where there are 8 quasiparticles with gap $\Delta$ and 1 quasiparticle with gap $2\Delta$. In this case, one finds 
\be
T_c = \frac{e^\gamma}{\pi} 2^{1/3} \Delta_0 \, .
\ee
Finally, we mention that the phase transition to the color-superconducting phase is only a second order transition at asymptotically large density, where 
gauge field fluctuations can be neglected. Taking these fluctuations into account turns the transition into a first order transition and induces an ${\cal O}(g)$ 
correction to the critical temperature \cite{Giannakis:2004xt}.

\chapter{Meissner effect in a superconductor}
\label{sec:meissner1}

 In chapter \ref{sec:cooper} we have discussed Cooper pairing and argued that this mechanism is valid 
in a superfluid as well as in a superconductor. However, we have not yet discussed the fundamental difference between a superfluid and a superconductor. 
The crucial ingredient in the theoretical description of a superconductor is a gauge symmetry. In this chapter, we will discuss what happens if we replace the global 
symmetry group that is broken spontaneously in a superfluid by a local symmetry group. We shall see that the Goldstone mode, which occurs in every system with 
spontaneously broken global symmetry, is not a physical excitation in a gauged system. In this case, Cooper pairing or Bose-Einstein condensation rather lead 
to a massive gauge boson. We shall discuss in Sec.\ \ref{sec:massive} how the disappearance of the Goldstone mode is related to the 
massiveness of the gauge boson, and then compute this mass explicitly in a field-theoretical calculation in Sec.\ \ref{sec:meissnermass}; for similar 
field-theoretical calculations in the context of quark matter, see Refs.\ \cite{Rischke:2000qz,Rischke:2000ra,Schmitt:2003aa,Alford:2005qw}. 

The meaning of this mass is actually very well known from the phenomenology of a superconductor. Superconductors expel externally applied magnetic fields. 
This is called the Meissner effect. More precisely, this means that the magnetic field is screened in the superconductor like $B\propto e^{-x/\lambda}$ with 
the {\it penetration depth} $\lambda$. In field-theoretical terms, the gauge boson acquires a magnetic mass, called the {\it Meissner mass} $m_M$,
which is nothing but the inverse penetration depth, $m_M=\lambda^{-1}$. Here, the gauge boson that becomes massive is the boson that ``sees''
the charge of the condensate. This is obviously the photon in an electronic superconductor because the electron Cooper pair carries electric charge. In a color
superconductor, (some of) the gluons and (possibly) the photon acquire a Meissner mass because the quark Cooper pairs carry color charge
and (depending on the particular phase) may also carry electric charge.

\section{Massive gauge boson}
\label{sec:massive}

In order to discuss the disappearance of the Goldstone mode in a gauge theory, we start with the Lagrangian 
\be \label{Leta}
{\cal L} = \partial_\mu\varphi^*\partial^\mu\varphi + \eta^2|\varphi|^2-\lambda|\varphi|^4 \, .
\ee
First, let us compare this Lagrangian to the one from chapter \ref{sec:phi4}, see Eq.\ (\ref{lagr}). 
There, we discussed a $\varphi^4$ model with a chemical potential $\mu$ 
and have seen that there is Bose-Einstein condensation when $\mu$ is larger than the boson mass $m$. Here, we revisit this model in a slightly simpler formulation: 
instead of introducing a mass $m$ and a chemical potential $\mu$ we work with the single parameter $\eta^2$ that plays the role of a negative mass squared, such that 
there is Bose-Einstein condensation for $\eta^2>0$. The Lagrangian (\ref{Leta}) is invariant under the global $U(1)$ symmetry  
\be
\varphi \to e^{-i\alpha} \varphi \, .
\ee
Let us introduce polar coordinates,
\be \label{polar}
\varphi = \frac{\rho}{\sqrt{2}}e^{i\psi} \, .
\ee
In this parametrization, the Lagrangian becomes
\be
{\cal L} = \frac{1}{2}\partial_\mu\rho\partial^\mu\rho + \frac{\rho^2}{2}\partial_\mu\psi\partial^\mu\psi + \frac{\eta^2}{2}\rho^2 -\frac{\lambda}{4}
\rho^4 \, .
\ee
Now, as in chapter \ref{sec:phi4}, we separate the condensate $\rho_0$ from the fluctuations and assume the condensate to be constant in 
space and time, $\rho(X) \to \rho_0 + \rho(X)$, where 
\be
\rho_0^2 = \frac{\eta^2}{\lambda} \, .
\ee
[Remember that in chapter \ref{sec:phi4} we had $\rho_0^2=(\mu^2-m^2)/\lambda$.] This yields
\be\label{Gold}
{\cal L} = \frac{1}{2}\partial_\mu\rho\partial^\mu\rho + \frac{(\rho_0+\rho)^2}{2}\partial_\mu\psi\partial^\mu\psi -\eta^2\rho^2-\sqrt{\lambda}\eta\rho^3
-\frac{\lambda}{4}\rho^4+\frac{\eta^4}{4\lambda} \, .
\ee
This shows, in a quick way, that there is a massive mode $\rho$ with mass term $-\eta^2\rho^2$, and a massless mode $\psi$ for which there is only the 
kinetic term. This is the Goldstone
mode, whose complete dispersion we have computed in chapter \ref{sec:phi4}. 
All other terms in the Lagrangian are interaction terms between $\psi$ and $\rho$ or self-interactions of $\rho$ (plus one constant term that
is independent of the dynamical fields $\rho$ and $\psi$).

Now let us extend the symmetry to a gauge symmetry, i.e., we extend the Lagrangian (\ref{Leta}) to 
\be
{\cal L} = (D_\mu\varphi)^*D^\mu\varphi + \eta^2|\varphi|^2-\lambda|\varphi|^4 -\frac{1}{4}F_{\mu\nu}F^{\mu\nu} \, , 
\ee
with the covariant derivative $D_\mu = \partial_\mu-igA_\mu$, the gauge field $A_\mu$, and the field strength tensor 
$F_{\mu\nu} = \partial_\mu A_\nu -\partial_\nu A_\mu$. Now the Lagrangian is invariant under {\it local} $U(1)$ transformations
\be
\varphi \to e^{-i\alpha(X)} \varphi \, , \qquad A_\mu\to A_\mu - \frac{1}{g}\partial_\mu\alpha \, .
\ee
With the parametrization of Eq.\ (\ref{polar}) we obtain
\bea 
{\cal L} &=& \frac{1}{2}\partial_\mu \rho\partial^\mu\rho + \frac{g^2\rho^2}{2}\left(A_\mu-\frac{1}{g}\partial_\mu\psi\right) 
\left(A^\mu-\frac{1}{g}\partial^\mu\psi\right) \non[2ex]
&&+ \frac{\eta^2}{2}\rho^2 -\frac{\lambda}{4}
\rho^4-\frac{1}{4}F_{\mu\nu}F^{\mu\nu} \, .
\eea
There are certain terms that couple the gauge field $A_\mu$ to the angular mode $\psi$. We may define the gauge invariant 
combination 
\be
B_\mu \equiv A_\mu-\frac{1}{g}\partial_\mu\psi \, , 
\ee
as our new gauge field (notice that the phase $\psi$ transforms as $\psi\to \psi - \alpha$). Then, we obtain with the same replacement 
$\rho(X)\to\rho_0+\rho(X)$ as above,
\bea \label{Gauge}
{\cal L} &=& \frac{1}{2}\partial_\mu\rho\partial^\mu\rho + \frac{g^2\rho_0^2}{2}B_\mu B^\mu + g^2 \rho_0\rho B_\mu B^\mu + \frac{g^2}{2}\rho^2B_\mu B^\mu 
-\eta^2\rho^2 \non[2ex]
&&-\sqrt{\lambda}\eta\rho^3-\frac{\lambda}{4}\rho^4+\frac{\eta^4}{4\lambda} -\frac{1}{4}F_{\mu\nu}F^{\mu\nu} \, .
\eea
(The $F_{\mu\nu}F^{\mu\nu}$ term has not changed because of $\partial_\mu A_\nu -\partial_\nu A_\mu =\partial_\mu B_\nu -\partial_\nu B_\mu$.) 
This result has to be compared to Eq.\ (\ref{Gold}): the Goldstone mode has disappeared! It has been ``eaten up'' by the gauge field, which has acquired a mass 
term with mass $g\rho_0$. 

It is instructive to count degrees of freedom in both cases. 
\begin{itemize}
\item {\it global $U(1)$ symmetry}: we start with 2 degrees of freedom, represented by the complex field $\varphi$. After spontaneous
symmetry breaking, we obtain 1 massive mode $\rho$ and 1 massless mode $\psi$.

\item {\it local $U(1)$ symmetry}: here we start with 2 degrees of freedom from the complex field $\varphi$ plus 2 degrees of freedom of the 
massless gauge field $A_\mu$. Spontaneous symmetry breaking leads to 1 massive mode $\rho$ plus 3 degrees of freedom of the now massive
gauge field $B_\mu$. So we end up with the same number of degrees of freedom, $2+2 = 1+3$, as it should be. There is no Goldstone mode.
\end{itemize}

This mechanism is very general and applies also to more complicated gauge groups, for instance in the context of electroweak symmetry 
breaking. The electroweak gauge group is $SU(2)\times U(1)$, and it is spontaneously broken to $U(1)$ through the Higgs mechanism.
There are 3+1 massless gauge fields to start with. Together with a complex doublet, the Higgs doublet, there are $2\times(3+1) + 4 = 12$ 
degrees of freedom. After symmetry breaking, 3 gauge fields
have become massive (eating up the 3 would-be Goldstone modes) and 1 remains massless. The massive gauge fields correspond to the $W^\pm$ and the $Z$ bosons, while
the massless gauge field is the photon. One massive scalar, the Higgs boson, is also left, i.e., 
there are $3\times 3 + 2\times 1 + 1 = 12$ degrees of freedom.

\section{Meissner mass from the one-loop polarization tensor}
\label{sec:meissnermass}

We now make the arguments of the previous subsection more concrete by computing the Meissner mass in a fermionic
superconductor. We thus go back to the formalism developed in chapter \ref{sec:cooper}. This formalism did not have 
a gauge boson which we now have to add. Remember that we had introduced a boson in order to account for the attractive interaction between the fermions.
This was a scalar boson, not a gauge boson. The boson that is responsible for the interaction between the fermions 
may or may not be identical to the gauge boson that becomes massive.  In an electronic superconductor it isn't, the two bosons are the phonon and the photon. 
In a color superconductor, however, the gluons that provide the interaction between the quarks are also the gauge bosons
that acquire a Meissner mass.

\subsection{Gauge boson propagator and screening masses}

Let us start with defining the Meissner mass via the gauge boson propagator. The gauge field contribution to the action is
\be \label{Fmunu}
-\frac{1}{4}\int_X F_{\mu\nu}F^{\mu\nu}  = -\frac{1}{2T^2}\sum_Q A_\mu(-Q)(Q^2g^{\mu\nu}-Q^\mu Q^\nu)A_\nu(Q) \, . 
\ee
Here we are working in the imaginary-time formalism of thermal field theory from the previous chapters, and we have used 
the Fourier transformation of the gauge field
\be
A(X) = \frac{1}{\sqrt{TV}}\sum_Q e^{-iQ\cdot X} A_\mu(Q) \, ,
\ee
with temperature $T$ and three-volume $V$. 
We can read off the inverse gauge boson propagator from Eq.\ (\ref{Fmunu}), 
\be
D_{0,\mu\nu}^{-1}(Q) = Q^2g_{\mu\nu}-\left(1-\frac{1}{\zeta}\right)Q_\mu Q_\nu \, ,
\ee
where we have added a gauge-fixing term in a covariant gauge $\partial_\mu A^\mu=0$ with gauge-fixing parameter $\zeta$. Physical observables must of course be 
independent of $\zeta$. Inversion gives 
\be
D_{0,\mu\nu}(Q) = \frac{g_{\mu\nu}}{Q^2}-(1-\zeta)\frac{Q_\mu Q_\nu}{Q^4} \, .
\ee
(One can easily check that $D_0^{\mu\nu}D_{0,\nu\sigma}^{-1}=g^\mu_{\;\;\;\sigma} = \delta^\mu_\sigma$.) This propagator describes the propagation of 
a gauge boson through vacuum. In a superconductor, we are of course interested in the propagation of the gauge boson through a medium. Therefore, the 
propagator must receive a correction, which is usually written in terms of a self-energy $\Pi_{\mu\nu}$, which is also called {\it polarization tensor},
\be \label{Dinvmunu}
D^{-1}_{\mu\nu}(Q) =D^{-1}_{0,\mu\nu}(Q)+\Pi_{\mu\nu}(Q) \, .
\ee
In our case, the self-energy is determined by the interaction of the gauge boson with the fermions of the superconductor. 
In a rotationally invariant system and due to the tranversality property of the self-energy $Q_\mu \Pi^{\mu\nu}=0$, the 
self-energy in an abelian gauge theory can be written as \cite{lebellac}
\be \label{Pimunu}
\Pi_{\mu\nu}(Q) = {\cal F}(Q)P_{L,\mu\nu} + {\cal G}(Q)P_{T,\mu\nu} \, , 
\ee
with scalar functions ${\cal F}$ and ${\cal G}$ and projection operators $P_L$, $P_T$ that are defined as follows.
The transverse projector is defined via
\begin{subequations}
\bea
P_T^{00}&=&P_T^{0i}=P_T^{i0}=0 \, , \\[2ex]
P_T^{ij}&=&\delta^{ij}-\hat{q}^i\hat{q}^j \, ,
\eea
\end{subequations}
and the longitudinal projector is
\be
P_L^{\mu\nu}=\frac{Q^\mu Q^\nu}{Q^2}-g^{\mu\nu}-P_T^{\mu\nu} \, .
\ee
To be more precise, by longitudinal and transverse we actually mean 3-longitudinal and 3-transverse. Both $P_T$ and $P_L$ are 4-transverse to $Q$, 
i.e., $Q_\mu P_L^{\mu\nu}=Q_\mu P_T^{\mu\nu}=0$, such that obviously $\Pi_{\mu\nu}$ is also 4-transverse to $Q$, as required. We will need the following 
relations, which can easily be checked from the definitions,
\be \label{PLPT}
P^{\mu\nu}_LP_{L,\nu\sigma}=-P^\mu_{L,\sigma} \, , \quad  P^{\mu\nu}_TP_{T,\nu\sigma}=-P^\mu_{T,\sigma}\,, \quad 
 P^{\mu\nu}_LP_{T,\nu\sigma}=P^{\mu\nu}_TP_{L,\nu\sigma}=0 \, .
\ee
From Eq.\ (\ref{Pimunu}) we can express the functions ${\cal F}$ and ${\cal G}$ in terms of certain components of the polarization tensor. 
To obtain an explicit form for ${\cal F}$ we may for instance consider the $\mu=\nu=0$ component, which yields
\be \label{FQ00}
{\cal F}(Q) = \frac{Q^2}{q^2}\Pi_{00}(Q) \, .
\ee
To obtain ${\cal G}$, we multiply 
Eq.\ (\ref{Pimunu}) with $P^{\sigma\mu}_T$ and take the $\sigma=i, \nu=j$ component. This yields 
\be \label{GPi}
{\cal G}(Q) = \frac{1}{2}(\delta_{jk}-\hat{q}_j\hat{q}_k)\Pi_{kj}(Q) \, .
\ee
After inserting Eq.\ (\ref{Pimunu}) into Eq.\ (\ref{Dinvmunu}), the inverse propagator can be written as
\be
D^{-1}_{\mu\nu}(Q) = [{\cal F}(Q)-Q^2]P_{L,\mu\nu}+[{\cal G}(Q)-Q^2]P_{T,\mu\nu} +\frac{1}{\zeta} Q_\mu Q_\nu \, . 
\ee
The formulation in terms of projectors makes the inversion of this expression very easy. The full boson propagator becomes
\be \label{photonprop}
D_{\mu\nu}(Q) = \frac{P_{L,\mu\nu}}{{\cal F}(Q)-Q^2} + \frac{P_{T,\mu\nu}}{{\cal G}(Q)-Q^2}+\zeta\frac{Q_\mu Q_\nu}{Q^4} \, .
\ee
With the help of the relations (\ref{PLPT}) one checks that $D^{-1}_{\mu\nu}D^{\nu\sigma}=g_{\mu}^{\;\;\;\sigma}$.  
Compare this propagator for instance to the propagator of a free scalar boson, $D_0 = (M^2-Q^2)^{-1}$. 
In this simple case, $M$ is obviously the mass of the boson. Similarly, the poles of the gauge boson propagator yield the masses that arise due to the 
interactions with the medium. There is a longitudinal and a transverse mass, corresponding to electric and magnetic screening. They are encoded in the 
functions ${\cal F}$ and ${\cal G}$ which, in turn, are related to the polarization tensor via Eqs.\ (\ref{FQ00}) and (\ref{GPi}). Therefore, we define 
the electric screening mass $m_D$ (Debye mass) and the magnetic screening mass $m_M$ (Meissner mass), 
\begin{subequations}
\bea
m_D^2 &=& - \lim_{{\bf q}\to 0} \Pi_{00}(0,{\bf q}) \, , \\[2ex]
m_M^2 &=& \frac{1}{2}\lim_{{\bf q}\to 0} (\delta_{ij}-\hat{q}_i\hat{q}_j)\Pi_{ij}(0,{\bf q}) \, . \label{mMdef}
\eea
\end{subequations}
While the electric screening mass becomes nonzero in any plasma with charged particles, the Meissner mass is nonzero only in a superconductor.
We now compute the Meissner mass in the one-loop approximation.

\subsection{Calculation of the Meissner mass}

\begin{figure} [t]
\begin{center}
\includegraphics[width=0.95\textwidth]{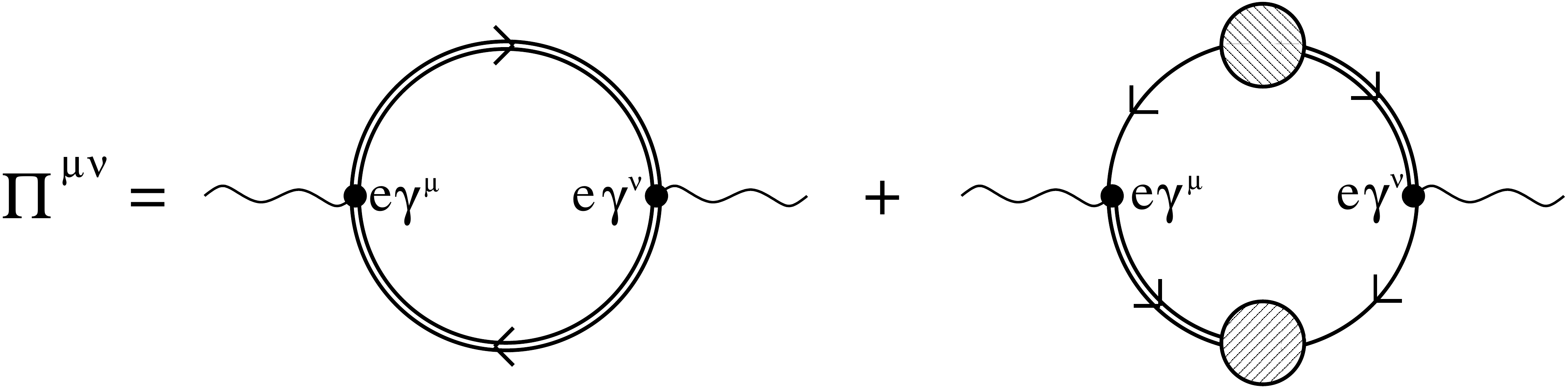}
\caption{Contributions to the one-loop polarization tensor $\Pi^{\mu\nu}$ from the propagators $G^\pm$ (left) and the anomalous propagators 
$F^\pm = -G_0^\mp\Phi^\pm G^\pm$ (right), see 
Eq.\ (\ref{NGtrace}). 
As in the diagram for the gap equation in Fig.\ \ref{figgapeq}, solid single lines represent the tree-level propagators $G_0^\pm$, solid double lines 
the full propagators $G^\pm$, and the hatched circles the gap matrices $\Phi^\pm$. The wavy lines represent the gauge boson propagator.
}
\label{figPimunu}
\end{center}
\end{figure}

At one-loop level, the polarization tensor is 
\be
\Pi^{\mu\nu}(Q) = \frac{1}{2}\frac{T}{V}\sum_K \Tr[\Gamma^\mu {\cal S}(K)\Gamma^\nu{\cal S}(K-Q)] \, , 
\ee
where the trace runs over Nambu-Gorkov and Dirac space, where ${\cal S}(K)$ is the fermion propagator in Nambu-Gorkov space from Eq.\ (\ref{SGF}), and 
where the vertex of the interaction between gauge boson and fermion in Nambu-Gorkov space is
\be
\Gamma^\mu = \left(\begin{array}{cc} e\gamma^\mu & 0 \\ 0 & -e\gamma^\mu \end{array}\right) \, .
\ee
It is convenient for the following to abbreviate 
\be
P\equiv K - Q \, .
\ee 
Using the explicit form of the Nambu-Gorkov propagator, the trace over Nambu-Gorkov space yields 
\bea \label{NGtrace}
\Pi^{\mu\nu}(Q) &=& \frac{e^2}{2}\frac{T}{V}\sum_K \Tr\Big[\gamma^\mu G^+(K)\gamma^\nu G^+(P)+\gamma^\mu G^-(K)\gamma^\nu G^-(P) \non[2ex]
&&-\gamma^\mu F^-(K)\gamma^\nu F^+(P)-\gamma^\mu F^+(K)\gamma^\nu F^-(P)\Big] \, . 
\eea
The contributions coming from the propagators $G^\pm$ and the anomalous propagators $F^\pm$ are shown diagrammatically in Fig.\ \ref{figPimunu}.

We use the propagators from  Eqs.\ (\ref{Gexp}) and (\ref{Fexp}),
\begin{subequations}
\bea
G^\pm(K) &=& \sum_e\frac{k_0\pm e\xi_k^e}{k_0^2-(\epsilon_k^e)^2}\gamma^0\Lambda_k^{\mp e} \, , \\[2ex]
F^\pm(K) &=& \pm \sum_e \frac{\Delta}{k_0^2-(\epsilon_k^e)^2}\gamma^5\Lambda_k^{\mp e} \, ,
\eea
\end{subequations}
with $\epsilon_k^e=\sqrt{(\xi_k^e)^2+\Delta^2}$, and 
\be
\xi_k^e \equiv k-e\mu \, , 
\ee
i.e., we work again in the ultrarelativistic limit for simplicity.

To compute the Meissner mass, we only need the spatial components $\mu=i$, $\nu=j$ of the polarization tensor. 
For these components, we need the Dirac traces 
\bea
\Tr[\gamma^i\gamma^0\Lambda_k^{\mp e_1}\gamma^j\gamma^0\Lambda_p^{\mp e_2}] 
&=& \Tr[\gamma^i\gamma^5\Lambda_k^{\pm e_1}\gamma^j\gamma^5\Lambda_p^{\mp e_2}] \non[2ex]
&=& \delta^{ij}(1-e_1e_2\uk\cdot\up)+e_1e_2(\hk^i\hp^j+\hk^j\hp^i) \, , 
\eea
where we have used $\Tr[\gamma^0\gamma^i\gamma^j]=0$, and 
\be
\Tr[\gamma^i\gamma^j] = -4\delta^{ij} \, , \qquad 
\Tr[\gamma^i\gamma^j\gamma^k\gamma^\ell] = 4(\delta^{ij}\delta^{k\ell}+\delta^{i\ell}\delta^{jk}-\delta^{ik}\delta^{j\ell}) \, .
\ee
With these results we compute 
\bea
\Pi^{ij}(Q) &=&  e^2\frac{T}{V}\sum_{e_1e_2}\sum_K [\delta^{ij}(1-e_1e_2\uk\cdot\up)+e_1e_2(\hk^i\hp^j+\hk^j\hp^i)] \non[2ex]
&& \hspace{1.5cm}\times \,\frac{k_0p_0+e_1e_2\xi_k^{e_1}\xi_p^{e_2}+\Delta^2}{[k_0^2-(\epsilon_k^{e_1})^2][p_0^2-(\epsilon_p^{e_2})^2]} \, .
\eea
We now perform the Matsubara sum over fermionic Matsubara frequencies $k_0 = -(2n+1)i\pi T$ and use that $q_0 = -2mi\pi T$ from the external 
four-momentum is a bosonic Matsubara frequency, $m,n\in \mathbb{Z}$. The explicit calculation in terms of a contour integral in the complex $k_0$ plane
is left as an exercise. The result is 
\bea
&&\Pi^{ij}(Q) = \frac{e^2}{4}\sum_{e_1e_2}\int\frac{d^3{\bf k}}{(2\pi)^3} \,
[\delta^{ij}(1-e_1e_2\uk\cdot\up)+e_1e_2(\hk^i\hp^j+\hk^j\hp^i)] \non[2ex]
&&\times\, \Bigg\{\frac{\epsilon_k^{e_1}\epsilon_p^{e_2}-e_1e_2\xi_k^{e_1}\xi_p^{e_2}-\Delta^2}{\epsilon_k^{e_1}\epsilon_p^{e_2}}
\left(\frac{1}{q_0-\epsilon_k^{e_1}-\epsilon_p^{e_2}}-\frac{1}{q_0+\epsilon_k^{e_1}+\epsilon_p^{e_2}}\right)\non[2ex]
&&\hspace{3cm}\times[1-f(\epsilon_k^{e_1})-f(\epsilon_p^{e_2})]\non[2ex]
&&+\frac{\epsilon_k^{e_1}\epsilon_p^{e_2}+e_1e_2\xi_k^{e_1}\xi_p^{e_2}+\Delta^2}{\epsilon_k^{e_1}\epsilon_p^{e_2}}
\left(\frac{1}{q_0+\epsilon_k^{e_1}-\epsilon_p^{e_2}}-\frac{1}{q_0-\epsilon_k^{e_1}+\epsilon_p^{e_2}}\right) \non[2ex]
&&\hspace{3cm}\times [f(\epsilon_k^{e_1})-f(\epsilon_p^{e_2})]\Bigg\} \, .
\eea
According to the definition of the Meissner mass (\ref{mMdef}), we can now set $q_0=0$,
\bea
\Pi^{ij}(0,{\bf q}) &=& \frac{e^2}{2}\sum_{e_1e_2}\int\frac{d^3{\bf k}}{(2\pi)^3} \,
[\delta^{ij}(1-e_1e_2\uk\cdot\up)+e_1e_2(\hk^i\hp^j+\hk^j\hp^i)] \allowdisplaybreaks \non[2ex]
&&\times\, \left(\frac{\epsilon_k^{e_1}\epsilon_p^{e_2}+e_1e_2\xi_k^{e_1}\xi_p^{e_2}+\Delta^2}{\epsilon_k^{e_1}\epsilon_p^{e_2}}
\frac{f(\epsilon_k^{e_1})-f(\epsilon_p^{e_2})}{\epsilon_k^{e_1}-\epsilon_p^{e_2}} \right. \allowdisplaybreaks\non[2ex]
&&\left. -\frac{\epsilon_k^{e_1}\epsilon_p^{e_2}-e_1e_2\xi_k^{e_1}\xi_p^{e_2}-\Delta^2}{\epsilon_k^{e_1}\epsilon_p^{e_2}}
\frac{1-f(\epsilon_k^{e_1})-f(\epsilon_p^{e_2})}{\epsilon_k^{e_1}+\epsilon_p^{e_2}} \right)\, .
\eea
Next, we are interested in the limit ${\bf q}\to 0$ which corresponds to ${\bf p}\to {\bf k}$. With 
\be
\int\frac{d\Omega}{4\pi} \,\hk^i\hk^j = \frac{\delta^{ij}}{3} \, ,
\ee
we obtain for the angular integral 
\be
\int\frac{d\Omega}{4\pi}\Big[\delta^{ij}(1-e_1e_2\uk\cdot\up)+e_1e_2(\hk^i\hp^j+\hk^j\hp^i)\Big]_{{\bf p}={\bf k}} 
= \left\{\begin{array}{cc} \displaystyle{\frac{2}{3}\delta^{ij}} & \mbox{for}\;
e_1=e_2 \\[2ex] \displaystyle{\frac{4}{3}\delta^{ij}} & \mbox{for}\;e_1\neq e_2 \end{array}\right. \, .
\ee
Consequently, we find
\bea
\Pi^{ij}(0,{\bf q}\to 0) 
&=& \frac{\delta^{ij}e^2}{3\pi^2}\int_0^\infty dk\,k^2\left[\frac{df(\epsilon_k^+)}{d\epsilon_k^+}
+\frac{df(\epsilon_k^-)}{d\epsilon_k^-} \right. \non[2ex]
&&\left.+ 2\,\frac{\epsilon_k^+\epsilon_k^--\xi_k^+\xi_k^-+\Delta^2}
{\epsilon_k^+\epsilon_k^-} \frac{f(\epsilon_k^+)-f(\epsilon_k^-)}{\epsilon_k^+-\epsilon_k^-} \right. \non[2ex]
&&\left.-2\,\frac{\epsilon_k^+\epsilon_k^-+\xi_k^+\xi_k^--\Delta^2}{\epsilon_k^+\epsilon_k^-} 
\frac{1-f(\epsilon_k^+)-f(\epsilon_k^-)}{\epsilon_k^++\epsilon_k^-} \right] \, .
\eea
At zero temperature, this becomes
\be \label{Piij0}
\Pi^{ij}(0,{\bf p}\to 0) 
= -\frac{2\delta^{ij}e^2}{3\pi^2}\int_0^\infty dk\,k^2\,
\frac{\epsilon_k^+\epsilon_k^-+\xi_k^+\xi_k^--\Delta^2}{\epsilon_k^+\epsilon_k^-(\epsilon_k^++\epsilon_k^-)} \,.
\ee
This integral can be performed exactly. We use a momentum cutoff $\Lambda$ for large momenta in order to discuss the ultraviolet divergences of the 
integral, 
\bea \label{integralUV}
\int_0^\Lambda dk\,k^2
\frac{\epsilon_k^+\epsilon_k^-+\xi_k^+\xi_k^--\Delta^2}{\epsilon_k^+\epsilon_k^-(\epsilon_k^++\epsilon_k^-)}
&=& \left[\frac{1}{4}(\xi_k^+\epsilon_k^- +\xi_k^-\epsilon_k^+)-\frac{\Delta^2}{2\mu}(\epsilon_k^+-\epsilon_k^-)\right. \non[2ex]
&&\hspace{-1cm}\left.-\frac{3\Delta^2}{4}\left\{
\ln[2(\xi_k^++\epsilon_k^+)] +\ln[2(\xi_k^-+\epsilon_k^-)]\right\}\right]_{k=0}^{k=\Lambda} \allowdisplaybreaks\non[2ex]
&=&\frac{\Lambda^2}{2} -\frac{3}{2}\Delta^2\ln\frac{2\Lambda}{\Delta} +\frac{5\Delta^2-2\mu^2}{4} \, .
\eea
Not surprisingly, there is an ultraviolet divergence $\propto \Lambda^2$ from the vacuum which we can subtract. However, there is another -- logarithmic -- cutoff 
dependence which depends on $\Delta$. 
We recall that the solution of the gap equation for a point-like interaction requires the introduction of an energy scale, see Sec.\ \ref{sec:BCSgap}. 
There, we restricted the momentum integral to a small vicinity around the Fermi surface by introducing a scale $\delta$. Had we worked with a simple momentum cutoff 
$\Lambda$ instead, as in the integral (\ref{integralUV}), we would have obtained the same weak-coupling result for the zero-temperature gap, 
with $\Lambda$ replacing $\delta$,
\be
\Delta = 2\Lambda e^{-\frac{2\pi^2}{\mu^2 G}} \, , 
\ee
such that the logarithm $\ln\frac{2\Lambda}{\Delta}$ goes like $(\mu^2 G)^{-1}$. This is a large factor for small coupling, but it is multiplied 
by $\Delta^2$, which is {\it exponentially} small for small coupling. Therefore, at weak coupling and after subtracting the vacuum contribution, 
the integral (\ref{integralUV}) can be approximated by $-\mu^2/2$.


Consequently, the result for the spatial components of the polarization tensor at zero temperature is
\be 
\Pi^{ij}(0,{\bf q}\to 0)=  \frac{\delta^{ij}e^2\mu^2}{3\pi^2} \, .
\ee
Inserting this into the definition of the Meissner mass (\ref{mMdef}), yields the final result
\be
m_M^2 = \frac{e^2\mu^2}{3\pi^2} \, .
\ee
One might wonder why this result is independent of the gap $\Delta$. After all, we expect the Meissner mass to be nonzero only in a superconductor, i.e., 
only for nonvanishing gap. The point is that we have taken the limit ${\bf p}\to 0$ for a fixed nonzero $\Delta$. If we had first taken the limit 
$\Delta\to 0$ we would have found $m_M=0$, as expected. This calculation is left as an exercise.

\chapter{BCS-BEC crossover}
\label{sec:BCSBEC}

We have discussed bosonic and fermionic superfluids. The underlying mechanisms were, on the one hand, Bose-Einstein condensation (BEC) and,
on the other hand, Cooper pairing according
to Bardeen-Cooper-Schrieffer (BCS) theory. We have also mentioned that fermionic superfluidity is also a form of BEC because a Cooper pair can, in some sense, 
be considered as a 
boson. However, this picture has of course to be taken with some care. When we discussed Cooper pairing, we were working in the weak-coupling limit. 
And we have seen that an arbitrarily weak interaction leads to Cooper pairing. Now, an infinitesimally small interaction between fermions does not  
create di-fermionic molecules which could be considered as a bosonic particle. Weakly coupled Cooper pairing is more subtle, it is a collective effect in 
which the constituents of a Cooper pair are spatially separated, typically over distances much larger than the average distance between the fermions in the system. 
But what if we increase the strength of the interaction? Is there a point where we can truly speak of di-fermionic molecules that undergo Bose-Einstein condensation?
Is there a justification for speaking of a Bose-Einstein condensation of Cooper pairs even at weak coupling? 

In this chapter, we shall see that BEC and BCS are indeed continuously connected, and the connection is made by varying the coupling strength 
between the fermions. The point is that there is no phase transition between BEC and BCS, and thus one speaks of the 
BCS-BEC crossover. Theoretical works showing that BEC is a limit of the very general BCS theory have been pioneered by 
D.\ Eagles in 1969 \cite{Eagles:1969zz} and P.\ Nozi\`{e}res and S.\ Schmitt-Rink in 1985 \cite{Nozieres:1985zz}. In principle,
various physical systems may show this crossover. To observe the crossover experimentally, one would ideally like to tune the interaction strength at will.
This is exactly what can be done in modern experiments with ultra-cold atomic gases. Therefore, despite the theoretical 
generality of the BCS-BEC crossover, we shall put our discussion in the context of ultra-cold fermionic atoms, where the crossover has first been demonstrated 
experimentally and which since then has remained an extremely active research field.

\section{Ultra-cold atomic gases}

Experiments with ultra-cold fermionic gases have been based on the experience gained from similar experiments with ultra-cold bosonic gases, which
has led to the first direct observation of Bose-Einstein condensation in 1995 \cite{1995Sci...269..198A,1995PhRvL..75.3969D}. Interest in 
the fermionic counterparts has begun in the 1990's and around 2003 several groups had established the creation of 
ultra-cold Fermi gases.
The fermions used in these experiments are usually $^{40}$K or $^6$Li. If you are interested in the details of these experiments I recommend 
the exhaustive review \cite{ketterle}, where also large parts of the theory are laid out. Other nice reviews, with more emphasis on theory, are
Refs.\ \cite{giorgini,2010AnPhy.325..233L,2012LNP...836.....Z}.

\begin{figure} [t]
\begin{center}
\includegraphics[width=0.75\textwidth]{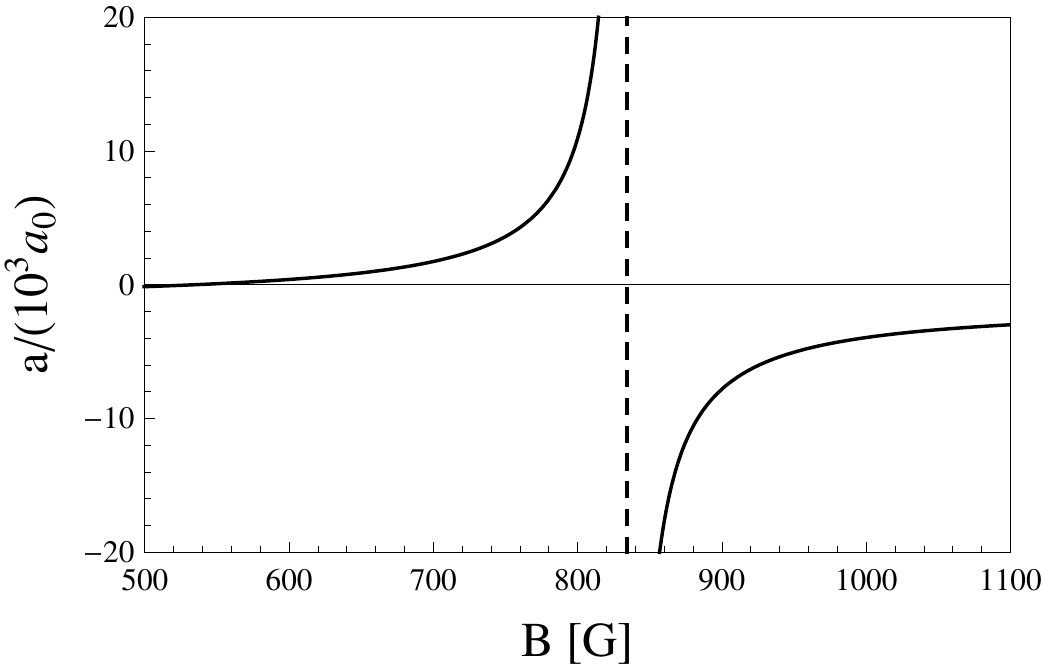}
\caption{Feshbach resonance. Scattering length $a$ in units of $10^3$ times the Bohr radius $a_0$ 
as a function of the applied magnetic field 
in Gauss according to the parametrization (\ref{aB}) with the numerical values for $^6$Li, $B_0 = 834.15\, {\rm G}$, $\Delta B = 300\, {\rm G}$, 
$a_{\rm bg} = -1405\,a_0$. 
}
\label{figscatter1}
\end{center}
\end{figure}

The main characteristics of the systems created in all these experiments are the 
low temperature and the diluteness. After several stages of different cooling techniques, the fermionic gases are brought down to 
temperatures of the order of $T\sim 50\, {\rm nK}$ at densities of the order of $n\sim 5\times 10^{12}\,{\rm cm}^{-3}$. 
The low temperature and diluteness are crucial for the properties of the gas. In this regime, both 
the thermal wavelength $\lambda = \sqrt{2\pi/mT}$, where $m$ is the mass of a single atom,
and the mean inter-particle distance $n^{-1/3}\sim k_F^{-1} \sim  10^4\, a_0$ are much larger than the spatial range of the inter-atomic potential 
$R_0\sim 50 \, a_0$, where $a_0$ is the Bohr radius and $k_F$ the Fermi wavevector,
\be
\lambda \gg R_0 \, , \qquad k_F^{-1} \gg R_0 \, .
\ee
As a consequence,
the complicated details of the short-range interaction potential become unimportant and the interaction is basically characterized by one
single quantity, the $s$-wave scattering length $a$. This scattering length is under experimental control and can be varied through a 
magnetic field,
\be \label{aB}
a(B) = a_{\rm bg}\left(1-\frac{\Delta B}{B-B_0}\right) \, ,
\ee   
see Fig.\ \ref{figscatter1}.
This parametrization describes the so-called {\it Feshbach resonance} at $B=B_0$ with a width $\Delta B$ and a background scattering length $a_{\rm bg}$
far away from the resonance. At $B=B_0$ the scattering length is infinite. This is called the {\it unitary limit}. The unitary limit
is particularly interesting since in this limit the only length scale that is left to characterize the interaction drops out, giving   
the unitary limit very general significance. For instance, in the very dense nuclear matter inside a neutron star, neutrons have a scattering length 
larger than their mean inter-particle distance, and 
parallels to the unitary limit in ultra-cold atoms may help to improve the theoretical understanding of this system.

Here we are not aiming at a detailed description of the atomic physics involved in the experiments with ultra-cold fermions because this is not the 
topic of the course. Nevertheless, let us give a very brief reminder of how the scattering length is defined, for more details about basic scattering theory 
see for instance Ref.\ \cite{landau}.
The Schr\"{o}dinger equation for scattering of two particles with masses $m_1$, $m_2$ and reduced mass $m_r\equiv m_1m_2/(m_1+m_2)$ 
can be written in the center-of-mass frame in terms of the scattering potential $V({\bf r})$,
\be \label{schro}
\left[-\frac{\nabla^2}{m}+V({\bf r})\right]\psi_k({\bf r}) = E \psi_k({\bf r}) \, ,
\ee
with $m\equiv 2m_r$ being the mass of a single atom in the case $m_1=m_2$. If the incoming particle moves along the 
$z$-axis and the angle between the $z$-axis and the scattered particle is denoted by $\theta$, the solution of the Schr\"{o}dinger equation
at large distances can be written as 
\be
\psi_k({\bf r}) \simeq e^{ikz} + f_k(\theta)\frac{e^{ikr}}{r} \, , 
\ee
with the {\it scattering amplitude} $f_k(\theta)$, which determines the differential cross section $d\sigma = |f_k(\theta)|^2 d\Omega$. From the 
general expression 
\be
f_k(\theta) = \frac{1}{2ik}\sum_{\ell=0}^\infty (2\ell + 1) [e^{2i\delta_\ell(k)}-1]P_\ell(\cos\theta) \, ,
\ee
where $\delta_\ell(k)$ is the {\it phase shift} of the collision and $P_\ell$ are the Legendre polynomials, we are only interested in the $s$-wave
scattering amplitude $f_s(k)$ because this is the dominant contribution in the context of cold fermionic gases, 
\bea
f_k(\theta)\simeq f_s(k) &=& \frac{1}{2ik} [e^{2i\delta_s(k)}-1] \allowdisplaybreaks \non[2ex] 
&=& \frac{1}{k\,\cot\delta_s(k)-ik} \simeq \frac{1}{-\frac{1}{a}
+R_0\frac{k^2}{2}-ik}\, .
\eea
Here we have introduced the scattering length $a$ and the effective range of the potential $R_0$ which appear as coefficients in the 
low-momentum expansion of $k\cot\delta_s(k)$. According to this expansion, the definition of the $s$-wave scattering length in terms of the phase shift is
\be
a = -\lim_{k\ll R_0^{-1}} \frac{\tan \delta_s(k)}{k} \, . 
\ee
We have discussed above that due to the diluteness of the system, the typical wavelengths are always very large, $R_0\ll k^{-1}$. Therefore, for very small scattering 
lengths, $k|a|\ll 1$, the scattering amplitude is $f_s\simeq -a$ while for large scattering lengths, i.e., in the unitary limit $k|a|\gg 1$, we can approximate 
$f_s\simeq i/k$. 

\begin{figure} [t]
\begin{center}
\includegraphics[width=0.75\textwidth]{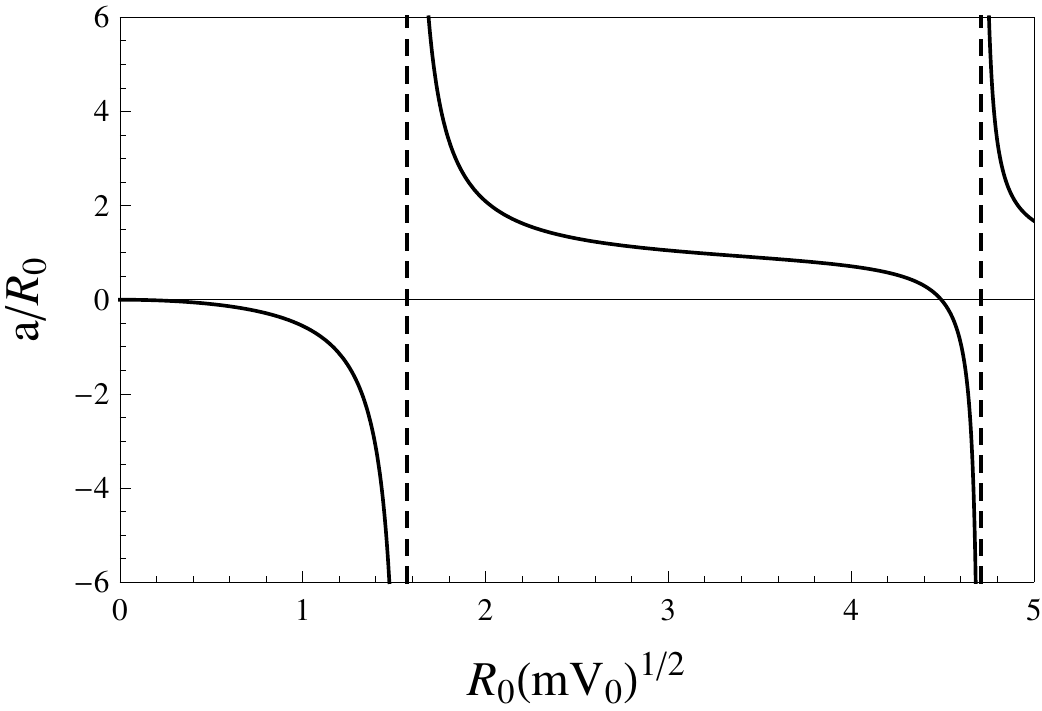}
\caption{
Scattering length $a$ for a square-well potential (\ref{sqwell}) according to Eq.\ (\ref{sc1}) as a function of the dimensionless combination of the 
parameters of the potential $R_0$, $V_0$ and the mass of the scattered particle $m$. 
The scattering length diverges whenever a new bound state appears. 
}
\label{figscatter2}
\end{center}
\end{figure}

To illustrate the meaning of the scattering length, it is useful to consider an attractive square-well scattering potential 
\be \label{sqwell}
V({\bf r}) = -V_0\Theta(R_0-r) \, .
\ee
[Even though below we shall rather work with a point-like potential $V({\bf r}) \propto \delta({\bf r})$.] 
In this case, one computes the scattering length \cite{landau}
\be \label{sc1}
a = R_0\left[1-\frac{\tan(R_0\sqrt{mV_0})}{R_0\sqrt{mV_0}}\right] \, .
\ee
As shown in Fig.\ \ref{figscatter2}, at very shallow potentials the scattering length starts off with small negative values. 
With increasing depth
of the potential, it becomes more and more negative, until it diverges at $R_0\sqrt{m V_0} = \pi/2$. This is the point where the first bound state 
develops. Then the scattering length is large and positive until the next bound state approaches etc.


Let us compute the energy $E$ of a shallow bound state $E = -\frac{\kappa^2}{m}$ with $\kappa\ll R_0^{-1}$. From the Schr\"{o}dinger equation (\ref{schro}) we 
obtain after 
Fourier transformation 
\bea
&&\int\frac{d^3{\bf q}}{(2\pi)^3} (q^2+\kappa^2) e^{i{\bf q}\cdot{\bf r}}\psi_\kappa({\bf q}) 
= -m\int\frac{d^3{\bf p}}{(2\pi)^3}\int\frac{d^3{\bf k}}{(2\pi)^3}e^{i({\bf p}+{\bf k})\cdot{\bf r}}v({\bf p})\psi_\kappa({\bf k}) \non[2ex]
&&\hspace{3cm}=-m\int\frac{d^3{\bf q}}{(2\pi)^3}\int\frac{d^3{\bf k}}{(2\pi)^3}e^{i{\bf q}\cdot{\bf r}}v({\bf q}-{\bf k})\psi_\kappa({\bf k})\, ,
\eea
where we have denoted the Fourier transform of $V({\bf r})$ by $v({\bf p})$, and thus
\bea
\psi_\kappa({\bf q}) &=& -\frac{m}{q^2+\kappa^2} \int\frac{d^3{\bf k}}{(2\pi)^3}v({\bf q}-{\bf k})\psi_\kappa({\bf k}) \non[2ex]
&\simeq& -\frac{mv_0}{q^2+\kappa^2} \int\frac{d^3{\bf k}}{(2\pi)^3}\psi_\kappa({\bf k}) \, , 
\eea
where we have used that, for small momenta, the scattering potential is approximated by a $\delta$-function, $V({\bf r})\propto \delta({\bf r})$
in position space and thus by a constant $v_0$ in 
momentum space. Integrating both sides over ${\bf q}$ then yields
\be \label{v0inv}
-\frac{1}{v_0} = m \int\frac{d^3{\bf q}}{(2\pi)^3} \frac{1}{q^2+\kappa^2} \, .
\ee
The integral on the right-hand side is ultraviolet divergent. This is due to our use of the point-like potential, where we did not care about 
large momenta. The physical potential is not constant in momentum space for all momenta. 
We thus need to renormalize our potential which can be done by the prescription 
\be\label{prescription}
\frac{1}{v_0} = \frac{m}{4\pi a} - \int\frac{d^3{\bf q}}{(2\pi)^3}\frac{m}{q^2} \, .
\ee
This can be viewed as going from the bare coupling $v_0$ to a physical coupling given by the scattering length $a$: if we ``switch on'' the 
divergent second term on the right-hand side we need to adjust the bare coupling in order to keep the physical coupling fixed. 
Replacing the bare coupling $v_0$ in Eq.\ (\ref{v0inv}) by the expression from Eq.\ (\ref{prescription}) yields
\bea
-\frac{m}{4\pi a} &=& \frac{m}{2\pi^2}\underbrace{\int_0^\infty dq \, \left(\frac{q^2}{q^2+\kappa^2}-1\right)}_{\displaystyle{-\kappa\pi/2}} 
= -\frac{\kappa m}{4\pi} \, .
\eea
We read off $\kappa = a^{-1}$. In particular, $a$ has to be positive for the bound state to exist. The energy of the bound state is
\be \label{bound}
E = -\frac{1}{ma^2} \, .
\ee
We shall come back to this result later in the interpretation of the BCS-BEC crossover.

\section{Crossover in the mean-field approximation}
 
At sufficiently small temperatures, the atoms in the optical trap become superfluid. In this subsection we are interested in their
behavior as a function of the scattering length $a$. The product $k_Fa$ with the Fermi momentum $k_F$ will play the role 
of an effective, dimensionless coupling constant. In this way we will generalize the weak-coupling solutions to the BCS gap equation 
from chapter \ref{sec:cooper} to arbitrary values of the coupling. Since we use the same framework given by the mean-field approximation,
the results will have to be taken with some care. Especially at nonzero temperature, we shall see that our approach does not provide
a correct description of the system. For zero temperature, however, the mean-field approximation is, at least qualitatively, correct. 

For an effective four-point coupling between the fermions we can write our gap equation (\ref{gapeq1}) as
\be \label{gapeqcoldat}
\Phi^+ = -v_0\frac{T}{V}\sum_KF^+(K) \, .
\ee
Now $v_0$ plays the role of the (bare) coupling strength, instead of $G$ in chapter \ref{sec:cooper}. Remember that $G$ and thus also $v_0$
have mass dimensions $-2$. Instead of the general Dirac fermions of chapter \ref{sec:cooper}, here we are interested in the non-relativistic case.
Therefore, we shall simply consider two fermion species with no additional structure. These species can be thought of 
as spin-up and spin-down fermions, but spin will nowhere appear in our calculation, so one can think more abstractly of species 1 and 2. 
The two fermion species may in general have different masses and chemical potentials. In this chapter, we restrict ourselves to fermions with 
equal masses and chemical potentials. We shall discuss the more complicated case of different chemical potentials, 
relevant for experiments with cold atoms as well as for quark matter, in chapter \ref{sec:mismatch}. 
The distinction of two species is necessary since the Cooper pair wave function has to be antisymmetric. This can only be achieved with at least one 
quantum number that distinguishes the constituents of a Cooper pair. In the experimental setup of ultra-cold atoms, the two species are provided by 
two hyperfine states of the respective fermionic atom or by two different atom species \cite{ketterle}.

Assuming equal masses and chemical potentials, the tree-level propagator is proportional to the unit matrix in this internal ``spin space'',
\be
[G_0^\pm(K)]^{-1} = (k_0\mp \xi_k)\cdot {\bf 1}_{\rm spin} \, , \qquad \xi_k\equiv \frac{k^2}{2m}-\mu \, . 
\ee
This propagator is obtained from the ultra-relativistic version (\ref{G0m1}) by dropping the anti-particle contribution, ignoring the Dirac structure, and replacing
the ultra-relativistic dispersion $k-\mu$ by the non-relativistic one. Our ansatz for the gap matrix is
\be
\Phi^+ = \Delta \sigma_2 \, , 
\ee
where the anti-symmetric Pauli matrix $\sigma_2$ takes into account that fermions of different species form Cooper pairs. Then, one can easily
compute the components of the Nambu-Gorkov propagator,
\begin{subequations}
\bea
G^\pm(K) &=& \left([G_0^\pm]^{-1}-\Phi^\mp G_0^\mp \Phi^\pm\right)^{-1} = \frac{k_0\pm\xi_k}{k_0^2-\epsilon_k^2} \, , \\[2ex]
F^\pm(K) &=& -G_0^\mp\Phi^\pm G^\pm = -\frac{\Delta\sigma_2}{k_0^2-\epsilon_k^2} \, , 
\eea
\end{subequations}
with the quasiparticle dispersion
\be
\epsilon_k = \sqrt{\xi_k^2 + \Delta^2} \, .
\ee
Consequently, the gap equation (\ref{gapeqcoldat}) becomes 
\be
-\frac{1}{v_0}=\int\frac{d^3{\bf k}}{(2\pi)^3}\frac{\tanh\frac{\epsilon_k}{2T}}{2\epsilon_k} \, ,
\ee
where we have used the Matsubara sum from Eq.\ (\ref{Matsusimple}). With the renormalization given in Eq.\ (\ref{prescription}) we obtain 
\be \label{gap3}
-\frac{m}{4\pi a} =  \int\frac{d^3{\bf k}}{(2\pi)^3} \left(\frac{\tanh\frac{\epsilon_k}{2T}}{2\epsilon_k}-\frac{m}{k^2}\right) \, .
\ee
Remember that for the solution of the gap equation in chapter 5 we needed a cutoff, for instance the Debye frequency in the case of an electronic superconductor. 
Here we are working with the same point-like approximation of the interaction. Therefore, the same problem arises, and we have solved it by expressing the 
coupling constant $v_0$ in terms of the scattering length. 

It is convenient to express the gap equation in terms of the Fermi momentum and the Fermi energy 
\be
k_F = (3\pi^2 n)^{1/3} \, , \qquad E_F = \frac{k_F^2}{2m} = \frac{(3\pi^2 n)^{2/3}}{2m} \, .
\ee
They are written in terms of the total charge density $n$ (= number density of the atoms) rather than the chemical potential since, in the experiment, 
the number of atoms is kept fixed.  
Then, taking the zero-temperature limit and changing the integration variable in the gap equation from $k$ to $x=k/\sqrt{2m\Delta}$, 
we can write the gap equation as
\be \label{gapcross}
-\frac{1}{k_Fa}=\frac{2}{\pi}\sqrt{\frac{\Delta}{E_F}}\,I_1\left(\frac{\mu}{\Delta}\right) \, , 
\ee
with the abbreviation
\be
I_1(z) \equiv \int_0^\infty dx\, x^2\left[\frac{1}{\sqrt{(x^2-z)^2+1}}-\frac{1}{x^2}\right] \, .
\ee
This gap equation shows that the combination $k_Fa$ plays the role of a dimensionless coupling constant. 

\begin{figure} [t]
\begin{center}
\includegraphics[width=0.75\textwidth]{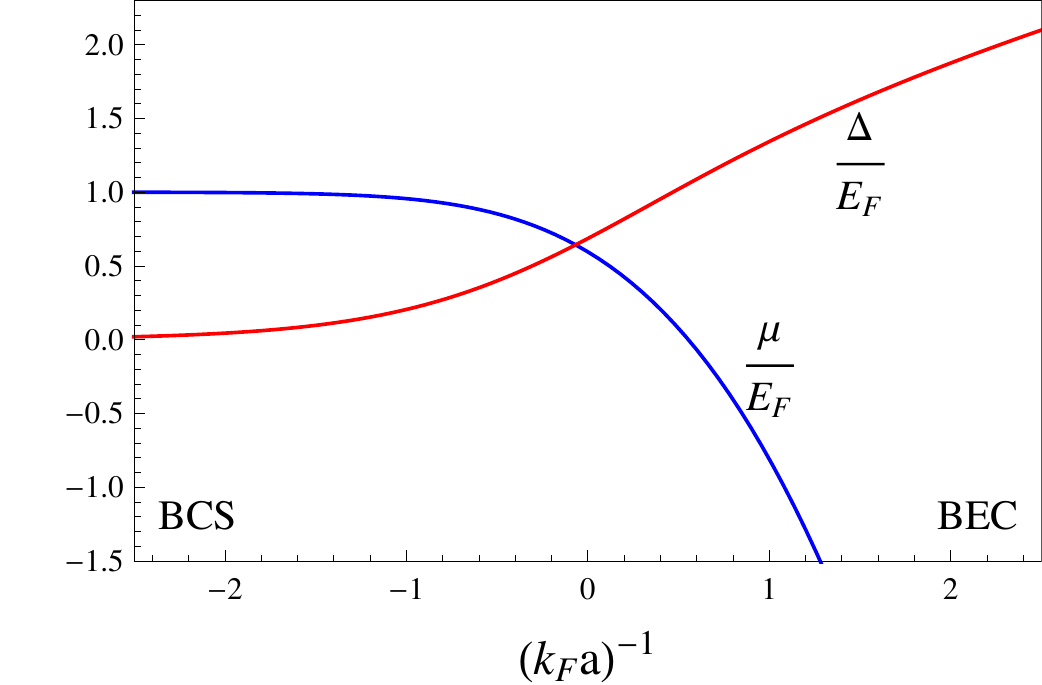}
\caption{(Color online) Zero-temperature results for $\Delta$ and $\mu$ in units of the Fermi energy throughout the BCS-BEC crossover, computed from Eqs.\ (\ref{gap12}).
}
\label{figcross1}
\end{center}
\end{figure}

\begin{figure} [t]
\begin{center}
\includegraphics[width=0.75\textwidth]{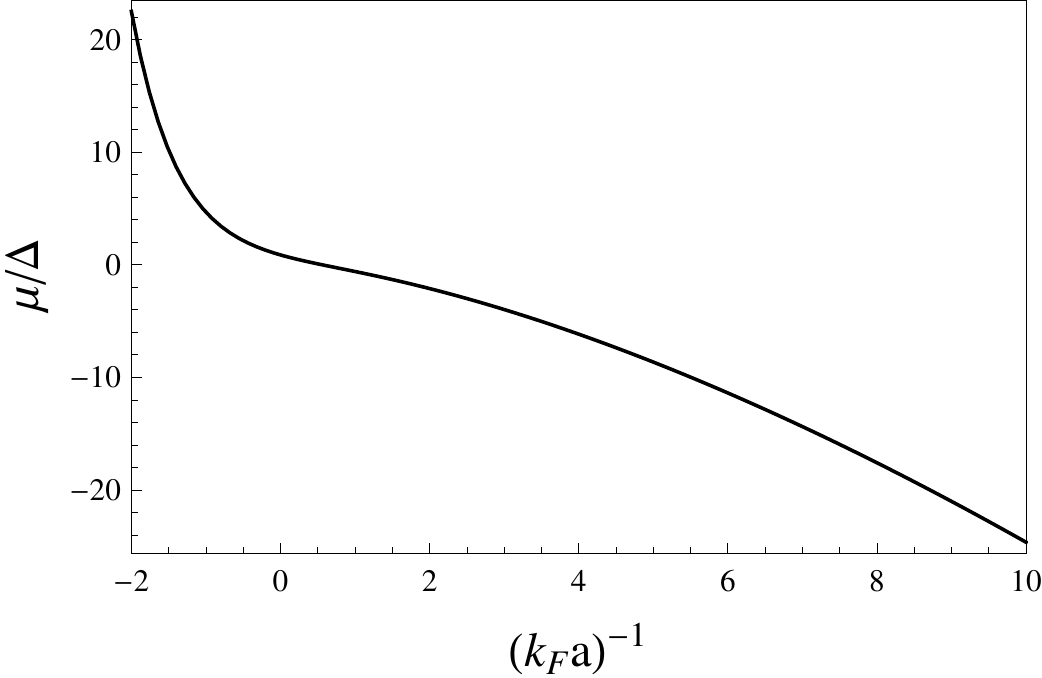}
\caption{Ratio $\mu/\Delta$ at zero temperature, computed from the solution shown in Fig.\ \ref{figcross1}.}
\label{figcross2}
\end{center}
\end{figure}

Besides the gap equation we have a second equation that arises from fixing the number density, and
we need to solve both equations for $\Delta$ and $\mu$. For the second equation we compute the number density in analogy to Eq.\ (\ref{nNG}),
\bea 
n &=& \frac{1}{2}\frac{T}{V}\sum_K\Tr\left[{\cal S}\frac{\partial {\cal S}^{-1}}{\partial \mu}\right] \non[2ex] 
&=&\frac{1}{2}\frac{T}{V}\sum_K\Tr[G^+-G^-] \non[2ex]
&=& 2\frac{T}{V}\sum_K\frac{\xi_k}{k_0^2-\epsilon_k^2} =
-\int\frac{d^3{\bf k}}{(2\pi)^3} \frac{\xi_k}{\epsilon_k}\,\tanh\frac{\epsilon_k}{2T} \, , \label{n3}
\eea
where the trace in the first line is taken over Nambu-Gorkov space and the internal $2\times 2$ space, and in the second line only over the internal space. 
At zero temperature and after subtracting the vacuum contribution $\mu=T=\Delta=0$, this becomes
\be 
n = \int\frac{d^3{\bf k}}{(2\pi)^3} \left(1-\frac{\xi_k}{\epsilon_k}\right) \, , 
\ee  
in agreement with Eq.\ (\ref{occupy}). Analogously to the gap equation, we rewrite this equation as 
\be\label{ncross}
1 = \frac{3}{2}\left(\frac{\Delta}{E_F}\right)^{3/2}\,I_2\left(\frac{\mu}{\Delta}\right) \, , 
\ee
with 
\be
I_2(z) = \int_0^\infty dx \, x^2\left[1-\frac{x^2-z}{\sqrt{(x^2-z)^2+1}}\right] \, .
\ee
We now have to solve the coupled equations (\ref{gapcross}) and (\ref{ncross}) for $\mu$ and $\Delta$ for given $E_F$ and $k_Fa$. The equations can be decoupled
by solving Eq.\ (\ref{ncross}) for $\Delta/E_F$ and inserting the result into Eq.\ (\ref{gapcross}), such that the two equations become
\begin{subequations} \label{gap12}
\bea
-\frac{1}{k_Fa} &=& \frac{2}{\pi}\left[\frac{2}{3I_2\left(\frac{\mu}{\Delta}\right)}\right]^{1/3}I_1\left(\frac{\mu}{\Delta}\right) \, , 
\label{gapcross2}\\[2ex]
\frac{\Delta}{E_F} &=& \left[\frac{2}{3I_2\left(\frac{\mu}{\Delta}\right)}\right]^{2/3} \, .
\label{ncross2}
\eea
\end{subequations}
In this form, the first equation only depends on the ratio $\mu/\Delta$. We may solve this equation for $\mu/\Delta$ 
and then insert the result into the second equation to obtain $\Delta$. The numerical 
evaluation yields the results shown in Figs.\ \ref{figcross1} and \ref{figcross2}.


From Fig.\ \ref{figcross2} we read off  
\be
\frac{\mu}{\Delta} \to \mp \infty \;\;\;\; \mbox{for} \;\;\;\; \frac{1}{k_Fa}\to\pm\infty \, .
\ee
Therefore, to obtain analytical approximations for these two limit cases, we need the asymptotic values of the integrals $I_1$ and $I_2$,
\begin{subequations}
\bea
I_1(z) &\to& \left\{\begin{array}{cc} \sqrt{z}(\ln 8z -2) & \mbox{for}\; z\to +\infty \\[2ex] -\frac{\pi}{2}\sqrt{|z|}
 & \mbox{for}\; z\to -\infty \end{array}\right. \, , \\[2ex]
I_2(z) &\to& \left\{\begin{array}{cc} \frac{2}{3}z^{3/2}  & \mbox{for}\; z\to +\infty \\[2ex] \frac{\pi}{8}|z|^{-1/2}
 & \mbox{for}\; z\to -\infty \end{array}\right. \, .
\eea
\end{subequations}
For $(k_Fa)^{-1}\to -\infty$ we thus find from Eq.\ (\ref{gapcross2})
\be
-\frac{1}{k_Fa} \simeq \frac{2}{\pi}\left(\ln \frac{8\mu}{\Delta}-2\right) \, , 
\ee
and from Eq.\ (\ref{ncross2})
\be
\mu\simeq E_F \, .
\ee
Putting both results together yields the gap as a function of the Fermi energy and the coupling strength,
\be \label{DeltakFaBCS}
\Delta \simeq \frac{8E_F}{e^2} e^{-\frac{\pi}{2k_F|a|}} \, .
\ee
For the opposite limit  $(k_Fa)^{-1}\to +\infty$, Eq.\ (\ref{gapcross2}) yields 
\be
\frac{\mu}{\Delta} \simeq -\sqrt{\frac{3\pi}{16}}\frac{1}{(k_Fa)^{3/2}} \, ,
\ee
and from inserting this result into Eq.\ (\ref{ncross2}) we obtain
\be \label{lim1Delta}
\Delta \simeq \sqrt{\frac{16}{3\pi}}\frac{E_F}{\sqrt{k_F a}} \, , 
\ee
from which we get 
\be \label{lim1mu}
\mu\simeq -\frac{E_F}{(k_F a)^2} = - \frac{1}{2ma^2} \, .
\ee
What do we learn from these results? Firstly, we recover the BCS results from 
Sec.\ \ref{sec:BCSgap} for small negative values of $k_Fa$: the gap is exponentially small, and 
the chemical potential is identical to the Fermi energy.
At small positive values of $k_Fa$, on the opposite side of the Feshbach resonance, we find that the chemical potential is one half of the 
energy of the bound state $E$ from Eq.\ (\ref{bound}). Consequently, in this limit, by adding a single fermion to the system one gains half of the binding energy. This 
suggests that the fermions are all bound in molecules of two fermions. In other words, the system has effectively become bosonic, and we may 
call this regime the BEC regime. We have thus continuously connected the BCS and BEC regimes; this is the BCS-BEC crossover. 

\begin{figure} [t]
\begin{center}
\includegraphics[width=0.85\textwidth]{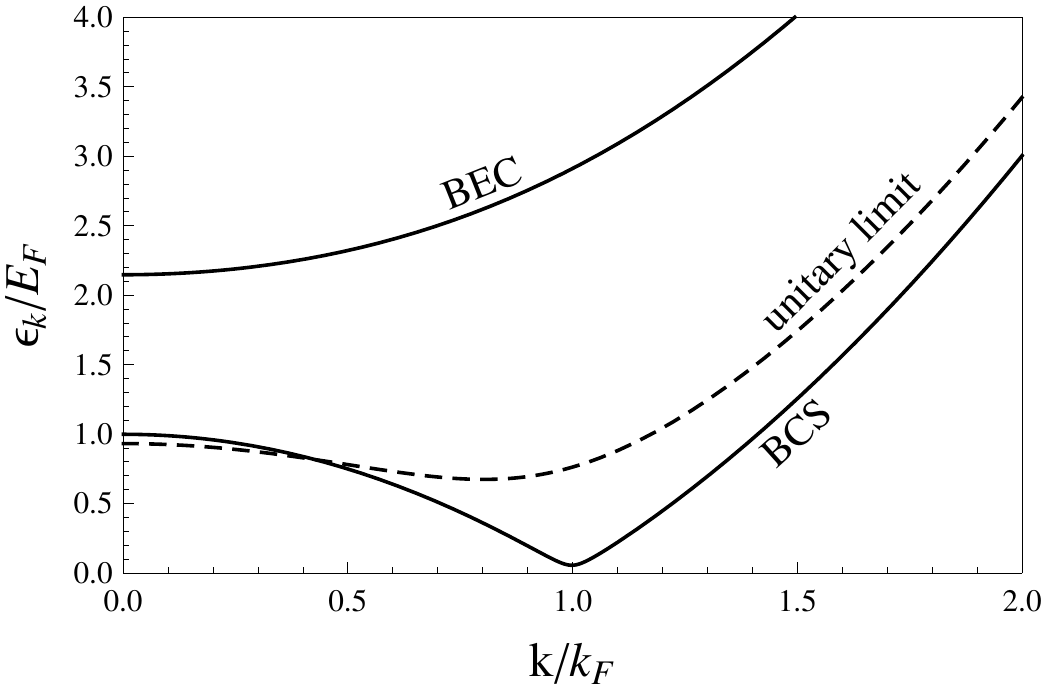}
\caption{Zero-temperature single-fermion excitation energies for $(k_Fa)^{-1}=-1.6$ (BCS), $(k_Fa)^{-1}=0$ (unitary limit), and $(k_Fa)^{-1}=1.2$ (BEC).
}
\label{figepsBECBCS}
\end{center}
\end{figure}

In Fig.\ \ref{figepsBECBCS}, we plot the quasiparticle dispersion $\epsilon_k$ for three different coupling strengths.
We know that in the BCS regime, $\Delta$ corresponds to 
the energy gap. In this case, the single-particle dispersion has a minimum at a certain nonzero momentum, and one needs the energy 
$2\Delta$ to excite single fermions in the system with this momentum. The corresponding curve is the non-relativistic analogue of the curve in Fig.\ \ref{figgap}.
Now our formalism goes beyond this situation since $\Delta$ becomes large 
while $\mu$ becomes negative. Fig.\ \ref{figepsBECBCS} shows that the minimum at a finite momentum disappears and in the BEC regime the 
minimum occurs at $k=0$. In this case, the gap is not $\Delta$, but
\be \label{muDelta}
\sqrt{\mu^2+\Delta^2} \simeq \frac{1}{2ma^2}\sqrt{1+\frac{16}{3\pi}(k_Fa)^3}  \, ,
\ee
where we have used Eqs.\ (\ref{lim1Delta}) and (\ref{lim1mu}). 

The physical picture of the BCS-BEC crossover is thus as follows. Without interactions, there is a well-defined Fermi surface at $\mu=E_F$. Now
we switch on a weak interaction. Weakly coupled Cooper pairs start to form due to the usual BCS mechanism. This is a pure Fermi surface 
phenomenon, i.e., everything happens in a small vicinity of the Fermi surface. But, the Fermi surface is gone because, by definition,
at a Fermi surface quasifermions can be excited with infinitesimally small energy and this is not possible after pairing because now the energy
$2\Delta$ is needed. Now we increase the interaction strength. The point is that we can understand the physics qualitatively by starting from 
our BCS picture: the Cooper pairs get bound stronger and stronger while at the same time we can no longer speak of a Fermi surface phenomenon
because the strong interaction is able to ``dig'' into the Fermi sphere. As a consequence, more fermions participate in Cooper pairing, namely the ones that, 
at weak coupling, were just sitting in the Fermi sphere, not contributing to any dynamics. Eventually, a bound state in the strict 
sense appears at the point where the scattering length diverges, and, going further to the regime where the scattering length goes to zero again,
this time from above, {\it all} fermions become paired in bosonic molecules. (The particle number of the two species has to be identical
in order for all fermions to find a partner.) Now, as Eq.\ (\ref{muDelta}) shows, one needs, to lowest order in $k_Fa$, half of the binding 
energy of a molecule to excite a single fermion.

\begin{figure} [t]
\begin{center}
\includegraphics[width=0.85\textwidth]{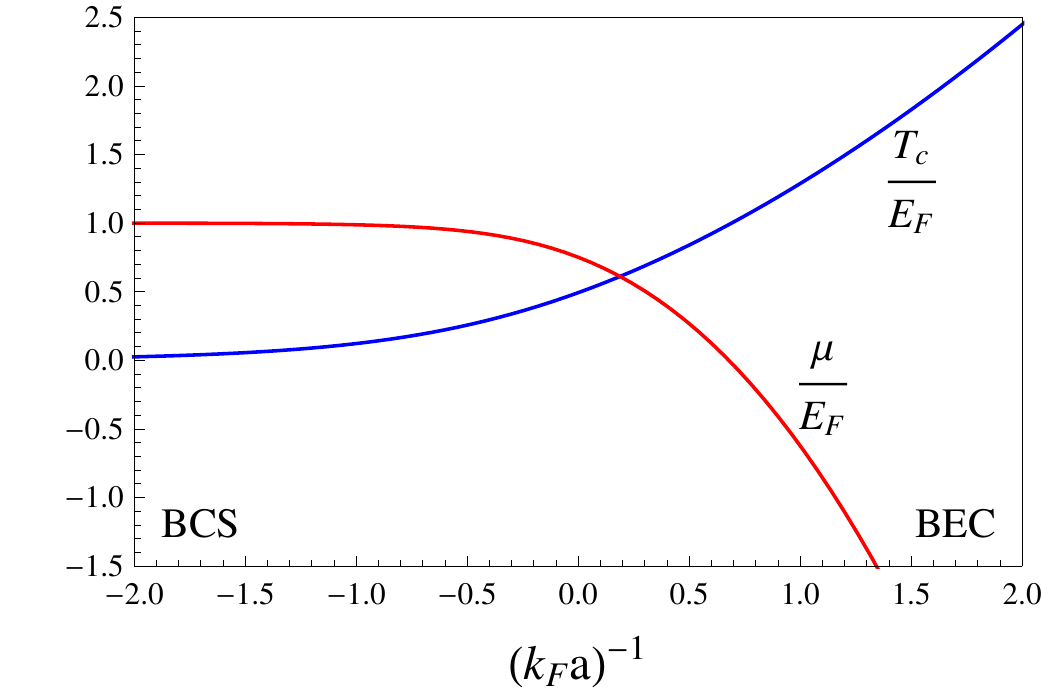}
\caption{(Color online) Critical temperature $T_c$ at which $\Delta=0$ and chemical potential at $T_c$ computed from Eqs.\ (\ref{Tc12}). 
}
\label{figTc1}
\end{center}
\end{figure}

\begin{figure} [t]
\begin{center}
\includegraphics[width=0.85\textwidth]{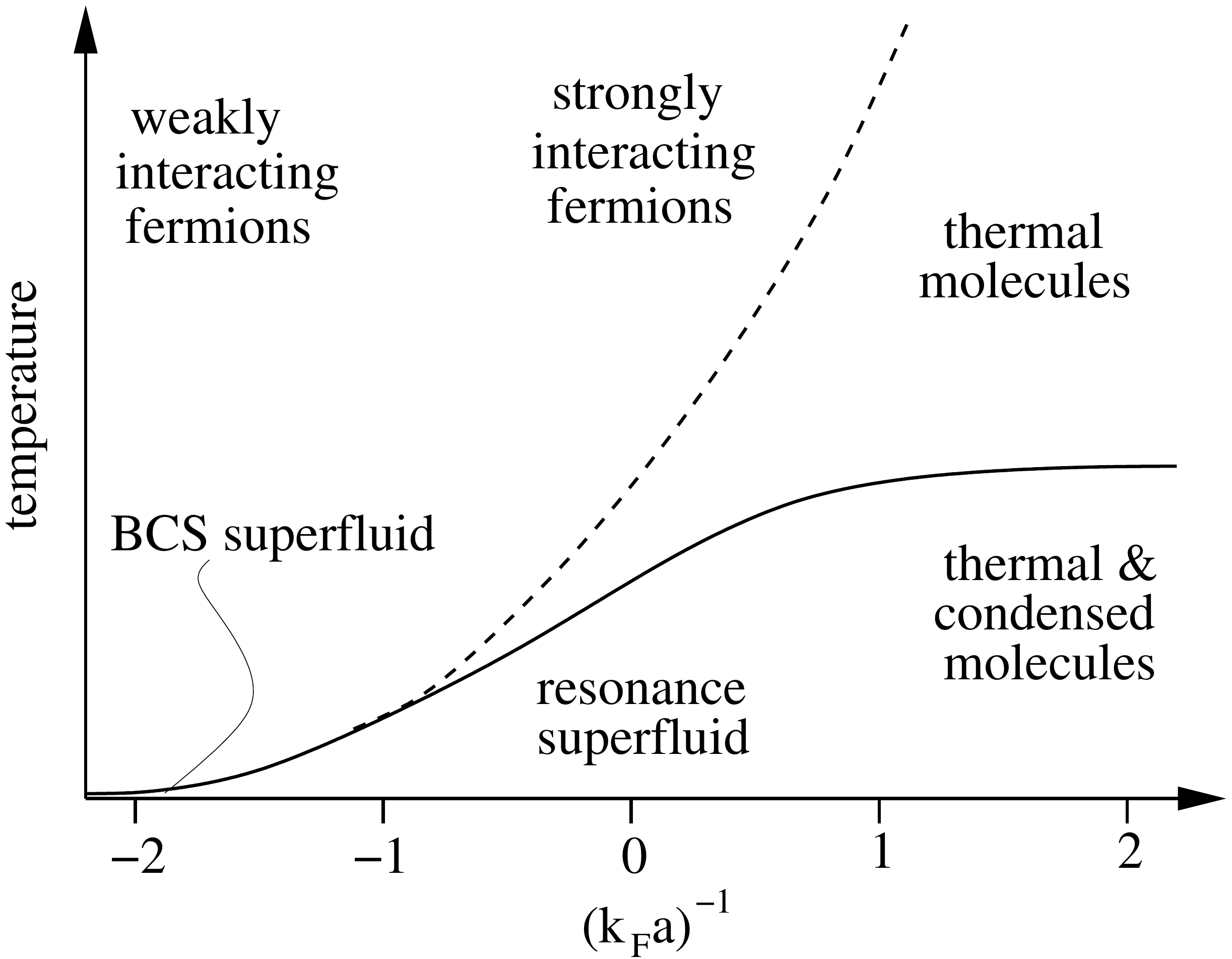}
\caption{
Schematic phase diagram according to Ref.\ \cite{ketterle}. The critical temperature from Fig.\ \ref{figTc1} is the temperature where difermions start to form
(dashed line), to be distinguished from the phase transition temperature where these difermions condense (solid line). 
Only in the BCS limit, these two temperatures coincide.
}
\label{figTc2}
\end{center}
\end{figure}

Finally, we consider nonzero temperatures, although our mean-field treatment becomes more questionable in this case. 
Nevertheless, let us try to determine the critical temperature $T_c$ from 
our gap equation and number equation. As before, we define $T_c$ as the temperature where $\Delta$ becomes 
zero. From Eqs.\ (\ref{gap3}) and (\ref{n3}) we find in this case 
\begin{subequations}
\bea
-\frac{m}{4\pi a} &=& \int\frac{d^3{\bf k}}{(2\pi)^3} \left(\frac{\tanh\frac{\xi_k}{2T_c}}{2\xi_k}-\frac{m}{k^2}\right) \, , \\[2ex]
n&=&\int\frac{d^3{\bf k}}{(2\pi)^3} \left(1-\tanh\frac{\xi_k}{2T_c}\right) \, ,
\eea
\end{subequations}
since $\epsilon_k = |\xi_k|$ for $\Delta=0$. Analogously to the $T=0$ case we can rewrite these equations as
\begin{subequations}\label{Tc12}
\bea
-\frac{1}{k_Fa} &=& \frac{2}{\pi}\left[\frac{2}{3J_2\left(\frac{\mu}{T_c}\right)}\right]^{1/3}J_1\left(\frac{\mu}{T_c}\right) \, , 
\\[2ex]
\frac{T_c}{E_F} &=& \left[\frac{2}{3J_2\left(\frac{\mu}{T_c}\right)}\right]^{2/3} \, ,
\eea
\end{subequations}
with
\begin{subequations}
\bea
J_1(z)&\equiv& \int_0^\infty dx\,x^2\left(\frac{\tanh\frac{x^2-z}{2}}{x^2-z}-\frac{1}{x^2}\right) \, , \\[2ex]
J_2(z)&\equiv& \int_0^\infty dx\,x^2\left(1-\tanh\frac{x^2-z}{2}\right) \, . 
\eea
\end{subequations}
The numerical evaluation is shown in Fig.\ \ref{figTc1}. The behavior in the BCS regime is as expected, it is left as an exercise to show that the BCS relation
(\ref{TcD0}) between the critical temperature and the zero-temperature gap is fulfilled. In the BEC regime, the critical temperature
seems to increase without boundary. Is this expected? If the picture of bosonic molecules is correct, one might think that the critical
temperature is given by the critical temperature of Bose-Einstein condensation. For non-interacting bosons with density $n/2$ (half the fermionic density) 
and mass $2m$ (twice the fermion mass), this temperature is (see any textbook about statistical mechanics, for instance \cite{huang})
\be
T_c^{\rm BEC} = \frac{\pi}{m}\left[\frac{n}{2\zeta(3/2)}\right]^{2/3} \, .
\ee
We might thus expect that the critical temperature should saturate at this value, or possibly at a slightly corrected value due to the interactions between the molecules. 
The reason for the discrepancy between $T_c$ and $T_c^{\rm BEC}$ is that they indeed describe two different transitions. 
In general, there is one temperature where fermions start forming pairs and one -- lower -- temperature 
where the fermion pairs start forming a Bose-Einstein condensate. 
Only in the weakly coupled BCS theory these temperatures are identical. 
This is shown schematically in the phase diagram in Fig.\ \ref{figTc2}.


In a nutshell, the important points to take away from this discussion are:

\begin{itemize}

\item At zero temperature, there is no phase transition between the BCS state at weak coupling -- where loosely bound Cooper pairs are formed
in a small vicinity around the Fermi surface --   and the BEC state at strong coupling -- where strongly bound di-fermions form a Bose condensate.
This implies that there is no qualitative difference between these two states. 

\item  This so-called BCS-BEC crossover can be realized experimentally in ultra-cold atoms in an optical trap. This is one of the 
few systems where the interaction strength is under complete experimental control and can be varied at will. Because of the very 
mild dependence on the details of the inter-atomic interaction, these experiments can give insight into a large class of physical systems.

\end{itemize}

\chapter{Low-energy excitations in a fermionic superfluid}
\label{sec:lowenergy}

In this chapter we discuss the Goldstone mode in a fermionic superfluid. So far, we have discussed the Goldstone mode in a bosonic superfluid, 
chapter \ref{sec:phi4}, and the absence of a Goldstone mode in a fermionic superconductor, chapter \ref{sec:meissner1}. In our field-theoretical discussion of
a fermionic superfluid in chapter \ref{sec:cooper} we have discussed {\it fermionic} excitations, in particular their energy gap. If this energy gap 
exists for all fermionic modes (all flavors and colors in quark matter, all directions in momentum space etc.), we can safely
ignore the fermions that reside in the Cooper pairs if we are interested in energies much smaller than this gap. 
Therefore, the low-energy excitations of such a ``conventional'' fermionic superfluid are dominated by {\it bosonic} excitations, the Goldstone mode. 
In fermionic superfluids that do have gapless modes (quark matter phases where
not all quark flavors form Cooper pairs, anisotropic phases where the gap vanishes in certain directions in momentum space etc.), 
fermionic {\it and} bosonic excitations coexist, even at arbitrarily low energies. In this chapter, however, we will only
consider a fully gapped superfluid where the Goldstone mode is dominant at low energies. 

Remember how the Goldstone mode was computed in the bosonic superfluid: we have computed the condensate from the classical equation of motion and 
have considered fluctuations around this condensate. From these fluctuations we have computed the elementary excitations of the system, the Goldstone 
mode and the massive mode. In the fermionic case, it was more complicated to compute the condensate, we have employed the mean-field approximation 
and have obtained the condensate via the gap equation. In order to discuss elementary excitations, again we need to consider fluctuations 
around the condensate. In this case, these are not fluctuations around a classical solution, but around the mean-field solution.

\section{Fluctuations around the mean-field background}

We consider a system of ultra-relativistic fermions with chemical potential $\mu$ and a point-like interaction, 
\be  \label{NJLLag}
{\cal L} = \overline{\psi}(i\gamma^\mu\partial_\mu+\gamma^0\mu)\psi + G (\overline{\psi}\psi)^2 \, . 
\ee
In chapter \ref{sec:cooper}, we have started from a more complicated interaction, including a bosonic propagator 
$D(X,Y)$. Only later, when we solved the gap equation, we have made use of the approximation of a point-like interaction, see Sec.\ \ref{sec:BCSgap}. 
The dimensionful coupling defined in this limit, $G=\frac{g^2}{2M^2}$, corresponds to the coupling that now appears in front of the interaction term of our 
Lagrangian. We can therefore follow the derivation in chapter \ref{sec:cooper} to obtain the mean-field Lagrangian in terms of the Nambu-Gorkov spinor 
$\Psi$, 
\bea \label{LMF}
{\cal L}_{\rm mf} &=&  \overline{\Psi}\left(\begin{array}{cc} [G_0^+]^{-1} & \Phi^- \\[2ex] \Phi^+&
[G_0^-]^{-1}\end{array}\right)\Psi + \frac{\Tr[\Phi^+\Phi^-]}{4G} \, , 
\eea
with the gap matrices $\Phi^\pm$, and where the trace is taken over Dirac space. This form of the mean-field Lagrangian can be read off from 
Eq.\ (\ref{ZNZZ}) with the Nambu-Gorkov propagator (\ref{fullSinv}) and the $\Tr[\Phi^+\Phi^-]$ term coming from $Z_0$ given in Eq.\ (\ref{Z0}).

As an ansatz for the gap matrix we choose again $\Phi^+=\Delta\gamma^5$, such that $\Phi^-=-\Delta^*\gamma^5$, to obtain the explicit form of the Lagrangian  
\be
{\cal L}_{\rm mf}=\overline{\Psi}\left(\begin{array}{cc} i\gamma^\mu\partial_\mu+\gamma^0\mu & -\Delta^*\gamma^5 \\[2ex] \Delta\gamma^5&
 i\gamma^\mu\partial_\mu-\gamma^0\mu\end{array}\right)\Psi - \frac{|\Delta|^2}{G} \, .  
\ee
In order to introduce fluctuations, it is crucial that we allow for the gap $\Delta$ to become complex.
Starting from this mean-field Lagrangian is particularly useful for our purpose because now we can proceed analogously to the bosonic superfluid. 
While in the bosonic case 
we have written the complex scalar field as a sum of its expectation value and fluctuations, we now write the gap $\Delta$ in the same way. Denoting 
the bosonic fluctuation field by $\eta(X)\in \mathbb{C}$, we can write 
\be  
\Delta(X) = \Delta e^{2i{\bf q}\cdot{\bf x}}+ \eta(X) \, , 
\ee
where $\Delta \in \mathbb{R}$ is the constant value of the gap, to be determined from the gap equation. In addition to the fluctuations
we have also introduced a topological mode, determined by the externally given three-vector ${\bf q}$. We have seen in chapter \ref{sec:phi4} that this topological mode
is necessary to introduce a superflow, and thus needed to describe the hydrodynamics of the superfluid. Even though we shall not 
work out the dynamics for the fermionic case in detail, it is instructive to keep the superflow in the derivation as long as possible.

The fluctuations $\eta(X)$ describe the Goldstone mode and a massive mode. They must be considered as a dynamical field, such that the partition function is
\be \label{Zfluc}
Z = \int {\cal D}\overline{\Psi}{\cal D}\Psi {\cal D}\eta^* {\cal D}\eta \, e^{S[\overline{\Psi},\Psi,\eta^*,\eta]} \, .
\ee
Notice that we now work beyond the mean-field approximation. We could have kept and bosonized the fluctuations from the beginning instead of throwing 
them away in Eq.\ (\ref{LMF}). As an exercise you can redo the derivation in that alternative way.

Due to the superflow, the inverse fermionic propagator depends on ${\bf x}$ in a non-trivial way. With a simple transformation of the fields, however, 
we can get rid of this dependence. This is analogous to the field transformation (\ref{vpprime}) in the bosonic $\varphi^4$ theory. Here we need to transform 
fermionic and bosonic fields,
\be
\psi'(X) = e^{i{\bf q}\cdot{\bf x}}\psi(X) \, , \qquad  \eta'(X) = e^{-2i{\bf q}\cdot{\bf x}}\eta(X) \, .
\ee
The factor 2 in the exponential of the transformation of the bosonic field indicates that two fermions form a Cooper pair. 
In terms of the transformed fields, the mean-field Lagrangian plus fluctuations can be written as\footnote{Note that $\psi = e^{-i{\bf q}\cdot{\bf x}}\psi'$, but
$\psi_C = e^{+i{\bf q}\cdot{\bf x}}\psi_C'$ because $\psi_C = C\overline{\psi}^T$.}
\be
{\cal L}_{{\rm mf}+{\rm fl}} = \bar{\Psi'}({\cal S}^{-1}+ h)\Psi' - \frac{\Delta^2}{G}-\frac{\Delta}{G}(\eta'^*+\eta')-\frac{|\eta'|^2}{G} \, ,
\ee
with the abbreviation
\be
h \equiv h[\eta'^*,\eta']\equiv \left(\begin{array}{cc} 0 & -\eta'^*\gamma^5 \\[2ex] \eta'\gamma^5 & 0 \end{array}\right) \, ,
\ee
and the fermionic propagator,
\be
{\cal S}^{-1} = \left(\begin{array}{cc} i\gamma^\mu\partial_\mu+\gamma^0\mu+\vg\cdot{\bf q} & -\Delta\gamma^5 \\[2ex] \Delta\gamma^5
& i\gamma^\mu\partial_\mu-\gamma^0\mu-\vg\cdot{\bf q}\end{array}\right) \, . 
\ee
In the basis of the field $\psi'$ the ${\bf x}$ dependence in the off-diagonal components of ${\cal S}^{-1}$ is gone, and the superflow appears in the form of the term
$\pm\vg\cdot{\bf q}$ in the diagonal components. Thus we could have introduced the superflow just like the spatial components of a 
gauge field from the beginning, formally promoting the space-time derivative to a covariant derivative with a background gauge field 
(the chemical potential plays the role of the temporal component of this gauge field). This is completely analogous to the bosonic field theory discussed in 
chapter \ref{sec:phi4}.

Since the fermionic fields only appear quadratically in the action, we can integrate them out. This leads to the partition function 
\be
Z = \int {\cal D}\eta'^* {\cal D}\eta' \, e^{S[\eta'^*,\eta']} \, ,
\ee
with an action that now only depends on the fluctuation fields,
\bea \label{Seffeta}
S[\eta'^*,\eta'] &=& \frac{1}{2}\int_X\Tr\ln({\cal S}^{-1}+h) - \int_X\left[\frac{\Delta^2}{G}+\frac{\Delta}{G}(\eta'^*+\eta')+\frac{|\eta'|^2}{G}\right] \, ,
\eea
where the trace is taken over Nambu-Gorkov and Dirac space.

\section{Expanding in the fluctuations}

Next, we expand the Tr ln term for small fluctuations. To this end, we write
\be
\Tr\ln ({\cal S}^{-1}+h) = \Tr\ln {\cal S}^{-1}(1+{\cal S}h) = \Tr\ln{\cal S}^{-1} + \Tr\ln (1+{\cal S}h) \, , 
\ee
and use the expansion of the logarithm, $\ln(1+x) = x-\frac{x^2}{2}+\frac{x^3}{3}-\ldots$.  
Consequently, keeping terms up to second order in $\eta$ and writing the space-time arguments explicitly, we obtain
\bea
\frac{1}{2}\int_X\Tr\ln({\cal S}^{-1}+h) &= & \frac{1}{2}\int_X\Tr\ln {\cal S}^{-1}(X,X)+\frac{1}{2}\int_X\Tr[{\cal S}(X,X)h(X)]\non[2ex]
&&\hspace{-2cm}-\frac{1}{4}\int_{X,Y}\Tr[{\cal S}(X,Y)h(Y){\cal S}(Y,X)h(X)] + {\cal O}(\eta^3) \, .
\eea
As a result, we can decompose the effective action (\ref{Seffeta}) into various contributions, according to their power of $\eta$,
\be
S[\eta'^*,\eta'] \simeq S^{(0)} + S^{(1)} + S^{(2)} \, . 
\ee
The various contributions are  
\begin{subequations}
\bea
S^{(0)} &\equiv& \frac{1}{2}\int_X\Tr\ln {\cal S}^{-1}(X,X) -\frac{V}{T}\frac{\Delta^2}{G}  \, , \\[2ex]
S^{(1)} &\equiv& \frac{1}{2}\int_X\Tr[{\cal S}^{-1}(X,X)h(X)] - \frac{\Delta}{G}\int_X(\eta'^*+\eta) \, , \\[2ex]
S^{(2)} &\equiv&  -\frac{1}{4}\int_{X,Y}\Tr[{\cal S}(X,Y)h(Y){\cal S}(Y,X)h(X)] - \int_X\frac{\eta'\eta'^*}{G} \, , \label{S2XY}
\eea
\end{subequations}
where we have used that the trivial space-time integral in the imaginary time formalism of thermal field theory gives $V/T$. 
Here, $S^{(0)}$ is the purely fermionic mean-field effective action that does not contain any fluctuations, i.e., the partition function 
can be written as 
\be \label{ZS0S1S2}
Z \simeq e^{S^{(0)}} \int {\cal D}\eta'^* {\cal D}\eta' \, e^{S^{(1)}+S^{(2)}} \, .
\ee
To evaluate the contributions $S^{(1)}$ and $S^{(2)}$, we
first write the propagator in Nambu-Gorkov space in terms of normal and anomalous propagators, as in chapter \ref{sec:cooper},
\be
{\cal S}(X,Y) = \left(\begin{array}{cc} G^+(X,Y) & F^-(X,Y) \\[2eX] F^+(X,Y) & G^-(X,Y) \end{array}\right) \, . 
\ee
Furthermore, we introduce the Fourier transforms for the propagators, 
\begin{subequations}\label{fourierFG}
\bea 
G^\pm(X,Y) &=& \frac{T}{V}\sum_K e^{-iK\cdot(X-Y)}G^\pm (K) \,  , \\[2ex]
F^\pm(X,Y) &=& \frac{T}{V}\sum_K e^{- iK\cdot(X-Y)}F^\pm (K) \, ,
\eea
\end{subequations}
where we have assumed translational invariance. 
Now, the terms linear in the fluctuations become 
\bea
S^{(1)} &=& \frac{1}{2}\Tr\int_X\Big[\frac{T}{V}\sum_K F^-(K)+\frac{\Phi^-}{2G}\Big]\gamma^5 \eta'(X) \non[2ex]
&&- \frac{1}{2}\Tr \int_X  \Big[\frac{T}{V}\sum_K F^+(K)+\frac{\Phi^+}{2G}\Big] \gamma^5\eta'^*(X) \, ,
\eea
where the trace is taken over Dirac space, and where we have reintroduced $\Phi^\pm = \pm \Delta \gamma^5$. The reason for this particular way of writing the 
result is that
we recover the mean-field gap equation (\ref{gapeq1}) (and its analogue for $F^+\to F^-$, $\Phi^+\to \Phi^-$). Therefore, at the point where the mean-field
gap equation is fulfilled, we have $S^{(1)}=0$. The reason is that the gap equation is obtained by minimizing the 
free energy with respect to $\Delta$, and this is nothing but looking for the point where the variation of the gap vanishes to linear order. Therefore, 
it is clear that $S^{(1)}$ must vanish at the mean-field solution. 
Again, this is analogous to the bosonic field theory discussed in chapter \ref{sec:phi4}: 
there, the contributions to the Lagrangian to linear order in the fluctuations vanished by 
using the equations of motion, see comments below Eq.\ (\ref{L1L2L3L4}). In this
sense, the classical equations of motion for the scalar field (that determine the Bose condensate) correspond to the mean-field gap equation
for the Cooper pair condensate in the fermionic theory. Here we consider fluctuations around a given mean-field background, while in 
chapter \ref{sec:phi4} we considered fluctuations around a given classical background. In both cases, the fluctuations themselves will, in general, back-react on the 
condensate. Therefore, our use of the mean-field gap equation is an approximation, just like the classical solution to the bosonic condensate is an approximation. 
In a more complete treatment, the fluctuations will give additional contributions
to the gap equation and thus a correction to the energy gap $\Delta$. 

Next, we need to evaluate $S^{(2)}$. For the fluctuation fields, we introduce real and imaginary parts, 
\be
\eta'(X) = \frac{1}{\sqrt{2}}[\eta_1'(X)+i\eta_2'(X)] \, , 
\ee
with $\eta_1',\eta_2'\in {\mathbb R}$, and their Fourier transforms,
\bea \label{fouriereta}
\eta_i'(X) &=& \frac{1}{\sqrt{TV}}\sum_K e^{-iK\cdot X}\eta_i'(K) \, , \qquad i=1,2 \, .
\eea
For the second term on the right-hand side of Eq.\ (\ref{S2XY}) this yields 
\be
-\int_X\frac{\eta'\eta'^*}{G} = -\frac{1}{2T^2}\sum_K\left[\frac{\eta_1'(K)\eta_1'(-K)}{G}+\frac{\eta_2'(K)\eta_2'(-K)}{G}\right] \, .
\ee
The trace over Nambu-Gorkov space in the first term on the right-hand side of Eq.\ (\ref{S2XY}) becomes 
\bea
\Tr[{\cal S}h{\cal S}h] &=& -\Tr[G^+\eta'^*\gamma^5 G^-\eta'\gamma^5]
-\Tr[G^-\eta'\gamma^5 G^+ \eta'^*\gamma^5] \non[2ex]
&&+\Tr[F^+\eta'^*\gamma^5 F^+ \eta'^*\gamma^5]+\Tr[F^-\eta'\gamma^5 F^- \eta'\gamma^5] \, , 
\eea
where we have omitted all space-time arguments for brevity. Going to momentum space and to the basis of $\eta_1'$, $\eta_2'$ yields for the first of these terms 
\bea
&&\int_{X,Y}\Tr[G^+(X,Y)\eta'^*(Y)\gamma^5 G^-(Y,X)\eta'(X)\gamma^5] \non[2ex]
&&= \frac{1}{2TV}\sum_{K,P}\Big[\eta_1'(K)\eta_1'(-K)+\eta_2'(K)\eta_2'(-K)-i\eta_1'(K)\eta_2'(-K)+i\eta_2'(K)\eta_1'(-K)\Big] \non[2ex]
&&\hspace{2cm}\times \Tr[G^+(P)\gamma^5 G^-(P+K)\gamma^5] \, ,
\eea
and analogously for the three other terms. To write the result in a compact and instructive way, we introduce the following abbreviations,
\begin{subequations} \label{PiSigmadef}
\bea
\Pi^\pm(K) &\equiv& \frac{1}{2}\frac{T}{V}\sum_P \Tr[G^\pm(P)\gamma^5 G^\mp(P+K)\gamma^5] \, , \\[2ex]
\Sigma^\pm(K) &\equiv& \frac{1}{2}\frac{T}{V}\sum_P \Tr[F^\pm(P)\gamma^5 F^\pm(P+K)\gamma^5] \, .
\eea
\end{subequations} 
Then, putting everything together, the quadratic contribution $S^{(2)}$ can be written as 
\be
S^{(2)} = -\frac{1}{2} \sum_K \eta'(K) \frac{D^{-1}(K)}{T^2} \eta'^\dagger(K) \, , 
\ee
where we have introduced the inverse bosonic propagator in momentum space 
\be
D^{-1} = \left(\begin{array}{cc} \displaystyle{\frac{1}{G}-\frac{\Pi^++\Pi^-}{2}+\frac{\Sigma^++\Sigma^-}{2}} & 
\displaystyle{i\frac{\Pi^+-\Pi^-}{2}-i\frac{\Sigma^+-\Sigma^-}{2}} 
\\[2ex] \displaystyle{-i\frac{\Pi^+-\Pi^-}{2}-i\frac{\Sigma^+-\Sigma^-}{2}} & \displaystyle{\frac{1}{G}-\frac{\Pi^++\Pi^-}{2}-\frac{\Sigma^++\Sigma^-}{2}} 
\end{array}\right) \, , 
\ee
and the two-component field
\be
\eta'(K)\equiv [\eta'_1(K),\eta'_2(K)] \, , 
\ee
using that $\eta_i(-K) = \eta_i^*(K)$. This property for the fluctuation fields in momentum space follows from the fact that the fields in 
position space $\eta_i(X)$ are real. 

\begin{figure} [t]
\begin{center}
\includegraphics[width=0.95\textwidth]{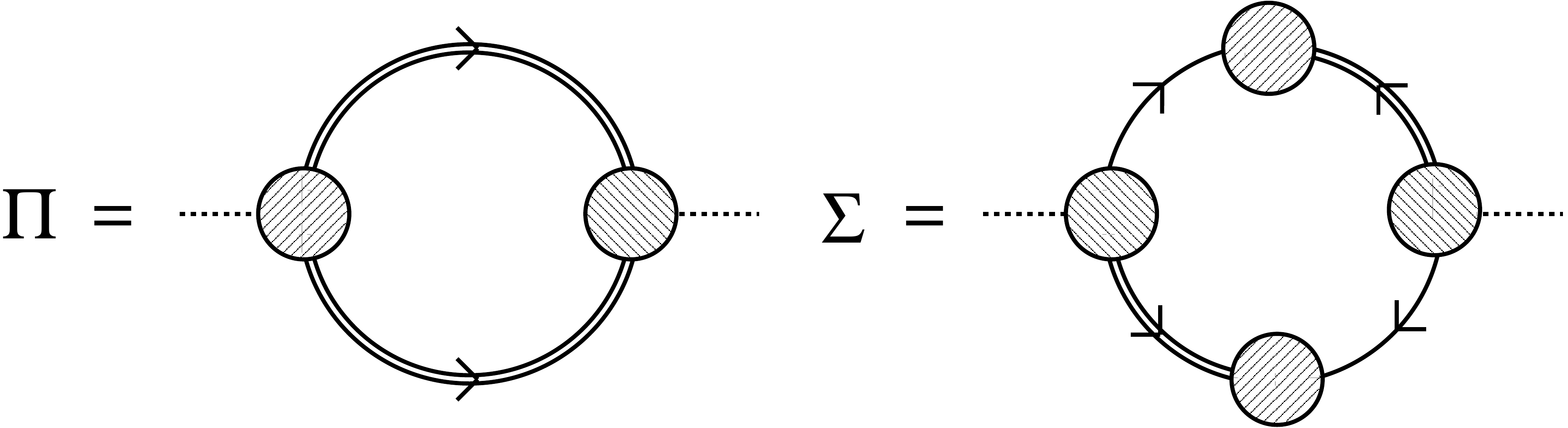}
\caption{Contributions to the inverse propagator (\ref{Dm1K}) of the bosonic excitations in a fermionic superfluid. The diagrammatic notation of the 
fermionic propagators and the 
condensate is as in Figs.\ \ref{figgapeq} and \ref{figPimunu}: hatched circles are the condensates $\Phi^\pm$ (the direction of the hatching  
distinguishes $\Phi^+$ and $\Phi^-$), solid single lines represent the tree-level
propagators $G_0^\pm$, and solid double lines the full propagators $G^\pm$, such that the loop in $\Sigma$ contains the anomalous propagators 
$F^+ = - G_0^- \Phi^+ G^+$. The dotted lines correspond to the elementary excitation, which couples to the fermions via the condensate.
For the algebraic expressions see Eqs.\ (\ref{PiSigmadef}), with $\Pi\equiv \Pi^+$, $\Sigma\equiv \Sigma^+$.}
\label{figPiSigma}
\end{center}
\end{figure}

The inverse bosonic propagator can by simplified a bit by using  
\be
G^+(-K) = -G^-(K) \, , \qquad F^+(-K) = -F^-(K) \, . 
\ee
This property of the propagators can be checked from their explicit form, see Eq.\ (\ref{propGF}). 
Then, by renaming the summation momentum, we see that $\Pi^+(K) = \Pi^-(-K)$ and $\Sigma^+(K)=\Sigma^-(K)$, 
and we can express the inverse propagator solely through $\Sigma^+$ and $\Pi^+$. For convenience, we can thus drop the superscript,
\be 
\Sigma\equiv \Sigma^+ \, , \qquad \Pi\equiv \Pi^+ \, , 
\ee
to write 
\be \label{Dm1K}
D^{-1}(K) = \left(\begin{array}{cc} \displaystyle{\frac{1}{G}-\bar{\Pi}(K)+\Sigma(K)} & i\delta\Pi(K) \\[2ex] 
-i\delta\Pi(K) 
& \displaystyle{\frac{1}{G}-\bar{\Pi}(K)-\Sigma(K)} \end{array}\right) \, ,
\ee
where we have abbreviated
\be
\bar{\Pi}(K) \equiv \frac{\Pi(K)+\Pi(-K)}{2} \, , \qquad \delta\Pi(K) \equiv \frac{\Pi(K)-\Pi(-K)}{2} \, .
\ee
The two momentum sums $\Pi$ and $\Sigma$ correspond to one-loop diagrams, see Fig.\ \ref{figPiSigma}. 

It is instructive to compare these diagrams to the 
similar, but not identical, diagrams that we have computed in the context of the Meissner mass in chapter \ref{sec:meissner1}, see Fig.\ \ref{figPimunu}. 
In both cases, we compute one-loop diagrams given by the fermionic propagators, each loop containing two propagators $G^\pm$ or two anomalous propagators $F^\pm$.
In the case of the photon polarization tensor, the vertex was given by the gauge coupling, i.e., by the interaction between the fermions and the gauge boson that 
appears in the Lagrangian through the covariant derivative. Now, in contrast, the {\it condensate} sits at each vertex. To see this notice that the matrices $\gamma^5$
in the expressions (\ref{PiSigmadef}) originate from the structure of the gap matrices $\Phi^\pm$. Because of the presence of the condensate at the vertex, 
the propagator $G^+$ can be coupled to the propagator $G^-$ (and $F^+$ to $F^+$) which, without the condensate, would simply violate charge conservation. Only
the condensate, where charge can be extracted from or deposited into, allows for such a coupling. In the case of the photon polarization tensor, the coupling was 
different, $G^+$ was coupled to $G^+$ (and $F^+$ to $F^-$). In the diagrams, this difference is manifest in the arrows of the propagators: here, in 
Fig.\ \ref{figPiSigma}, the propagators hit the vertex from both sides while in Fig.\ \ref{figPimunu} charge flows through the vertex in one direction, as usual.  

Before we evaluate the boson propagator explicitly for small momenta, we can write down an expression for the free energy density of the system. By 
performing the integration over the fluctuations in the partition function (\ref{ZS0S1S2}) and using $\Omega = - \frac{T}{V}\ln Z$ we obtain 
\be \label{OmDS}
\Omega = \frac{\Delta^2}{G} -\frac{1}{2}\frac{T}{V}\sum_K\Tr\ln \frac{{\cal S}^{-1}(K)}{T}  
+\frac{1}{2}\frac{T}{V}\sum_K \Tr\ln \frac{D^{-1}(K)}{T^2} \, . 
\ee

\section{Goldstone mode and low-energy expansion }

In chapter \ref{sec:phi4} we have already discussed a bosonic propagator in a superfluid, see Eq.\ (\ref{nondiag}). In the current fermionic formalism 
we have arrived at a much more complicated form of this propagator because we have started on a different microscopic level. Nevertheless, from 
general principles, we expect similar low-energy properties in both systems. In particular, the Goldstone theorem tells us that there should be a gapless mode
also in the fermionic superfluid, which we should obtain by computing the poles of the propagator $D(K)$. We also expect that, in the low-energy 
limit, the dispersion of the Goldstone mode is linear in the momentum. For higher energies, for example 
at temperatures of the order of the gap $\Delta$, the fermionic nature of the superfluid matters, and the physics compared to the bosonic scenario
of chapter \ref{sec:phi4} must be different. This behavior is also encoded in $D(K)$. 

In the following, we set the superflow to zero for simplicity, ${\bf q}=0$, such that we 
can work with the fermionic propagators in momentum space that we already know from chapter \ref{sec:cooper},
\begin{subequations} \label{propGF}
\bea
G^\pm(P) &=& \sum_{e=\pm} \frac{p_0\pm e\xi_p^e}{p_0^2-(\epsilon_p^e)^2}\gamma^0 \Lambda^{\mp e}_p\, , \allowdisplaybreaks \\[2ex] 
F^\pm(P) &=& \pm \sum_{e=\pm} \frac{\Delta\gamma^5\Lambda_p^{\mp e}}{p_0^2-(\epsilon_p^{e})^2} \, ,  
\eea
\end{subequations}
where we restrict ourselves to ultra-relativistic fermions, $m=0$, in which case the energy projectors are given by $\Lambda_p^e = \frac{1}{2}(1+e\gamma^0\vg\cdot\up)$, 
and
\be
\xi_p^e\equiv p-e\mu \, , \qquad \epsilon_p^e = \sqrt{(\xi_p^e)^2+\Delta^2} \, .
\ee
We will also need the explicit form of the gap equation, 
\be \label{1g}
\frac{\Delta}{G} =\sum_{e=\pm}\int\frac{d^3{\bf p}}{(2\pi)^3}\frac{\Delta}{2\epsilon_p^e}[1-2f(\epsilon_p^e)] \, .
\ee
In order to compute $\Pi(K)$ and $\Sigma(K)$, it is convenient to abbreviate $Q\equiv P+K$ and
\begin{subequations}
\bea
\epsilon_1 &\equiv& \epsilon_p^{e_1} \, , \qquad \epsilon_2\equiv \epsilon_q^{e_2} \, , \\[2ex]
\xi_1&\equiv& p-e_1\mu \, , \qquad \xi_2\equiv q-e_2\mu \, .
\eea
\end{subequations}
(Here, $q = |{\bf p}+{\bf k}|$.) With these abbreviations and the traces over Dirac space, 
\be
-\Tr[\gamma^0\Lambda_p^{-e_1}\gamma^5\gamma^0\Lambda_q^{e_2}\gamma^5]=\Tr[\gamma^5\Lambda_p^{-e_1}\gamma^5\gamma^5\Lambda_q^{-e_2}\gamma^5]=1+e_1e_2\up\cdot\uq \, ,  
\ee
we obtain 
\bea
\Pi(K) = -\frac{1}{2}\frac{T}{V}\sum_P\sum_{e_1e_2}(1+e_1e_2\up\cdot\uq)\frac{p_0+e_1\xi_1}{p_0^2-\epsilon_1^2}\frac{q_0-e_2\xi_2}{q_0^2-\epsilon_2^2} \, . 
\eea
To compute $\Pi(-K)$, we change $P\to -P$ in the momentum sum and notice that $\up\cdot\uq$, $\xi_1$, $\xi_2$, $\epsilon_1$, $\epsilon_2$ are invariant under
the simultaneous sign change ${\bf p}\to -{\bf p}$, ${\bf k}\to -{\bf k}$. Therefore, the expression for $\Pi(-K)$ can be written as the one for $\Pi(K)$, only
with opposite signs in front of $e_1\xi_1$ and $e_2\xi_2$. Consequently, for (half of) the sum of $\Pi(K)$ and $\Pi(-K)$, we obtain 
\bea
&&\bar{\Pi}(K)= 
-\frac{1}{2}\frac{T}{V}\sum_P\sum_{e_1e_2}(1+e_1e_2\up\cdot\uq)\frac{p_0q_0-e_1e_2\xi_1\xi_2}{(p_0^2-\epsilon_1^2)(q_0^2-\epsilon_2^2)} \non[2ex]
&&= \frac{1}{2}\sum_{e_1e_2}\int\frac{d^3{\bf p}}{(2\pi)^3}\frac{1+e_1e_2\up\cdot\uq}{4\epsilon_1\epsilon_2} \non[2ex]
&&\times \left\{(\epsilon_1\epsilon_2+e_1e_2\xi_1\xi_2)\left(\frac{1}{k_0+\epsilon_1+\epsilon_2}-\frac{1}{k_0-\epsilon_1-\epsilon_2}\right)
[1-f(\epsilon_1)-f(\epsilon_2)] \right.\non[2ex]
&&\left.-(\epsilon_1\epsilon_2-e_1e_2\xi_1\xi_2)\left(\frac{1}{k_0+\epsilon_1-\epsilon_2}
-\frac{1}{k_0-\epsilon_1+\epsilon_2}\right)[f(\epsilon_1)-f(\epsilon_2)]\right\} \, , 
\eea
where, in the second step, we have performed the Matsubara sum over fermionic Matsubara frequencies $p_0=-(2n+1)i\pi T$, using that $k_0=-2mi\pi T$ are bosonic
Matsubara frequencies, $n,m\in {\mathbb Z}$. All distribution functions $f$ are Fermi 
distributions. For (half of) the difference between $\Pi(K)$ and $\Pi(-K)$, we compute  
\bea \label{OffPi}
&&\delta\Pi(K)= 
\frac{1}{2}\frac{T}{V}\sum_P\sum_{e_1e_2}(1+e_1e_2\up\cdot\uq)\frac{p_0 e_2\xi_2-q_0 e_1 \xi_1}{(p_0^2-\epsilon_1^2)(q_0^2-\epsilon_2^2)} \allowdisplaybreaks
\non[2ex]
&&= \frac{1}{2}\sum_{e_1e_2}\int\frac{d^3{\bf p}}{(2\pi)^3}\frac{1+e_1e_2\up\cdot\uq}{4\epsilon_1\epsilon_2} \non[2ex]
&&\times \left\{(\epsilon_1 e_2\xi_2+\epsilon_2 e_1\xi_1)\left(\frac{1}{k_0+\epsilon_1+\epsilon_2}+\frac{1}{k_0-\epsilon_1-\epsilon_2}\right)
[1-f(\epsilon_1)-f(\epsilon_2)] \right.\non[2ex]
&&\left.+(\epsilon_1 e_2\xi_2-\epsilon_2 e_1\xi_1)\left(\frac{1}{k_0+\epsilon_1-\epsilon_2}
+\frac{1}{k_0-\epsilon_1+\epsilon_2}\right)[f(\epsilon_1)-f(\epsilon_2)]\right\} \, . 
\eea
Finally, the loop containing the anomalous propagators becomes
\bea
\Sigma(K) &=& \frac{1}{2}\frac{T}{V}\sum_P\sum_{e_1e_2}(1+e_1e_2\up\cdot\uq)\frac{\Delta^2}{(p_0^2-\epsilon_1^2)(q_0^2-\epsilon_2^2)} \non[2ex]
&=& \frac{1}{2}\sum_{e_1e_2}\int\frac{d^3{\bf p}}{(2\pi)^3}\frac{1+e_1e_2\up\cdot\uq}{4\epsilon_1\epsilon_2} \non[2ex]
&&\times \Delta^2 \left\{\left(\frac{1}{k_0+\epsilon_1+\epsilon_2}-\frac{1}{k_0-\epsilon_1-\epsilon_2}\right)[1-f(\epsilon_1)-f(\epsilon_2)]\right.
\non[2ex]
&& \left.+\left(\frac{1}{k_0+\epsilon_1-\epsilon_2}-\frac{1}{k_0-\epsilon_1+\epsilon_2}\right)
[f(\epsilon_1)-f(\epsilon_2)]\right\} \, .
\eea
We can now check whether there is indeed a Goldstone mode. A Goldstone mode means that the propagator $D$ has a pole at $K=0$, i.e., the determinant of the 
inverse propagator (\ref{Dm1K}) has a zero at $K=0$. We can set $k_0=0$ directly and immediately read off from Eq.\ (\ref{OffPi}) 
that the off-diagonal elements of $D^{-1}$ vanish,
\be
\delta\Pi(0,{\bf k}) = 0 \,.
\ee
Regarding the spatial momentum ${\bf k}$, we have to be a bit 
more careful because, once we have set $k_0=0$, there are denominators that become zero for ${\bf k}=0$. In the limit ${\bf k}\to 0$ we have ${\bf q}\to{\bf p}$ and 
thus $\up\cdot\uq=1$. Consequently, only the terms where $e_1=e_2$ contribute, and we find 
\begin{subequations}
\bea
\bar{\Pi}(0,0) &=& \frac{1}{2}\sum_e\int \frac{d^3{\bf p}}{(2\pi)^3} \left\{\frac{(\epsilon_p^e)^2+(\xi_p^e)^2}{2(\epsilon_p^e)^3}[1-2f(\epsilon_p^e)]
-\frac{\Delta^2}{(\epsilon_p^e)^2}\frac{\partial f}{\partial\epsilon_p^e}\right\} \, , \hspace{1cm} \allowdisplaybreaks\\[2ex] 
\Sigma(0,0) &=& \frac{1}{2}\sum_e\int \frac{d^3{\bf p}}{(2\pi)^3} \left\{\frac{\Delta^2}{2(\epsilon_p^e)^3}[1-2f(\epsilon_p^e)]+\frac{\Delta^2}{(\epsilon_p^e)^2}
\frac{\partial f}{\partial\epsilon_p^e}\right\} \, , 
\eea
\end{subequations}
where we have used $(\epsilon_p^e)^2-(\xi_p^e)^2 = \Delta^2$. With these results and the gap equation (\ref{1g}) we find that one of the 
diagonal components of the inverse propagator vanishes too,
\be
\frac{1}{G} - \bar{\Pi}(0,0) -\Sigma(0,0) = 0 \, .
\ee
As a result, the determinant of $D^{-1}$ vanishes at $(k_0,{\bf k}) = (0,0)$, and we have thus shown that, for all temperatures below the critical temperature, 
there exists a Goldstone mode.

A general calculation of the bosonic excitations is only possible numerically. Let us therefore discuss the zero-temperature, low-energy limit where 
an analytical evaluation is possible. At zero temperature, we can set all distribution 
functions to zero, $f=0$, because their arguments are positive, $\epsilon_1,\epsilon_2>0$. At low energies, only the Goldstone mode is relevant, and 
we expect its dispersion to be linear, $k_0 = u k$. The corresponding slope $u$ can be computed analytically. To this end, 
it is sufficient to expand $\bar{\Pi}(K)$, $\delta\Pi(K)$, and $\Sigma(K)$ up to quadratic order in $k_0$ and $k$, such that we can write the inverse propagator as 
\be
D^{-1}(K) = \left(\begin{array}{cc} a_1 k_0^2 + b_1 k^2 + c  & -ik_0 d  \\[2ex] ik_0 d & a_2 k_0^2 + b_2 k^2 \end{array}\right) + {\cal O}(K^3)\, ,
\ee
where we have already dropped all potential contributions that do not actually appear. In particular, we already know that the lower right component 
has to vanish for $k_0=k=0$. Setting $k_0=uk$, the determinant of this matrix becomes a polynomial in $k$ with quadratic and quartic terms. The quartic terms
are of course not determined consistently because we have already truncated the expansion of the matrix elements at order $k^2$. 
Requiring the coefficient of the $k^2$ term to vanish yields the slope of the Goldstone dispersion as a function of the various coefficients,
\be \label{uabcd}
u^2 = \frac{b_2c}{d^2-a_2c} \, .
\ee
It remains to compute the coefficients from the loop integrals. This is straightforward, but very tedious, so it is best done with the help of a computer. 
We introduce a momentum cutoff $\Lambda$ and abbreviate the integral over the modulus of the momentum by
\be \label{cutoffp}
\int_p\equiv \int_0^\Lambda \frac{dp\,p^2}{2\pi^2} \, .
\ee
Then, after performing the angular integral over the angle between the momentum of the Goldstone mode ${\bf k}$ and the fermion momentum ${\bf p}$ and
using the gap equation (\ref{1g}), the results become  
\begin{subequations}\label{abcdcoeff}
\bea
a_1 &=& -\sum_e\int_p \frac{(\xi_p^e)^2}{8(\epsilon_p^e)^5} = -\frac{\mu^2}{24\pi^2\Delta^2} + \frac{1}{6\pi^2} -\frac{1}{8\pi^2}\ln\frac{2\Lambda}{\Delta}
  \, , \allowdisplaybreaks\\[2ex] 
a_2 &=& -\sum_e\int_p \frac{1}{8(\epsilon_p^e)^3} = -\frac{\mu^2}{8\pi^2\Delta^2} +\frac{1}{8\pi^2}-\frac{1}{8\pi^2}\ln\frac{2\Lambda}{\Delta}  
\, , \allowdisplaybreaks\\[2ex]
b_1&=& \frac{1}{12}\sum_e\int_p \left[\frac{1}{p^2\epsilon_p^e}+\frac{p^2-2e\mu p-2\Delta^2}{2p^2(\epsilon_p^e)^3}
+\frac{(2p+3e\mu)\Delta^2}{p(\epsilon_p^e)^5}-\frac{5\Delta^4}{(\epsilon_p^e)^7}\right] \non[2ex]
&&- \int_p\frac{\epsilon_p^+\epsilon_p^--\xi_p^+\xi_p^-
-\Delta^2}{6p^2\epsilon_p^+\epsilon_p^-(\epsilon_p^++\epsilon_p^-)} \non[2ex] 
&=& \frac{\mu^2}{72\pi^2\Delta^2} -\frac{5}{24\pi^2}+\frac{1}{8\pi^2}\ln\frac{2\Lambda}{\Delta}\non[2ex] 
&&+\frac{\Delta^2}{24\pi^2\mu\sqrt{\mu^2+\Delta^2}}\ln\frac{\sqrt{\mu^2+\Delta^2}+\mu}{\sqrt{\mu^2+\Delta^2}-\mu}
  \, , \allowdisplaybreaks\\[2ex]
b_2&=& \frac{1}{12}\sum_e\int_p \left[\frac{1}{p^2\epsilon_p^e}+\frac{p-2e\mu}{2p(\epsilon_p^e)^3}
+\frac{3\Delta^2}{2(\epsilon_p^e)^5}\right] - \int_p\frac{\epsilon_p^+\epsilon_p^--\xi_p^+\xi_p^-
+\Delta^2}{6p^2\epsilon_p^+\epsilon_p^-(\epsilon_p^++\epsilon_p^-)} \non[2ex]
&=& \frac{\mu^2}{24\pi^2\Delta^2}-\frac{1}{12\pi^2}+\frac{1}{8\pi^2}\ln\frac{2\Lambda}{\Delta} \, , \allowdisplaybreaks\\[2ex]
c &=& \sum_e\int_p \frac{\Delta^2}{2(\epsilon_p^e)^3} = \frac{\mu^2}{2\pi^2}-\frac{\Delta^2}{2\pi^2}+\frac{\Delta^2}{2\pi^2}\ln\frac{2\Lambda}{\Delta} \, , 
\allowdisplaybreaks\\[2ex] 
d &=& \sum_e\int_p \frac{e\xi_p^e}{4(\epsilon_p^e)^3} = -\frac{3\mu}{4\pi^2}+\frac{\mu}{2\pi^2}\ln\frac{2\Lambda}{\Delta} \, .
\eea
\end{subequations} 
We find the same logarithmic cutoff dependence 
as in the calculation of the Meissner mass, see Eq.\ (\ref{integralUV}). Let us abbreviate
\be
\alpha\equiv \frac{\Delta}{\mu} \, , \qquad \beta \equiv \left(\ln \frac{2\Lambda}{\Delta}\right)^{-1} \, .
\ee
Then, inserting the coefficients (\ref{abcdcoeff}) into 
Eq.\ (\ref{uabcd}) yields
\be \label{ualphabeta}
u^2 = \frac{1}{3}\frac{\beta^2(1-3\alpha^2+2\alpha^4)+\beta\alpha^2(4-5\alpha^2)+3\alpha^4}{\beta^2(1+7\alpha^2+\alpha^4)-2\beta\alpha^2(5+\alpha^2)+\alpha^2(4+\alpha^2)}
\, .
\ee
The reason we have chosen this way of writing the result is as follows. As we know from chapter \ref{sec:cooper}, the cutoff dependent logarithm appears in the 
gap equation. More precisely, if we use the momentum cutoff as in (\ref{cutoffp}) in the gap equation (\ref{1g}), the weak-coupling solution for the gap is
[see also discussion below Eq.\ (\ref{integralUV})]
\be
\Delta = 2\Lambda e^{-\frac{2\pi^2}{\mu^2 G}} \, , 
\ee
such that the logarithm can be expressed in terms of the coupling strength
\be
\beta  = \frac{\mu^2 G}{2\pi^2} \, . 
\ee
(Notice that this dimensionless quantity is not only a measure for the coupling strength, but includes the density of states at the Fermi surface.)   
Therefore, at weak coupling, and taking the cutoff to be larger than, but of the order of, the chemical potential, we have
\be
\alpha \ll \beta \ll 1  \, , 
\ee 
because $\alpha$ is exponentially suppressed compared to $\beta$, while $\beta$ is small because we assume the coupling to be small.

\begin{figure} [t]
\begin{center}
\includegraphics[width=0.85\textwidth]{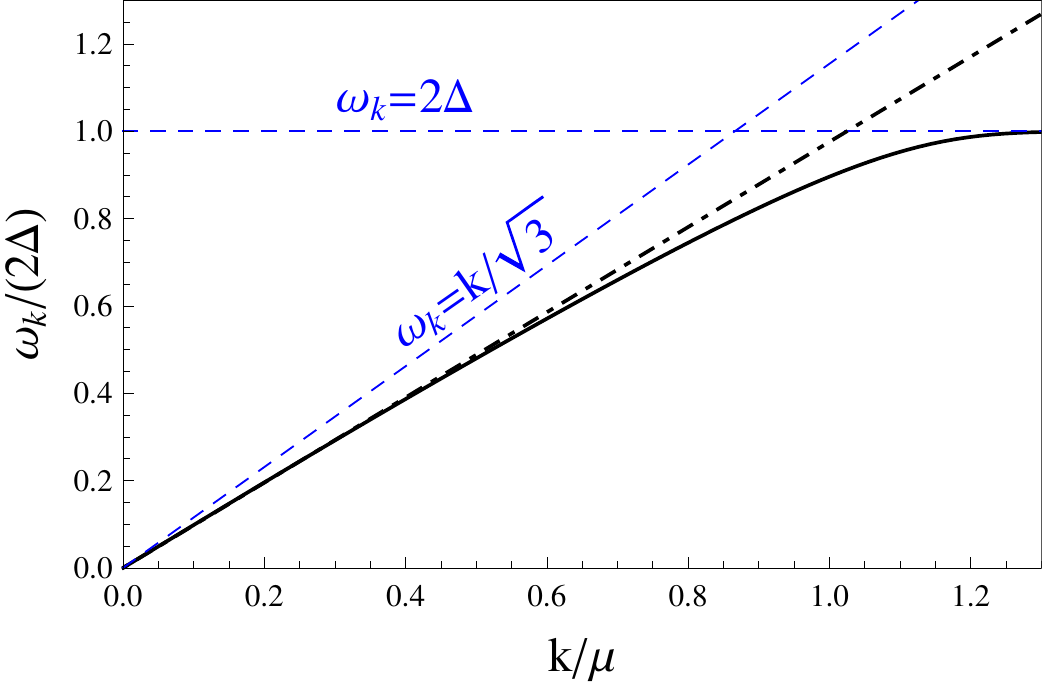}
\caption{(Color online) Zero-temperature Goldstone mode dispersion $\omega_k$ as a function of momentum $k$. The dashed-dotted line is the result from Eq.\ (\ref{ualphabeta}) 
that gives the linear low-energy behavior. The plot was obtained with the parameters $\mu=0.2\Lambda$, $\Delta = 0.05 \Lambda$, where $\Lambda$ is the 
momentum cutoff.}
\label{figomegak}
\end{center}
\end{figure}

We see from Eq.\ (\ref{ualphabeta}) that, in the limit of infinitesimally small coupling, the slope of the Goldstone mode approaches $u=1/\sqrt{3}$. The lowest-order
correction is given by the $\alpha^2$ term in the denominator and reduces the slope, 
\be \label{uapprox}
u\simeq\frac{1}{\sqrt{3}}\left[1+\left(\frac{2\Delta}{\mu}\frac{2\pi^2}{\mu^2 G}\right)^2\right]^{-1/2} \, .
\ee
In Fig.\ \ref{figomegak} we plot the zero-temperature dispersion of the Goldstone mode by evaluating the momentum integrals numerically
and compare the result with the analytic expression (\ref{ualphabeta}). In order to see a sizable deviation from $u=1/\sqrt{3}$ in the linear regime,
we have chosen a relatively large $\Delta$, too large for the approximation (\ref{uapprox}) to hold. Nevertheless, the deviation goes in the same direction, 
i.e., reduces the slope compared to $u=1/\sqrt{3}$.
One can check that there is a parameter regime where this is no longer true and the slope becomes larger than that value. 
We also see from the plot that there is a special energy given by $2\Delta$, which is smoothly approached from below by the dispersion for large momenta. 
We know that beyond this energy fermionic 
excitations become important. As a consequence, it turns out that the excitations in this energy regime are no longer given by stable quasiparticles. 
It is beyond the scope of this course to discuss this regime, for related discussions see Ref.\ \cite{Fukushima:2005gt} for superfluid quark matter and 
Refs.\ \cite{2008PhRvA..77b3626D,Gubankova:2008ya,1997PhRvB..5515153E} for ultra-cold fermionic atoms.

\chapter{Cooper pairing with mismatched Fermi momenta}
\label{sec:mismatch}

In all previous chapters where we have discussed fermionic superfluids we have assumed that the two Fermi momenta of the fermions that form a Cooper pair, 
say fermion A and fermion B, are identical. This is the simplest form of Cooper pairing. It is an interesting question what happens if we release this constraint. 
The general expectation is that 
it becomes
more difficult for the fermions to form Cooper pairs because Cooper pairing, at least at weak coupling, occurs in a small vicinity of the Fermi
surface, as we have seen. Now, when there are two different Fermi surfaces, can the fermions from Fermi surface A pair with fermions from 
Fermi surface B? If they do so, it seems that they would have to form Cooper pairs with nonzero momentum while, in the standard Cooper pairing, 
fermions on opposite sides of the same Fermi sphere pair, such that the total momentum of a Cooper pair vanishes. 
Although nonzero-momentum Cooper pairing is indeed one possibility, we shall see that, if the mismatch in Fermi momenta is 
sufficiently small, standard Cooper pairing with zero-momentum Cooper pairs is still possible. 

The question of mismatched Cooper pairing was first discussed theoretically for an electronic superconductor. In this case, the two fermion species 
are simply electrons distinguished by their spin, and a mismatch in Fermi momenta can in principle be created by Zeeman splitting in an external 
magnetic field. This situation has already been envisioned a few years after the development of BCS theory in 1962 by B.S.\ Chandrasekhar \cite{Chandrasekhar:1962}
and A.M.\ Clogston \cite{Clogston:1962}. However, a clean experimental study of pairing with mismatched Fermi momenta is difficult in 
this case, because an external magnetic field is obviously in conflict with electronic superconductivity due to the Meissner effect.
 
In recent years, the question of mismatched Cooper pairing has regained interest in the contexts of quark matter and ultra-cold fermionic atoms.
In dense quark matter inside a compact star, the Fermi momenta of the quarks of different flavors are necessarily different. The reason is essentially the
difference in masses between the light up and down quarks and the heavier strange quark. Together with the conditions of electric neutrality and chemical 
equilibrium with respect to the weak interactions, this leads to three different Fermi momenta for unpaired quark matter, for a pedagogical discussion see
Ref.\ \cite{arXiv:1001.3294}. An overview over the multitude of possible phases that can arise in Cooper-paired quark matter with mismatched Fermi momenta 
can be found in Ref.\ \cite{Alford:2007xm}. While in dense quark matter the various Fermi momenta at a given density are unambiguously determined by QCD (even though 
poorly known due to theoretical difficulties), in experiments with ultra-cold fermionic atoms (see chapter \ref{sec:BCSBEC}) this mismatch can be controlled 
at will. There, an imbalance between two fermion species can be created by unequal populations of two hyperfine 
states \cite{2006Sci...311..492Z,2006Sci...311..503P}, for reviews see Refs.\ \cite{sheehy,2010RPPh...73k2401C}. Another interesting application of the physics discussed 
in this chapter is chiral symmetry breaking in QCD. In this case, pairs between fermions and anti-fermions form a chiral condensate, and a mismatch is created by a finite 
baryon chemical potential. Since this kind of pairing does not take place at the Fermi surface, it seems very different from Cooper pairing. Interestingly, however,
in the presence of a strong magnetic field, the dynamics of the system are completely analogous to BCS Cooper pairing \cite{Gusynin:1994xp}, and 
in the presence of a chemical potential there is an analogue of the Chandrasekhar-Clogston limit discussed in this 
chapter \cite{Preis:2012fh,Gorbar:2009bm,Preis:2010cq,Gorbar:2011ya}.

It should be emphasized that the physics of mismatched Cooper pairing is by no means universal. In other words, if you impose a mismatch on 
quark matter, it will react very differently compared to fermionic atoms under the same constraint. It is not our goal to elaborate on the complications that 
arise in each specific system. We will focus on a two-species system with mismatch in chemical potentials in a field-theoretical treatment and compute 
the quasiparticle excitations and the free energy of the paired state. We shall derive the so-called Chandrasekhar-Clogston limit, 
beyond which the paired state becomes unstable, and discuss whether and how a superfluid can accommodate a difference in charge densities of the two fermion species
that pair.

\section{Quasiparticle excitations}

We work in the relativistic field-theoretical formalism of chapters \ref{sec:cooper}, \ref{sec:meissner1}, and \ref{sec:lowenergy}.
Additionally, we introduce a two-dimensional space for two fermion species. You can think of these species as two different atoms or two hyperfine states, or two 
quark flavors etc. As a consequence, the fermion propagator will now be a $16\times 16$ matrix: 
2 degrees of freedom from the two fermion species, 2 from fermions/charge-conjugate fermions (Nambu-Gorkov space), and 4 from spin $\frac{1}{2}$ and particle/antiparticle 
degrees of freedom (Dirac space). Notice that in three-flavor quark matter, this space is even larger; due to 3 color and 3 flavor degrees of freedom, the fermion
propagator is a $72\times 72$ matrix. For convenience, we will sometimes refer to the two fermion species as ``flavors''. We require fermions of different flavors to form 
Cooper pairs, which we implement by the following ansatz for the gap matrix,
\be \label{Phim}
\Phi^\pm = \pm \Delta \sigma_1 \gamma_5 \, ,
\ee
with the off-diagonal and symmetric Pauli matrix $\sigma_1$. As mentioned below Eq.\ (\ref{Phiansatz}), the ansatz must lead to an overall anti-symmetric
Cooper pair. Using the same Dirac structure as for the single-species system in chapter \ref{sec:cooper}, we therefore must add a symmetric structure in the 
internal flavor space. 
  
We also have to promote the inverse tree-level propagator from Eq.\ (\ref{G0m1}) to a matrix in flavor space,
\bea \label{G0m1m}
[G_0^\pm]^{-1} &=& \left(\begin{array}{cc} \gamma^\mu K_\mu \pm \mu_1 \gamma^0  & 0  \\ 0 & \gamma^\mu K_\mu \pm \mu_2 \gamma^0 \end{array}\right) \non[2ex]
&=& \sum_{e=\pm} \gamma^0 \Lambda_k^{\pm e} \left(\begin{array}{cc} k_0\pm(\mu_1-ek) & 0  \\ 0 & k_0\pm(\mu_2-ek) \end{array}\right) \, ,
\eea
where we have introduced different chemical potentials $\mu_1$ and $\mu_2$ for the two flavors. Since we work with vanishing fermion
masses, the chemical potentials are identical to the Fermi momenta, $\mu_i = k_{F,i}$. Below it will be convenient to work with 
the average chemical potential $\bar{\mu}$ and (half of) the difference between the chemical potentials $\delta\mu$,
\be
\bar{\mu}\equiv \frac{\mu_1+\mu_2}{2} \, , \qquad \delta\mu \equiv \frac{\mu_1-\mu_2}{2} \, .
\ee
Without loss of generality we may assume $\mu_1>\mu_2$, such that $\delta\mu>0$.

The tree-level propagator is easily obtained by inverting $[G_0^\pm]^{-1}$,
\be \label{G0m}
G_0^\pm = \sum_{e=\pm} \gamma^0 \Lambda_k^{\mp e} \left(\begin{array}{cc} \frac{1}{k_0\pm(\mu_1-ek)} & 0 \\ 0 & \frac{1}{k_0\pm(\mu_2-ek)} 
\end{array}\right) \, .
\ee
Our first goal is to compute the quasiparticle excitations. To this end we need to compute the full propagator from Eq.\ (\ref{Gpm}),
\be
G^\pm =   \left([G_0^\pm]^{-1}-\Phi^\mp G_0^\mp \Phi^\pm\right)^{-1} \, .
\ee
With
\bea
\Phi^\mp G_0^\mp \Phi^\pm  
&=& \sum_e\gamma^0\Lambda_k^{\pm e}\left(\begin{array}{cc} \frac{\Delta^2}{k_0\mp(\mu_2-ek)} & 0 \\ 0 & \frac{\Delta^2}{k_0\mp(\mu_1-ek)} 
\end{array}\right)
\eea
(note the flip of the chemical potentials due to the matrix multiplication in flavor space), we find
\bea 
G^\pm 
&=& \sum_e \gamma^0\Lambda_k^{\mp e}\left(\begin{array}{cc} \frac{k_0\mp(\mu_2-ek)}{(k_0\pm\delta\mu)^2
-(\epsilon_k^e)^2} & 0 \\ 0 & \frac{k_0\mp(\mu_1-ek)}{(k_0\mp\delta\mu)^2-(\epsilon_k^e)^2} 
\end{array}\right) \, , \label{Gpmm}
\eea
where we have denoted
\be
\epsilon_k^e \equiv \sqrt{(\bar{\mu}-ek)^2+\Delta^2} \, ,
\ee
which, without mismatch, would be the single-particle fermionic excitations in the superfluid. 
In the derivation of Eq.\ (\ref{Gpmm}) we have used the relation 
\be
[k_0\pm(\mu_1-ek)][k_0\mp(\mu_2-ek)] = (k_0\pm\delta\mu)^2-(\bar{\mu}-ek)^2 \, .
\ee
This relation is very useful for the following and will be used multiple times. 

\begin{figure} [t]
\begin{center}
\includegraphics[width=0.85\textwidth]{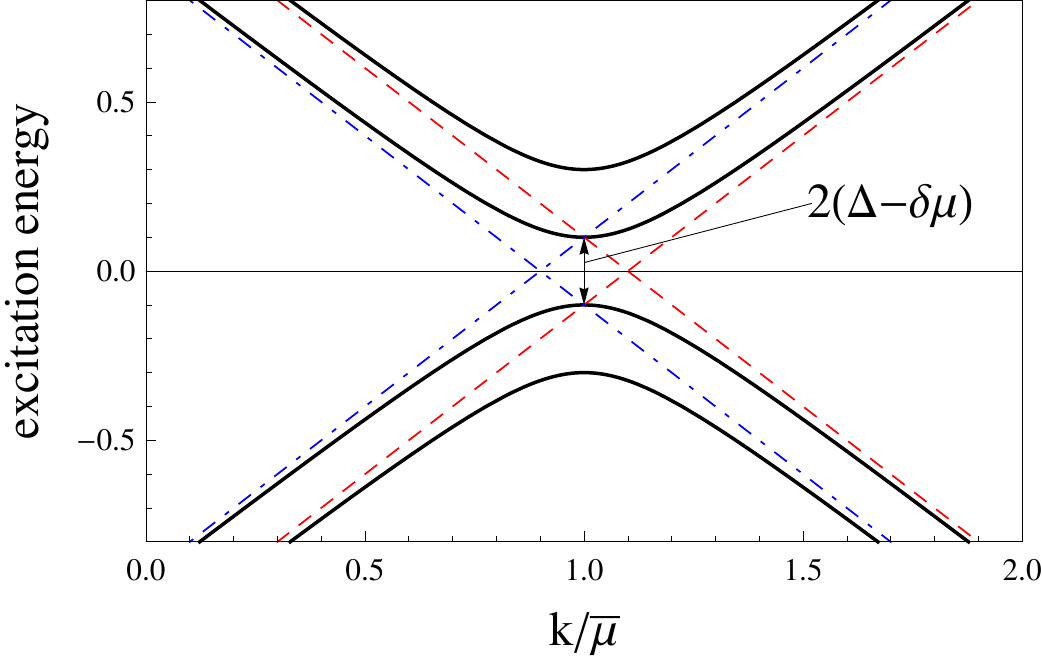}
\caption{(Color online) 
Fermionic quasiparticle excitations (in units of $\bar{\mu}$) in an ultra-relativistic superfluid where fermions with different chemical potentials form 
Cooper pairs. The solid lines show the dispersion relations in the superfluid state, while the dashed and dashed-dotted lines are the dispersions of fermions
and fermion-holes of species 1 and 2 in the absence of Cooper pairing. The parameters for this plot are  $\delta\mu = 0.1 \bar{\mu}$, $\Delta = 0.2\bar{\mu}$.
}
\label{figepsm}
\end{center}
\end{figure}

We see that the poles of the propagator are 
\be
\epsilon_k^e+\delta\mu \, , \qquad \epsilon_k^e-\delta\mu \, , \qquad -\epsilon_k^e+\delta\mu \, , \qquad -\epsilon_k^e-\delta\mu \, , 
\ee
i.e., including anti-particles ($e=-$) there are 8 poles. 
In the absence of pairing, the 8 poles correspond to fermions of species 1 and 2
and fermion-holes of species 1 and 2, and the same for the anti-fermions. In the case of pairing, the new quasiparticles are mixtures
of the original ones, but of course the number of excitation branches is still 8. The excitations  with and without pairing are 
shown in Fig.\ \ref{figepsm}, where we do not include the anti-fermions. 
We see that, at low momenta, two of the excitation branches are fermions (fermion holes) of species 1 while, at large momenta, they become fermion holes (fermions)
of species 2; in between, they are mixtures thereof. For the other two quasiparticles, exchange 1 and 2 in this sentence.  
We see that the mismatch leads to a reduction of the effective energy gap, and for $\delta\mu>\Delta$ the effective gap is gone. 
It turns out, however, that this gapless scenario corresponds to an unstable state, as we shall discuss below.

\section{Free energy}
\label{sec:clogston}

When we have discussed Cooper pairing in the previous chapters, we have always assumed without proof that the paired state has lower free energy than the 
non-superfluid state. In principle, this needs to be checked, because the non-superfluid state $\Delta=0$ is always a solution of the gap equation. 
In the case of pairing with mismatch, this free energy comparison will turn out to be very important. Therefore, in this section, we compute the free energy
and, as a side result, we shall prove that without mismatch the superfluid state is indeed preferred over the non-superfluid state. 

As a starting point, let us go back to the free energy density (\ref{OmDS}), 
\be  \label{Omega01}
\Omega =  -\frac{1}{2}\frac{T}{V}\sum_K\Tr\ln \frac{{\cal S}^{-1}}{T}  + \frac{\Delta^2}{G} \, , 
\ee
where ${\cal S}$ is the propagator in Nambu-Gorkov space (\ref{SGF}), where we have neglected the contribution of the fluctuations, 
and where the trace is taken over Nambu-Gorkov, Dirac, and the internal flavor space. 
For a rigorous derivation in the presence of a mismatch, let us 
use a more general form of the free energy,  
\be \label{Omega}
\Omega =  -\frac{1}{2}\frac{T}{V}\sum_K\Tr\ln \frac{{\cal S}^{-1}}{T} + \frac{1}{4}\frac{T}{V}\sum_K\Tr[1-{\cal S}_0^{-1}{\cal S}] \, ,
\ee
where ${\cal S}_0^{-1}$ is the inverse tree-level propagator in Nambu-Gorkov space (\ref{inversetree}).
This form of the free energy requires some explanation. It results from the so-called Cornwall-Jackiw-Tomboulis (CJT) or two-particle irreducible (2PI) 
formalism \cite{Luttinger:1960ua,Baym:1962sx,Cornwall:1974vz}. This is a self-consistent formalism which is particularly suited for
systems with spontaneously broken symmetry. For applications to superfluids and superconductors, see for instance Sec.\ IV in Ref.\ \cite{Alford:2007xm} 
and references therein. Without getting into the details of the formalism, let us briefly motivate the form of the free energy (\ref{Omega}).
The CJT effective action density is a functional of the Nambu-Gorkov propagator ${\cal S}$,
\be \label{CJT}
\Gamma[{\cal S}] = \frac{1}{2}\frac{T}{V}\sum_K\Tr\ln \frac{{\cal S}^{-1}}{T} - \frac{1}{2}\frac{T}{V}\sum_K\Tr[1-{\cal S}_0^{-1}{\cal S}] +\Gamma_2[{\cal S}] \, , 
\ee
where $\Gamma_2[{\cal S}]$ is the set of all two-particle irreducible diagrams (= diagrams that do not become disconnected by cutting any two of its lines). 
Extremizing the effective action with respect to the propagator yields 
the Dyson-Schwinger equation ${\cal S}^{-1} = {\cal S}_0^{-1}+\Sigma$ (\ref{DysonS}), which, in our context, is the gap equation for the superfluid energy gap. 
For a concrete calculation, $\Gamma_2[{\cal S}]$ has to be truncated at a certain number of loops. To derive the gap equation at one-loop level, 
one has to go to two-loop level in the effective action.  The free energy (\ref{Omega}) is (the negative of) the effective action density $\Gamma[{\cal S}]$ 
at the stationary point. This is seen by using $\Gamma_2[{\cal S}] = \frac{1}{4}\frac{T}{V}\sum_K \Tr[\Sigma{\cal S}]$ and expressing $\Sigma$ in terms of 
${\cal S}_0$ and ${\cal S}$ via the Dyson-Schwinger equation. The free energy density (\ref{Omega}), in turn, is 
identical to the one from Eq.\ (\ref{Omega01}), if the latter is evaluated at the stationary point too, i.e., if we replace $\frac{\Delta^2}{G}$ by a momentum integral
with the help of the gap equation (\ref{1g}). This is confirmed by the explicit evaluation of Eq.\ (\ref{Omega}), for the result see Eq.\ (\ref{OmegafinalT}). 

We now compute the two terms in the free energy (\ref{Omega}) separately. 
In the $\Tr\ln$ term, we perform the trace in Nambu-Gorkov space with the help of $\Tr\ln = \ln{\rm det}$ and 
\be
{\rm det}\left(\begin{array}{cc} A & B \\ C & D \end{array}\right) = {\rm det}(AD-BD^{-1}CD) \, .
\ee
Then, with the full inverse propagator from Eq.\ (\ref{fullSinv}), 
\be 
{\cal S}^{-1} = \left(\begin{array}{cc} [G_0^+]^{-1} & \Phi^- \\[2ex] \Phi^+ & [G_0^-]^{-1} \end{array}\right) \, ,
\ee
we find 
\be
\Tr\ln \frac{{\cal S}^{-1}}{T} = \Tr \ln \frac{[G_0^+]^{-1}[G_0^-]^{-1}-\Phi^-G_0^-\Phi^+[G_0^-]^{-1}}{T^2} \, .
\ee
With the inverse tree-level propagators (\ref{G0m1m}), we compute
\be
[G_0^+]^{-1}[G_0^-]^{-1} = \sum_e \Lambda_k^{-e} \left(\begin{array}{cc} k_0^2-(\mu_1-ek)^2 & 0 \\ 0 & k_0^2-(\mu_2-ek)^2\end{array}\right) \, ,
\ee
and, with the help of Eqs.\ (\ref{Phim}), (\ref{G0m1m}), (\ref{G0m}), 
\be
\Phi^-G_0^-\Phi^+[G_0^-]^{-1} = \Delta^2\sum_e\Lambda_k^{-e}
\left(\begin{array}{cc} \frac{k_0-(\mu_1-ek)}{k_0-(\mu_2-ek)} & 0 \\[2ex] 0 & \frac{k_0-(\mu_2-ek)}{k_0-(\mu_1-ek)}\end{array}\right) \, ,
\ee
such that
\bea
&& [G_0^+]^{-1}[G_0^-]^{-1}-\Phi^-G_0^-\Phi^+[G_0^-]^{-1} \non[2ex]
&&= \sum_e \Lambda_k^{-e}\left(\begin{array}{cc} \frac{k_0-(\mu_1-ek)}{k_0-(\mu_2-ek)}
[(k_0+\delta\mu)^2 -(\epsilon_k^e)^2]\hspace{-0.8cm} & 0 \\[2ex] 0 & 
\hspace{-0.8cm}\frac{k_0-(\mu_2-ek)}{k_0-(\mu_1-ek)}[(k_0-\delta\mu)^2 -(\epsilon_k^e)^2]\end{array}\right) \, . 
\eea
In Dirac space, this matrix has the form $a_+{\cal P}_++a_-{\cal P}_-$ with complete, orthogonal projectors ${\cal P}_\pm$. For such a matrix, $a_\pm$ are eigenvalues 
with degeneracy $\Tr[{\cal P}_\pm]$, hence we have $\Tr\ln(a_+{\cal P}_++a_-{\cal P}_-) = \Tr[{\cal P}_+]\,\ln a_+ + \Tr[{\cal P}_-]\,\ln a_-$. In our case, the 
degeneracy of each eigenvalue is $\Tr[\Lambda_k^{-e}]=2$. Therefore, we obtain 
\bea \label{term1}
&&-\frac{1}{2}\frac{T}{V}\sum_K \Tr\ln {\cal S}^{-1} = -\frac{T}{V}\sum_K\sum_e\left[\ln\frac{k_0-(\mu_1-ek)}{k_0-(\mu_2-ek)}
\frac{(k_0+\delta\mu)^2-(\epsilon_k^e)^2}{T^2}\right. \non[2ex]
&& \left. \hspace{4.4cm}+ \ln\frac{k_0-(\mu_2-ek)}{k_0-(\mu_1-ek)}
\frac{(k_0-\delta\mu)^2-(\epsilon_k^e)^2}{T^2}\right] \allowdisplaybreaks \non[2ex]
&&=-\frac{T}{V}\sum_K\sum_e \left[\ln\frac{(\epsilon_k^e+\delta\mu)^2-k_0^2}{T^2}+\ln\frac{(\epsilon_k^e-\delta\mu)^2-k_0^2}{T^2} \right] \non[2ex]
&&= -2 \sum_e \int\frac{d^3 {\bf k}}{(2\pi)^3}\left[\epsilon_k^e
+T\ln\left(1+e^{-\frac{\epsilon_k^e+\delta\mu}{T}}\right) +T\ln\left(1+e^{-\frac{\epsilon_k^e-\delta\mu}{T}}\right)\right] \, ,
\eea
where, in the last step, we have performed the Matsubara sum over fermionic Matsubara frequencies $k_0=-(2n+1)i\pi T$ (dropping an infinite constant).

Now we turn to the second term of the free energy (\ref{Omega}). Performing the trace over Nambu-Gorkov space yields
\be
\Tr[1-{\cal S}_0^{-1}{\cal S}]=\Tr\left[2-[G_0^+]^{-1}G^+-[G_0^-]^{-1}G^-\right] \, . 
\ee
We see that the anomalous propagators $F^\pm$ drop out. Now, with 
\be
[G_0^\pm]^{-1}G^\pm = \sum_e \Lambda_k^{\mp e} \left(\begin{array}{cc} 1+ \frac{\Delta^2}{(k_0\pm\delta\mu)^2-(\epsilon_k^e)^2}& 0 \\[2ex]
 0 & 1+ \frac{\Delta^2}{(k_0\mp\delta\mu)^2-(\epsilon_k^e)^2}\end{array}\right) \, , 
\ee
we find, after performing the traces over Dirac and flavor space, and performing the Matsubara sum,
\bea \label{term2}
\frac{1}{4}\frac{T}{V}\sum_K\Tr[1-{\cal S}_0^{-1}{\cal S}] &=& 
-\frac{T}{V}\sum_K \sum_e \left[\frac{\Delta^2}{(k_0+\delta\mu)^2-(\epsilon_k^e)^2}+\frac{\Delta^2}{(k_0-\delta\mu)^2-(\epsilon_k^e)^2}\right] \non[2ex]
&=&\sum_e \int\frac{d^3 {\bf k}}{(2\pi)^3} \frac{\Delta^2}{\epsilon_k^e} \left[1-f(\epsilon_k + \delta\mu)-f(\epsilon_k - \delta\mu)\right] \, .
\eea
Inserting the results (\ref{term1}) and (\ref{term2}) into the free energy density (\ref{Omega}) yields 
\bea \label{OmegafinalT}
\Omega &=& -2\sum_e\int\frac{d^3 {\bf k}}{(2\pi)^3}\,\left\{\epsilon_k^e+T\ln\left(1+e^{-\frac{\epsilon_k^e-\delta\mu}{T}}\right) +
T\ln\left(1+e^{-\frac{\epsilon_k^e+\delta\mu}{T}}\right) \right.\non[2ex]
&&\left.\hspace{2.2cm} - \frac{\Delta^2}{2\epsilon_k^e}[1-f(\epsilon_k^e-\delta\mu)-f(\epsilon_k^e+\delta\mu)]
\right\} \, . 
\eea

\subsection{Chandrasekhar-Clogston limit}

Let us evaluate the free energy (\ref{OmegafinalT}) at zero temperature. For this limit, we remember that we have assumed $\delta\mu>0$, and 
we use 
\be
\lim_{T\to 0} T\ln(1+e^{-x/T}) =  -x\Theta(-x) \, , \qquad \lim_{T\to 0} f(x)= \Theta(-x) \, , 
\ee
to obtain\footnote{This can be written in an alternative, maybe more instructive, way as 
\be
\Omega = - \sum_e\int\frac{d^3 {\bf k}}{(2\pi)^3} \left\{|\epsilon_k^e+\delta\mu|+|\epsilon_k^e-\delta\mu|
-\frac{\Delta^2}{2\epsilon_k^e}[{\rm sgn}\,(\epsilon_k^e+\delta\mu)+{\rm sgn}\,(\epsilon_k^e-\delta\mu)]\right\} \, , \nonumber
\ee
where no assumption about the sign of $\delta\mu$ has been made.
}
\bea
\Omega = -2 \sum_e\int\frac{d^3 {\bf k}}{(2\pi)^3} \left[\epsilon_k^e + (\delta\mu -\epsilon_k^e)\Theta(\delta\mu -\epsilon_k^e)
-\frac{\Delta^2}{2\epsilon_k^e}\Theta(\epsilon_k^e-\delta\mu)\right] \, .\label{OmT0}
\eea
By allowing $\epsilon_k^e-\delta\mu$ to become negative, we are allowing $\delta\mu$ to become larger than $\Delta$. This is the case where there is no energy
gap left in the excitation spectrum, as already mentioned above.

First we check that we reproduce the non-superfluid free energy $\Omega_0$. This is not completely obvious 
because setting $\Delta=0$ in the single-particle energies $\epsilon_k^e$ yields $|k-e\bar{\mu}|$, not $k-e\bar{\mu}$.  
It is left as an exercise to show that setting $\Delta=0$ in Eq.\ (\ref{OmT0}) yields 
\bea
\Omega_0 = 2\int\frac{d^3 {\bf k}}{(2\pi)^3} [(k-\mu_1)\Theta(\mu_1-k) + (k-\mu_2)\Theta(\mu_2-k) -2k] \, .
\eea
In this form, one recovers the zero-temperature expression for the free energy $\Omega_0 = \epsilon - \mu_1 n_1 - \mu_2 n_2$ with the energy density $\epsilon$ and
the charge densities of the two fermion species $n_1$, $n_2$. After subtracting the vacuum contribution one obtains the expected result,
\be \label{Omnonsf}
\Omega_0 = - \frac{\mu_1^4}{12\pi^2} - \frac{\mu_2^4}{12\pi^2} \, .
\ee
Next, we turn to the free energy of the superfluid state. For the following argument, let us abbreviate the integrand in the free energy (\ref{OmT0}) 
by $I_\Delta^e$, 
\be
\Omega = \sum_e\int_0^\infty I_\Delta^e \, , 
\ee
where the integral stands for the integral over the modulus of the three-momentum $k$. 
As in Sec.\ \ref{sec:BCSgap}, we assume the gap $\Delta$ to vanish everywhere in momentum space except for a small vicinity around the Fermi surface, in this case 
around the average Fermi surface,  $k\in [\bar{\mu}-\delta,\bar{\mu}+\delta]$, where it is assumed to be constant with \mbox{$\Delta\ll\delta\ll\bar{\mu}$}. 
We also assume that $\delta\mu$ is of the order of the gap $\Delta$, such that $\delta\mu\ll \delta$. 
Then we can write
\bea
\Omega &=& \Omega_0 +\sum_e\int_{\bar{\mu}-\delta}^{\bar{\mu}+\delta}I_\Delta^e - \sum_e\int_{\bar{\mu}-\delta}^{\bar{\mu}+\delta}I_0^e
\non[2ex]
&\simeq &  \Omega_0 + \int_{\bar{\mu}-\delta}^{\bar{\mu}+\delta}(I_\Delta^+-I_0^+) \, , 
\eea
where, in the second step, we have set the antiparticle gap to zero. This is possible because at zero temperature and positive chemical potentials the 
anti-particles play no role in the physics of the system. Consequently, we have  
\be
\Omega = \Omega_0 + \Delta\Omega \, , 
\ee
where 
\bea
\Delta\Omega &=& -\frac{1}{\pi^2}\int_{\bar{\mu}-\delta}^{\bar{\mu}+\delta} dk\,k^2\Bigg\{\epsilon_k^+-\frac{\Delta^2}{2\epsilon_k^+}
+\left(\delta\mu - \epsilon_k^++\frac{\Delta^2}{2\epsilon_k^+}\right)\Theta(\delta\mu-\epsilon_k^+)  \non[2ex] 
&& \hspace{1.5cm} -\,\Big[|k-\bar{\mu}| 
+ (\delta\mu -|k-\bar{\mu}|)\,\Theta(\delta\mu-|k-\bar{\mu}|)\Big]
\Bigg\} \,  
\eea
is the free energy difference between the superfluid and non-superfluid phases.
We compute the integrals of the three contributions separately. The first contribution is
\bea
\int_{\bar{\mu}-\delta}^{\bar{\mu}+\delta} dk\,k^2\left(\epsilon_k^+-\frac{\Delta^2}{2\epsilon_k^+}\right) 
&=& \delta\sqrt{\delta^2+\Delta^2}\left(\bar{\mu}^2+\frac{\delta^2}{2}-\frac{\Delta^2}{4}\right)+\frac{\Delta^4}{8}
\ln\frac{\sqrt{\delta^2+\Delta^2}+\delta}{\sqrt{\delta^2+\Delta^2}-\delta} \non[2ex]
&=& \bar{\mu}^2\delta^2 + \frac{\bar{\mu}^2\Delta^2}{2} + \frac{\delta^4}{2} + 
{\cal O}\left(\Delta^4\right) \, .
\eea
Using $\delta>\delta\mu$, the second contribution becomes
\bea
&&\int_{\bar{\mu}-\delta}^{\bar{\mu}+\delta} dk\,k^2\left(\delta\mu - \epsilon_k^++\frac{\Delta^2}{2\epsilon_k^+}\right)
\Theta(\delta\mu-\epsilon_k^+) \allowdisplaybreaks\non[2ex]
&&=\Theta(\delta\mu-\Delta)\int_{\bar{\mu}-\sqrt{\delta\mu^2-\Delta^2}}^{\bar{\mu}+\sqrt{\delta\mu^2-\Delta^2}} 
dk\,k^2\left(\delta\mu - \epsilon_k^++\frac{\Delta^2}{2\epsilon_k^+}\right) \allowdisplaybreaks\non[2ex]
&&=\Theta(\delta\mu-\Delta)\left[\delta\mu \sqrt{\delta\mu^2-\Delta^2}\left(\bar{\mu}^2+\frac{\delta\mu^2}{6}+
\frac{\Delta^2}{12}\right)-\frac{\Delta^4}{8}\ln\frac{\delta\mu+\sqrt{\delta\mu^2-\Delta^2}}{\delta\mu-\sqrt{\delta\mu^2-\Delta^2}}\right]
\allowdisplaybreaks\non[2ex] 
&&= \Theta(\delta\mu-\Delta)\delta\mu\,\bar{\mu}^2\sqrt{\delta\mu^2-\Delta^2}
+{\cal O}(\Delta^4) \, .
\eea
The neglected terms of order $\Delta^4$ and higher include the terms proportional to $\delta\mu^4$ and $\delta\mu^2\Delta^2$.
Finally, the third contribution is
\bea \label{int3}
&&\int_{\bar{\mu}-\delta}^{\bar{\mu}+\delta} dk\,k^2 \Big[|k-\bar{\mu}| + (\delta\mu -|k-\bar{\mu}|)\,\Theta(\delta\mu-|k-\bar{\mu}|)\Big] \non[2ex]
&&= \bar{\mu}^2\delta^2 + \frac{\delta^4}{2} + \bar{\mu}^2\delta\mu^2+\frac{\delta\mu^4}{6}  \, .
\eea
To be consistent with the expansions of the previous two terms, we need to omit the $\delta\mu^4$ term in this result. 
Then, putting everything together yields
\be \label{OmOm0}
\Delta \Omega \simeq \frac{\bar{\mu}^2\delta\mu^2}{\pi^2}-\frac{\bar{\mu}^2\Delta^2}{2\pi^2} 
- \Theta(\delta\mu-\Delta)\frac{\bar{\mu}^2\delta\mu\sqrt{\delta\mu^2-\Delta^2}}{\pi^2}
 \, .
\ee 
Let us first consider the case without mismatch, $\delta\mu=0$. In this case, with $\mu\equiv \bar{\mu}=\mu_1=\mu_2$, the free energy difference is
\be
\Delta\Omega \simeq  -\frac{\mu^2\Delta^2}{2\pi^2} \, .
\ee
This contribution to the free energy density is called {\it condensation energy} and shows that the free energy is lowered by the gap, i.e., 
the superfluid state wins over the non-superfluid state for all nonzero values of $\Delta$, which we had assumed without proof in the previous chapters. 

Next, we switch on the mismatch, but keep it smaller than the gap, $0<\delta\mu<\Delta$. In this case,  
\be \label{Om12}
\Delta \Omega \simeq \frac{\bar{\mu}^2\delta\mu^2}{\pi^2}-\frac{\bar{\mu}^2\Delta^2}{2\pi^2}  \, .
\ee 
The mismatch induces an additional, positive, contribution to the free energy: now you not only gain free energy from pairing 
but also have to pay a price in free energy. The superfluid state is now  
only preferred over the non-superfluid state for 
\be
\delta\mu < \frac{\Delta}{\sqrt{2}} \, .
\ee
This is called the {\it Chandrasekhar-Clogston limit}. For $\delta\mu$ beyond this limit, the superfluid state breaks down. It depends on the specific system 
under consideration whether $\Delta$ or $\delta\mu$ is larger. For instance, in experiments with ultra-cold fermionic atoms, both quantities can be more or less controlled
independently, since $\Delta$ is basically a measure of the interaction strength while $\delta\mu$ (or rather the mismatch in atom number densities 
$\delta n$) can be tuned directly. Therefore, the whole phase diagram in the $\Delta$-$\delta \mu$ plane can in principle be explored. 
In quark matter inside a compact star, on the other hand, both $\Delta$ and $\delta\mu$ are complicated functions of a single parameter, the quark 
chemical potential. Of course, the situation in quark matter is even more complicated because there are several different $\delta\mu$'s, and possibly several 
different $\Delta$'s because of the larger number of fermion species. 

\begin{figure} [t]
\begin{center}
\includegraphics[width=0.95\textwidth]{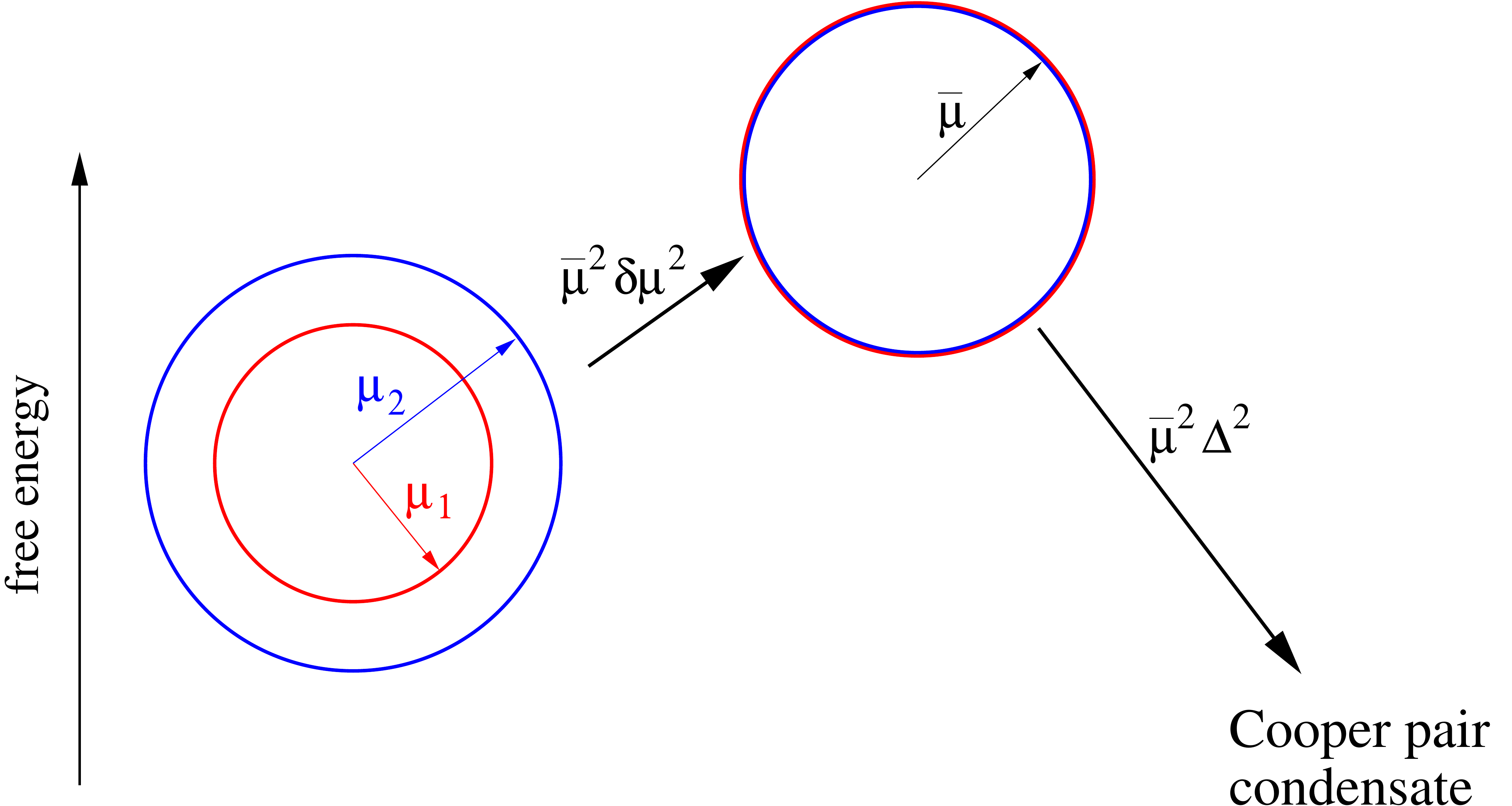}
\caption{(Color online) Illustration of the free energy balance for Cooper pairing in a system where Cooper pairs are formed by fermions which, in the absence of pairing, 
have different Fermi momenta. In the ultra-relativistic limit, these Fermi momenta are given by the chemical potentials $\mu_1$ and $\mu_2$. 
For sufficiently small values of the mismatch $\delta\mu=(\mu_1-\mu_2)/2$, the cost in free energy $\propto \bar{\mu}^2\delta\mu^2$ 
is equivalent to the cost needed to create an ``intermediate'', fictitious state, where both 
species have the same Fermi momentum $\bar{\mu}=(\mu_1+\mu_2)/2$. If the condensation energy $\propto \bar{\mu}^2\Delta^2$ compensates this cost, pairs will form. 
}
\label{figbalance}
\end{center}
\end{figure}

From Eq.\ (\ref{Om12}) we can read off a nice illustrative picture for mismatched Cooper pairing. To this end, let us first write the 
free energy of the non-superfluid state (\ref{Omnonsf}) in terms of $\delta\mu$ and $\bar{\mu}$. For $\delta\mu\ll\bar{\mu}$, we have 
\be
\Omega_0 = - \frac{\mu_1^4}{12\pi^2} - \frac{\mu_2^4}{12\pi^2} =  -\frac{\bar{\mu}^4}{6\pi^2} - \frac{\bar{\mu}^2\delta\mu^2}{\pi^2}
+ {\cal O}(\delta\mu^4)\, . 
\ee
Adding the difference in free energies (\ref{Om12}) to this expression gives the free energy of the superfluid state in the form
\be
\Omega \simeq -\frac{\bar{\mu}^4}{6\pi^2} - \frac{\bar{\mu}^2\Delta^2}{2\pi^2} \, .
\ee
This is nothing but the free energy of a superfluid where both fermion species have the {\it same} Fermi momentum $\bar{\mu}$. 
Therefore, the free energy of the superconducting state in the presence of a mismatch can be understood by first creating a (fictitious) state
where both flavors have one common Fermi surface -- paying a cost in free energy $\propto \bar{\mu}^2\delta\mu^2$ -- and then by forming Cooper pairs in the usual BCS 
way at this common Fermi surface -- which yields an energy gain $\propto \bar{\mu}^2 \Delta^2 $. If the gain exceeds the cost, pairing will happen.
This is illustrated in Fig.\ \ref{figbalance}. One immediate consequence of this picture is that the paired state ``locks'' the two species together, i.e., 
their charge densities in the paired state are identical.  

\begin{figure} [t]
\begin{center}
\includegraphics[width=0.85\textwidth]{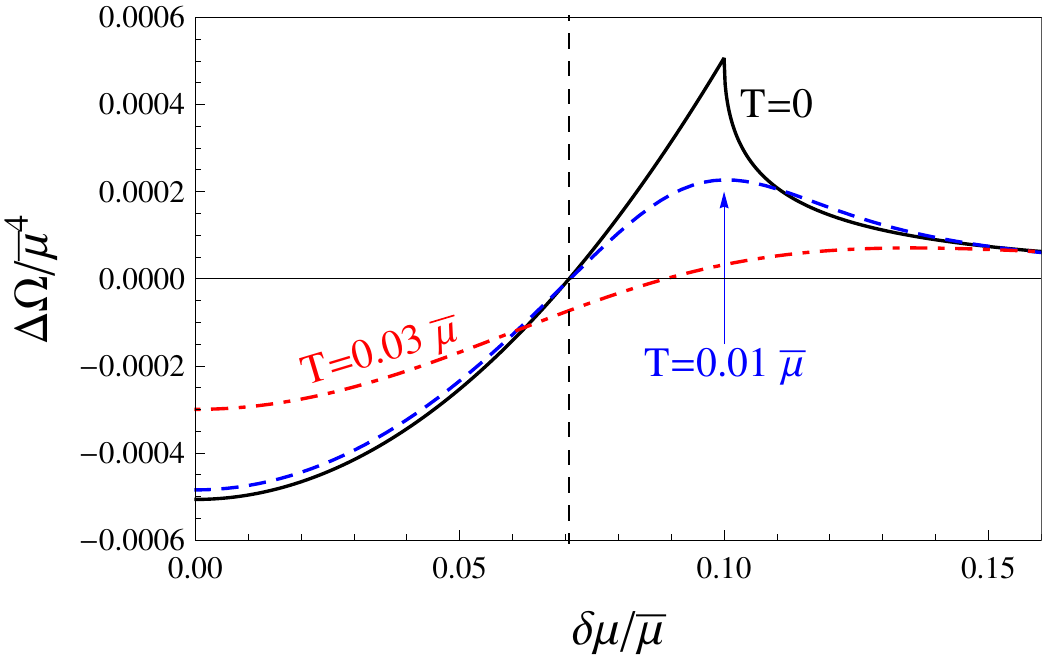}
\caption{(Color online) Difference in free energies $\Delta\Omega$ between superfluid and non-superfluid states as a function of the mismatch $\delta\mu$.
For $\Delta\Omega<0$, the superfluid state is preferred. The zero-temperature result (\ref{OmOm0}) is compared to the general result, using (\ref{OmegafinalT}).
In this plot, the energy gap is kept fixed for all three temperatures, $\Delta = 0.1\bar{\mu}$. The vertical dashed line indicates the Chandrasekhar-Clogston limit
$\delta\mu = \Delta/\sqrt{2}$.   
 }
\label{figdOm}
\end{center}
\end{figure}

We plot the difference between the free energies of the superfluid and normal states in Fig.\ \ref{figdOm}. Here we have included the 
term with the step function in Eq.\ (\ref{OmOm0}), i.e., we have allowed for $\delta\mu>\Delta$. However, this scenario occurs beyond the 
Chandrasekhar-Clogston limit, where the non-superfluid state is always preferred over the superfluid state. We have also plotted two curves 
for nonzero temperatures, by evaluating the general result (\ref{OmegafinalT}) numerically. The nonzero-temperature results are shown for the 
same value of the gap. One can think of increasing the coupling constant $G$ such that, 
even though the temperature is increased, the gap has remained the same. 
By plotting the result in this way, we do not illustrate the effect of the melting of the Cooper pair condensate, but we see that nonzero temperatures 
can support a larger mismatch $\delta\mu$ {\it relative to the gap} $\Delta$ in the superfluid phase.

\section{Superfluids with mismatched charge densities}

So far we have asked the question whether the superfluid state is favored over the non-superfluid state in the presence of a given mismatch in chemical potentials. 
A related question is whether and when the superfluid
state is favored for a given mismatch in the charge density. This is the more relevant question in the context of cold atoms, where 
experiments with different number densities of the two fermion species can be performed. In quark matter, the 
situation is more complicated. In this case, there are different fermions with different electric and color charges and one requires
the system to be color and electrically neutral. This is also a constraint on the different charge densities, but not all of the charge densities are fixed by this
constraint. There are 9 different species (3 colors, 3 flavors), but only two constraints (color \& electric neutrality). One might think that in a system of 
up, down and strange quarks, Cooper pairing can occur in the standard way because the same number of up, down and strange quarks form a neutral system. 
However, due to the heaviness of the strange quark, it is energetically very costly to fill up the strange Fermi sea as much as the Fermi seas of the light 
up and down quarks. Therefore, the system wants to become less strange, and to keep electric neutrality with fewer strange quarks, it wants to increase the number 
of down quarks. This leads to the situation where all three number densities of up, down and strange quarks are different from each other. Therefore, 
pairing between quarks of different flavors (which is the favorite pairing channel in QCD) is challenged in a similar way as if we consider two atom species
with fixed and unequal number densities in the laboratory. Therefore, the following question is of general interest. 
\begin{itemize}
\item Can a fermionic superfluid accommodate different charge densities of the fermion species that form Cooper pairs? And if yes, how?
Put differently: if I impose a difference in number densities, can the system be superfluid?
\end{itemize}
A complete answer to this question is very difficult because, besides the fully gapped state and the non-superfluid state, there are more exotic, partially gapped,
superfluids which may form in the case of imbalanced fermion populations. We can only briefly touch these complications, and first will 
approach the problem by computing the charge densities in the superfluid state. They can be obtained by taking the derivative of the 
effective action (\ref{CJT}) with respect to $\mu_1$ and $\mu_2$.
To this end, notice that the derivative with respect to the full propagator ${\cal S}$ vanishes at the stationary point, i.e., we only need to 
take the {\it explicit} derivatives with respect to the chemical potentials. Since they only appear explicitly in the tree-level propagator ${\cal S}_0^{-1}$,
we obtain 
\bea
n_i &=& \frac{1}{2}\frac{T}{V}\sum_K\Tr\left[\frac{\partial {\cal S}_0^{-1}}{\partial\mu_i}\,{\cal S}\right] \non[2ex]
&=& \frac{1}{2}\frac{T}{V}\sum_K\Tr\left[\frac{\partial [G_0^+]^{-1}}{\partial\mu_i}G^+ + \frac{\partial [G_0^-]^{-1}}{\partial\mu_i}G^-\right] \, , 
\eea
where, in the second step, we have performed the trace in Nambu-Gorkov space. This is the generalization
of Eq.\ (\ref{nNG}) to two flavors. 
With the propagators (\ref{G0m1m}) and (\ref{Gpmm}) we obtain after performing the trace in flavor and Dirac space,
\bea\label{n1start}
n_1 &=& \frac{T}{V}\sum_K\sum_e\left[\frac{k_0-(\mu_2-ek)}{(k_0+\delta\mu)^2-(\epsilon_k^e)^2}
-\frac{k_0+(\mu_2-ek)}{(k_0-\delta\mu)^2-(\epsilon_k^e)^2}\right] \allowdisplaybreaks\non[2ex]
&=& \frac{T}{V}\sum_K\sum_e \frac{2k_0^2(ek-\mu_1)+2(\mu_2-ek)[(\epsilon_k^e)^2-\delta\mu^2]}
{[(k_0+\delta\mu)^2-(\epsilon_k^e)^2][(k_0-\delta\mu)^2-(\epsilon_k^e)^2]} \allowdisplaybreaks\non[2ex]
&=& -\sum_e \int \frac{d^3{\bf k}}{(2\pi)^3} \left[\frac{1}{2}\left(1+e\frac{\xi_k^e}{\epsilon_k^e}\right)\tanh\frac{\epsilon_k^e-\delta\mu}{2T} \right.\non[2ex]
&&\left.\hspace{1.9cm}-\frac{1}{2}\left(1-e\frac{\xi_k^e}{\epsilon_k^e}\right)\tanh\frac{\epsilon_k^e+\delta\mu}{2T}\right]  \, , 
\eea
where, in the last step, we have performed the Matsubara sum and abbreviated $\xi_k^e\equiv k-e\bar{\mu}$. 
Since $n_2$ is obtained from $n_1$ by exchanging $1\leftrightarrow 2$ and thus by $\delta\mu \to -\delta\mu$, we can write the charge densities as 
\begin{subequations}
\bea
n_1 &=& 
2\sum_e e \int \frac{d^3{\bf k}}{(2\pi)^3} \Bigg[\frac{1}{2}\left(1-\frac{\xi_k^e}{\epsilon_k^e}\right)+
\frac{\xi_k^e}{\epsilon_k^e}\frac{f(\epsilon_k^e-\delta\mu)+f(\epsilon_k^e+\delta\mu)}{2}
\non[2ex]
&&\hspace{3.5cm}+e\frac{f(\epsilon_k^e-\delta\mu)-f(\epsilon_k^e+\delta\mu)}{2}\Bigg] \, ,  
\\[2ex]
n_2 &=& 
2\sum_e e \int \frac{d^3{\bf k}}{(2\pi)^3} \Bigg[\frac{1}{2}\left(1-\frac{\xi_k^e}{\epsilon_k^e}\right)+
\frac{\xi_k^e}{\epsilon_k^e}\frac{f(\epsilon_k^e-\delta\mu)+f(\epsilon_k^e+\delta\mu)}{2}
\non[2ex]
&&\hspace{3.5cm}-e\frac{f(\epsilon_k^e-\delta\mu)-f(\epsilon_k^e+\delta\mu)}{2}\Bigg] 
\, .
\eea
\end{subequations}
For $\delta\mu=0$ we have $n_1=n_2\equiv n$ and recover the charge density $n$ from Eq.\ (\ref{nbog}). Let us discuss the charge densities in more detail 
at zero temperature. In this case, using $\delta\mu>0$, 
\begin{subequations} \label{densbreach}
\bea
n_1 &=& 
2\sum_e e \int \frac{d^3{\bf k}}{(2\pi)^3} \left[\frac{1}{2}\left(1-\frac{\xi_k^e}{\epsilon_k^e}\right) 
+\frac{e}{2}\left(1+e\frac{\xi_k^e}{\epsilon_k^e}\right) \Theta(\delta\mu-\epsilon_k^e)\right]\, , 
\\[2ex]
n_2 &=& 
2\sum_e e \int \frac{d^3{\bf k}}{(2\pi)^3} \left[\frac{1}{2}\left(1-\frac{\xi_k^e}{\epsilon_k^e}\right) -\frac{e}{2}
\left(1-e\frac{\xi_k^e}{\epsilon_k^e}\right) \Theta(\delta\mu-\epsilon_k^e)\right]
\, .
\eea
\end{subequations}
For $\delta\mu<\Delta$, the step functions do not give a contribution, and the occupation numbers of the two fermion species are both given by 
the result without mismatch, see Fig.\ \ref{figocc}. This confirms the observation 
we have made in the context of the free energy: as long as $\delta\mu<\Delta$, the system behaves almost like a usual superfluid where Cooper pairs at the
common Fermi surface $\bar{\mu}$ are formed. It does not behave exactly like a usual superfluid because, even though not obvious from the charge densities, 
the mismatch $\delta\mu$ does matter for some properties of the system. For instance, as we have seen in Fig.\ \ref{figepsm}, the energy gap of the quasifermions
is effectively reduced from $\Delta$ to $\Delta-\delta\mu$.

\begin{figure} [t]
\begin{center}
\includegraphics[width=0.75\textwidth]{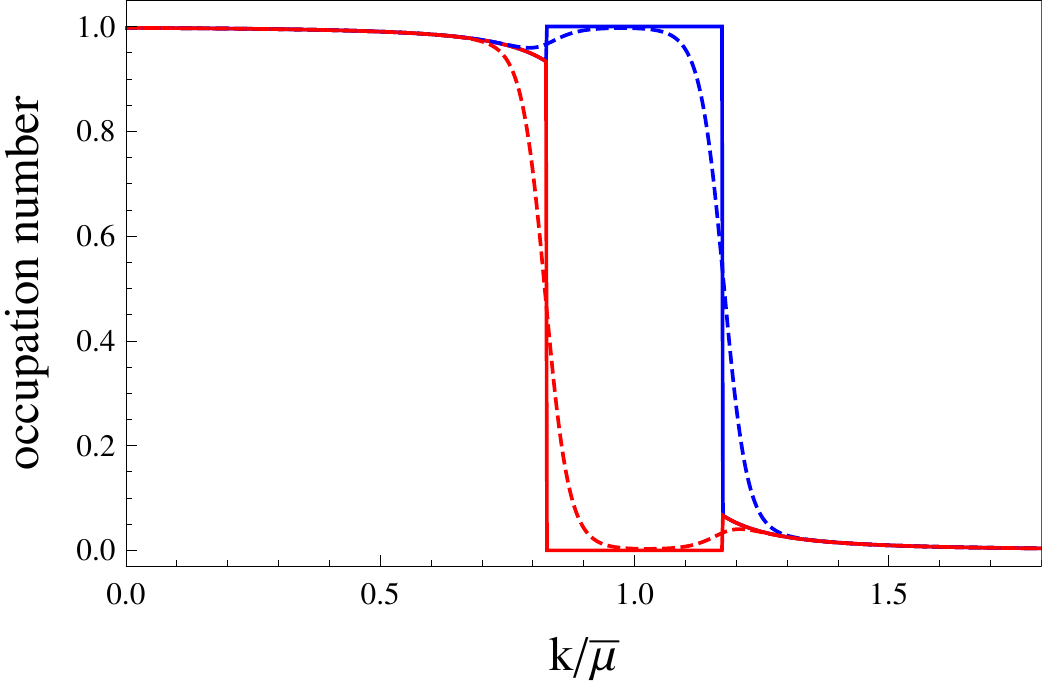}
\caption{(Color online) 
Occupation numbers for the two fermion species that form Cooper pairs with $\delta\mu>\Delta$ at zero (solid lines) and nonzero (dashed lines) temperatures. 
At $T=0$, only one fermion species resides in a  certain region in momentum space around the average Fermi surface $\bar{\mu}$. 
This ``breach'' is less pronounced at nonzero temperatures, here $T=0.02\bar{\mu}$. For all curves, $\Delta=0.1\bar{\mu}$, $\delta\mu=0.2\bar{\mu}$.
}
\label{figbreach}
\end{center}
\end{figure}

For $\delta\mu>\Delta$, the step functions in Eqs.\ (\ref{densbreach}) become nonzero for certain momenta. 
Let us focus on the particle contribution $e=+$, which is the dominant 
contribution for $\bar{\mu}\gg \Delta, \delta \mu$. (In the relativistic BCS-BEC crossover, $\bar{\mu}$ approaches zero in the BEC limit, and anti-particle contributions
become important \cite{Nishida:2005ds,Abuki:2006dv,Deng:2006ed}.) The step functions yield a contribution from a certain 
shell in momentum space,
\be
\int_0^\infty dk\,\Theta(\delta\mu-\epsilon_k^+) = \Theta(\delta\mu - \Delta) \int_{k_-}^{k_+} dk \, , 
\ee
where 
\be \label{kpm}
k_\pm \equiv \bar{\mu}\pm \sqrt{\delta\mu^2-\Delta^2} \, .
\ee
This shell between $k_-$ and $k_+$, sometimes called ``breach'' \cite{Gubankova:2003uj}, is only populated by fermions of the majority species, 
see Fig.\ \ref{figbreach}. Therefore, at sufficiently large $\delta\mu$, the superfluid seems to allow for a difference in charge densities. 
We have to remember, however, that our calculation of the free energy has shown that this state is energetically less favorable than the non-superfluid state, 
see Fig.\ \ref{figdOm}. In other words, if we want more particles of one flavor than of the other flavor, the system will choose not to be 
superfluid. 

In fact, there are situations where the free energy comparison {\it  does} indicate the breached phase
to be preferred. This occurs for instance in dense quark matter under the neutrality constraint \cite{Shovkovy:2003uu,Huang:2003xd,Alford:2003fq}. 
In this case, however, one finds a more subtle instability: if one computes the Meissner masses for the gluons in such a state -- in analogy to our calculation of 
the photon Meissner mass without mismatch in chapter \ref{sec:meissner1} -- one finds imaginary masses for $\delta\mu>\Delta$, 
which is unphysical \cite{Huang:2004bg,Casalbuoni:2004tb,Fukushima:2005cm,Gubankova:2006gj}. (As a -- rather long -- exercise, you may repeat the calculation of 
chapter \ref{sec:meissner1} with a nonzero $\delta\mu$ to verify this statement.) 
This indicates an instability. In other words, suppose that among the phases whose free energy we have compared (non-superfluid and superfluid), 
the breached superfluid turns out to be preferred. Then the imaginary Meissner mass indicates that we have not included the true ground state into our comparison; 
there must be a state with even lower free energy. This state is a superfluid that breaks rotational and/or translational invariance.

Although the breached state has not been found to be a viable option to the ground state in any case (be it quark matter or cold atoms or nuclear 
matter etc.), it helps us to understand possible other superfluid states that {\it can} accommodate excess particles of one fermion species:
instead of the unstable isotropic breach in momentum space, there are stable phases in which the excess, unpaired,  particles sit in ``caps'' at the 
north and/or south pole of the Fermi sphere \cite{Alford:2000ze,Schafer:2005ym,2006PhRvA..74a3614S}. In general,
these phases exhibit counter-propagating currents and a crystalline structure of the order parameter, i.e., a periodically 
varying gap function $\Delta({\bf r})$. They are variants of the so-called FFLO phases (also called LOFF phases, if you prefer a more pronounceable version), 
which have been originally suggested in the context of 
solid state physics by P.\ Fulde, R.A.\ Ferrell \cite{Fulde:1964zz}, and A.\ Larkin, Y.\ Ovchinnikov \cite{larkin:1964zz}. The details of these
phases are beyond the scope of this course, see Refs.\ \cite{Casalbuoni:2003wh,Anglani:2013gfu} for reviews.   

Another -- maybe less spectacular -- option for a superfluid with a mismatch in charge
densities is phase separation, as observed in experiments with ultra-cold atoms \cite{2006PhRvL..97c0401S,2006PhRvL..97s0407P}. 
In such a state, certain regions in position space 
are filled with a usual superfluid (the center of the trap in the atomic experiments), while others contain the non-superfluid state where excess 
particles of one species can easily be accommodated. 

We conclude this chapter with a brief discussion of the charge densities in the non-relativistic BCS-BEC crossover in the presence of a mismatch.
They are given by the $e=+$ contributions of Eqs.\ (\ref{densbreach}),    
\bea
n_{1/2} &=& \int \frac{d^3{\bf k}}{(2\pi)^3} \left[\frac{1}{2}\left(1-\frac{\xi_k}{\epsilon_k}\right)
\pm\frac{1}{2}\left(1\pm\frac{\xi_k}{\epsilon_k}\right)
\Theta(\delta\mu-\epsilon_k)\right] \, , 
\eea
with the ultra-relativistic dispersion relations replaced by
\be
\epsilon_k = \sqrt{\xi_k^2+\Delta^2} \, , \qquad \xi_k = \frac{k^2}{2m} -\bar{\mu} \, .
\ee
Instead of Eq.\ (\ref{kpm}) we now have
\be
k_\pm^2 = 2m(\bar{\mu} \pm \sqrt{\delta\mu^2-\Delta^2}) \, .
\ee
Now remember from chapter \ref{sec:BCSBEC} that in the BCS-BEC crossover the chemical potential $\bar{\mu}$ is in general not large compared to the gap $\Delta$. 
This opens up a third qualitatively different situation besides $\delta\mu<\Delta$ (usual pairing) and the breached pairing shown in Fig.\ \ref{figbreach}. 
Namely, if $\delta\mu >\Delta$, we need to distinguish between $\bar{\mu}>\sqrt{\delta\mu^2-\Delta^2}$ and $\bar{\mu}<\sqrt{\delta\mu^2-\Delta^2}$.
The first case corresponds to the breach.  In the second case, there is no $k_-$ (it formally becomes imaginary). Therefore, what was a shell in momentum space 
between $k_-$ and $k_+$ in the breached phase, now simply becomes a sphere with radius $k_+$. This superfluid state is only possible in the 
BEC regime, where the Cooper pairs have become
bosonic molecules. The physical picture of this state is rather simple: all fermions of the minority species (including the ones deep in the Fermi sea)
take a partner from the majority species to form a molecule. The remaining unpaired fermions of the majority species form a Fermi sphere with Fermi momentum 
$k_+$. 

In the presence of a mismatch, the BCS-BEC crossover is not really a crossover anymore because phase transitions occur. For a given value of the scattering length,
one starts from the usual superfluid at vanishing mismatch. Then, upon increasing the mismatch, some stress is put on the usual Cooper pairing, and some
kind of unusual superfluid occurs, be it a phase-separated state or a LOFF state etc. Then, for a sufficiently large mismatch, Cooper pairing becomes impossible and
the non-superfluid state is the ground state. Since these phase transitions occur at values of $\delta\mu$ that depend on the scattering length, there are also
phase transitions in the other direction, i.e., by varying the scattering length at fixed $\delta\mu$. Therefore, the path from the BCS to the BEC regime
is less smooth than in the case without mismatch. The resulting phase diagrams can be found for instance in 
Ref.\ \cite{sheehy}, see Ref.\ \cite{Deng:2006ed} for a relativistic version.

\bibliographystyle{spphys}
\bibliography{refs2}

\begin{thebibliography}{100}
\providecommand{\url}[1]{{#1}}
\providecommand{\urlprefix}{URL }
\expandafter\ifx\csname urlstyle\endcsname\relax
  \providecommand{\doi}[1]{DOI \discretionary{}{}{}#1}\else
  \providecommand{\doi}{DOI \discretionary{}{}{}\begingroup
  \urlstyle{rm}\Url}\fi

\bibitem{khal}
I.~Khalatnikov, \emph{An Introduction to the Theory of Superfluidity}
  (Addison-Wesley, New York, 1989)

\bibitem{pines}
P.~Nozi\`{e}res, D.~Pines, \emph{The Theory of Quantum Liquids} (Perseus Books,
  Cambridge, Massachusetts, 1999)

\bibitem{landau1}
L.~Landau, E.~Lifshitz, \emph{Statistical Physics} (Pergamon Press, Oxford,
  1980)

\bibitem{annett}
J.~Annett, \emph{Superconductivity, Superfluids, and Condensates} (Oxford
  Univ.\ Press, New York, 2004)

\bibitem{srednicki}
M.~Srednicki, \emph{Quantum Field Theory} (Cambridge University Press,
  Cambridge, 2007)

\bibitem{pokorsky}
S.~Pokorsky, \emph{Gauge Field Theories} (Cambridge University Press,
  Cambridge, 2000)

\bibitem{kapusta}
J.~Kapusta, C.~Gale, \emph{Finite-temperature field theory: Principles and
  Applications} (Cambridge University Press, New York, 2006)

\bibitem{lebellac}
M.~Le~Bellac, \emph{Thermal Field Theory} (Cambridge University Press,
  Cambridge, 2000)

\bibitem{thermal}
A.~Schmitt,
  {http://hep.itp.tuwien.ac.at/\raisebox{0.2mm}{\texttildelow}aschmitt/thermal%
13.pdf}  (2013)

\bibitem{1982PhLA...91...70K}
I.M. {Khalatnikov}, V.V. {Lebedev}, Physics Letters A \textbf{91}, 70 (1982)

\bibitem{1982ZhETF..83.1601L}
V.V. {Lebedev}, I.M. {Khalatnikov}, Zh.\ Eksp.\ Teor.\ Fiz. \textbf{83}, 1601
  (1982).
\newblock [Sov.\ Phys.\ JETP, {\bf 56}, 923 (1982)]

\bibitem{carter89}
B.~Carter, in \emph{Relativistic Fluid Dynamics (Noto 1987)}, ed. by A.~Anile,
  M.~Choquet-Bruhat (Springer-Verlag, 1989), pp. 1--64

\bibitem{1992PhRvD..45.4536C}
B.~{Carter}, I.M. {Khalatnikov}, Phys.Rev. \textbf{D45}, 4536 (1992)

\bibitem{Alford:2012vn}
M.G. Alford, S.K. Mallavarapu, A.~Schmitt, S.~Stetina, Phys.Rev. \textbf{D87},
  065001 (2013)

\bibitem{Pisarski:1999av}
R.D. Pisarski, D.H. Rischke, Phys.Rev. \textbf{D60}, 094013 (1999)

\bibitem{fetter}
A.~Fetter, J.~Walecka, \emph{Quantum theory of many-particle systems}
  (McGraw-Hill, New York, 1971)

\bibitem{tinkham}
M.~Tinkham, \emph{Introduction to Superconductivity} (McGraw-Hill, New York,
  1996)

\bibitem{vollhardt}
D.~Vollhardt, P.~W{\"o}lfle, \emph{The superfluid phases of helium 3} (Taylor
  \& Francis, London, 1990)

\bibitem{Alford:2007xm}
M.G. Alford, A.~Schmitt, K.~Rajagopal, T.~Sch{\"a}fer, Rev.Mod.Phys.
  \textbf{80}, 1455 (2008)

\bibitem{Page:2006ud}
D.~Page, S.~Reddy, Ann.Rev.Nucl.Part.Sci. \textbf{56}, 327 (2006)

\bibitem{arXiv:1001.3294}
A.~Schmitt, Lect.Notes Phys. \textbf{811}, 1 (2010)

\bibitem{2013arXiv1302.6626P}
D.~{Page}, J.M. {Lattimer}, M.~{Prakash}, A.W. {Steiner}, eprint
  arXiv:1302.6626  (2013)

\bibitem{Rischke:2000qz}
D.H. Rischke, Phys.Rev. \textbf{D62}, 034007 (2000)

\bibitem{Rischke:2000ra}
D.H. Rischke, Phys.Rev. \textbf{D62}, 054017 (2000)

\bibitem{Schmitt:2003aa}
A.~Schmitt, Q.~Wang, D.H. Rischke, Phys.Rev. \textbf{D69}, 094017 (2004)

\bibitem{Alford:2005qw}
M.~Alford, Q.h. Wang, J.Phys. \textbf{G31}, 719 (2005)

\bibitem{bcs}
J.~Bardeen, L.~Cooper, J.~Schrieffer, Phys.Rev. \textbf{106}, 162 (1957)

\bibitem{giorgini}
S.~{Giorgini}, L.P. {Pitaevskii}, S.~{Stringari}, Reviews of Modern Physics
  \textbf{80}, 1215 (2008)

\bibitem{ketterle}
W.~{Ketterle}, M.W. {Zwierlein}, Nuovo Cimento Rivista Serie \textbf{31}, 247
  (2008)

\bibitem{2010AnPhy.325..233L}
K.~{Levin}, Q.~{Chen}, C.C. {Chien}, Y.~{He}, Annals of Physics \textbf{325},
  233 (2010)

\bibitem{2012LNP...836.....Z}
W.~{Zwerger} (ed.), \emph{{The BCS-BEC Crossover and the Unitary Fermi Gas}},
  \emph{Lecture Notes in Physics, Berlin Springer Verlag}, vol. 836 (2012)

\bibitem{Nishida:2005ds}
Y.~Nishida, H.~Abuki, Phys.Rev. \textbf{D72}, 096004 (2005)

\bibitem{Abuki:2006dv}
H.~Abuki, Nucl.Phys. \textbf{A791}, 117 (2007)

\bibitem{Deng:2006ed}
J.~Deng, A.~Schmitt, Q.~Wang, Phys.Rev. \textbf{D76}, 034013 (2007)

\bibitem{2009NuPhA.823...83G}
H.~{Guo}, C.C. {Chien}, Y.~{He}, Nuclear Physics A \textbf{823}, 83 (2009)

\bibitem{He:2013gga}
L.~He, S.~Mao, P.~Zhuang, Int.J.Mod.Phys. \textbf{A28}, 1330054 (2013)

\bibitem{2008PhRvA..77b3626D}
R.B. {Diener}, R.~{Sensarma}, M.~{Randeria}, Phys.Rev. \textbf{A77}, 023626
  (2008)

\bibitem{Gubankova:2008ya}
E.~Gubankova, M.~Mannarelli, R.~Sharma, Annals Phys. \textbf{325}, 1987 (2010)

\bibitem{2011AnPhy.326..193S}
A.M.J. {Schakel}, Annals of Physics \textbf{326}, 193 (2011)

\bibitem{Fukushima:2005gt}
K.~Fukushima, K.~Iida, Phys.Rev. \textbf{D71}, 074011 (2005)

\bibitem{Chandrasekhar:1962}
B.S. Chandrasekhar, Appl.Phys.Lett. \textbf{1}, 7 (1962)

\bibitem{Clogston:1962}
A.M. Clogston, Phys.Rev.Lett. \textbf{9}, 266 (1962)

\bibitem{2006Sci...311..492Z}
M.W. {Zwierlein}, A.~{Schirotzek}, C.H. {Schunck}, W.~{Ketterle}, Science
  \textbf{311}, 492 (2006)

\bibitem{2006Sci...311..503P}
G.B. {Partridge}, W.~{Li}, R.I. {Kamar}, Y.a. {Liao}, R.G. {Hulet}, Science
  \textbf{311}, 503 (2006)

\bibitem{sheehy}
L.~{Radzihovsky}, D.E. {Sheehy}, Reports on Progress in Physics \textbf{73},
  076501 (2010)

\bibitem{Alford:2002kj}
M.~Alford, K.~Rajagopal, JHEP \textbf{0206}, 031 (2002)

\bibitem{Rajagopal:2005dg}
K.~Rajagopal, A.~Schmitt, Phys.Rev. \textbf{D73}, 045003 (2006)

\bibitem{Preis:2012fh}
F.~Preis, A.~Rebhan, A.~Schmitt, Lect.Notes Phys. \textbf{871}, 51 (2013)

\bibitem{1938Natur.141...74K}
P.~{Kapitza}, Nature \textbf{141}, 74 (1938)

\bibitem{1938Natur.141...75A}
J.F. {Allen}, A.D. {Misener}, Nature \textbf{141}, 75 (1938)

\bibitem{1924ZPhy...26..178B}
S.N. Bose, Zeitschrift fur Physik \textbf{26}, 178 (1924)

\bibitem{einstein}
A.~Einstein, Sitzungsber.\ Kgl.\ Preuss.\ Akad.\ Wiss. \textbf{261} (1924)

\bibitem{1938Natur.141..643L}
F.~{London}, Nature \textbf{141}, 643 (1938)

\bibitem{1941PhRv...60..356L}
L.~{Landau}, Physical Review \textbf{60}, 356 (1941)

\bibitem{1938Natur.141..913T}
L.~{Tisza}, Nature \textbf{141}, 913 (1938)

\bibitem{balibar}
S.~{Balibar}, Journal of Low Temperature Physics \textbf{146}, 441 (2007)

\bibitem{griffin}
A.~{Griffin}, Journal of Physics Condensed Matter \textbf{21}, 164220 (2009)

\bibitem{1935RSPSA.151..342W}
J.O. {Wilhelm}, A.D. {Misener}, A.R. {Clark}, Royal Society of London
  Proceedings Series A \textbf{151}, 342 (1935)

\bibitem{peshkov1946determination}
V.~Peshkov, J. Phys. USSR \textbf{10}, 389 (1946)

\bibitem{1947PhRv...71..600L}
C.T. {Lane}, H.A. {Fairbank}, W.M. {Fairbank}, Physical Review \textbf{71}, 600
  (1947)

\bibitem{donnelly}
R.J. {Donnelly}, Physics Today \textbf{62}, 34 (2009)

\bibitem{2013arXiv1302.2871S}
L.A. {Sidorenkov}, M.~{Khoon Tey}, R.~{Grimm}, Y.H. {Hou}, L.~{Pitaevskii},
  S.~{Stringari}, Nature \textbf{498}, 78 (2013)

\bibitem{landauhydro}
L.~Landau, E.~Lifshitz, \emph{Fluid mechanics} (Pergamon Press, Oxford, 1987)

\bibitem{2009PhRvA..80e3601T}
E.~{Taylor}, H.~{Hu}, X.J. {Liu}, L.P. {Pitaevskii}, A.~{Griffin},
  S.~{Stringari}, Phys.Rev. \textbf{A80}, 053601 (2009)

\bibitem{2010NJPh...12d3040H}
H.~{Hu}, E.~{Taylor}, X.J. {Liu}, S.~{Stringari}, A.~{Griffin}, New Journal of
  Physics \textbf{12}, 043040 (2010)

\bibitem{Alford:2013koa}
M.G. Alford, S.K. Mallavarapu, A.~Schmitt, S.~Stetina, Phys.Rev. \textbf{D89},
  085005 (2014)

\bibitem{Schmitt:2013nva}
A.~Schmitt, Phys.Rev. \textbf{D89}, 065024 (2014)

\bibitem{gross1961structure}
E.P. Gross, Il Nuovo Cimento Series 10 \textbf{20}, 454 (1961)

\bibitem{pitaevskii1961vortex}
L.~Pitaevskii, Sov. Phys. JETP \textbf{13}, 451 (1961)

\bibitem{donnellybook}
R.~Donnelly, \emph{Quantized Vortices in Helium II} (Cambridge University
  Press, Cambridge, 1991)

\bibitem{2012PhLB..716....1A}
{ATLAS Collaboration}, Physics Letters B \textbf{716}, 1 (2012)

\bibitem{2012PhLB..716...30C}
{CMS Collaboration}, Physics Letters B \textbf{716}, 30 (2012)

\bibitem{1976NuPhB.105..445N}
H.B. {Nielsen}, S.~{Chadha}, Nuclear Physics B \textbf{105}, 445 (1976)

\bibitem{Miransky:2001tw}
V.~Miransky, I.~Shovkovy, Phys.Rev.Lett. \textbf{88}, 111601 (2002)

\bibitem{Brauner:2010wm}
T.~Brauner, Symmetry \textbf{2}, 609 (2010)

\bibitem{Watanabe:2011ec}
H.~Watanabe, T.~Brauner, Phys.Rev. \textbf{D84}, 125013 (2011)

\bibitem{2012PhRvL.108y1602W}
H.~{Watanabe}, H.~{Murayama}, Phys.Rev.Lett. \textbf{108}, 251602 (2012)

\bibitem{2014arXiv1402.7066W}
H.~{Watanabe}, H.~{Murayama}, eprint arXiv:1402.7066  (2014)

\bibitem{Haber:1981fg}
H.E. Haber, H.A. Weldon, Phys.Rev.Lett. \textbf{46}, 1497 (1981)

\bibitem{Andersson:2006nr}
N.~Andersson, G.~Comer, Living Rev.Rel. \textbf{10}, 1 (2007)

\bibitem{Carter:1995if}
B.~Carter, D.~Langlois, Phys.Rev. \textbf{D51}, 5855 (1995)

\bibitem{Herzog:2008he}
C.~Herzog, P.~Kovtun, D.~Son, Phys.Rev. \textbf{D79}, 066002 (2009)

\bibitem{Son:2002zn}
D.~Son, eprint arXiv:hep-ph/0204199  (2002)

\bibitem{Comer:2002dm}
G.~Comer, R.~Joynt, Phys.Rev. \textbf{D68}, 023002 (2003)

\bibitem{Nicolis:2011cs}
A.~Nicolis, eprint arXiv:1108.2513  (2011)

\bibitem{2001cond.mat..1299S}
S.~{Stringari}, eprint arXiv:cond-mat/0101299  (2001)

\bibitem{oai:arXiv.org:cond-mat/0305138}
J.O. Andersen, Rev.Mod.Phys. \textbf{76}, 599 (2004)

\bibitem{2011PPN....42..460Y}
V.I. {Yukalov}, Physics of Particles and Nuclei \textbf{42}, 460 (2011)

\bibitem{unruh}
W.~Unruh, Phys.Rev.Lett. \textbf{46}, 1351 (1981)

\bibitem{2005LRR.....8...12B}
C.~{Barcel{\'o}}, S.~{Liberati}, M.~{Visser}, Living Reviews in Relativity
  \textbf{8}, 12 (2005)

\bibitem{Mannarelli:2008jq}
M.~Mannarelli, C.~Manuel, Phys.Rev. \textbf{D77}, 103014 (2008)

\bibitem{1911KNAB...14..113K}
H.~{Kamerlingh Onnes}, Koninklijke Nederlandse Akademie van Wetenschappen
  Proceedings Series B Physical Sciences \textbf{14}, 113 (1911)

\bibitem{vandelft}
D.~van Delft, P.~Kes, Physics Today \textbf{63}, 38 (2010)

\bibitem{1933NW.....21..787M}
W.~{Meissner}, R.~{Ochsenfeld}, Naturwissenschaften \textbf{21}, 787 (1933)

\bibitem{1950PhRv...79..845F}
H.~{Fr{\"o}hlich}, Phys. Rev. \textbf{79}, 845 (1950)

\bibitem{1986ZPhyB..64..189B}
J.G. {Bednorz}, K.A. {M{\"u}ller}, Zeitschrift f{\"u}r Physik B Condensed
  Matter \textbf{64}, 189 (1986)

\bibitem{1972PhRvL..28..885O}
D.D. {Osheroff}, R.C. {Richardson}, D.M. {Lee}, Physical Review Letters
  \textbf{28}, 885 (1972)

\bibitem{RevModPhys.69.645}
D.M. Lee, Rev. Mod. Phys. \textbf{69}, 645 (1997)

\bibitem{RevModPhys.69.667}
D.D. Osheroff, Rev. Mod. Phys. \textbf{69}, 667 (1997)

\bibitem{2005Natur.435.1047Z}
M.W. {Zwierlein}, J.R. {Abo-Shaeer}, A.~{Schirotzek}, C.H. {Schunck},
  W.~{Ketterle}, Nature \textbf{435}, 1047 (2005)

\bibitem{bogol}
N.N. {Bogoliubov}, Doklady Akad.\ Nauk SSSR \textbf{119}, 52 (1958)

\bibitem{1959NucPh..13..655M}
A.B. {Migdal}, Nucl.\ Phys. \textbf{13}, 655 (1959)

\bibitem{oai:arXiv.org:1011.6142}
D.~Page, M.~Prakash, J.M. Lattimer, A.W. Steiner, Phys.Rev.Lett. \textbf{106},
  081101 (2011)

\bibitem{Shternin:2010qi}
P.S. Shternin, D.G. Yakovlev, C.O. Heinke, W.C. Ho, D.J. Patnaude,
  Mon.Not.Roy.Astron.Soc. \textbf{412}, L108 (2011)

\bibitem{Ivanenko:1969gs}
D.D. Ivanenko, D.F. Kurdgelaidze, Lett. Nuovo Cim. \textbf{IIS1}, 13 (1969)

\bibitem{Barrois:1977xd}
B.C. Barrois, Nucl. Phys. \textbf{B129}, 390 (1977)

\bibitem{Frautschi:1978rz}
S.C. Frautschi, Presented at Workshop on Hadronic Matter at Extreme Energy
  Density, Erice, Italy, Oct 13-21, 1978

\bibitem{Barrois:1979pv}
B.C. Barrois, Non-perturbative effects in dense quark matter.
\newblock Ph.D. thesis, California Institute of Technology, Pasadena,
  California (1979)

\bibitem{Bailin:1979nh}
D.~Bailin, A.~Love, J. Phys. \textbf{A12}, L283 (1979)

\bibitem{Son:1998uk}
D.T. Son, Phys.Rev. \textbf{D59}, 094019 (1999)

\bibitem{Bailin:1983bm}
D.~Bailin, A.~Love, Phys.Rept. \textbf{107}, 325 (1984)

\bibitem{Nambu:1961tp}
Y.~Nambu, G.~Jona-Lasinio, Phys.Rev. \textbf{122}, 345 (1961)

\bibitem{Nambu:1961fr}
Y.~Nambu, G.~Jona-Lasinio, Phys.Rev. \textbf{124}, 246 (1961)

\bibitem{Buballa:2003qv}
M.~Buballa, Phys.Rept. \textbf{407}, 205 (2005)

\bibitem{Schafer:2000tw}
T.~Sch{\"a}fer, Phys.Rev. \textbf{D62}, 094007 (2000)

\bibitem{Alford:2002rz}
M.G. Alford, J.A. Bowers, J.M. Cheyne, G.A. Cowan, Phys.Rev. \textbf{D67},
  054018 (2003)

\bibitem{Schmitt:2004et}
A.~Schmitt, Phys.Rev. \textbf{D71}, 054016 (2005)

\bibitem{Gross:1973id}
D.J. Gross, F.~Wilczek, Phys.Rev.Lett. \textbf{30}, 1343 (1973)

\bibitem{Politzer:1973fx}
H.D. Politzer, Phys.Rev.Lett. \textbf{30}, 1346 (1973)

\bibitem{Alford:1998mk}
M.G. Alford, K.~Rajagopal, F.~Wilczek, Nucl.Phys. \textbf{B537}, 443 (1999)

\bibitem{Pisarski:1999tv}
R.D. Pisarski, D.H. Rischke, Phys.Rev. \textbf{D61}, 074017 (2000)

\bibitem{Brown:1999aq}
W.E. Brown, J.T. Liu, H.c. Ren, Phys.Rev. \textbf{D61}, 114012 (2000)

\bibitem{Schmitt:2002sc}
A.~Schmitt, Q.~Wang, D.H. Rischke, Phys.Rev. \textbf{D66}, 114010 (2002)

\bibitem{Giannakis:2004xt}
I.~Giannakis, D.f. Hou, H.c. Ren, D.H. Rischke, Phys.Rev.Lett. \textbf{93},
  232301 (2004)

\bibitem{Eagles:1969zz}
D.~Eagles, Phys.Rev. \textbf{186}, 456 (1969)

\bibitem{Nozieres:1985zz}
P.~Nozi{\`e}res, S.~Schmitt-Rink, J.Low.Temp.Phys. \textbf{59}, 195 (1985)

\bibitem{1995Sci...269..198A}
M.H. {Anderson}, J.R. {Ensher}, M.R. {Matthews}, C.E. {Wieman}, E.A. {Cornell},
  Science \textbf{269}, 198 (1995)

\bibitem{1995PhRvL..75.3969D}
K.B. {Davis}, M.O. {Mewes}, M.R. {Andrews}, N.J. {van Druten}, D.S. {Durfee},
  D.M. {Kurn}, W.~{Ketterle}, Phys.Rev.Lett. \textbf{75}, 3969 (1995)

\bibitem{landau}
L.~Landau, E.~Lifshitz, \emph{Quantum Mechanics: Non-relativistic Theory}
  (Pergamon Press, New York, 1987)

\bibitem{huang}
K.~Huang, \emph{Statistical Mechanics} (John Wiley \& Sons, New York, 1987)

\bibitem{1997PhRvB..5515153E}
J.R. {Engelbrecht}, M.~{Randeria}, C.A.R. {S{\'a} de Melo}, Phys.Rev.
  \textbf{B55}, 15153 (1997)

\bibitem{2010RPPh...73k2401C}
F.~{Chevy}, C.~{Mora}, Reports on Progress in Physics \textbf{73}, 112401
  (2010)

\bibitem{Gusynin:1994xp}
V.~Gusynin, V.~Miransky, I.~Shovkovy, Phys.Lett. \textbf{B349}, 477 (1995)

\bibitem{Gorbar:2009bm}
E.~Gorbar, V.~Miransky, I.~Shovkovy, Phys.Rev. \textbf{C80}, 032801 (2009)

\bibitem{Preis:2010cq}
F.~Preis, A.~Rebhan, A.~Schmitt, JHEP \textbf{1103}, 033 (2011)

\bibitem{Gorbar:2011ya}
E.~Gorbar, V.~Miransky, I.~Shovkovy, Phys.Rev. \textbf{D83}, 085003 (2011)

\bibitem{Luttinger:1960ua}
J.M. Luttinger, J.C. Ward, Phys. Rev. \textbf{118}, 1417 (1960)

\bibitem{Baym:1962sx}
G.~Baym, Phys. Rev. \textbf{127}, 1391 (1962)

\bibitem{Cornwall:1974vz}
J.M. Cornwall, R.~Jackiw, E.~Tomboulis, Phys. Rev. \textbf{D10}, 2428 (1974)

\bibitem{Gubankova:2003uj}
E.~Gubankova, W.V. Liu, F.~Wilczek, Phys.Rev.Lett. \textbf{91}, 032001 (2003)

\bibitem{Shovkovy:2003uu}
I.~Shovkovy, M.~Huang, Phys.Lett. \textbf{B564}, 205 (2003)

\bibitem{Huang:2003xd}
M.~Huang, I.~Shovkovy, Nucl.Phys. \textbf{A729}, 835 (2003)

\bibitem{Alford:2003fq}
M.~Alford, C.~Kouvaris, K.~Rajagopal, Phys.Rev.Lett. \textbf{92}, 222001 (2004)

\bibitem{Huang:2004bg}
M.~Huang, I.A. Shovkovy, Phys.Rev. \textbf{D70}, 051501 (2004)

\bibitem{Casalbuoni:2004tb}
R.~Casalbuoni, R.~Gatto, M.~Mannarelli, G.~Nardulli, M.~Ruggieri, Phys.Lett.
  \textbf{B605}, 362 (2005)

\bibitem{Fukushima:2005cm}
K.~Fukushima, Phys.Rev. \textbf{D72}, 074002 (2005)

\bibitem{Gubankova:2006gj}
E.~Gubankova, A.~Schmitt, F.~Wilczek, Phys.Rev. \textbf{B74}, 064505 (2006)

\bibitem{Alford:2000ze}
M.G. Alford, J.A. Bowers, K.~Rajagopal, Phys.Rev. \textbf{D63}, 074016 (2001)

\bibitem{Schafer:2005ym}
T.~Sch{\"a}fer, Phys.Rev.Lett. \textbf{96}, 012305 (2006)

\bibitem{2006PhRvA..74a3614S}
D.T. {Son}, M.A. {Stephanov}, Phys.Rev. \textbf{A74}, 013614 (2006)

\bibitem{Fulde:1964zz}
P.~{Fulde}, R.A. {Ferrell}, Physical Review \textbf{135}, 550 (1964)

\bibitem{larkin:1964zz}
A.~Larkin, Y.~Ovchinnikov, Zh.Eksp.Teor.Fiz. \textbf{47}, 1136 (1964).
\newblock [Sov.\ Phys.\ JETP {\bf 20}, 762 (1965)]

\bibitem{Casalbuoni:2003wh}
R.~Casalbuoni, G.~Nardulli, Rev.Mod.Phys. \textbf{76}, 263 (2004)

\bibitem{Anglani:2013gfu}
R.~Anglani, R.~Casalbuoni, M.~Ciminale, N.~Ippolito, R.~Gatto, et~al.,
  Rev.Mod.Phys. \textbf{86}, 509 (2014)

\bibitem{2006PhRvL..97c0401S}
Y.~{Shin}, M.W. {Zwierlein}, C.H. {Schunck}, A.~{Schirotzek}, W.~{Ketterle},
  Phys.Rev.Lett. \textbf{97}, 030401 (2006)

\bibitem{2006PhRvL..97s0407P}
G.B. {Partridge}, W.~{Li}, Y.A. {Liao}, R.G. {Hulet}, M.~{Haque}, H.T.C.
  {Stoof}, Phys.Rev.Lett. \textbf{97}, 190407 (2006)

\end{thebibliography}

\end{document}